\documentclass [PhD] {uclathes}
\usepackage{amsmath,amssymb}
\usepackage{hyperref}
\usepackage{graphicx}
\usepackage{threeparttable}
\usepackage{url}
\usepackage{mwe}
\usepackage[dvipsnames]{xcolor}
\usepackage{tcolorbox}
\usepackage{amsfonts} 
\usepackage{physics}
\usepackage{subcaption}
\usepackage{mathtools}
\usepackage{appendix}
\AtBeginEnvironment{subappendices}{%
\chapter*{Appendix}
\addcontentsline{toc}{chapter}{Appendices}
\counterwithin{figure}{section}
\counterwithin{table}{section}
}
\usepackage{times,lipsum}

\usepackage{diagbox}

\newcommand{\rm}{\mathrm}
\newcommand{\bm}{\mathsymbol}
\newcommand{\sla}{\slashed}
\newcommand{\new}{\textcolor{black}}
\usepackage[textwidth=6.5in,textheight=9.0in]{geometry}
\def\bea#1\eea{\begin{align}#1\end{align}}
\newcommand{\bef}{\begin{figure}[hbt]\centering}
\newcommand{\eef}{\end{figure}}

\usepackage{pdfpages}
\usepackage{bbm}
\usepackage{enumitem}
\usepackage{xspace}
\usepackage{slashed}
\usepackage{caption}
\usepackage{lipsum}
\newcommand{\GG}{{\mathcal G}}
\newcommand{\nnu}{\nonumber\\}
\usepackage{mathrsfs}
\usepackage{appendix}
\usepackage{cleveref}
\usepackage{bm}

\def\C{{\widetilde C}}

\title          {3D Imaging via Polarized Jet Fragmentation Functions\\
                 and Quantum Simulation of the QCD Phase Diagram}
\author         {Fanyi Zhao}
\department     {Physics}
\degreeyear     {2023}

\chair          {Zhong-Bo\ Kang}
\member         {Zvi Bern}
\member         {Huan Z.\ Huang}
\member         {E.\ T.\ Tomboulis}

\acknowledgments {
I would like to express my sincere gratitude to my supervisor, Zhongbo Kang, for his guidance, support, and encouragement throughout my PhD journey. His expertise, insight, and unwavering dedication to my research have been invaluable.

I would also like to extend my thanks to my collaborators, Leonard P. Gamberg, Kyle Lee, Ding Yu Shao, and John Terry, for their contributions to the research. Their insights and expertise have enriched the project and helped to achieve its goals.

I am grateful for the support and encouragement of my friends and colleagues, both within and outside the department. Their advice, feedback, and encouragement have been instrumental in shaping my research and in helping me to navigate the challenges of graduate school.

I would also like to thank the support from UC consortium, Mani L. Bhaumik Institute for Theoretical Physics and Center for Quantum Science and Engineering (CQSE) at UCLA as well as the staff and faculty of the Department of Physics and Astronomy at UCLA, who have provided a welcoming and stimulating environment for research and learning.

Finally, I want to express my gratitude to my family for their unwavering support and encouragement. Their love, patience, and understanding have been the bedrock of my academic and personal growth.

Without the contributions, support, and encouragement of these individuals, this research would not have been possible. Thank you all.
}

\vitaitem   {2014--2018}
                {B.S.~Physics,
                School of Physical Sciences, University of Science and Technology of China (USTC).}
\vitaitem   {2017--2018}
                {Teaching Assistant, School of Physical Sciences, USTC.}
\vitaitem   {2018--2021}
                {Teaching Assistant, Department of Physics and Astronomy, UCLA.}
\vitaitem   {2021--2023}
                {Research Assistant, Department of Physics and Astronomy, UCLA.}
                
\publication{
\begin{enumerate}
\item 
``Collins-type Energy-Energy Correlators and Nucleon Structures''\\
Z.~B.~Kang, K.~Lee, D.~Y.~Shao and F.~Zhao, \textit{Proceedings of \textbf{DIS2023}: XXX International Workshop on Deep-Inelastic Scattering and Related Subjects, Michigan State University, USA, 27-31 March 2023}, [arXiv:2307.06935 [hep-ph]]

\item 
``Predictions for the sPHENIX physics program''\\
R. Belmont et al., submitted to Nucl. Phys. A, [arXiv:2305.15491 [nucl-ex]]

\item
``Neutrino-tagged jets at the Electron-Ion Collider''\\
M.~Arratia, Z.~B.~Kang, S.~J.~Paul, A.~Prokudin, F.~Ringer, and F.~Zhao, Phys. Rev. D \textbf{107}, 094036 (2023) [arXiv:2212.02432 [hep-ph]]

\item
``Transverse-momentum-dependent factorization at next-to-leading power''\\
L.~Gamberg, Z.~B.~Kang, D.~Y.~Shao, J.~Terry and F.~Zhao, [arXiv:2211.13209 [hep-ph]]

\item 
``Studying chirality imbalance with quantum algorithms''\\
A.~M.~Czajka, Z.~B.~Kang, H.~Ma, Y.~Tee and F.~Zhao, [arXiv:2210.03062 [hep-ph]]

\item
``Quantum Simulation of Chiral Phase Transitions''\\
A.~M.~Czajka, Z.~B.~Kang, H.~Ma and F.~Zhao, JHEP \textbf{08} 209 (2022) [arXiv:2112.03944 [hep-ph]].

\item
``Spin asymmetries in electron-jet production at the EIC'' \\
Z.~B.~Kang, K.~Lee, D.~Y.~Shao and F.~Zhao, \textit{Proceedings of the 24th International Spin Symposium \textbf{(SPIN2021)}}, 10.7566/JPSCP.37.020128 [arXiv:2201.04582 [hep-ph]].

\item 
``Jet azimuthal anisotropy in $ep$ collisions''\\
R.~Esha, Z.~B.~Kang, K.~Lee, D.~Y.~Shao and F.~Zhao, \textit{Proceedings of \textbf{DIS2022}: XXIX International Workshop on Deep-Inelastic Scattering and Related Subjects}, Zenodo (2022) 7193209

\item
``Spin asymmetries in electron-jet production at the future electron ion collider''\\
Z.~B.~Kang, K.~Lee, D.~Y.~Shao and F.~Zhao, JHEP \textbf{11}, 005 (2021), [arXiv:2106.15624 [hep-ph]].

\item
``Transverse $\Lambda$ polarization in $e^+e^-$ collisions''\\
L.~Gamberg, Z.~B.~Kang, D.~Y.~Shao, J.~Terry and F.~Zhao, Phys. Lett. B \textbf{818}, 136371 (2021) [arXiv:2102.05553 [hep-ph]].

\item 
``QCD resummation on single hadron transverse momentum distribution with the thrust axis''\\
Z.~B.~Kang, D.~Y.~Shao and F.~Zhao, JHEP \textbf{12}, 127 (2020) [arXiv:2007.14425 [hep-ph]].

\item
``Polarized jet fragmentation functions''\\
Z.~B.~Kang, K.~Lee and F.~Zhao, Phys. Lett. B \textbf{809}, 135756 (2020) [arXiv:2005.02398 [hep-ph]].
\end{enumerate}
}

\abstract{Understanding the interactions between elementary particles and mapping out the internal structure of the hadrons are of fundamental importance in high energy nuclear and particle physics. This thesis concentrates on the strong interaction, described by Quantum Chromodynamics (QCD). We introduce a novel concept called ``polarized jet fragmentation functions'' and develop the associated theory framework known as QCD factorization which allows us to utilize jet substructure to probe spin dynamics of hadrons, especially nucleon's three-dimensional imaging. 
Furthermore, non-perturbative QCD studies, particularly of the QCD phase diagram, are important for understanding the properties of hadrons. The development of quantum computing and simulators can potentially improve the accuracy of finite-temperature simulations and allow researchers to explore extreme temperatures and densities in more detail. In this thesis, I present my work in two aspects of QCD studies: (1) investigating the nucleon structure using polarized jet fragmentation functions and (2) illustrating how to apply quantum computing techniques for studying phase diagram of a low energy QCD model. The first category investigates phenomena such as hadron production inside jets, spin asymmetries, etc., providing valuable insight into the behavior of quarks and gluons in hadrons. The second category provides potential applications of quantum computing in QCD and explores the non-perturbative nature of QCD. }

\begin{document}
\makeintropages
%

\chapter{Introduction}\label{ch1:intro}
\section{Motivation}
The study of high energy collisions is an essential aspect of nuclear and particle physics that aim to understand the fundamental interactions between elementary particles, now described by the well-known Standard Model of particle physics~\cite{ParticleDataGroup:2022pth}. Quantum Chromodynamics (QCD) is one of the pillars of the Standard Model, describing the strong interaction - one of the four fundamental forces of nature. This force holds quarks and gluons - collectively known as partons - together in hadrons such as the proton, and protons and neutrons together in atomic nuclei. QCD was developed and defined about fifty years ago~\cite{Gross:2022hyw}. One hallmark of QCD is asymptotic freedom, which states that the strong force between quarks and gluons decreases with increasing energy. The asymptotic freedom of strong interactions was discovered in 1973 by David Gross, Frank Wilczek, and David Politzer~\cite{Gross:1973id,Politzer:1973fx}, who shared the Nobel Prize in physics in 2004. Asymptotic freedom enables us to compute the partonic cross sections within the framework of perturbative QCD (pQCD), i.e. the expansion in terms of the strong coupling constant order by order. Although this pQCD paradigm has gained enormous success e.g. in computing the total hadronic cross section in $e^+e^-$ annihilation, it has ultimate difficulties in understanding physical processes involving hadrons in which the relevant energy scale is such that the strong coupling is too strong for pQCD computations to be applicable and is thus dubbed as the non-perturbative QCD regime.  

However, understanding hadron structure is of fundamental importance to science. The exploration of the internal structure of the hadrons (e.g. the proton and neutron) in terms of quarks and gluons, the degrees of freedom of QCD, has been and still is at the frontier of high energy nuclear physics research. Concurrent advances in the experimental use of high energy scattering processes and theoretical breakthroughs in understanding ``asymptotic freedom'' and developing the perturbation theory of strong interactions have provided a way of mapping out the internal landscape of nucleons. Specifically, perturbative QCD allows one to prove ``factorization theorems''~\cite{Collins:1989gx} for high-energy processes, which state that the physical observables involving hadrons can be written as a convolution of short-distance partonic cross sections and long-distance parton distribution functions (PDFs) that encode the bound state properties, or structure, of colliding hadrons. Armed with these theorems, theorists are then able to extract the low-energy properties of the hadron structure from the experimental data.

In past decades, a one-dimensional picture of nucleons has emerged, in the sense that we could learn about the longitudinal motion of partons in fast moving nucleons, as encoded in the so-called collinear PDFs. In recent years, theoretical breakthroughs in the community have paved the way to extending this simple picture in the transverse as well as longitudinal momentum space, i.e. three dimensions (3D). This new information is encoded in the novel concept of ``Transverse Momentum Dependent parton distributions'' (TMDs), which helps address long-standing questions concerning the confined motion of quarks and gluons inside the nucleon. How do they move in the transverse plane? Do they orbit, and carry orbital angular momentum? What are the quantum correlations between the motion of quarks, their spin and the spin of the nucleon? TMDs provide new and much richer information on the nucleon structure and they allow for the first time to carry out 3D imaging of the nucleon~\cite{Accardi:2012qut,AbdulKhalek:2021gbh}. 

The traditional processes to access TMDs are the semi-inclusive deep inelastic scattering (SIDIS), Drell-Yan production, and $e^+e^-$ collisions. In this thesis, we introduce the new concept called ``polarized jet fragmentation functions'' that utilize jet substructure to study TMD physics and spin dynamics. We develop the QCD factorization formalism for the relevant processes that can be measured in the experiments, especially at the future Electron-Ion Collider (EIC). It is important to realize that the detailed information on the nucleon structure as encoded in such more differential parton distribution functions are also crucial for the physics program at the Large Hadron Collider and for the search for signs of new physics beyond the Standard Model. 


Another very important areas of research in non-perturbative QCD is the study of the QCD phase diagram~\cite{Fukushima:2010bq}. This diagram maps out the behavior of QCD matter at different temperatures and densities. At high temperatures and densities, it is believed that quarks and gluons exist in a new state of matter called the quark-gluon plasma (QGP). Studying the properties of the QGP and mapping out the QCD phase diagram is very important for understanding the early Universe, the properties of neutron stars, and the behavior of heavy-ion collisions. In recent years there has been much progress on the investigation of the QCD phase diagram with lattice QCD simulations. However, studying QCD at finite baryon density is challenging, as the sign problem in lattice QCD simulations becomes increasingly severe as the density increases~\cite{Ratti:2021ubw}.


The development of quantum computing has opened up new possibilities for studying QCD. Quantum simulators have the potential to overcome the sign problem in lattice QCD simulations, allowing researchers to study the finite-temperature behavior of QCD at finite density more accurately. This would enable researchers to explore the QCD phase diagram in more detail, providing new insights into the behavior of QCD matter at extreme temperatures and densities. Since simulating QCD is not possible at the moment, i.e. at the noisy intermediate-scale quantum (NISQ) era, in this thesis, we illustrate how a quantum algorithm can be used to study phase diagram for a low energy model of QCD, the Nambu–Jona-Lasinio model in 1+1 dimension, at finite temperature and finite chemical potential. 


\section{Structure of this thesis}
This thesis includes two parts of studies in QCD: (1) utilizing polarized jet fragmentation functions for quantum 3D imaging of hadrons, and (2) understanding phase diagram of a low energy model of QCD via quantum simulation.

The first part is provided in~\cref{ch1:qcdintro} - \cref{sec:ajff}, where we start with an introduction of theoretical background of QCD in~\cref{ch1:qcdintro}, provide a comprehensive review of Semi-Inclusive Deep Inelastic Scattering and TMD factorization in~\cref{ch3:tmd}, then introduce the concept of polarized Jet Fragmentation Functions in~\cref{ch4:pjff}. In~\cref{sec:ajff}, we investigate promising observables using polarized jet fragmentation functions in $pp$ collisions, electron-jet production at the future EIC, etc. These studies provide insight into the behavior of quarks and gluons inside hadrons, and theoretical predictions can be compared with experimental results to validate the theory. For the second category, we illustrate how quantum algorithm is used for simulating phase diagram of a low energy model of QCD, the Nambu–Jona-Lasinio model in 1+1 dimension, in particular chiral phase diagram and chirality imbalance in~\cref{sec:npqcd}. Finally, we conclude the thesis in~\cref{sec:conclude}. 

Overall, this thesis aims to provide a comprehensive overview of exploring polarized jet fragmentation functions for quantum imaging and applying quantum simulation for chiral phase transitions, which offer unique perspectives on the behavior of quarks and gluons in high-energy collisions. The results of these studies may lead to a better understanding of the fundamental interactions between particles and the development of new theoretical frameworks for high-energy physics.

%
%




% ...

     

\begin{part}{Quantum Imaging via Polarized Jet Fragmentation Functions}
                 
\chapter{QCD background}\label{ch1:qcdintro}
\begin{quote}
\rule{0.875\textwidth}{0.5pt}\\
This chapter provides an introduction to Quantum Chromodynamics (QCD), a fundamental theory of the strong interaction. QCD is based on the properties of asymptotic freedom and color confinement, which are of great importance. The concept and formalism of QCD factorization is discussed and how this would allow us to extract information on the structure of hadrons.
\\
\rule{0.875\textwidth}{0.5pt}
\end{quote}
\section{Introduction}
Hadrons, including protons and neutrons (known as nucleons), are the predominant constituents of visible matter in the universe. Thus, comprehending their internal structure holds paramount importance. The nucleon forms a frontier of subatomic physics and has been under intensive study for the last several decades. Significant progress have been made in characterizing the one-dimensional momentum distribution of nucleon constituents through Feynman parton distribution functions (PDFs)~\cite{Lin:2017snn}. These investigations not only reveal the partonic composition of nucleons but also provide a valuable avenue for probing strong interactions. Therefore, unsolved fundamental questions such as how the spin and orbital properties of quarks and gluons within the nucleon combine to form its total spin, how quarks and gluons are spatially distributed within nucleons, and so on, are intriguing and have stimulated further theoretical and experimental endeavors in the field of hadronic physics, leading to the construction of major facilities aimed at addressing them~\cite{Achenbach:2023pba}.
 
Every several years (around seven or so), the nuclear physics community would get together to conduct a study of the opportunities and priorities for United States nuclear physics research and to recommend a long range plan that will provide
a framework for the coordinated advancement of the Nation's nuclear science program over the next decade. The most recent one is the Long Range Plan in 2015~\cite{Aprahamian:2015qub} and starting from the end of 2022, the nuclear physics community is in the process of developing a new Long Range Plan. Since the last Long Range Plan in 2015, significant progress has been achieved in the so-called cold QCD research, for which exploring the nucleon structure is one of the main goals. The successful completion of the CEBAF 12 GeV upgrade has enabled a full-fledged experimental program. Also, various hadron physics facilities, including CEBAF at JLab, RHIC at BNL, and the LHC at CERN, have yielded fruitful and exciting new results. These advancements encompass static properties and partonic structure of hadrons, nuclear modifications of structure functions, many-body physics of nucleons in nuclear structures, and the effects of dense cold matter. These recent findings not only scrutinize fundamental aspects of QCD, such as its chiral structure and predictions for novel hadronic states but also offer a glimpse into the future prospects of nucleon tomography, facilitating a deeper understanding of mass and spin origins.

In the subsequent sections of this chapter, we will review the theoretical foundations of QCD, encompassing the widely known factorization formalism which enables us to investigate the structure of hadrons and their hadronization process by combining perturbative QCD techniques and global analysis in QCD phenomenology. This would serve as a starting point for the more differential parton distribution functions such as the transverse-momentum dependent parton distribution functions to be studied in details in the later chapters of this thesis.

\section{Quantum Chromodynamics}
In theoretical physics, quantum chromodynamics (QCD) is the theory of the strong interaction between quarks mediated by gluons. The QCD Lagrangian is given by
\begin{equation}
\mathcal{L}_{\text{QCD}}=-\frac{1}{4} F_{\mu \nu}^a F^{a \mu \nu}+\sum_q\bar{\psi}_q^i\left[i \gamma^\mu\left(D_\mu\right)_{i j}-m_q\delta_{ij}\right] \psi_q^j\,,\label{eq:(1.10)}
\end{equation}
where $m_q$ is the quark mass, $\psi_q^i$ is the quark field with $i$ denoting the color index, that can be given by Red, Green, and Blue in SU(3) gauge group, namely $\psi_q=\left(\psi_{q R}, \psi_{q G}, \psi_{q B}\right)^T$. The Dirac matrix $\gamma^\mu$ is used to express the vector nature of the strong interaction, where $\mu$ is a Lorentz vector index. The gluon field strength tensor $F_{\mu \nu}^a$ with a color index $a$ ranging from 1 to 8 is given by
\begin{equation}
F_{\mu \nu}^a=\partial_\mu A_\nu^a-\partial_\nu A_\mu^a+g_sf^{abc}A_\mu^b A_\nu^c\,,
\end{equation}
with the gluon field denoted by $A_\mu^a$. The covariant derivative $D_\mu$ is given by $\left(D_\mu\right)_{i j}=\delta_{i j} \partial_\mu-i g_s t_{i j}^a A_\mu^a$. The parameter $g_s$ is the interaction strength and it is related to the more conventional strong coupling constant $\alpha_s$ as follows
\begin{equation}
    \alpha_s = \frac{g_s^2}{4\pi}\,.
\end{equation}
On the other hand, $t_{i j}^a$ are the standard generating matrices of the SU(3) group and $f^{abc}$ are the fully antisymmetric structure constants of the gauge group, defined so that $\left[t^a,\, t^b\right]=if^{abc}t^c$.



QCD has a very special feature, called ``asymptotic freedom'' and the discovery of asymptotic freedom was acknowledged with the Nobel Prize in Physics awarded to D. Gross, H. Politzer, and F. Wilczek in 2004~\cite{Gross:1973ju,Politzer:1974fr}. It means that the interaction between quarks and gluons decreases in strength at progressively higher energies. Specifically, the strong coupling constant $\alpha_s(\mu)$ satisfies the following renomralization group equation:
\begin{equation}
\mu^2 \frac{\partial \alpha_s}{\partial \mu^2}=\frac{\partial \alpha_s}{\partial \ln \mu^2}=\beta\left(\alpha_s\right)\,,\quad
\beta\left(\alpha_s\right)=-\alpha_s^2\left(b_0+b_1 \alpha_s+b_2 \alpha_s^2+\ldots\right)\,,
\label{eq:19}
\end{equation}
where $b_0$ is the 1-loop $\beta$-function coefficient and likewise, $\beta_{1,2}$ are the 2-loop and 3-loop coefficients. We have 
\begin{equation}
    \beta_0 = \frac{11C_A - 4n_f T_R}{12\pi} = \frac{33 - 2n_f}{12\pi}\,,
\end{equation}
where we have used $C_A = 3$ and $T_R=1/2$, and $n_f$ is the number of quark flavors. Since we have $u,d,s,c,b,t$ in total six quark flavors, we have $n_f = 6$ and thus $b_0 > 0$. The minus sign in Eq.~\eqref{eq:19} is the origin of ``asymptotic freedom'' and it would lead to $\alpha_s(\mu)$ decreases as $\mu$ increases. For a given physical process, one would take the renormalization scale $\mu$ to be the scale of the momentum transfer $Q$ in that process, then $\alpha_s(\mu^2=Q^2)$ is indicative of the effective strength of the strong interaction in that process. 

\begin{figure}[htb]
    \centering
    \includegraphics[width=3.6in]{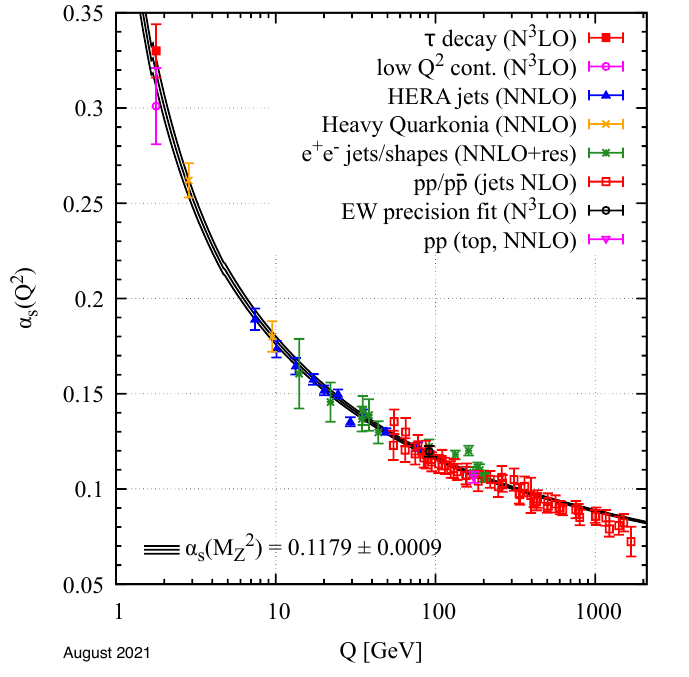}   
    \caption{Summary of running coupling $\alpha_s$ measured as a function of the energy scale $Q$ provided in~\cite{ParticleDataGroup:2022pth}. In brackets, the respective degree of perturbative QCD applied for the extraction of $\alpha_s$ is indicated (for example, NLO: next-to-leading order; NNLO+res.: next-to-next-to-leading order matched to a resummed calculation; etc.)}
    \label{fig:alphas}
\end{figure}

Summary of running coupling $\alpha_s$ measured as a function of the energy scale $Q$ is given in \cref{fig:alphas}, taken from the recent Particle Data Group~\cite{ParticleDataGroup:2022pth}. For $Q>100$ GeV, $\alpha_s\sim 0.1$, i.e. becomes relatively weak for processes involving large momentum transfer, which are often referred to as ``hard processes''. On the other hand, the theory is strongly interacting for scales around and below 1 GeV.

\section{QCD factorization}
Asymptotic freedom enables us to compute the partonic cross sections between quarks and gluons in the hard processes via the perturbation theory, i.e. the expansion in terms of $\alpha_s$ order by order. This has led to remarkable progress. Probably one of the most well-known example is the total hadronic cross section in $e^+e^-$ annihilation, $e^+e^-\to \mathrm{hadrons}$, which is generated at leading order by $e^+e^-\to q\bar{q}$. The conventional $R$ ratio is defined as the hadronic cross section over the cross section of $e^+e^-\to \mu^+\mu^-$,
\begin{equation}
    R(s) = \frac{\sigma(e^+e^-\to \mathrm{hadrons})}{\sigma(e^+e^-\to \mu^+\mu^-)}\,.
\end{equation}
This $R$ ratio has been measured for a wide range of center-of-mass energy $\sqrt{s}$. \cref{fig:ee_R} shows the $R$ ratio is compared with the 3-loop perturbative QCD computations as a function of $\sqrt{s}$, taken from the Particle Data Group~\cite{ParticleDataGroup:2022pth}. As one can see, the theory agrees very well with the experimental data. 

\begin{figure}[htb]
    \centering
    \includegraphics[width=4.2in]{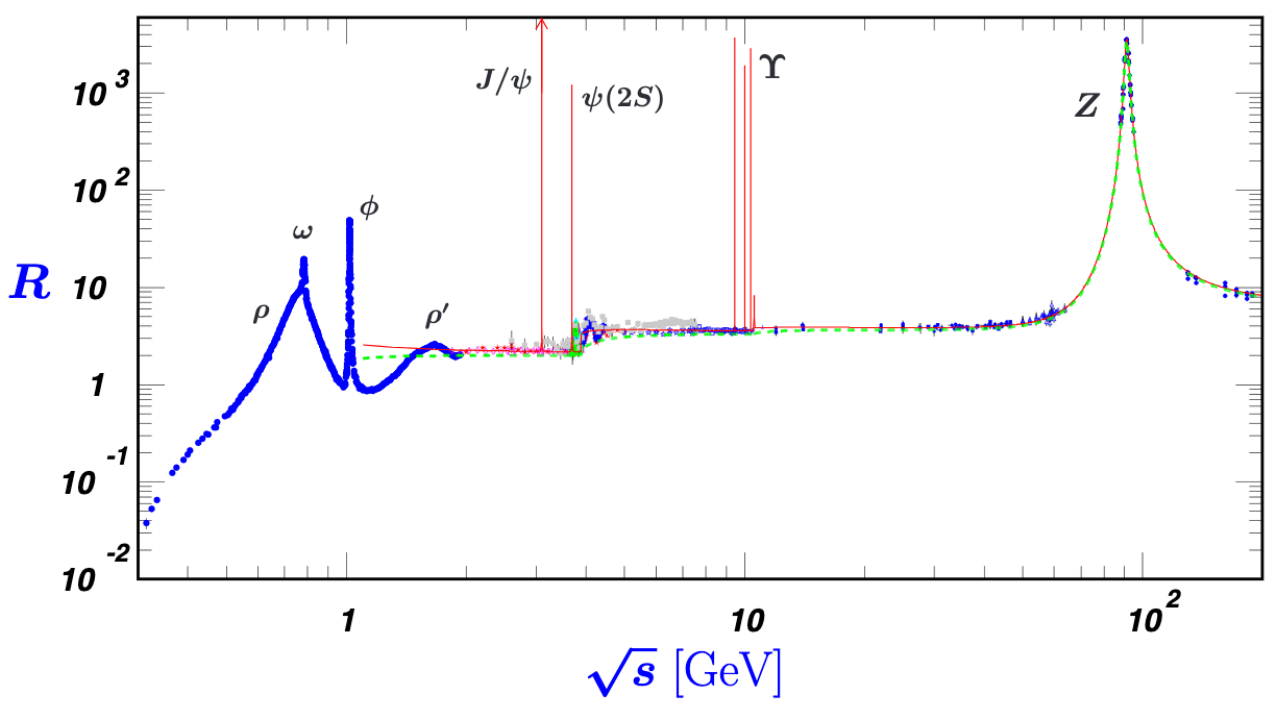}   
    \caption{World data on the ratio $R(s)$ as a function of the center-of-mass energy, $\sqrt{s}$, in $e^+e^-$ annihilation, taken from the Particle Data Group~\cite{ParticleDataGroup:2022pth}. }
    \label{fig:ee_R}
\end{figure}

Even with this remarkable success, perturbative QCD had difficulties in describing the scattering processes involving hadrons. For example, the total cross section of deep inelastic lepton-proton ($\ell\,p$) scattering (DIS), $\ell+p\to \ell'+X$, involves the incoming proton. Since the proton is a composite object made up of quarks and gluons, which interact strongly with each other inside the proton, it is unclear how one would compute the DIS cross section. 

\begin{figure}[hbt]
    \centering
    \includegraphics[width=2.5in]{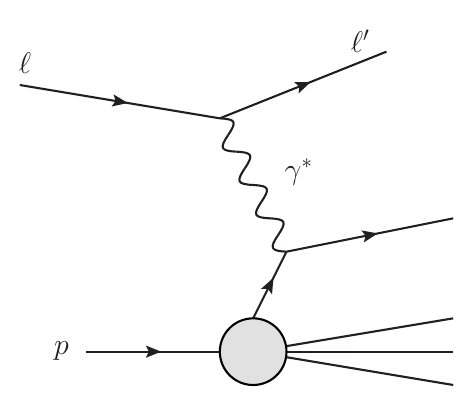}   
    \caption{The standard DIS process, where an incoming lepton $\ell$ is colliding with the proton $p$ by exchanging a virtual photon $\gamma^*$.}
    \label{fig:DIS}
\end{figure}

QCD factorization formalism~\cite{Collins:1989gx} came to rescue. It states that the cross section involving the hadron can be written as a convolution of short-distance partonic cross sections and long-distance parton distribution functions (PDFs) that encode the bound state properties, or structure, of colliding nucleons. To make it concrete, let us use the DIS process, $\ell(l)+p(P)\to \ell'(l')+X$, as an example. We define the usual variables
\begin{equation}
    x = \frac{Q^2}{2P\cdot q}\,,
    \qquad
    y  = \frac{P\cdot q}{P\cdot l}\,,
    \label{eq:DIS-variable}
\end{equation}
where $q = l - l'$ is the momentum of the exchanged virtual photon and $Q^2 = -q^2$. Thus the DIS differential cross section can be written as
\begin{equation}
    \frac{d^2\sigma}{dx\,dQ^2} = \frac{4\pi\alpha^2}{2x Q^4}\left[\left(1+(1-y)^2\right)F_2(x, Q^2) - y^2 F_L(x, Q^2)\right]\,,
\end{equation}
where $\alpha$ is the electromagnetic coupling. On the other hand, $F_2(x, Q^2)$ and $F_L(x, Q^2)$ are the proton structure functions, which encode the interaction between the virtual photon and the proton. See the illustration of the DIS process in \cref{fig:DIS}.

\begin{figure}[hbt]
    \centering
    \includegraphics[width=3in]{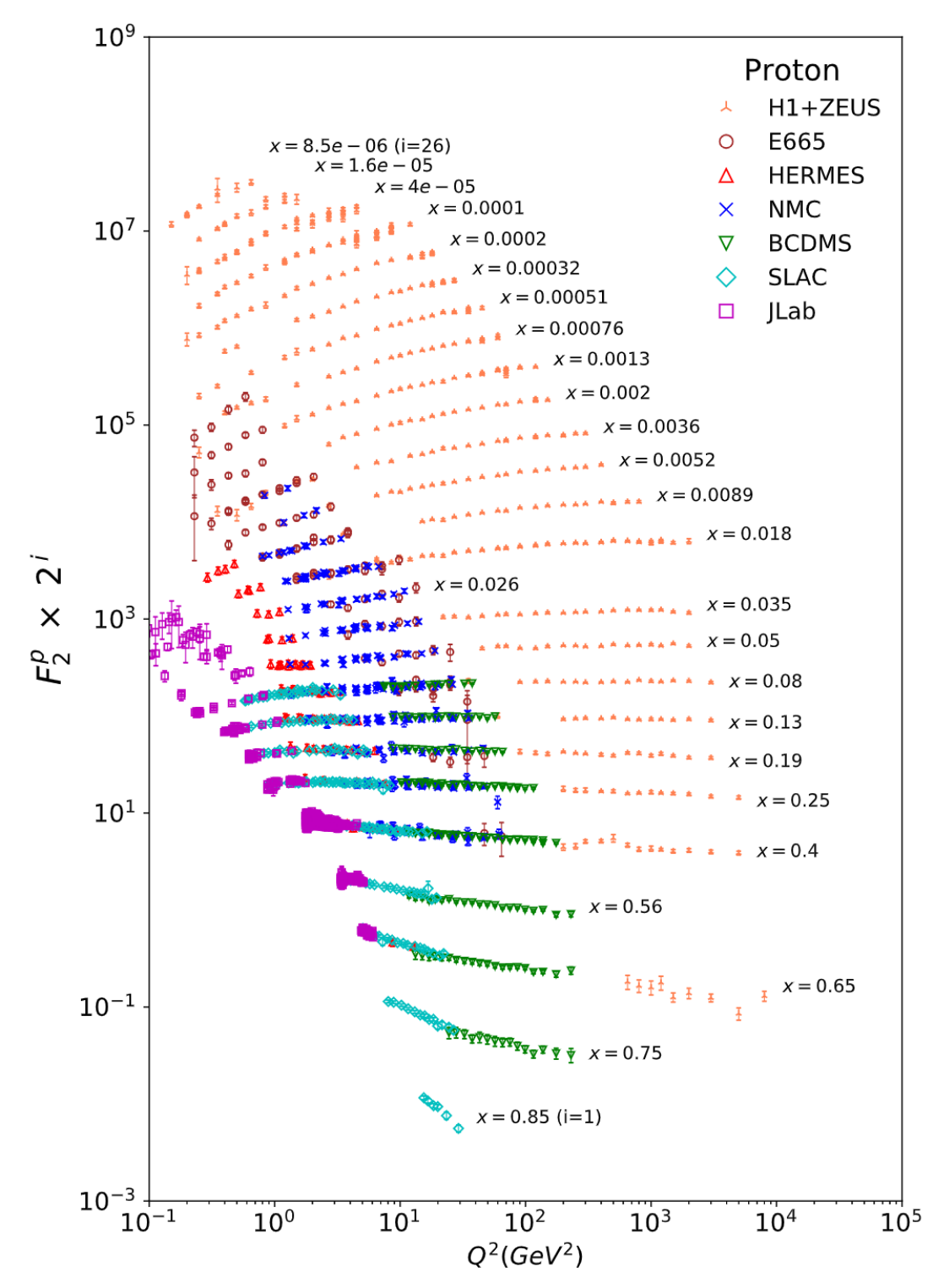}   
    \caption{The proton structure function $F_2^p$ as a function of $Q^2$ for different Bjorken $x$, taken from the Particle Data Group~\cite{ParticleDataGroup:2022pth}. }
    \label{fig:F_2-ep}
\end{figure}

Following QCD factorization theorem, one can write as $F_2(x, Q^2)$ as follows
\begin{equation}
    F_2(x, Q^2) = x \sum_{n=0} \frac{\alpha_s^n(\mu_R^2)}{(2\pi)^n} \sum_{i=q,g}
    \int_{x}^1\frac{d\hat x}{\hat x} C_{2,i}^{(n)}(\hat x, Q^2, \mu_R^2, \mu_F^2) f_{i/p}\left(\frac{x}{\hat x}, \mu_F^2\right) + \mathcal{Q}\left(\frac{\Lambda^2_{\mathrm{QCD}}}{Q^2}\right)\,,
    \label{eq:fac-DIS}
\end{equation}
similarly for $F_L(x, Q^2)$. Here, $\mu_R$ and $\mu_F$ are the so-called renormalizataion and factorization scales, respectively. The structure function is expanded as a series in powers of $\alpha_s(\mu_R^2)$, while each term involves a short-distance coefficient $C_{2, i}^{(n)}$ that can be computed order by order with Feynman diagram techniques. The coefficient functions $C_{2, i}^{(n)}$ are known up to $\mathcal{O}(\alpha_s^3)$, i.e. next-to-next-to-next-to-leading order (N$^3$LO)~\cite{Vermaseren:2005qc}. On the other hand, $f_{i/p}\left(x, \mu_F^2\right)$ is the aforementioned collinear PDF, which is non-perturbative and describes the probability density of finding a parton $i$ inside the proton that carries a fraction $x$ of its longitudinal momentum. Note that the above factorization is valid up to the power corrections of $\mathcal{O}(\Lambda_{\mathrm{QCD}}^2/Q^2)$. This is precisely the physical reasoning for the QCD factorization: the quark and gluon dynamics inside the proton (that is associated with the PDFs) happens at the physical scale $\Lambda_{\mathrm{QCD}}\sim 1/R_{p}\sim 200$ MeV where $R_p\sim 1$ fm is the proton size. The electron-parton scattering happens at a high-energy scale $\sim Q$. When $\Lambda_{\mathrm{QCD}}\ll Q$, the physics happening in these two widely separated scales should not interfere with each other and this leads to the factorization in Eq.~\eqref{eq:fac-DIS}. The PDFs follow the so-called Dokshitzer–Gribov–Lipatov–Altarelli–Parisi (DGLAP) evolution equations~\cite{Altarelli:1977zs,Gribov:1972ri,Dokshitzer:1977sg}, which have the following form
\begin{equation}
     \frac{\partial}{\partial\ln\mu^2}f_{i/p}(x, \mu^2) = \sum_{j=g,q,\bar q} \int_{x}^1\frac{dz}{z} P_{ij}\left(z, \alpha_s\right) f_{j/p}\left(\frac{x}{z}, \mu^2\right)\,.
\end{equation}
Here $P_{ij}\left(z, \alpha_s\right)$ are the splitting functions or evolution kernels which can be computed in the perturbation theory order by order. 
\begin{figure}[hbt]
    \centering
    \includegraphics[width=5.2in]{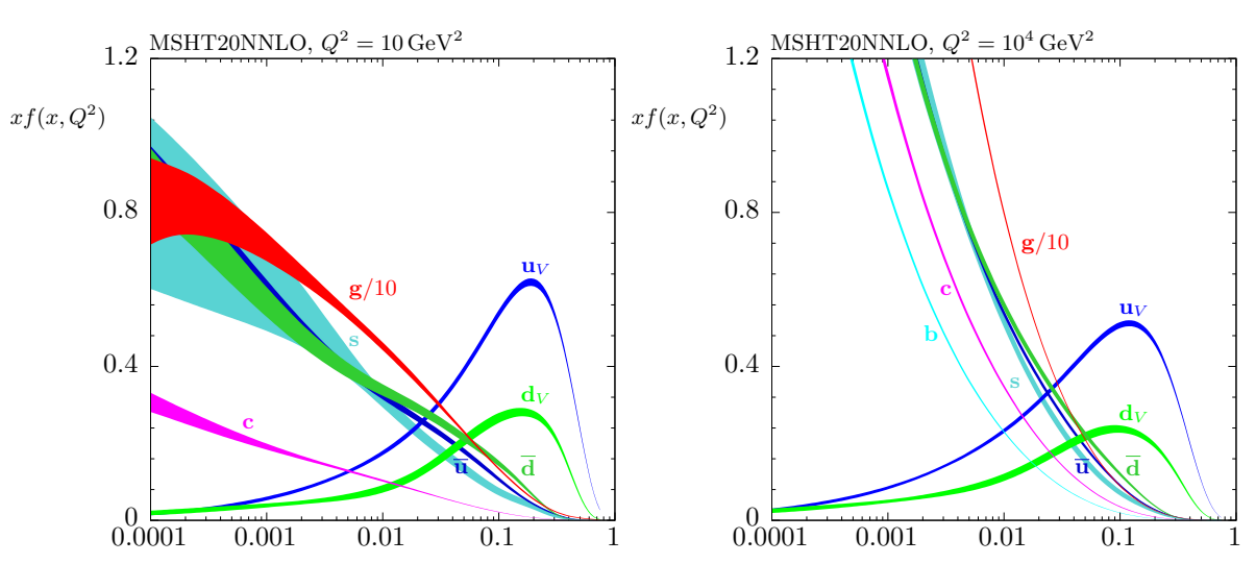}   
    \caption{The Parton Distribution Functions (PDFs) $f_{i/p}(x, Q^2)$ as a function of $x$ at $Q^2 = 10$ GeV$^2$ (left) and $Q^2 = 10^4$ GeV$^2$ (right).}
    \label{fig:PDFs}
\end{figure}

Armed with QCD factorization theorems, theorists are then able to extract the collinear PDFs from the experimental data. For example, a lot of data have been collected for DIS process in lepton-proton collisions shown in \cref{fig:F_2-ep}, which is taken from the Particle Data Group~\cite{ParticleDataGroup:2022pth}. From these data, one can extract the PDFs through the procedure of ``global analysis''. For example, one of the modern PDFs from MSHT20 group~\cite{Bailey:2020ooq} is shown in \cref{fig:PDFs}, where the PDFs $f_{i/p}(x, Q^2)$ are plotted as a function of $x$ at $Q^2 = 10$ GeV$^2$ (left) and $Q^2 = 10^4$ GeV$^2$ (right).

Similarly, one can introduce the collinear fragmentation function to describe the hadronization process for a parton fragmenting into a hadron. For example, if one measures a specific hadron $h$ in the $e^+e^-$ collisions,
\begin{equation}
    e^+e^-\to \left(\gamma^*, \, Z\right)\to h + X\,,
\end{equation}
the cross section for this single inclusive hadron production can be written as
\begin{equation}
    \frac{1}{\sigma_{\mathrm{tot}}}\frac{d\sigma^h}{dz}
    =\frac{\sigma_0}{\sum_q e_q^2}\left[2F_1^h(z, Q^2) + F_L^h(z, Q^2)\right]\,.
\end{equation}
Here the energy $E_h$ of the observed hadron scaled to the beam energy $Q/2$ is denoted by the variable $z=2p_h\cdot q/Q^2 = 2E_h/Q$ with $q$ being the momentum of the intermediate $\gamma^*$ or $Z$ boson and $q^2=Q^2$. Just like the DIS process, the unpolarized structure functions $F_1^{h}$ and $F_L^{h}$ can also be studied within the collinear QCD factorization, where they can be written as the convolution of short-distance partonic results and the long-distance non-perturbative functions. 
\begin{figure}[hbt]
    \centering
    \includegraphics[width=3.6in]{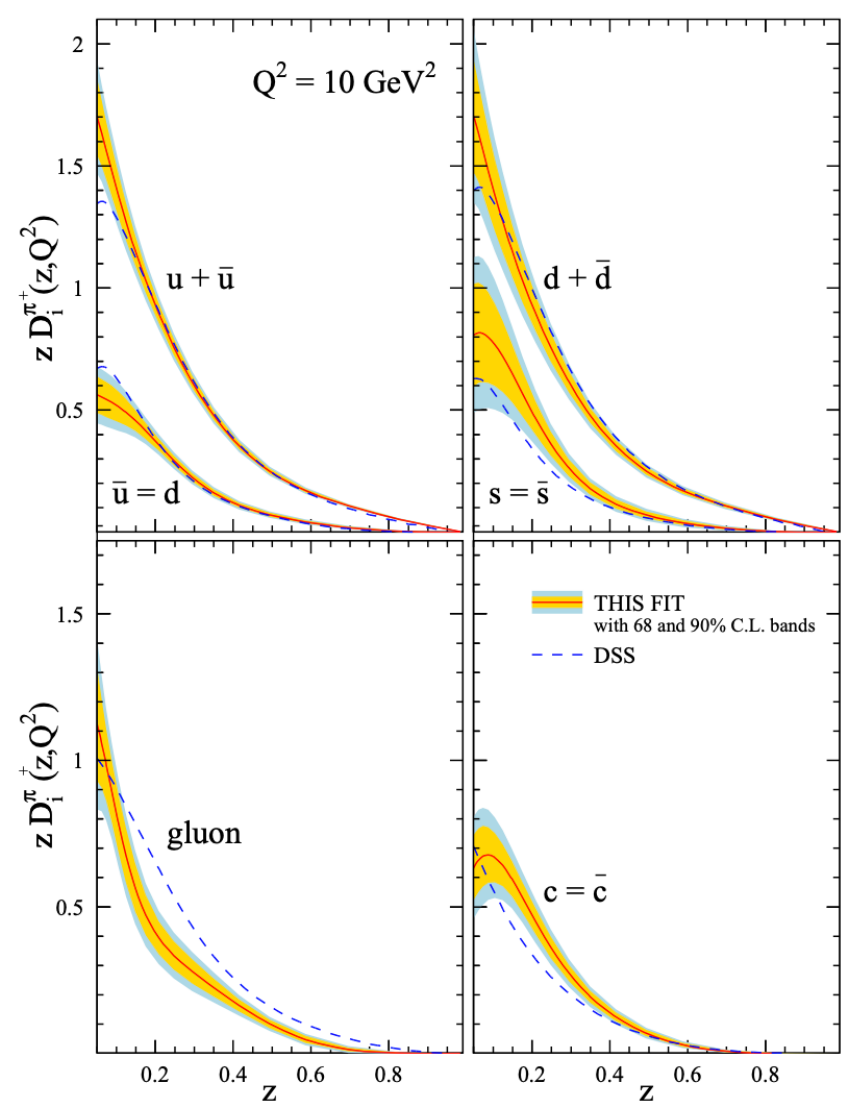}   
    \caption{The Fragmentation Functions (PDFs) $D_{h/i}(z, Q^2)$ for the $h=\pi^+$ hadron is plotted as a function of $z$ at $Q^2 = 10$ GeV$^2$, taken from~\cite{deFlorian:2014xna}.}
    \label{fig:FFs}
\end{figure}
For example, to the NLO accuracy, they  are given by~\cite{deFlorian:2007aj}
\begin{eqnarray}
\label{eq:f1nlo}
\nonumber
2 F_1^{h}(z,Q^2) &=& \sum_{q} e_q^2\;
\Bigg[  D_{h/q}(z,Q^2)  + \frac{\alpha_s(Q^2)}{2\pi} \left( C_q^1 \otimes D_{h/q} + C_g^1 \otimes D_{h/g} \right) (z,Q^2) \Bigg]\,, \\
\nonumber
\label{eq:flnlo}
F_L^{H}(z,Q^2) &=& \frac{\alpha_s(Q^2)}{2\pi} \sum_{q} e_q^2 
\left[ C_q^L \otimes D_{h/q} +C_g^L \otimes D_{h/g}
\right](z,Q^2)\,.
\end{eqnarray}
Here $C^1_{q,g}$ and $C^L_{q,g}$ are the corresponding short-distance coefficient functions that can be computed in perturbation theory order by order. On the other hand, $D_{h/q,g}$ is the collinear fragmentation function that gives the probability density for the quark $q$ or gluon $g$ fragmenting into the hadron~$h$. The collinear fragmentation functions follow a ``time-like''~\footnote{It is called ``time-like'' since $q^2 = Q^2>0$, while for DIS it would be ``space-like'' since $q^2 = - Q^2 < 0$.} DGLAP evolution equations. They can also be extracted from the experimental data. See a recent extraction in~\cite{deFlorian:2014xna}
\begin{figure}[htb]
    \centering
    \includegraphics[width=2.6in]{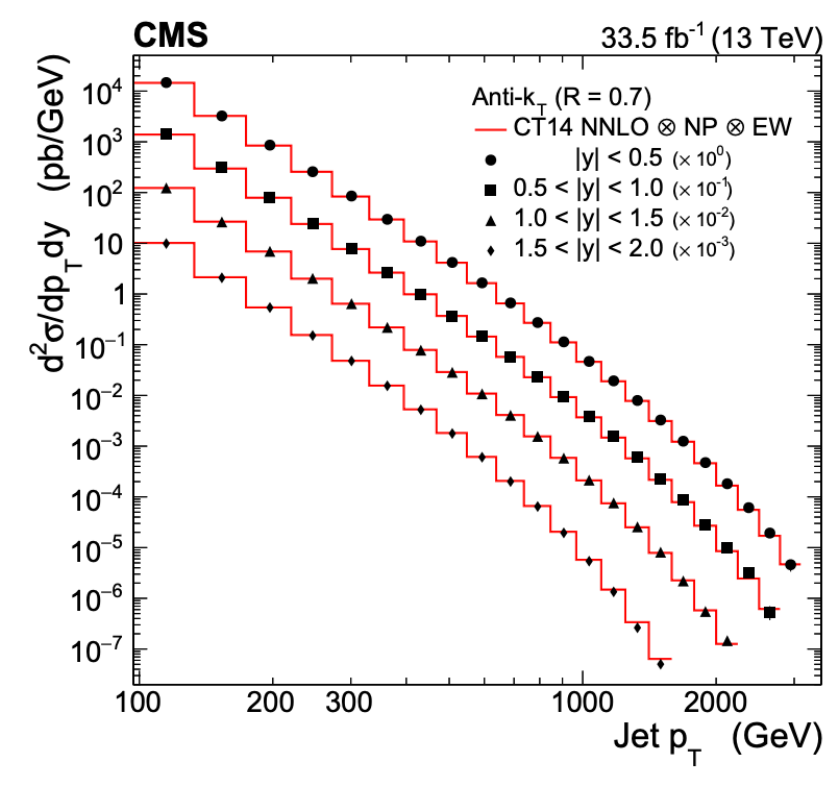} 
    \caption{Single inclusive jet cross section in $pp$ collisions (taken from the CMS collaboration at the LHC~\cite{CMS:2021yzl}) is plotted as a function of jet $p_T$, compared with the theory calculation at the next-to-next-to-leading order. }
    \label{fig:pp_data1}
\end{figure}
\begin{figure}[htb]
    \centering
    \includegraphics[width=2.6in]{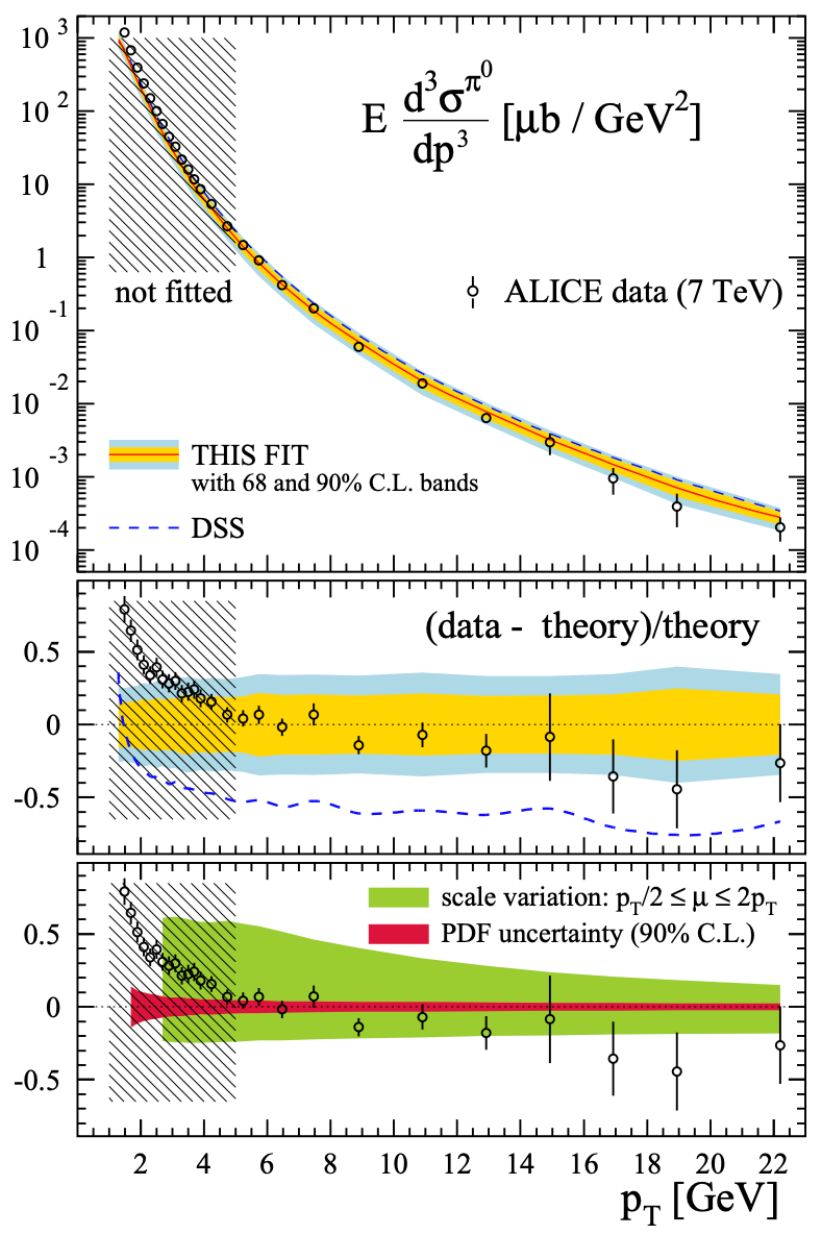}   
    \caption{The same fit of the fragmentation function in $e^+e^-$ collisions can also describe the single inclusive hadron cross section in $pp$ collisions (taken from the ALICE collaboration~\cite{ALICE:2012wos} at the LHC and figure from~\cite{deFlorian:2014xna}).}
    \label{fig:pp_data2}
\end{figure}
One of the key concept in QCD factorization is the universality of these collinear PDFs and FFs. In other words, the same set of collinear PDFs and/or FFs can be used for other scattering processes, e.g. single inclusive jet in proton-proton ($pp$) collisions, $pp\to \mathrm{jet}+X$, or single inclusive hadron production in $pp$ collisions, $pp\to h+X$. See \cref{fig:pp_data1,fig:pp_data2} for the comparison between theory and experimental data for single inclusive jet (\cref{fig:pp_data1}) and single inclusive hadron (\cref{fig:pp_data2}) in $pp$ collisions. It demonstrates that QCD factorization works remarkably well. 

With QCD collinear factorization well established, progress have been made in the field in terms of computing the partonic hard scattering cross sections and splitting functions with high precision at higher orders of perturbation theory. For instance, the evolution kernels of longitudinal momentum distribution functions, both spin-dependent and spin-independent, are now fully known to next-to-next-to-leading order (NNLO)~\cite{Moch:2004pa,Vogt:2004mw,Moch:2014sna} and beyond~\cite{Falcioni:2023luc,Hekhorn:2023gul,Falcioni:2023vqq}. Significant computations have been carried out for partonic cross sections in processes such as electron-proton scattering, extending beyond NNLO for inclusive DIS~\cite{Zijlstra:1992qd,Zijlstra:1993sh,Borsa:2022irn} and jet production in DIS~\cite{Currie:2017tpe,Currie:2018fgr,Boughezal:2018azh,Borsa:2020ulb}. Additional advancements include the calculation of heavy quark and quarkonium production in various hard scattering processes~\cite{Bodwin:1994jh,Brambilla:2010cs,Ma:2016exq,Cheung:2017loo,Cheung:2017osx,Cheung:2018tvq,Cheung:2018upe,Cheung:2021epq,Vogt:2018oje,Vogt:2019xmm}. 

Besides the progress in perturbative computations for partonic cross sections, in the last decade, we have also seen important progress in understanding the low-energy properties of the nucleon structure, encoded in the more differential parton distribution functions, e.g. the transverse momentum dependent parton distribution functions and/or fragmentation functions. We will now discuss in details the progress the community made along this direction and put my thesis in the proper context for introducing the contribution we made.


\chapter{TMD Factorization and SIDIS Process}\label{ch3:tmd}
\begin{quote}
\rule{0.875\textwidth}{0.5pt}\\
We review the Semi-Inclusive Deep Inelastic Scattering (SIDIS) in detail, a fundamental process for studying the structure of hadrons. We begin by discussing the kinematics and structure functions of SIDIS, which are essential for understanding the experimental measurements. We also present an in-depth introduction of the Transverse Momentum Dependent (TMD) factorization, which is a theoretical framework for describing the SIDIS cross-section. Also crucial ingredients in TMD factorization like TMD parton distribution function, TMD fragmentation function, hard and soft function are introduced. The chapter aims to provide a comprehensive understanding of SIDIS and TMD factorization, which are crucial for studying the structure of hadrons.
\\
\rule{0.875\textwidth}{0.5pt}
\end{quote}
\section{Introduction: 3D momentum tomography of hadrons}
The collinear parton distribution functions, $f_{i/p}(x, \mu)$, introduced in the previous chapter, provide the information for quarks and gluons inside the proton, specifically the longitudinal motion of the partons. This is because we consider the parent proton momentum $P$ as in the $+z$ or longitudinal direction as shown in \cref{fig:1d_3d} (left), while the parton carries the momentum fraction $x$ of the proton and thus its momentum is given by $k\approx xP$. In this sense, the collinear PDFs are usually considered to be providing the 1D structure of the proton in the momentum space. However, the parton inside the proton would also have the momentum component that is transverse to the parent proton, denoted as ${\bm k}_T$  in \cref{fig:1d_3d} (right). Writing the parton momentum as $k = xP+{\bm k}_T$, one would have the natural question - what role does this transverse momentum would play? In the last decade or so, theoretical breakthroughs~\cite{Boussarie:2023izj,Collins:2011zzd} have paved the way to extending the 1D structure in the longitudinal as well as transverse momentum space, providing 3D structure of the proton. This new information is encoded in the concept of ``Transverse Momentum Dependent parton distribution functions'', or simply called TMDs. 

\begin{figure}[htb]
    \centering
\includegraphics[width=0.7\textwidth]{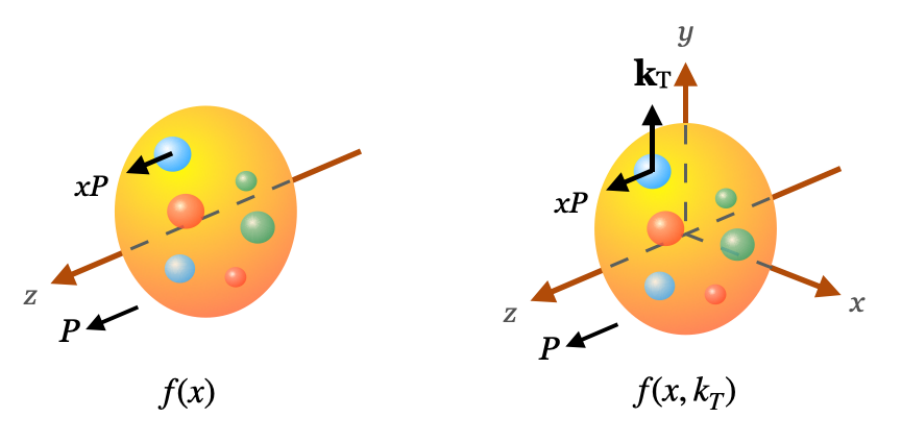}   
    \caption{Illustration of the standard 1D parton distribution functions $f_{i/p}(x)$ (left) and the 3D transverse momentum dependent parton distribution functions $f_{i/p}(x, k_T)$ (right).}
    \label{fig:1d_3d}
\end{figure}

The TMDs provide not only an intuitive illustration of nucleon tomography, but also the important opportunities to investigate the specific nontrivial QCD dynamics associated with their physics: QCD factorization, universality of the parton distributions and fragmentation functions, and their scale evolution. For example, one has to generalize the so-called QCD collinear factorization introduced in the previous chapter to deal with the TMDs. This new factorization named ``TMD factorization'' has been well established for the semi-inclusive hadron production in deep inelastic $ep$ scattering (SIDIS)~\cite{Ji:2004wu,Ji:2004xq}, Drell-Yan production in $pp$ collisions~\cite{Collins:1984kg,Echevarria:2011epo}, and back-to-back hadron pair production in $e^+e^-$ collisions~\cite{Collins:1981uk}. For the modern reviews, see~\cite{Boussarie:2023izj,Collins:2011zzd}. At the moment, all the 3D structure of the proton as encoded in the TMDs are extracted from these three standard processes: SIDIS, Drell-Yan, and $e^+e^-$ collisions. In the next section, we will provide a detailed review for the SIDIS process and its TMD factorization. 

A recent global extraction of unpolarized TMD parton distribution functions (TMD PDFs) and TMD fragmentation functions (TMD FFs) have been performed in~\cite{Bacchetta:2022awv} at the next-to-next-to-next-to-leading logarithmic (N$^3$LL) accuracy. This extraction is  based on more than two thousand data points from several experiments for both SIDIS and Drell-Yan production. For example, the Drell-Yan production data include those from earlier Fermilab~\cite{Ito:1980ev}, the RHIC~\cite{PHENIX:2018dwt}, CDF~\cite{CDF:2012brb} and D0~\cite{D0:2007lmg} collaobrations at the Tevatron, and LHCb~\cite{LHCb:2016fbk}, ATLAS~\cite{ATLAS:2019zci} and CMS~\cite{CMS:2019raw} collaborations at the LHC. On the other hand, the SIDIS data are collected by the HERMES~\cite{HERMES:2012uyd} and COMPASS~\cite{COMPASS:2017mvk} collaborations. Another recent work~\cite{Moos:2023yfa} extracted the unpolarized TMD PDFs from the Drell-Yan production process alone but with higher precision (N$^4$LL accuracy). 

When the experimental data are collected for the polarized scattering, one would be able to measure various spin asymmetries from which the spin-dependent TMD PDFs and/or TMD FFs can be extracted. Two of the spin-dependent TMDs have attracted most attentions in the past decade: the Sivers function~\cite{Sivers:1989cc,Sivers:1990fh} and the Collins function~\cite{Collins:1992kk}. The quark Sivers function describes the distribution of unpolarized quark inside the tranversely polarized proton through a correlation between the transverse momentum of the quark with respect to the proton and the transvese spin of the proton. On the other hand, the Collins fragmentation function describes a tranversely polarized quark fragmenting into an unpolarized hadron while the hadron's transverse momentum with respect to the quark is correlated with the quark's transverse spin. For recent global analysis of the Sivers functions, see~\cite{Cammarota:2020qcw,Bury:2020vhj,Bury:2021sue,Gamberg:2022kdb,Echevarria:2020hpy,Bacchetta:2020gko}. On the other hand, for recent global analysis of the Collins functions, see~\cite{Kang:2015msa,Gamberg:2022kdb,Cammarota:2020qcw}, where the Collins functions are extracted from the Collins spin asymmetry in SIDIS and the Collins azimuthal asymmetry in two hadron production in $e^+e^-$ collisions.

\section{Semi-Inclusive Deep Inelastic Scattering}\label{sec:th-sidis}
Semi-Inclusive Deep Inelastic Scattering (SIDIS) is a fundamental process that provides invaluable insights into the inner structure of hadrons and the distribution of their constituents. SIDIS is one of the most important processes for probing TMD PDFs and TMD FFs and will be the key process at the future EIC. In this section, we provide the general form of the cross section for polarized SIDIS and parameterize it in terms of suitable structure functions. For completeness, we also review the full parameterization of quark-quark correlation functions at the leading power. Note that this is well established in the community and we review the material to set up the notations and framework for our work in the next two chapters. The relation of the structure functions given below are consistent with the parameterization in~\cite{Diehl:2005pc,Bacchetta:2006tn}. 

\subsection{Kinematics}\label{sec:sec2}
In SIDIS, a lepton is interacting with a nucleon, with a scattering lepton and one of the produced hadrons detected. The interaction occurs through a virtual photon of virtuality $Q$. The cross section depends on the azimuthal angles of the final state hadron relative to the virtual photon axis and the target polarization, as shown in Fig.~\ref{fig:0611265-2}. In the low transverse momentum region of the outgoing hadron compared to $Q$, the cross section can be described using TMD PDFs and TMD FFs as shown below. 


\begin{figure}[htb]
    \centering
\includegraphics[width=0.8\textwidth]{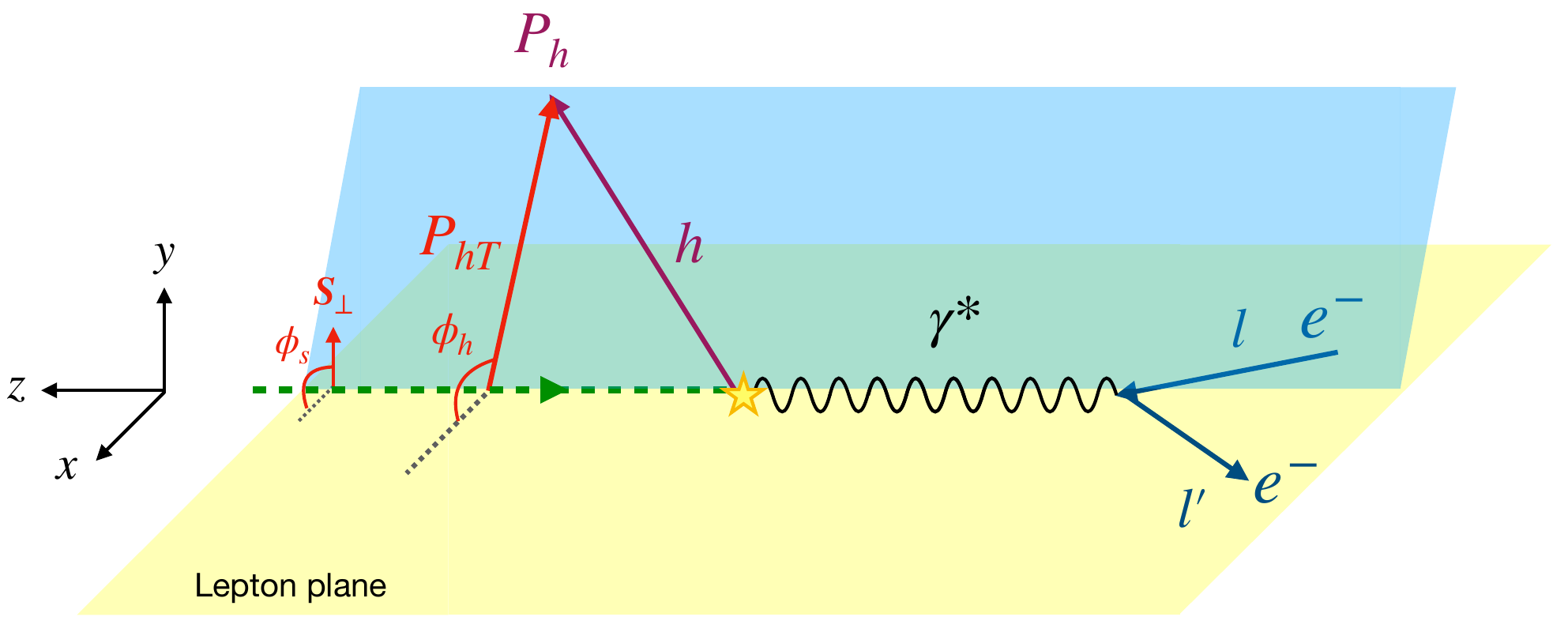}  
    \caption{Definition of azimuthal angles for semi-inclusive deep inelastic scattering in the target
rest frame~\cite{Bacchetta:2004jz}. $P_{h\perp}$ and $S_{\perp}$ are the transverse parts of $P_h$ and $S$ with respect to the photon momentum.}
    \label{fig:0611265-2}
\end{figure}

By defining the beam lepton $\ell$, the nucleon target $N$, and the produced hadron $h$ with their four-momenta in the following process
\begin{align}
\ell(l)+N(P) \rightarrow \ell\left(l^{\prime}\right)+h\left(P_h\right)+X\,,
\end{align}
and define $M$ and $M_h$ as the masses of the nucleon $N$ and the hadron $h$ respectively, one has the same DIS variables $q^2 = -Q^2$, $x$, and $y$ in Eq.~\eqref{eq:DIS-variable}, we also introduce 
\begin{align}
\gamma=\frac{2 M x}{Q}\,,
\qquad
z_h=\frac{P \cdot P_h}{P \cdot q}\,.
\end{align}
In the target rest frame, following the Trento conventions~\cite{Bacchetta:2004jz}, the azimuthal angle $\phi_h$ of the outgoing hadron are given by
\begin{align}
\cos \phi_h=-\frac{l_\mu P_{h \nu} g_{\perp}^{\mu \nu}}{\sqrt{l_{\perp}^2 P_{h \perp}^2}},&\quad g_{\perp}^{\mu \nu}  =g^{\mu \nu}-\frac{q^\mu P^\nu+P^\mu q^\nu}{P \cdot q\left(1+\gamma^2\right)}+\frac{\gamma^2}{1+\gamma^2}\left(\frac{q^\mu q^\nu}{Q^2}-\frac{P^\mu P^\nu}{M^2}\right)\,,\label{eq:2.31}\\
\sin \phi_h=-\frac{l_\mu P_{h \nu} \epsilon_{\perp}^{\mu \nu}}{\sqrt{l_{\perp}^2 P_{h \perp}^2}}\,,&\quad \epsilon_{\perp}^{\mu \nu}  =\epsilon^{\mu \nu \rho \sigma} \frac{P_\rho q_\sigma}{P \cdot q \sqrt{1+\gamma^2}}\,.\label{eq:2.32}
\end{align}
Here the transverse momentum with respect to the photon momentum $l_{\perp}^\mu=g_{\perp}^{\mu \nu} l_\nu$ and $P_{h \perp}^\mu=g_{\perp}^{\mu \nu} P_{h \nu}$ are defined and with the convention of antisymmetric tensor $\epsilon^{0123}=1$, one has the relations $g_{\perp}^{11}=g_{\perp}^{22}=-1$ and $\epsilon_{\perp}^{12}=-\epsilon_{\perp}^{21}=1$. The helicity of the lepton beam is represented by $\lambda_e$ and the covariant spin vector $S$ of the target can be decomposed as~\footnote{Note that the sign convention for the longitudinal spin component is such that the target spin is parallel to the virtual photon momentum for $\lambda_N=-1$.}
\begin{align}
S^\mu=\lambda_N \frac{P^\mu-q^\mu M^2 /(P \cdot q)}{M \sqrt{1+\gamma^2}}+S_{\perp}^\mu, \quad \lambda_N=\frac{S \cdot q}{P \cdot q} \frac{M}{\sqrt{1+\gamma^2}}, \quad S_{\perp}^\mu=g_{\perp}^{\mu \nu} S_\nu\,.\label{eq:2.6}
\end{align}
And accordingly, one can define the azimuthal angle $\phi_S$ of the spin $S$
\begin{align}
\cos \phi_S=-\frac{l_\mu S_{\nu} g_{\perp}^{\mu \nu}}{\sqrt{l_{\perp}^2 S_{\perp}^2}},&\quad \sin \phi_S=-\frac{l_\mu S_{\nu} \epsilon_{\perp}^{\mu \nu}}{\sqrt{l_{\perp}^2 S_{\perp}^2}}\,.
\end{align}
For the discussions in this section, we only consider the production of unpolarized hadron $h$.

\subsection{Hadronic tensor and Leptonic tensor}\label{sec:light-cone}
Next, in the investigation of parton distribution and fragmentation functions, it is convenient to utilize light-cone coordinates for effective manipulations. Specifically, for an arbitrary four-vector $v$, one can express $v^{ \pm}=\left(v^0 \pm v^3\right) / \sqrt{2}$ and $\boldsymbol{v}_T=\left(v^1, v^2\right)$ in a specified reference frame. All components are then represented as $\left[v^{+}, v^{-}, \boldsymbol{v}_T\right]$. Additionally, we employ the transverse tensors $g_T^{\alpha \beta}$ and $\epsilon_T^{\alpha \beta}$ given by
\begin{align}
g_T^{\alpha \beta}=g^{\alpha \beta}-n_a^\alpha n_b^\beta-n_b^\alpha n_a^\beta, \quad \quad \epsilon_T^{\alpha \beta}=\epsilon^{\alpha \beta \rho \sigma} n_{a,\rho} n_{b,\sigma}\, .
\end{align}
where only the components $g_T^{11} = g_T^{22} = -1$ and $\epsilon_T^{12} = -\epsilon_T^{21} = 1$ are nonzero. The light-cone decomposition of a vector is formulated in a Lorentz covariant manner, involving two light-like vectors $n_a=\left[1,0, \mathbf{0}_T\right]$ and $n_b=\left[0,1, \mathbf{0}_T\right]$ with $n_a^2 = n_b^2 = 0$ and $n_a\cdot{n_b} = 1$. Additionally, $\boldsymbol{v}_T$ can be promoted to a four-vector, denoted as $v_T=\left[0,0, \boldsymbol{v}_T\right]$. This light-cone representation facilitates the description of vectors in our analysis and any four-vector $v^\mu$ can be decomposed as
\begin{align}
v^\mu=v^{+} n_a^\mu+v^{-} n_b^\mu+v_T^\mu\,,
\end{align}
where $v^{+}=v \cdot n_b, v^{-}=v \cdot n_a$ and $v_T \cdot n_a=v_T \cdot n_b=0$. 

Note that scalar products of transverse four-vectors $v_T \cdot w_T$ are in Minkowski space and they are related to the transverse two-vectors by $v_T \cdot w_T=-\boldsymbol{v}_T \cdot \boldsymbol{w}_T$. Moreover, when discussing about the distribution functions, we apply the light-cone coordinates where momentum $P$ has no transverse component, namely
\begin{align}
P^\mu=P^{+} n_a^\mu+\frac{M^2}{2 P^{+}} n_b^\mu\,,\label{eq:3.3}
\end{align}
and the spin vector $S$ of the target is decomposed in the following form
\begin{align}
S^\mu=S_L \frac{\left(P \cdot n_b\right) n_a^\mu-\left(P \cdot n_a\right) n_b^\mu}{M}+S_T^\mu\,,\label{eq:3.4}
\end{align}
where one can easily find that $S_L=M\left(S \cdot n_b\right) /\left(P \cdot n_b\right)$. As for the fragmentation functions, we choose the coordinate where $P_h^-$ is the large component,
\begin{align}
P_h^\mu=P_h^{-} n_b^\mu+\frac{M_h^2}{2 P_h^{-}} n_a^\mu\,.\label{eq:3.5}
\end{align}

In~\cite{Bacchetta:2006tn}, the semi-inclusive deep inelastic scatterings are investigated under the condition where $Q^2$ becomes large while keeping $x$, $z_h$, and $P_{h\perp}^2$ fixed. To facilitate the calculation, one can choose a specific frame that satisfies both \cref{eq:3.3} and \cref{eq:3.5}, and in this frame, we have $xP^+ = P_h^-/z_h = Q/\sqrt{2}$. It is important to note that this choice of frame differs from the one in \cref{sec:sec2}, where the transverse direction was defined with respect to the momenta of the target and the virtual photon, rather than the momenta of the target and the produced hadron. The relationship between these two choices is elaborated in~\cite{Mulders:1995dh,Boer:2003cm}, indicating that $S_L$ and $S_T$, as defined by \cref{eq:3.4} with  \cref{eq:3.3} and \cref{eq:3.5}, deviate from $\lambda_N$ and $S_\perp$ in \cref{eq:2.6} by terms of order $1/Q^2$ and $1/Q$, respectively.

\begin{figure}[h!bt]
    \centering
    \includegraphics[width=0.35\textwidth]{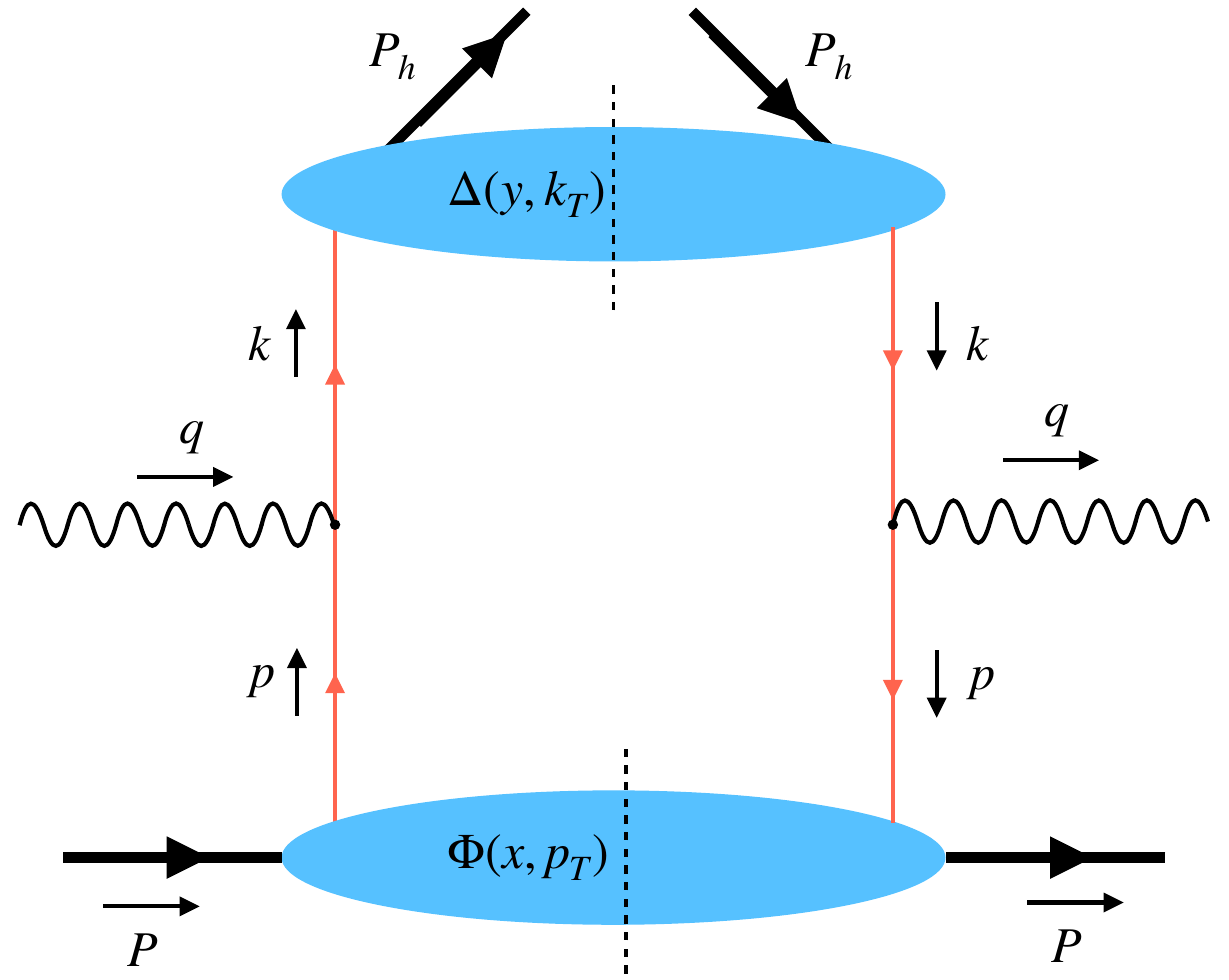} \\
    \includegraphics[width=0.8\textwidth]{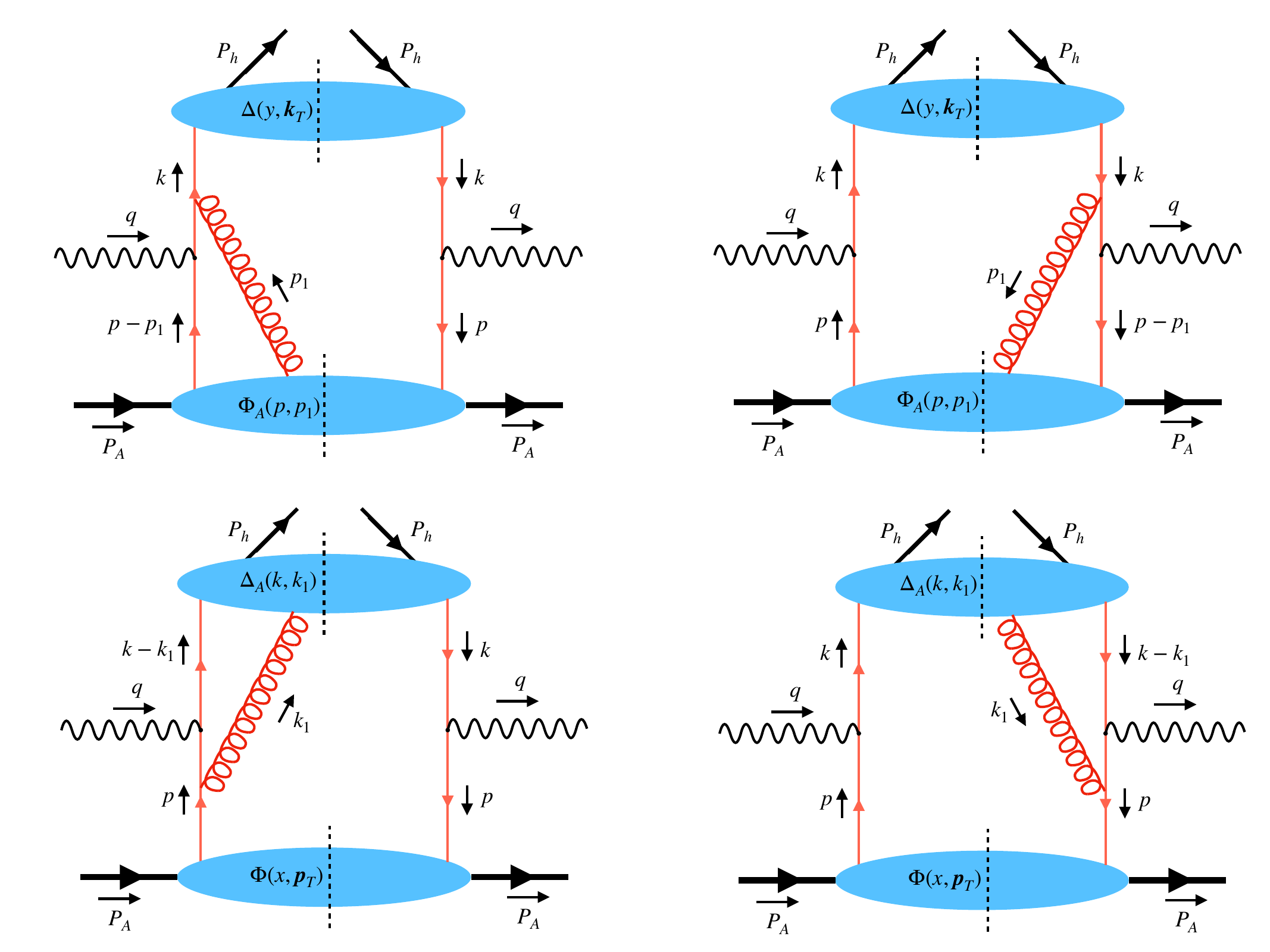}   
    \caption{Examples of graphs contributing to semi-inclusive DIS at low transverse momentum of
the produced hadron.}
    \label{fig:0611265-1}
\end{figure}

Next one can write down the lepton-production cross section described by a contraction of a hadronic tensor and a leptonic tensor,
\begin{align}
\frac{d \sigma}{d x d y d \psi d z_h d \phi_h d P_{h \perp}^2}=\frac{\alpha^2 y}{8 z_h Q^4} 2 M W^{\mu \nu} L_{\mu \nu}\,,\label{eq:3.6}
\end{align}
where the leptonic tensor and the hadronic tensor are respectively given as
\begin{align}
L_{\mu \nu}=&2\left(l_\mu l_\nu^{\prime}+l_\mu^{\prime} l_\nu-l \cdot l^{\prime} g_{\mu \nu}\right)+2 i \lambda_e \epsilon_{\mu \nu \rho \sigma} l^\rho l^{\sigma \sigma}\,,\\
2 M W^{\mu \nu}=&\frac{1}{(2 \pi)^3} \sum_X \int \frac{d^3 \boldsymbol{P}_X}{2 P_X^0} \delta^{(4)}\left(q+P-P_X-P_h\right)\left\langle P\left|J^\mu(0)\right| h, X\right\rangle\left\langle h, X\left|J^\nu(0)\right| P\right\rangle\,.
\end{align}
Here the summation runs over the polarizations of all hadrons in the final state and one has $J^\mu$ representing the electromagnetic current divided by the elementary charge.

The following calculations are based on the factorization of the cross section, breaking it down into a hard photon-quark scattering process and non-perturbative functions that describe the distribution of quarks in the target or the fragmentation of a quark into the observed hadron. For our analysis, we focus on the leading terms in the $1/Q$ expansion of the cross section~\footnote{More comprehensive details on higher twists can be found in~\cite{Bacchetta:2006tn,Ebert:2021jhy,Gamberg:2022lju}.} and consider graphs with the hard scattering at tree level. Loops are allowed only in the form shown in \cref{fig:0611265-1}, with gluons serving as external legs of the non-perturbative functions. The corresponding expression of the hadronic tensor can be found in~\cite{Mulders:1995dh,Boer:2003cm} and we have
\begin{align}
 &2 M W^{\mu \nu}=\,2 z_h \sum_a e_a^2 \int d^2 \boldsymbol{p}_T d^2 \boldsymbol{k}_T \delta^2\left(\boldsymbol{p}_T+\boldsymbol{q}_T-\boldsymbol{k}_T\right) \operatorname{Tr}\left[\Phi^a\left(x, p_T\right) \gamma^\mu \Delta^a\left(z_h, k_T\right) \gamma^\nu\right]
\,,\label{eq:3.9}
\end{align}
where the sum is carried out over quark and antiquark flavors $a$, with $e_a$ representing the fractional charge of the struck quark or antiquark. The correlation functions $\Phi$ for quark distributions, $\Delta$ for quark fragmentation can be parametrized into the leading-twist TMDs. 
In the subsequent subsections, we will provide a detailed discussion of the correlation functions $\Phi$ for quark PDFs and $\Delta$ for quark FFs. It is important to realize that the diagrams with one attached gluon in~\cref{fig:0611265-1} contribute to the cross section at the leading power when the gluon field is either parallel to the incoming nucleon $N$ or parallel to the outgoing hadron $h$, in which case it will become part of the Wilson line for $\Phi$ or $\Delta$ to make them gauage invariant. With this consideration, all the diagrams in~\cref{fig:0611265-1} would be cast into the same form as in \cref{eq:3.9}.

\section{TMD factorization}\label{sec:th-fac}
In this section, we embark on a comprehensive exploration of transverse momentum distributions (TMDs) and their underlying factorization in the context of this thesis. TMDs provide crucial insights into the spatial distribution and motion of quarks and gluons within hadrons, constituting a fundamental component of our understanding of QCD dynamics. By investigating the intricate interplay between the intrinsic transverse momenta of partons and the hard scattering processes, we aim to unravel the rich phenomenology associated with TMDs. 

\subsection{Transverse-momentum dependent distributions}\label{sec:th-pdf}
Before introducing the TMD factorization formalism, we first provide a short review for the definitions of the Transverse-momentum dependent parton distribution functions (TMDPDFs) for later convenience. TMDPDFs are defined through the so-called quark-quark correlation function~\cite{Mulders:1995dh}, $\Phi(x,\bm{k}_T;S)$, 
\begin{align}
\Phi(x, \bm{k}_T; S) = \int \frac{d\xi^- d^2 \bm \xi_T}{(2\pi)^3} e^{i k\cdot \xi}
\left.\langle PS| \bar\psi(0) \psi(\xi)|PS\rangle\right|_{\xi^+ = 0}\,,
\end{align}
where $k^ += xp^+$ with $p^+$ is the large light-cone component of the proton, and $\bm{k}_T$ is the quark transverse momentum with respect to the parent proton. Here we have suppressed the relevant gauge link for our process, which is the same as that for SIDIS process and renders the expression on the right-hand side gauge invariant. In different processes, the structure of the gauge link can change which leads to the important and nontrivial process-dependence of the TMDPDFs~\cite{Collins:2002kn,Boer:2003cm,Bomhof:2004aw,Bacchetta:2005rm,Collins:2011zzd,Kang:2020xez,Buffing:2018ggv}. The correlation function
$\Phi(x, \bm{k}_T; S)$ can be parametrized by TMDPDFs at leading twist accuracy~\cite{Mulders:1995dh,Goeke:2005hb,Bacchetta:2006tn} as
\begin{align}
\label{eq:TMDPDFs}
\Phi(x,\bm{k}_T;S)=&\frac{1}{2}\Bigg[\left(f_{1} - \frac{\epsilon_{T}^{ij} k_{T}^i S_{T}^j}{M}f_{1 T}^{\perp}\right) \slashed{n}_a+\left(\lambda_p g_{1L} + \frac{{\bm k}_T\cdot{\bm S}_T}{M}g_{1T}\right)\gamma_{5} \slashed{n}_a\\
& -i\sigma_{i\mu}n_a^\mu \left(h_1{S_{T}^i}\gamma_5 -ih_1^{\perp}\frac{k_{T}^i}{M}+  h_{1L}^\perp\frac{\lambda_p k_{T}^i}{M} \gamma_5+h_{1T}^\perp\frac{{\bm k}_T\cdot{\bm S}_T k_{T}^i - \frac{1}{2}k_T^2S_{T}^i}{M^2}\gamma_5\right)\Bigg]\,, \notag
\end{align}
where $\sigma_{\mu\nu}=\frac{i}{2}\left[\gamma_\mu,\gamma_\nu\right]$. We have eight quark TMDPDFs $f_1(x, k_T^2)$, $f_{1T}^{\perp}(x, k_T^2)$, $g_{1L}(x, k_T^2)$, $g_{1T}(x, k_T^2)$, $h_1(x, k_T^2)$, $h_1^\perp(x, k_T^2)$, $h_{1L}^\perp(x, k_T^2)$, and $h_{1T}^\perp(x, k_T^2)$, and their physical interpretations are summarized in \cref{intpdf}. For details, see~\cite{Mulders:1995dh,Goeke:2005hb,Bacchetta:2006tn,Bacchetta:2004jz,Boer:2011fh,Accardi:2012qut}.

As usual, we find it convenient to work in the Fourier or $\bm{b}$-space. Taking the Fourier transformation of the correlation function, we have 
\begin{align}
\tilde{\Phi}\left(x, \bm{b}; S\right)=\int d^2\bm{k}_T\,e^{-i\bm{k}_T\cdot \bm{b}}\, \Phi(x,\boldsymbol{k}_T;S)\,,
\end{align}
and the $\bm{b}$-space correlation function $\tilde{\Phi}(x, \bm{b}; S)$ at leading twist is given by~\cite{Boer:2011xd}
\begin{align}
\label{eq:TMDPDFsb}
\tilde{\Phi}(x,\bm{b};S)=&\frac{1}{2}\Bigg[\left(\tilde{f}_{1} + {i\epsilon_{T}^{ij} b^i S_{T}^j}{M}\tilde{f}_{1 T}^{\perp(1)}\right) \slashed{n}_a+\left(\lambda_p \tilde{g}_{1L} - {i{\bm b}\cdot{\bm S}_T}{M}\tilde{g}^{(1)}_{1T}\right)\gamma_{5} \slashed{n}_a\nonumber\\
& \hspace{0.7cm} -i\sigma_{i\mu}n_a^\mu \bigg({S_{T}^i}\tilde{h}_1\gamma_5 -{b^{i}}{M}\tilde{h}_1^{\perp(1)}- i{\lambda_p b^{i}}{M}\tilde{h}_{1L}^{\perp(1)} \gamma_5\nnu
& \hspace{2.7cm}-\frac{1}{2}\left({\bm b}\cdot{\bm S}_T b^{i}-\frac{1}{2} b^2S_{T}^i\right) M^{2} \tilde{h}_{1 T}^{\perp(2)}\gamma_5\bigg)\Bigg]\,,
\end{align}
where $b = |\bm{b}|$ denotes the magnitude of the vector $\bm{b}$. Here, the TMDPDFs in $\bm{b}$-space are defined as
\begin{align}\label{btilde}
\tilde{f}^{(n)}(x,b^2)=\frac{2 \pi n !}{\left({M^{2}}\right)^{n}} \int dk_T \, k_T\left(\frac{k_T}{b}\right)^{n} J_{n}\left(k_T b\right) f\!\left(x, k_T^2\right)\,,
\end{align}
where $n=0$ by default when denoted without a superscript. For simplicity, we have suppressed the additional scale-dependence in both $f\left(x, k_T^2\right)$ and $\tilde{f}^{(n)}(x,b^2)$, and we will specify these scale-dependence explicitly below when we present the factorization formula. 
\begin{table} 
    \centering
\begin{tabular}{ |c|c|c|c| } 
 \hline
 \diagbox[width=3em]{$H$}{$q$} & $U$ & $L$ & $T$ \\ 
  \hline
 $U$ & $f_1$ &  & $h_1^\perp$  \\ 
  \hline
$L$ &  &  $g_{1L}$& $h_{1L}^\perp$ \\ 
  \hline
$T$ & $f_{1T}^\perp$ & $g_{1T}$ & $h_1,\ h_{1T}^\perp$\\ 
  \hline
\end{tabular}
  \caption{TMDPDFs for quarks. We have quark polarizations in the row with $U=\,$unpolarized, $L=\,$longitudinal polarized, and $T=\,$transversely polarized quarks. On the other hand, the column represents polarization of the hadron $H$ (i.e. the proton in our case).}
  \label{intpdf}
\end{table}

\subsection{Transverse-momentum dependent fragmentation functions}\label{sec:th-ff}
We start with writing the parametrization of the transverse-momentum dependent fragmentation functions (TMDFFs) correlator~\cite{Metz:2016swz} in the momentum space. 
\bea
{\Delta}\left(z_h, \boldsymbol{k}_\perp,S_h\right) =& \sum_X \int \frac{d \xi^{+}d^2{\bm \xi}_T}{(2 \pi)^3} e^{i (k^{-} \xi^{+}+{\bm k}_{\perp}\cdot{\bm \xi}_T)/z_h}\left\langle 0\left| \psi_{q}\left(\xi^{+}, 0^{-}, {\bm \xi}_{T}\right)\right| p_h, S_{h} ; X\right\rangle \nnu
& \times\left\langle p_h, S_{h} ; X\left|\bar{\psi}_{q}\left(0^{+}, 0^{-}, {\bm 0}_{T}\right) \right| 0\right\rangle \,,
\eea
where $\boldsymbol{k}_\perp$ is the transverse momentum of the final hadron $h$ with respect to the fragmenting quark $q$ and we suppress the Wilson lines that make the correlator gauge invariant. To the leading twist accuracy, the parametrization is given as
\bea
\label{eq:TMDFF}
\Delta(z_h,\bm{k}_\perp,S_{h})=&\frac{1}{2}\Bigg\{\left({{D}}_{1}- \frac{\epsilon_{T}^{ij} k_{\perp}^i S_{h\perp}^j}{z_hM_h} {{D}}_{1 T}^{\perp}\right)\slashed{n}_b+\left(\lambda_h{{G}}_{1L}- \frac{\boldsymbol{k}_\perp\cdot\boldsymbol{S}_{h\perp}}{z_hM_h} {{G}}_{1 T}\right)\slashed{n}_b\gamma_5
\nnu
&
-i\sigma_{i\mu}n_b^\mu\left(H_1S_{h\perp}^i-iH_1^\perp\frac{k_\perp^i}{z_hM_h}-H_{1L}^\perp\frac{\lambda_h k_\perp^i}{z_hM_h}\gamma_5 \right.
\nnu
&\left.\hspace{15mm}+H_{1T}^\perp\frac{\boldsymbol{k}_\perp\cdot \boldsymbol{S}_{h\perp}k_\perp^i  - \frac{1}{2} {k}_\perp^{2}S_{h\perp}^i}{z_h^2M_h^2}\gamma_5\right)\Bigg\}\,,
\eea
where $n_b$ is the light-cone vector defined by the outgoing quark direction.

Just as in \cref{unp_JFF_FF}, we find it more convenient to derive the relations between the TMDJFFs and TMDFFs using the Fourier space expressions of the TMDFFs. The Fourier transformation for the TMDFF correlator is defined as
\bea
\tilde{\Delta}(z_h,\bm{b},S_h)=\frac{1}{z_h^2}\int d^2\bm{k}_\perp e^{-i\bm{k}_\perp\cdot \bm{b}/z_h}\Delta(z_h,\bm{k}_\perp,S_h)\,.
\eea
The TMDFF correlator in $\bm{b}$-space is then given as
\bea
\label{eq:TMDFFb}
\tilde{\Delta}(z_h,\bm{b},S_h)=&\frac{1}{2}\Bigg\{\left({\tilde{D}}_{1}(z_h,b^2)+ i{\epsilon_{T}^{ij} b^i S_{h\perp}^j}{z_hM_h} {\tilde{D}}_{1 T}^{\perp(1)}(z_h,b^2)\right)\slashed{n}_b\nnu
&\hspace{0.5cm}+\left(\lambda_h{\tilde{G}}_{1L}(z_h,b^2) + i{\boldsymbol{b}\cdot\boldsymbol{S}_{h\perp}}{z_hM_h} {\tilde{G}}_{1 T}^{(1)}(z_h,b^2)\right)\slashed{n}_b\gamma_5\nnu
&\hspace{0.5cm} -i\sigma_{i\mu}n_b^\mu\bigg[\tilde{H}_1(z_h,b^2)S_{h\perp}^i-\tilde{H}_1^{\perp(1)}(z_h,b^2){b^i}{z_hM_h}+i\tilde{H}_{1L}^{\perp(1)}(z_h,b^2){\lambda_h b^i}{z_hM_h}\gamma_5 \nnu
&\hspace{2.3cm}-\tilde{H}_{1T}^{\perp(2)}(z_h,b^2)\frac{1}{2}\left({\boldsymbol{b}\cdot \boldsymbol{S}_{h\perp}b^i - \frac{1}{2}{b}^{2}S_{h\perp}^i}\right){z_h^2M_h^2}\gamma_5\bigg]\Bigg\}\,,
\eea
where we defined
\bea\label{btilde2}
\tilde{{F}}^{(n)}(z_h,b^2)=&\frac{1}{z_h^2}\frac{2 \pi n ! }{\left(z_h^2M_h^{2}\right)^{n}} \int d{k}_\perp {k}_\perp\left(\frac{{k}_\perp}{b}\right)^{n} J_{n}\left(\frac{{b}{k}_\perp}{z_h}\right) {F}^{h/q}\left(z_h, {k}_\perp^2\right)\,.
\eea
Note that $F$ stands generally for all TMDFFs with appropriate $n$ value and by default $n=0$. We then begin with unsubtracted TMDFFs, which follow the same parametrization, and make the scale explicit by replacing
\bea
\tilde{{F}}^{(n)}(z_h,b^2) \to \tilde{{F}}^{(n),{\rm unsub}}(z_h,b^2,\mu,\zeta/\nu^2)\,,
\eea
where $\mu$ is the usual renormalization scale, $\nu$ is a rapidity scale, and $\zeta$ is the so-called Collins-Soper scale~\cite{Boussarie:2023izj}.

\subsection{Results of the structure functions}
By substituting the parameterizations of the various PDF and FF correlators into equation \eqref{eq:3.9}, one can compute the lepton-hadron production cross section for SIDIS and extract the forms of all the structure functions $F_{AB,(C)}$. 

With the assumption of single photon exchange, the differential cross section can be described by a set of structure functions that are model-independent~\cite{Gourdin:1973qx,Kotzinian:1994dv,Diehl:2005pc} and one obtains~\cite{Diehl:2005pc,Bacchetta:2006tn},
\begingroup
\allowdisplaybreaks
\begin{align}
 \frac{d \sigma}{d x d y d \psi d z d \phi_h d P_{h \perp}^2}= & \frac{\alpha^2}{x y Q^2} \frac{y^2}{2(1-\varepsilon)}\left(1+\frac{\gamma^2}{2x}\right)\bigg\{F_{U U, T}+\varepsilon F_{U U, L}+\varepsilon \cos \left(2 \phi_h\right) F_{U U}^{\cos 2 \phi_h}\nnu
 &+\lambda_N\varepsilon \sin \left(2 \phi_h\right) F_{U L}^{\sin 2 \phi_h}+\lambda_N \lambda_e\sqrt{1-\varepsilon^2} F_{L L}\nnu
&  +\left|\boldsymbol{S}_{\perp}\right|\left[\sin \left(\phi_h-\phi_S\right)\left(F_{U T, T}^{\sin \left(\phi_h-\phi_S\right)}+\varepsilon F_{U T, L}^{\sin \left(\phi_h-\phi_S\right)}\right)\right. \nnu
&+\varepsilon \sin \left(\phi_h+\phi_S\right) F_{U T}^{\sin \left(\phi_h+\phi_S\right)}+\varepsilon \sin \left(3 \phi_h-\phi_S\right) F_{U T}^{\sin \left(3 \phi_h-\phi_S\right)}\bigg]\nnu
&+\left|\boldsymbol{S}_{\perp}\right|\lambda_e \sqrt{1-\varepsilon^2}\cos\left(\phi_h-\phi_S\right)F_{L T}^{\cos \left(\phi_h-\phi_S\right)}\bigg\}\,,\label{eq:2.7}
\end{align}
\endgroup
where the ratio $\varepsilon$ is defined by
\begin{align}
\varepsilon=\frac{1-y-\frac{1}{4} \gamma^2 y^2}{1-y+\frac{1}{2} y^2+\frac{1}{4} \gamma^2 y^2}\approx \frac{1-y}{1-y+\frac{1}{2} y^2}\,.
\end{align}
Thus one has the overall depolarization factor given by
\begin{align}
\frac{y^2}{2(1-\varepsilon)} &\approx\left(1-y+\frac{1}{2} y^2\right)\,,
\end{align}
and the rest factors are written as
\begin{eqnarray}
&\frac{y^2}{2(1-\varepsilon)} \,\varepsilon \approx(1-y)\,, &\quad \frac{y^2}{2(1-\varepsilon)} \sqrt{1-\varepsilon^2}  \approx y\left(1-\frac{1}{2} y\right)\,,\\
&\frac{y^2}{2(1-\varepsilon)} \sqrt{2 \varepsilon(1+\varepsilon)} \approx(2-y) \sqrt{1-y}\,, &\quad\frac{y^2}{2(1-\varepsilon)} \sqrt{2 \varepsilon(1-\varepsilon)}  \approx y \sqrt{1-y}\,.
\end{eqnarray}
In the context of our study, the structure functions on the right-hand side of the equation depend on several parameters, including $x$, $Q^2$, $z$, and $P_{h \perp}^2$. The angle $\psi$ represents the azimuthal angle of $\ell^{\prime}$ (the scattered lepton) around the lepton beam axis, with respect to an arbitrary fixed direction. In the case of a transversely polarized target, we specifically choose this fixed direction to align with the direction of $S$ (the target polarization vector). The relationship between $\psi$ and $\phi_S$ is detailed in~\cite{Diehl:2005pc}, where, in the context of deep inelastic kinematics, $d \psi$ is approximately equal to $d \phi_S$.

The subscripts of the structure functions signify the respective polarizations of the beam and target, while an additional subscript in $F_{U U, T}$, $F_{U U, L}$, $F_{U T, T}^{\sin \left(\phi_h-\phi_S\right)}$, and $F_{U T, L}^{\sin \left(\phi_h-\phi_S\right)}$ specifies the polarization of the virtual photon. Here, the terms "longitudinal" and "transverse" target polarization refer to the photon direction. However, converting to experimentally relevant longitudinal or transverse polarizations with respect to the lepton beam direction is a straightforward process, and details can be found in~\cite{Diehl:2005pc}.

To simplify the notation of these structure functions, we introduce the unit vector $\hat{\bm{h}}=\bm{P}_{h \perp}/|\bm{P}_{h \perp}|$. As an example, we first write down the factorization formalism for $F_{UU,T}$,
\begin{align}
F_{UU,T}=&x \sum_i \sigma_0 H\left(Q^2, \mu\right) \int_0^\infty \frac{db}{2\pi}\, b\, b^{m+n} J_{m+n}(q_Tb) \nnu
&\times\tilde{f}^{\rm unsub.}_{1,i/p}\left(x, b,\mu,\zeta_1/\nu^2\right) \tilde{D}^{\rm unsub.}_{1,h/i}\left(z_h, b,\mu,\zeta_2/\nu^2\right)\tilde{S}_{n_a\,n_b}(b,\mu,\nu)\,,\label{eq:fuutnew}
\end{align}
where $\displaystyle\sigma_0=\frac{\alpha_{\mathrm{em}} \alpha_s}{s Q^2} \frac{2\left(\hat{u}^2+\hat{s}^2\right)}{\hat{t}^2}$ is the Born cross section for the unpolarized electron and quark scattering process. $H\left(Q^2, \mu\right)$ is the hard function that encodes physics at the hard scale $Q$ and at the next-to-leading order, it is given by~\cite{Kang:2021ffh}
\begin{align}
H\left(Q^2, \mu\right) & =1+\frac{\alpha_s(\mu)}{2 \pi}C_F\left(- \ln^2 \frac{Q^2}{\mu^2}-3\ln \frac{Q^2}{\mu^2}-8+\frac{\pi^2}{6}\right)+\mathcal{O}\left(\alpha_s^2\right)\,,
\end{align}
In \cref{eq:fuutnew}, $\tilde{f}^{\rm unsub.}_{1,i/p}$ is the renormalized beam function defined in the SCET literature~\cite{Stewart:2009yx} (also know as unsubtracted TMD PDF), describing collinear radiation close to the proton, $\tilde{D}^{\rm unsub.}_{1,h/i}$ is the unsubtracted TMD FF. And $\tilde{S}_{n_a\,n_b}$ is the soft function that encodes soft gluon radition between the colliding partons. 
Up to NLO, the soft function is given as
\begin{align}
    \tilde{S}_{n_a\,n_b}(b,\mu,\nu)=1+\frac{\alpha_s(\mu)C_F}{2\pi}\left[-L_b^2+4L_b \ln\frac{\mu}{\nu}-\frac{\pi^2}{6}\right]+\mathcal{O}(\alpha_s^2)\,,
\end{align}
with $\displaystyle L_b=\ln\frac{b^2\mu^2}{b_0^2}$ with $b_0=2e^{-\gamma_E}$. Note that both soft function and the unsubtracted TMDs contain additional divergence called ``rapidity divergence''. In order to regularize them, we use the rapidity regulator method introduced in~\cite{Chiu:2012ir}, which is why we have a new rapidity scale $\nu$ similar to the normal renormalization scale $\mu$. 

Note the dependence of rapidity divergence scale $\nu$ cancels between the unsubtracted function and soft function~\cite{Chiu:2012ir}, leaving only the Collins-Soper scale $\zeta_1$, $\zeta_2$
\begin{align}
\tilde{f}_{i/p}\left(x, b,\mu,\zeta_1\right)=\tilde{f}^{\rm unsub.}_{i/p}\left(x, b,\mu,\zeta_1/\nu^2\right)\sqrt{\tilde{S}_{n_a\,n_b}(b,\mu,\nu)}\,,\\
\tilde{D}_{h/i}\left(z_h, b,\mu,\zeta_2\right)=\tilde{D}^{\rm unsub.}_{h/i}\left(z_h, b,\mu,\zeta_2/\nu^2\right)\sqrt{\tilde{S}_{n_a\,n_b}(b,\mu,\nu)}\,.
\end{align}
The method of including soft function into the unsubtracted TMDs was first introduced by Collins \cite{Collins:2011zzd}. With this new definition, one can view that the decomposition of TMD PDF $\tilde{f}_{i/p}\left(x, b, \mu, \zeta_1\right)$ and TMD FF $\tilde{D}_{h/i}\left(z, b, \mu, \zeta_2\right)$ into collinear and soft matrix elements~\cite{Becher:2010tm,Becher:2011xn,Becher:2012yn,Echevarria:2011epo,Echevarria:2012js,Echevarria:2014rua,Chiu:2012ir,Li:2016axz}. Now we can simplify the factorization in \cref{eq:fuutnew} and define the notation~\cite{Bacchetta:2006tn,Boer:2011xd,Boussarie:2023izj}
\begin{align}
\mathcal{B}[\tilde{f}^{(m)} \tilde{D}^{(n)}]=&x \sum_i H_{ii}(Q^2,\mu) \int_0^\infty \frac{db}{2\pi}\, b\, b^{m+n} J_{m+n}(q_Tb)  \tilde{f}^{(m)}_{i/p}\left(x, b,\mu,\zeta_1\right) \tilde{D}^{(n)}_{h/i}\left(z_h, b,\mu,\zeta_2\right)\,,
\end{align}
where the Fourier-transformed TMD PDFs $\tilde{f}^{(m)}_{i/p}$ and TMD FFs $\tilde{D}^{(n)}_{h/i}$ have been defined in~\cref{btilde,btilde2}.
Finally, one arrives at the structure functions shown in \cref{eq:2.7} written in terms of TMDs~\cite{Boer:2011xd}:
\begingroup
\allowdisplaybreaks
\begin{align}
F_{U U,T}\left(x, z_h, P_{h T}, Q^2\right) & =\mathcal{B}\left[\tilde{f}_1^{(0)} \tilde{D}_1^{(0)}\right]\,, \\
F_{U U,L}\left(x, z_h, P_{h T}, Q^2\right) & =0\,,\\
F_{U U}^{\cos 2 \phi_h}\left(x, z_h, P_{h T}, Q^2\right) & =M_N M_h \mathcal{B}\left[\tilde{h}_1^{\perp(1)} \tilde{H}_1^{\perp(1)}\right]\,, \\
F_{U L}^{\sin 2 \phi_h}\left(x, z_h, P_{h T}, Q^2\right) & =M_N M_h \mathcal{B}\left[\tilde{h}_{1 L}^{\perp(1)} \tilde{H}_1^{\perp(1)}\right]\,, \\
F_{L L}\left(x, z_h, P_{h T}, Q^2\right) & =\mathcal{B}\left[\tilde{g}_1^{(0)} \tilde{D}_1^{(0)}\right]\,, \\
F_{U T,T}^{\sin \left(\phi_h-\phi_S\right)}\left(x, z_h, P_{h T}, Q^2\right) & =-M_N \mathcal{B}\left[\tilde{f}_{1 T}^{\perp(1)} \tilde{D}_1^{(0)}\right]\,, \\
F_{U T,L}^{\sin \left(\phi_h-\phi_S\right)}\left(x, z_h, P_{h T}, Q^2\right) & =0\,, \\
F_{U T}^{\sin \left(\phi_h+\phi_S\right)}\left(x, z_h, P_{h T}, Q^2\right) & =M_h \mathcal{B}\left[\tilde{h}_1^{(0)} \tilde{H}_1^{\perp(1)}\right]\,, \\
F_{U T}^{\sin \left(3 \phi_h-\phi_S\right)}\left(x, z_h, P_{h T}, Q^2\right) & =\frac{M_N^2 M_h}{4} \mathcal{B}\left[\tilde{h}_{1 T}^{\perp(2)} \tilde{H}_1^{\perp(1)}\right]\,,\\
F_{L T}^{\cos \left(\phi_h-\phi_S\right)}\left(x, z_h, P_{h T}, Q^2\right) & =M_N \mathcal{B}\left[\tilde{g}_{1 T}^{\perp(1)} \tilde{D}_1^{(0)}\right]\,, 
\end{align}
\endgroup

\section{Summary}
So far we have provided a comprehensive review of Semi-Inclusive Deep Inelastic Scattering (SIDIS), a fundamental process in studying the structure of hadrons. We begin by discussing the essential aspects of SIDIS, including its kinematics and structure functions, which are crucial for interpreting experimental measurements. In this chapter, we also introduce the Transverse Momentum Dependent (TMD) factorization framework, which provides a theoretical description of the SIDIS cross-section. Key components of TMD factorization, such as TMD parton distribution functions and TMD fragmentation functions, are presented. The main goal of this chapter is to provide readers with a thorough understanding of SIDIS and TMD factorization, as these concepts are essential for investigating the structure of hadrons and introducing the studies of jets.

The subsequent chapter will introduce a novel concept called Polarized Jet Fragmentation Functions (JFFs), developed by the author of this dissertation. It will discuss the motivation and significance of polarized JFFs, emphasizing their importance in understanding the spin structure of hadrons. Detailed calculations of polarized JFFs for both Collinear and Transverse Momentum Dependent (TMD) cases will be presented. The objective of this chapter is to provide a comprehensive insight into polarized JFFs and their significance in studying the spin structure of hadrons.


\chapter{Polarized Jet Fragmentation Functions (JFFs)}\label{ch4:pjff}

\begin{quote}
\rule{0.875\textwidth}{0.5pt}\\
The Polarized Jet Fragmentation Functions (JFFs) is a newly developed concept by the author of this dissertation. The chapter starts by discussing the motivation and significance of polarized JFFs, which is a crucial ingredient for understanding the spin structure of hadrons. The focus then shifts to the detailed calculations of polarized JFFs for both Collinear and Transverse Momentum Dependent (TMD) cases. The chapter aims to provide a comprehensive understanding of the polarized JFFs and its importance for the study of the spin structure of hadrons.
\\
\rule{0.875\textwidth}{0.5pt}
\end{quote}

Over the last few years, the study of hadron distributions inside jets has received increasing attention as an effective tool to understand the fragmentation process, describing how the color carrying partons transform into color-neutral particles such as hadrons. Understanding such a fragmentation process is important as it will provide us with a deep insight into the elusive mechanism of hadronization. Theoretical objects which describe the momentum distribution of hadrons inside a fully reconstructed jet is called jet fragmentation functions (JFFs). The usefulness of studying the longitudinal momentum distribution of the hadron in the jet rather than the hadron production itself stems from the former process being differential in the momentum fraction $z_h \equiv p_{hT}/ p_{JT}$, where $p_{hT}$ and $p_{JT}$ are the transverse momenta of the hadron and the jet with respect to the beam axis, respectively. Collinear JFFs in the first process can be matched onto the standard collinear fragmentation functions (FFs), enabling us to extract the usual universal FFs more directly by ``scanning'' the differential $z_h$ dependence. The theoretical developments on the JFFs were first studied in the context of exclusive jet production ~\cite{Procura:2009vm,Jain:2011xz,Jain:2011iu,Chien:2015ctp} and was later extended to the inclusive jet production case ~\cite{Arleo:2013tya,Kaufmann:2015hma,Kang:2016ehg,Dai:2016hzf,Kang:2019ahe}. 

At the same time, the transverse momentum distribution of the hadrons within jets can be sensitive to the transverse momentum dependent fragmentation, described by transverse momentum dependent jet fragmentation functions (TMDJFFs). In ~\cite{Kang:2017glf}, it was demonstrated that such TMDJFFs are closely connected to the standard transverse momentum dependent FFs (TMDFFs) ~\cite{Bacchetta:2000jk,Mulders:2000sh,Metz:2016swz} when the transverse momentum of the hadron is measured with respect to the standard jet axis. For the TMD study of the hadron with respect to the Winner-Take-All jet axis, see ~\cite{Neill:2016vbi,Neill:2018wtk}. As for the TMD study inside the groomed jet, see ~\cite{Makris:2017arq,Makris:2018npl,Gutierrez-Reyes:2019msa}. For the recent works on resummation of $\ln z_h$ and $\ln(1-z_h)$, see \cite{Neill:2020bwv,Kaufmann:2019ksh}.

Because of its phenomenological relevance and effectiveness, study of the JFFs has become a very important topic over recent years at the LHC and RHIC, producing measurements for a wide range of identified particles within the jet. Calculations for the JFFs have been performed for single inclusive jet production in unpolarized proton-proton collisions in the context of light charged hadrons ~\cite{Chien:2015ctp,Kaufmann:2015hma,Kang:2016ehg}, heavy-flavor mesons~\cite{Chien:2015ctp,Bain:2016clc,Anderle:2017cgl}, heavy quarkonium~\cite{Kang:2017yde,Bain:2017wvk}, and photons~\cite{Kaufmann:2016nux}. For the relevant experimental results for the LHC and RHIC, see~\cite{Aaboud:2019oac,Aad:2011td,Aad:2011sc,Chatrchyan:2012gw,Chatrchyan:2014ava,Aad:2014wha,Aaboud:2017tke,Aaij:2017fak,Acharya:2019zup,Aaij:2019ctd,Acharya:2018eat,Aad:2019onw,Sirunyan:2019vlp} and~\cite{Adamczyk:2017wld,Adamczyk:2017ynk}, respectively. Study of JFFs is not only important at the LHC and RHIC as already proven to be, but also provides novel insights at the future Electron-Ion Collider~(EIC)~\cite{Accardi:2012qut,Aschenauer:2017jsk,Liu:2018trl,Arratia:2019vju} as we will show below.  

In this chapter, we provide the general theoretical framework for studying the distribution of hadrons inside a jet by taking full advantage of the polarization effects. We introduce polarized jet fragmentation functions, where the parton that initiates the jet and the hadron that is inside the jet can both be polarized. We do this in the context of both $pp$ collider like LHC and RHIC, as well as $ep$ collider like the future EIC. Analogous to the standard FFs, we find a slew of different JFFs that have close connection with the corresponding standard FFs. 

When a proton with a general polarization collides with an unpolarized proton or lepton, different JFFs appear with different parton distribution functions (PDFs) and characteristic modulation in the azimuthal angles measured with respect to the scattering plane. Therefore, these observables are not only useful in exploring the spin-dependent FFs, but also in understanding the polarized PDFs. For instance, with the extra handle in $z_h$, we would be able to reduce uncertainties coming from the final state fragmentation functions by restricting to a well-determined $z_h$ region. Alternatively, with well-determined polarized PDFs at hand, we can directly probe spin-dependent FFs through a study of different JFFs. Some applications of spin-dependent JFFs relevant for the RHIC were considered in~\cite{Adamczyk:2017wld,Yuan:2007nd,DAlesio:2010sag,DAlesio:2011kkm,Kang:2017btw,DAlesio:2017bvu}, but other applications are far and wide. To demonstrate this, we consider two phenomenological applications in detail. We demonstrate how one can use spin-dependent JFFs to study the collinear helicity FFs and so-called TMD polarizing fragmentation functions (TMD PFFs). There are, of course, many more possible applications of studying other polarized JFFs which we also list in this paper and will present the details in a forthcoming long paper. Other potential applications include probing the polarization of heavy quarkonium inside the jet~\cite{Kang:2017yde}, which is very promising at the LHC and RHIC. 

\section{Kinematics}\label{sec:4.1}
To properly define the momentum and the spin vectors, we apply a light-cone vector $n_a=\left[1, 0,\mathbf{0}_T\right]$ with its conjugate vector $n_b=\left[0,1, \mathbf{0}_T\right]$ that have been introduced in \cref{sec:light-cone}. Accordingly, we decomposes any four-vector $p^\mu$ as $p = \left[p^+, p^-,\bm{p}_T\right]$. Namely,
\bea
p^\mu =p^+ n_a^{\mu}+ p^- n_b^{\mu} + p_T^\mu\,,
\eea
where $p^+ = n_b\cdot p = \frac{1}{\sqrt{2}}\left(p^0 + p^z\right)$ and $p^- = {n_a}\cdot p = \frac{1}{\sqrt{2}}\left(p^0 - p^z\right)$. Let us specify the kinematics of the hadron inside the jet. If the hadron is in a reference frame in which it moves along the $+z$-direction and has no transverse momentum, then the $p_h^-$ component of its momentum would be very large while the $p_h^+$ component is small, $p_h^+\ll p_h^-$. We can parameterize the momentum $p_h$ and the spin vector $S_h$ of the hadron, respectively, as
\begin{align}
p_h = \left(\frac{M_h^2}{2p_h^-},p_h^-,0\right)\,,
\qquad
S_h = \left(-\Lambda_h \frac{M_h}{2p_h^-},\Lambda_h \frac{p_h^-}{M_h},{\bm S}_{h\perp}\right)\,,
\end{align}
where $M_h$ is the mass of the hadron, and $\Lambda_h$ and ${\bm S}_{h\perp}$ describe the longitudinal and transverse polarization of the hadron inside the jet, respectively. It is evident that they satisfy the relation $p_h\cdot S_h =0$ as required. 

\section{Polarized collinear JFFs}\label{sec:2}

In this section, we introduce the definition of the exclusive and semi-inclusive jet fragmentation function in SCET with both unpolarized and polarized fragmenting hadron, which are used in the description of longitudinal momentum fraction distribution within jets in $pp$ collisions.
The siJFFs describe the fragmentation of a hadron $h$ within a jet that is initiated by a parton $c$.
We first provide their operator definitions, perform its calculation to NLO, derive and solve its RG evolution equation.

\subsection{Collinear JFFs in semi-inclusive jet productions}\label{ss.collinear_JFF}
The general correlators that define the collinear jet fragmentation functions in such a hadron frame are given by~\cite{Kang:2016ehg}
\begin{align}
\Delta^{h/q}(z, z_h,\omega_J, S_h) =& \frac{z}{2N_c} \delta\left(z_h - \frac{\omega_h}{\omega_J}\right)\nnu
&\times\langle 0| \delta\left(\omega - n_a\cdot {\mathcal P} \right) \chi_n(0) |(Jh)X\rangle\langle (Jh)X|\bar \chi_n(0) |0\rangle,\label{eq:correlators1}\,,\\
\Delta^{h/g,\ \mu\nu}(z, z_h, \omega_J, S_h) =& \frac{z\,\omega}{(d-2)(N_c^2-1)} \delta\left(z_h - \frac{\omega_h}{\omega_J}\right) \nnu
&\times \langle 0|  \delta\left(\omega - n_a\cdot {\mathcal P} \right){\mathcal B}_{n\perp}^\mu(0)|(Jh)X\rangle \langle (Jh)X|{\mathcal B}_{n\perp}^\nu(0)  |0\rangle\,,\label{eq:correlators2}
\end{align}
for quark and gluon jets, respectively. Here $\omega_J$ and $\omega_h$ are the energy of the jet and that of the identified hadron inside the jet, respectively and they are related to the momenta of the jet $p_J$, and the hadron $p_h$ by $\omega_J=p_J^-$ and $\omega_h=p_h^-$. The energy fractions $z$ and $z_h$ are defined as 
\begin{align}
z = \frac{\omega_J}{\omega}, \qquad z_h =  \frac{\omega_h}{\omega_J}\,,\label{eq:definez}
\end{align}
where $\omega$ is the energy of the parton that initiates the jet. Thus $z$ is the momentum fraction of the parton carried by the jet, while $z_h$ is the momentum fraction of the jet carried by the hadron. 

Note that the state $|(Jh)X\rangle$ represents the final-state unobserved particles $X$ and the observed jet $J$ with an identified hadron $h$ inside, denoted collectively by $(Jh)$. Because of this, the equations above also contain the kinematics of the jet, such as the jet radius $R$. We will suppress them here for simplicity, but express them out explicitly when we discuss their evolution equation in the following subsection. Also note here, we have used the gauge invariant quark and gluon fields, $\chi_n$ and $\mathcal{B}_{n\perp}^\mu$, in the Soft Collinear Effective Theory~\cite{Bauer:2000ew,Bauer:2000yr,Bauer:2001ct,Bauer:2001yt},  
\begin{equation}\label{eq:gauge}
\chi _n = W_{n}^{\dagger} \xi _n ,
\quad
\mathcal{B}_{n \perp}^{\mu} = \frac{1}{g} \bqty{W_{n}^{\dagger} i D_{n \perp}^{\mu} W_n} ,
\end{equation}
where $n$ in the subscript denotes the light-cone vector and has its spatial component aligned with the jet axis. In \cref{eq:gauge}, the covariant derivative is $i D_{n \perp}^{\mu}=\mathcal{P}_{n \perp}^{\mu}+g A_{n \perp}^{\mu}$, with $\mathcal{P}^{\mu}$ the label momentum operator.
On the other hand, $W_{n}$ is the Wilson line of collinear gluons:
\begin{equation}
W_n \pqty{x}
=
\sum_{\textrm{perms}} \exp(-g \frac{1}{n_a \cdot \mathcal{P}} n_a\cdot A_n \pqty{x}) .
\end{equation}
With these collinear quark and gluon fields at hand, one can define the correlator definitions for the quark semi-inclusive JFFs with different polarizations as \cite{Kang:2016ehg}
\begingroup
\allowdisplaybreaks
\begin{align}
\GG_q^h(z, z_h, \omega_J, \mu) =& {\rm Tr}\left[\frac{\slashed{n}_a}{2}\Delta^{h/q}(z, z_h,\omega_J, S_h)\right]\,,\label{eq:gq}\\
\Delta\GG_q^h(z, z_h, \omega_J, \mu) =& {\rm Tr}\left[\frac{\slashed{n}_a}{2}\gamma_5\Delta^{h/q}(z, z_h,\omega_J, S_h)\right]\,,\label{eq:dgq}\\
\Delta_T\GG_q^h(z, z_h, \omega_J, \mu) =& {\rm Tr}\left[\frac{\slashed{n}_a}{2}\gamma_\perp^i\gamma_5\Delta^{h/q}(z, z_h,{\bm j}_\perp, S_h)\right]\,,\label{eq:dtgq}
\end{align}
\endgroup
and the gluon semi-inclusive JFFs are given as \cite{Kang:2016ehg}
\begingroup
\allowdisplaybreaks
\begin{align}
\GG_g^h(z, z_h, \omega_J, \mu) =& - \frac{z\,\omega}{(d-2)(N_c^2-1)} \delta\left(z_h - \frac{\omega_h}{\omega_J}\right) \nnu
&\times \langle 0|  \delta\left(\omega - n_a\cdot {\mathcal P} \right) {\mathcal B}_{n\perp \mu}(0) |(Jh)X\rangle \langle (Jh)X|{\mathcal B}_{n\perp}^\mu(0)  |0\rangle,\\
\Delta\GG_g^h(z, z_h, \omega_J, \mu) = & - \frac{z\,\omega}{(d-2)(d-3)(N_c^2-1)} \delta\left(z_h - \frac{\omega_h}{\omega_J}\right)\nnu
&\times \langle 0| \delta \left(\omega - n_a\cdot \mathcal{P} \right) \mathcal{B}_{n\perp \mu}(0)|(Jh)X\rangle \langle (Jh)X| \mathcal{B}_{n \perp}^\mu(0)  |0\rangle\,,
\end{align}
\endgroup
where $(d-2)$ is the number of polarizations for gluons in $d$ space-time dimensions and $N_c$ is the number of colors for quarks.
As given in \cref{eq:gq,eq:dgq,eq:dtgq}, to obtain the helicity and transversity distributions of hadron in quark JFFs, we replace the ${\slashed{n}_a}$ in unpolarized JFF $\GG_q^h(z, z_h, \omega_J, \mu)$ by ${\slashed{n}_a}\gamma_5$ and ${\slashed{n}_a}\gamma_\perp^i\gamma_5$ respectively. In \cref{t.col_JFF}, we categorize $\GG_q^h$, $\Delta\GG_q^h$ and $\Delta_T\GG_q^h$. Note that we only consider massless quark flavors.

In addition, we would like to point out that the semi-inclusive jet fragmentation functions can also depend on the jet radius $R$, namely in general we have $\mathcal{G}_i^h \pqty{z, z_h, \omega _J, R, \mu}$. However, in the remainder of this thesis, we leave this dependence implicit to shorten our notation.
\begin{table}
\begin{center}
\begin{tabular}{ |c|c|c|c| } 
\hline
$h/q$ & U & L & T \\ 
\hline
U & $\mathcal{G}_q^h$ &  &  \\ 
\hline
L &  &  $\Delta\mathcal{G}_q^h$& \\ 
\hline
T &  &  & $\Delta_T\mathcal{G}_q^h$\\ 
\hline
\end{tabular}
\hspace{1.0 cm}
\begin{tabular}{ |c|c|c|c| } 
\hline
$h/q$ & U & L & T \\ 
\hline
U & $D^{h/q}$ &  &  \\ 
\hline
L &  &  $\Delta D^{h/q}$& \\ 
\hline
T &  &  & $\Delta_T D^{h/q}$\\ 
\hline
\end{tabular}
\end{center}
\caption{Collinear quark semi-inclusive JFFs (left) and collinear FFs (right). The labels ``U", ``L" and ``T" in the header row represent unpolarized, longitudinally polarized and transversely polarized fragmenting quarks. And the labels in the header column represent the corresponding polarizations of the produced hadrons.}\label{t.col_JFF}
\end{table}

\subsubsection{NLO calculation}
\label{sss.collinear_siJFFs_calculation}

Since the semi-inclusive JFFs $\mathcal{G}_i^h \pqty{z, z_h, \omega _J, \mu}$ describe the distribution of hadrons inside the jet, which contains hadronization/non-perturbative information, they are not perturbatively calculable.
In this respect, they are different from the purely perturbative semi-inclusive jet functions introduced in \cite{Kang:2016mcy}.
However, we can still follow the standard perturbative QCD methodology and obtain the renormalization properties by evaluating the partonic jet fragmentation functions.
Hence, we will replace the hadron $h$ by a parton $j$, and compute perturbatively compute $\mathcal{G}_i^j \pqty{z, z_h, \omega _J, \mu}$ as an expansion of the strong coupling constant $\alpha _s$.

Here we outline the calculation of the semi-inclusive JFFs for quark and gluon initiated jets $\mathcal{G}_{q, g}^h \pqty{z, z_h, \omega _J, \mu}$, which have close relations to the conventional collinear FFs. For the unpolarized case, the collinear unpolarized JFFs $\mathcal{G}_{i}^h \pqty{z, z_h, \omega _J, \mu}$ is related to the collinear unpolarized FFs $D^{h/i}(z_h,\mu)$ as follows
\begin{align}
\mathcal{G}_{i}^h \pqty{z, z_h, \omega _J, \mu} = \sum_j\int_{z_h}^1 \frac{dz_h'}{z_h'}\mathcal{J}_{ij}(z,z_h',\omega_J,\mu)\,D^{h/j}\left(\frac{z_h}{z_h'},\mu\right)\,,
\end{align}
where the coefficient functions $\mathcal{J}_{ij}$ can be found in~\cite{Kang:2016ehg}. Note that we have selected $\omega_J=p_{JT} R$ with the jet transverse momentum $p_{JT}$ and the jet radius $R$, which are dependent on the jet kinematics. We have also included the renormalization scale $\mu$ in the matching to collinear FFs. By studying the perturbative behavior of these JFFs, one can derive their renormalization group (RG) equations, which are the same as the usual time-like DGLAP evolution equations, 
\begin{align}
\mu\frac{d}{d\mu}\mathcal{G}_{i}^h \pqty{z, z_h, \omega _J, \mu} = \frac{\alpha_s(\mu)}{\pi} 
 \sum_j \int_z^1  \frac{dz'}{z'}P_{ji}\left(\frac{z}{z'}, \mu \right) \mathcal{G}_{i}^h \pqty{z', z_h, \omega _J, \mu}\,,
\end{align}
where $P_{ji}$ are the splitting functions for unpolarized fragmentation functions~\cite{Altarelli:1977zs,Stratmann:1996hn}. 

Next we turn to the polarized JFFs. The leading order polarized bare semi-inclusive JFF in the $\overline{\textrm{MS}}$ scheme can be written as:
\begin{equation}
\Delta _{(T)} \GG _i^{j, (0)} (z, z_h, \omega _J,\mu) = \delta (1 - z) \delta (1 - z_h) ,
\end{equation}
notice that $z$ is equal to 1 because at LO the total energy of the initiating parton is transferred to the jet, and $z_h$ is equal to one because the fragmenting parton inside the jet carries entire jet energy.

\begin{figure}[htb]
\centering
\includegraphics[width = 0.6 \textwidth]{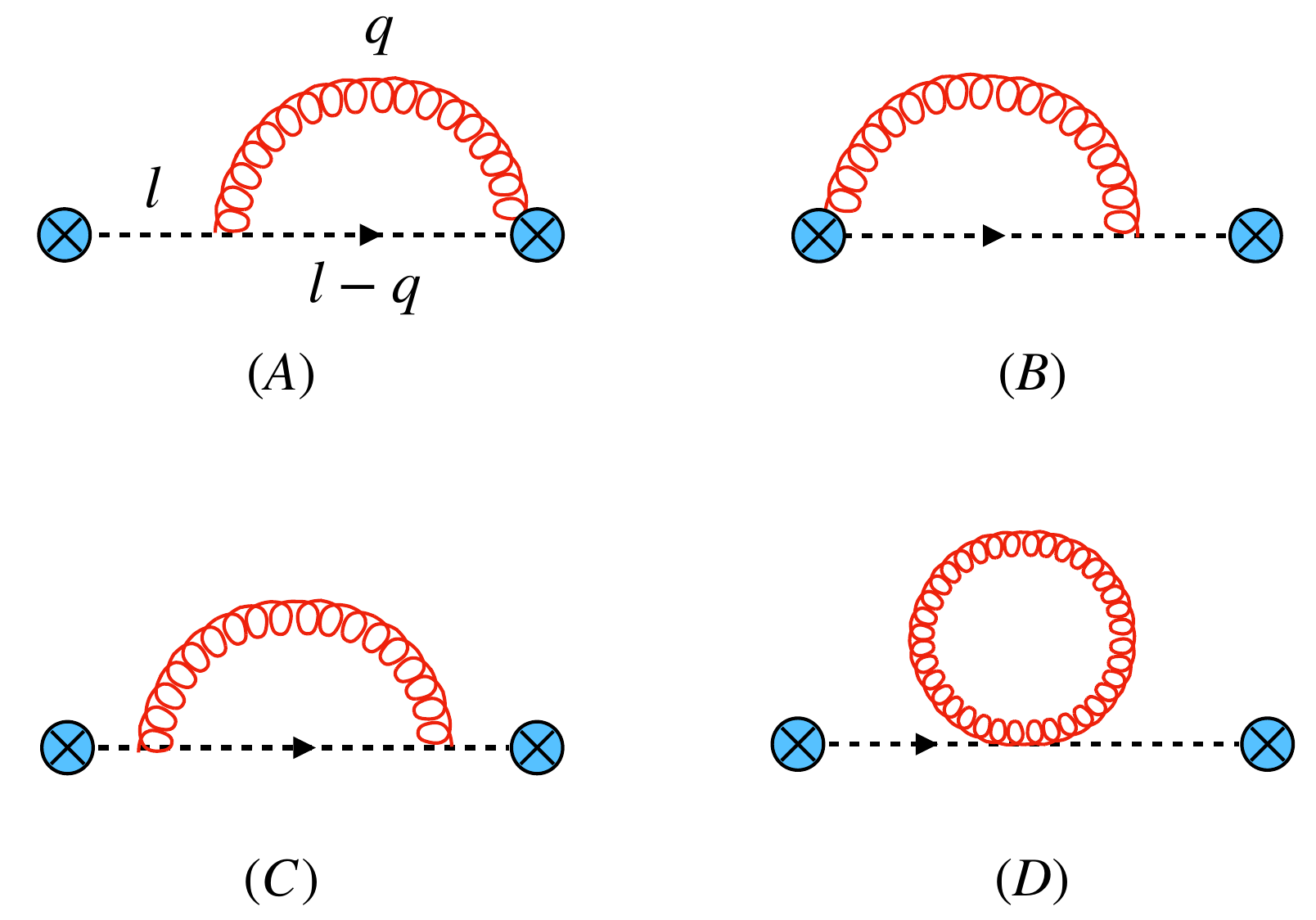}
\caption{
Feynman diagrams that contribute to the polarized semi-inclusive quark JFFs. As illustrated in (A), the quark initiating the jet has momentum $l=(l^+,l^-=\omega,\bm{0}_\perp)$ with quark energy $\omega$ as introduced in~\cref{eq:definez}. The black dashed lines are collinear quarks and red curly lines represent collinear gluons.
}
\label{f.jet_NLO_Feynman}
\end{figure}
At the next-to-leading order (NLO) in SCET, the collinear processes will contribute, and the corresponding Feynman diagrams can be found in \cite{Kang:2016mcy, Kang:2016ehg}. The semi-inclusive jet fragmentation functions (JFFs) are obtained by considering all possible final state cuts of the Feynman diagrams depicted in \cref{f.jet_NLO_Feynman}. To carry out the calculations, we work in pure dimensional regularization with $d = 4 - 2 \epsilon$ dimensions, focusing only on cuts through loops, where there are two final-state partons. The remaining cuts result in virtual contributions, leading to scaleless integrals that vanish in dimensional regularization. These virtual contributions primarily change infrared (IR) poles to ultraviolet (UV) poles, except for the IR poles that will eventually be matched onto the standard collinear fragmentation functions. Thus, in the end, we are left with UV poles only, which will be addressed through renormalization.

Considering the quark semi-inclusive JFF, we have two contributions to consider: \cref{f.jet_NLO_Feynman} (A) and \cref{f.jet_NLO_Feynman} (B)+(C), similar to the semi-inclusive jet function analyzed in a previous work \cite{Jain:2011xz,Chien:2015ctp}. Specifically, we focus on $\Delta _{(T)} \mathcal{G} _q^{q, (1)}$, where the incoming quark has momentum $l^-$, and the final-state quark has momentum $l^--q^-$. Based on these momenta, we define the branching fraction $x=(l^--q^-)/l^-$. At this perturbative order, there are only two possibilities. First, if both the quark and the gluon are inside the jet, as illustrated in \cref{f.jet_NLO} (A), we have $z = 1$ and $z_h = x$. Second, if the gluon exits the jet, as shown in \cref{f.jet_NLO} (B), we have $z = x$ and $z_h = 1$. Extending the discussions from \cite{Kang:2016ehg}, we consider both cases where both partons are in the jet and only one parton is in the jet:
\begin{figure}[htb]
\centering
\includegraphics[width = 0.76 \textwidth]{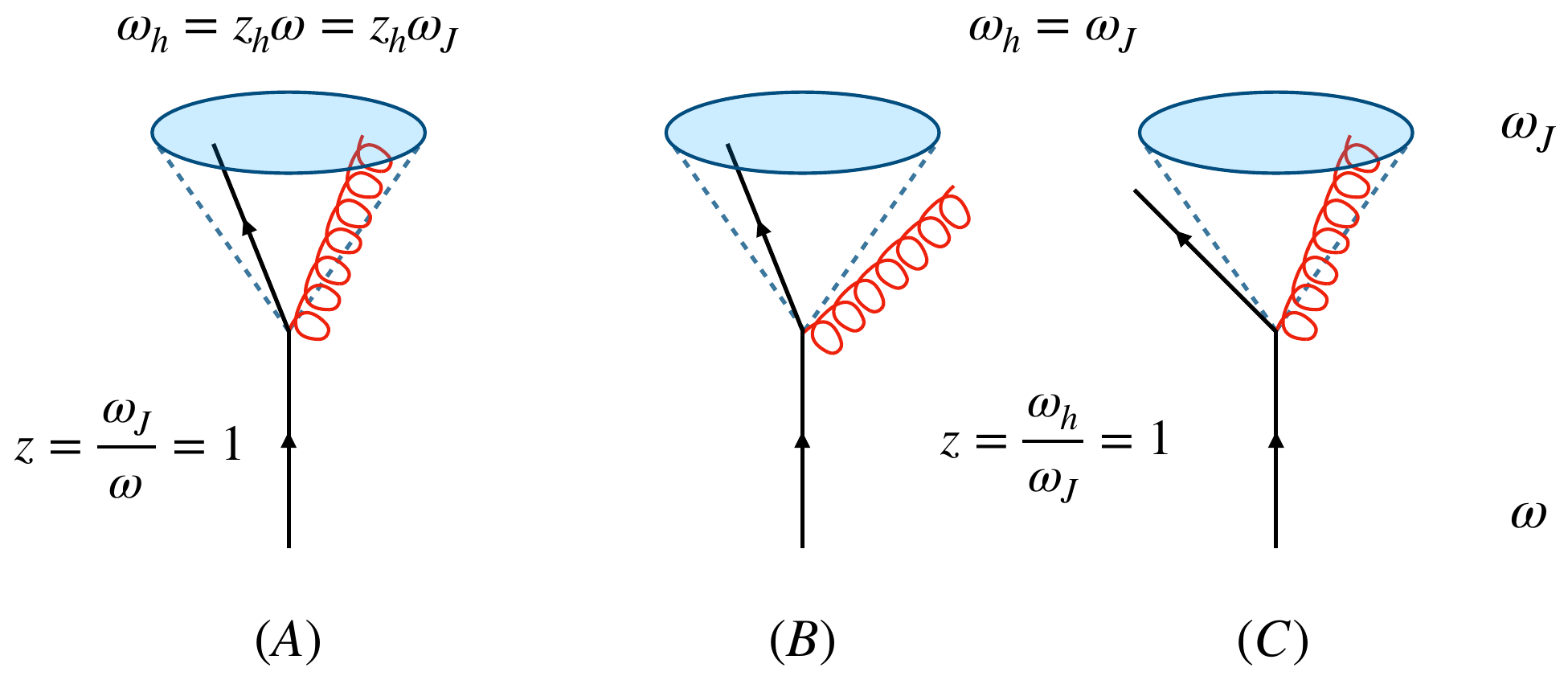}
\caption{
Contributions required for studying semi-inclusive JFFs. In (A), one observes both the quark and the gluon inside the jet. In (B) and (C) only the quark or gluon is inside the jet, respectively.
}
\label{f.jet_NLO}
\end{figure}
\begin{enumerate}
    \item Both partons are inside the jet\\
    The scenario depicted in \cref{f.jet_NLO} (A) corresponds to a quark-initiated jet. In this case, all the initial quark energy $\omega$ is fully transferred to the jet, leading to $z = \omega_J /\omega = 1$. However, the energy of the fragmenting parton, denoted by $\omega_h$, can be less than the jet energy, resulting in a general ratio $z_h = \omega_h/\omega_J < 1$.

For the splitting process $i \rightarrow jk$, where $j$ denotes the fragmenting parton, the one-loop bare semi-inclusive Jet Fragmentation Fragmentation Function (JFF) in the $\overline{MS}$ scheme can be expressed as follows:
\begin{align}
\Delta _{(T)} \GG _{i}^{jk, (1)} (z, z_h, \omega_J,\mu) = & \frac{\alpha_s}{2 \pi} \frac{(e^{\gamma _E} \mu^2)^{\epsilon}}{\Gamma (1-\epsilon)}
\left(\delta(1-z)\Delta_{(T)}\hat{P}_{ji}(z_h,\epsilon)\int\frac{dq_\perp^2}{(q_\perp^2)^{1+\epsilon}}\Theta^{\text{anti}-k_T}_{\textrm{both}}\right)\,.\label{eq:gboth}
\end{align}
The superscript ``$jk$" indicates that this is the contribution proportional to $\mathcal{\alpha_s}$ where both partons $j$ and $k$ remain inside the jet. The term $\Theta^{\text{anti}-k_T}{\textrm{both}}$ represents the constraints of the anti-$k_T$ jet algorithm with both partons remaining inside the jet, and it is expressed using Heaviside functions:
\begin{align}
\Theta ^{\text{anti}-k_T}_{\textrm{both}} & = \theta \pqty{z_h (1 - z_h) \omega _J \tan{\frac{\mathcal{R}}{2}} - q_{\perp}} \,,\quad \mathcal{R} \equiv R / \cosh(\eta)\,.\label{eq:(2.9)}
\end{align}
\item Only one parton is inside the jet\\
The scenario where one parton remains inside the jet while another parton exits the jet is depicted in \cref{f.jet_NLO} (B) and (C) for a quark-initiated jet. In this case, the final-state quark (or gluon) forms the jet with a jet energy $\omega_J = l^- - q^- = zl^-$. This means that only a fraction $z$ of the incoming quark energy $\omega$ is transferred to the jet energy. At this perturbative order, all the jet energy is fully transferred to the fragmenting parton inside the jet, leading to an overall delta function that enforces $z_h = \omega_h/\omega_J = 1$.

It is crucial to distinguish this situation from earlier work~\cite{Kang:2016mcy} that considered the exclusive limit of the JFFs. In that case, an upper cut $\Lambda$ was imposed on the total energy outside the measured jets to ensure an exclusive $n$-jet configuration. However, it was demonstrated explicitly in the context of angularities~\cite{Ellis:2010rwa} that for the exclusive case, this contribution is power suppressed as $\mathcal{O}(\Lambda/Q)$, where $Q$ is the large scale of the process. In the present inclusive cross-section calculation, such constraints are not necessary, as we need to integrate over all momentum configurations similar to the case of fragmentation functions. Consequently, there are no power corrections of the form $\mathcal{O}(\Lambda/Q)$.

The constraints imposed by the jet algorithms require that one of the partons must be outside the jet, which can be formulated in the following manner for both cone and anti-$k_T$ algorithms:
\begin{align}
\Theta ^{\text{anti}-k_T}_{j} & = \theta \pqty{q_{\perp} - (1 - z) \omega _J \tan\frac{\mathcal{R}}{2}}  \,,\quad \mathcal{R} \equiv R / \cosh(\eta)\,.\label{eq:algone} 
\end{align}
Note that this restriction is formulated in terms of the variable $z$, whereas the constraints in \cref{eq:(2.9)} are related to the variable $z_h$. Once again, we consider the splitting process $i \rightarrow jk$, where only parton $j$ remains inside the jet and eventually fragments into the observed hadron.

This part of the bare semi-inclusive jet fragmentation functions can be expressed as follows:
\begin{align}
\Delta _{(T)} \GG _i^{j(k), (1)} (z, z_h, \omega_J,\mu) = &\frac{\alpha_s}{2 \pi} \frac{(e^{\gamma _E} \mu^2)^{\epsilon}}{\Gamma (1-\epsilon)}\left(\delta (1-z_h) \Delta_{(T)} \hat{P}_{ji}(z,\epsilon) \int\frac{dq_\perp^2}{(q_\perp^2)^{1+\epsilon}} \Theta^{\text{anti}-k_T}_{j} \right)\,. \label{eq:gone}
\end{align}
The superscript ``$j(k)$" indicates that parton $k$ exits the jet. The structure of this expression is similar to \cref{eq:gboth}, except for the presence of a different overall delta function and a distinct jet algorithm constraint. $\Theta^{\text{anti}-k_T}_{j}$.
\end{enumerate}
Finally, by combining~\cref{eq:gboth} and~\cref{eq:gone}, one obtains
\begin{align}
\Delta _{(T)} \GG _i^{j, (1)} (z, z_h, \omega_J,\mu) = & \frac{\alpha_s}{2 \pi} \frac{(e^{\gamma _E} \mu^2)^{\epsilon}}{\Gamma (1-\epsilon)}
\bigg(\delta(1-z)\Delta_{(T)}\hat{P}_{ji}(z_h,\epsilon)\int\frac{dq_\perp^2}{(q_\perp^2)^{1+\epsilon}}\Theta^{\text{anti}-k_T}_{\textrm{both}} \nnu
& +
\delta (1-z_h) \Delta_{(T)} \hat{P}_{ji}(z,\epsilon) \int\frac{dq_\perp^2}{(q_\perp^2)^{1+\epsilon}} \Theta^{\text{anti}-k_T}_{j}\bigg)\,, \label{e.bare_collinear_JFF_1}
\end{align}
where $\Theta^{\text{anti}-k_T}_{\textrm{both}}$ and $\Theta^{\text{anti}-k_T}_{j}$ are the anti-$k_T$ jet algorithm constraints with both partons in jet and only one parton in jet as introduced in~\cref{eq:(2.9)} and~\cref{eq:algone}.

The longitudinally polarized splitting functions $\Delta  \hat{P}_{ji} (x , \epsilon)$ in \cref{e.bare_collinear_JFF_1} are given in~\cite{Vogelsang:1996im}:
\begingroup
\allowdisplaybreaks
\begin{align}
\Delta\hat{P}_{qq}(x,\epsilon) & = C_F\left[\frac{1+x^2}{1-x}-\epsilon(1-x)\right]\,,\label{e.Delta_P_qq}\\
\Delta\hat{P}_{gq}(x,\epsilon) & = C_F\left[2-x+2\epsilon(1-x)\right]\,,\label{e.Delta_P_gq}\\
\Delta\hat{P}_{qg}(x,\epsilon) & = T_F\left[2x-1-2\epsilon(1-x)\right]\,,\label{e.Delta_P_qg}\\
\Delta\hat{P}_{gg}(x,\epsilon) & = 2 C_A\left[\frac{1}{1-x}-2x+1+2\epsilon(1-x)\right]\label{e.Delta_P_gg}\,,
\end{align}
\endgroup
and the transversely polarized splitting functions $\Delta _{T} \hat{P}_{ji} (x , \epsilon)$ only exist for $\Delta _{T} \hat{P}_{qq}$ and have been provided in~\cite{Vogelsang:1997ak}:
\begin{align}
\Delta_T\hat{P}_{qq}(x,\epsilon) & = C_F\left(\frac{2x}{1-x}\right)\label{e.Delta_T_P_qq}\,.
\end{align}

After inserting the $\Theta$ functions for anti-$k_T$ algorithm and carrying out the integration in \cref{e.bare_collinear_JFF_1}, one obtains the bare results
for longitudinally polarized semi-inclusive JFFs $\Delta\mathcal{G}^{j}_{j,\text{bare}}(z,z_h,\omega_J,\mu)$ with $i, j\in \Bqty{q, g}$,
\begingroup
\allowdisplaybreaks
\begin{align}
\Delta\mathcal{G}^{q}_{q,\text{bare}}&(z,z_h,\omega_J,\mu)=\,\delta(1-z)\delta(1-z_h)\nonumber\\
&+\frac{\alpha_s}{2\pi}\left[\left(\frac{1}{\epsilon}+L\right)\Delta P_{qq}(z)\delta(1-z_h)-\left(\frac{1}{\epsilon}+L\right)\Delta P_{qq}(z_h)\delta(1-z)\right]\nonumber\\
&+\delta(1-z)\frac{\alpha_s}{2\pi}\left[2C_F(1+z_h^2)\left(\frac{\ln{(1-z_h)}}{1-z_h}\right)_++C_F(1-z_h)+2\Delta P_{qq}(z_h)\ln z_h \right]\nonumber\\
&-\delta(1-z_h)\frac{\alpha_s}{2\pi}\left[2C_F(1+z^2)\left(\frac{\ln{(1-z)}}{1-z}\right)_++C_F(1-z)\right]\,,\label{e.bare_collinear_G_qq}\\
\Delta\mathcal{G}^{g}_{q,\text{bare}}&(z,z_h,\omega_J,\mu)=\,\frac{\alpha_s}{2\pi}\left[\left(\frac{1}{\epsilon}+L\right)\Delta P_{gq}(z)\delta(1-z_h)-\left(\frac{1}{\epsilon}+L\right)\Delta P_{gq}(z_h)\delta(1-z)\right]\nonumber\\
&+\delta(1-z)\frac{\alpha_s}{2\pi}\left[2\Delta P_{gq}(z_h)\ln (z_h(1-z_h))-2C_F(1-z_h) \right]\nonumber\\
&-\delta(1-z_h)\frac{\alpha_s}{2\pi}\left[2\Delta P_{gq}(z)\ln (1-z)-2C_F(1-z)\right]\,,\label{e.bare_collinear_G_qg}\\
\Delta\mathcal{G}^{q}_{g,\text{bare}}&(z,z_h,\omega_J,\mu)=\,\frac{\alpha_s}{2\pi}\left[\left(\frac{1}{\epsilon}+L\right)\Delta P_{qg}(z)\delta(1-z_h)-\left(\frac{1}{\epsilon}+L\right)\Delta P_{qg}(z_h)\delta(1-z)\right]\nonumber\\
&+\delta(1-z)\frac{\alpha_s}{2\pi}\left[2\Delta P_{qg}(z_h)\ln (z_h(1-z_h))+2T_F(1-z_h) \right]\nonumber\\
&-\delta(1-z_h)\frac{\alpha_s}{2\pi}\left[2\Delta P_{qg}(z)\ln (1-z)+2T_F(1-z)\right]\,,\label{e.bare_collinear_G_gq}\\
\Delta\mathcal{G}^{g}_{g,\text{bare}}&(z,z_h,\omega_J,\mu)=\,\delta(1-z)\delta(1-z_h)\nonumber\\
&+\frac{\alpha_s}{2\pi}\left[\left(\frac{1}{\epsilon}+L\right)\Delta P_{gg}(z)\delta(1-z_h)-\left(\frac{1}{\epsilon}+L\right)\Delta P_{gg}(z_h)\delta(1-z)\right]\nonumber\\
&+\delta(1-z)\frac{\alpha_s}{2\pi}\bigg[4C_A \left(2(1-z_h)^2+z_h\right)\left(\frac{\ln (1-z_h)}{1-z_h}\right)_+\bigg]\nonumber\\
&+\delta(1-z)\frac{\alpha_s}{2\pi}\bigg[2\Delta P_{gg}(z_h)\ln z_h-4C_A(1-z_h)\bigg]\nonumber\\
&-\delta(1-z_h)\frac{\alpha_s}{2\pi}\left[4 C_A\left(2(1-z)^2+z\right)\left(\frac{\ln (1-z)}{1-z}\right)_+-4C_A(1-z)\right]\,,\label{e.bare_collinear_G_gg}
\end{align}
\endgroup
where $L \equiv \ln(\frac{\mu ^2}{\omega _J^2 \tan[2](\mathcal{R}/2)})$.
The transversely polarized semi-inclusive JFFs are given by:
\begingroup
\allowdisplaybreaks
\begin{align}
\Delta _T\mathcal{G}^{q}_{q,\text{bare}}&(z,z_h,\omega_J,\mu)=\delta(1-z)\delta(1-z_h)\nonumber\\
&+\frac{\alpha_s}{2\pi}\left[\left(\frac{1}{\epsilon}+L\right)\Delta_T P_{qq}(z)\delta(1-z_h)-\left(\frac{1}{\epsilon}+L\right)\Delta_T P_{qq}(z_h)\delta(1-z)\right]\nonumber\\
&+\delta(1-z)\frac{\alpha_s}{2\pi}\left[4C_Fz_h\left(\frac{\ln{(1-z_h)}}{1-z_h}\right)_++2\Delta_T P_{qq}(z_h)\ln z_h \right]\nonumber\\
&-\delta(1-z_h)\frac{\alpha_s}{2\pi}\left[4C_Fz\left(\frac{\ln{(1-z)}}{1-z}\right)_+\right]\, .\label{e.bare_collinear_G_qq_T}
\end{align}
\endgroup
Here the functions $\Delta_{(T)} P_{ji}(z_h)$ are the longitudinally (transversely) polarized Altarelli-Parisi splitting kernels
\begin{align}
\Delta{P}_{qq}(x)
& =
C_F \bqty{\frac{2}{(1-x)_+}-1-x+\frac{3}{2}\delta(1-x)} , \label{e.polarized_AP_splittings1}\\
\Delta{P}_{gq}(x)
& =
C_F \bqty{2-x} ,\label{e.polarized_AP_splittings2} \\
\Delta{P}_{qg}(x)
& =
T_F \bqty{2x-1} , \label{e.polarized_AP_splittings3}\\
\Delta{P}_{gg}(x)
& =
2 C_A \bqty{\frac{1}{(1-x)_+}-2x+1}
+
\frac{\beta _0}{2} \delta(1-x) , \label{e.polarized_AP_splittings4}\\
\Delta_T{P}_{qq}(x)
& =
C_F \bqty{\frac{2x}{(1-x)_+} + \frac{3}{2} \delta(1-x)} ,\label{e.polarized_AP_splittings5}
\end{align}
where $\beta _0 \equiv \frac{11}{3} C_A - \frac{4}{3} T_F n_f$ and $n_f$ is number of flavors.
The ``plus'' distributions are defined as usual by:
\begin{equation}
\int _0^1 \dd{z} f \pqty{z} \bqty{g \pqty{z}}_+
=
\int _0^1 \dd{z} \pqty{f \pqty{z} - f \pqty{1}} g \pqty{z}.
\end{equation}
Note that there is no gluon involved splitting for $\Delta _T$ since there is no gluonic transversity distribution at leading twist.

It is important to point out that the $1 / \epsilon$ poles with a factor of $\Delta _{\pqty{T}} \pqty{z_h} \delta \pqty{1 - z}$ in \cref{e.bare_collinear_G_qq,e.bare_collinear_G_qg,e.bare_collinear_G_gq,e.bare_collinear_G_gg,e.bare_collinear_G_qq_T} are IR poles that will be matched onto the standard longitudinally (transversely) polarized collinear FFs.
The poles with a factor of $\Delta _{\pqty{T}} \pqty{z} \delta \pqty{1 - z_h}$, on the other hand, are the UV poles which will be taken care of by renormalization.
Since the UV poles do not involve the variable $z_h$, we should expect that $z_h$ is merely a parameter when doing the renormalization.
In matching onto the collinear FFs, however, $z_h$ will become a relevant variable.
Both the renormalization and matching onto collinear FFs will be discussed in \cref{sss.collinear_siJFFs_renormalization}.

\subsubsection{Renormalization and matching onto collinear FFs}
\label{sss.collinear_siJFFs_renormalization}

The subsequent action we will take involves the renormalization of the bare semi-inclusive JFFs obtained previously, followed by matching them onto the renormalized partonic fragmentation functions to address the IR poles.
The relationship between the bare and renormalized semi-inclusive JFFs is as follows:
\begin{equation}
\Delta _{\pqty{T}} \mathcal{G} _{i, \mathrm{bare}}^j \pqty{z, z_h, \omega _J, \mu}
=
\sum _k \int _z^1 \frac{\dd{z'}}{z'}
\Delta _{\pqty{T}} Z_{ik} \pqty{\frac{z}{z'}, \mu}
\Delta _{\pqty{T}} \mathcal{G} _i^j \pqty{z', z_h, \omega _J, \mu} ,
\end{equation}
where $\Delta _{\pqty{T}} Z_{ik} \pqty{\frac{z}{z'}, \mu}$ is the renormalization matrix and $\Delta _{\pqty{T}} \mathcal{G} _i^j \pqty{z', z_h, \omega _J, \mu}$ are the renormalized semi-inclusive JFFs.
As pointed out in \cref{sss.collinear_siJFFs_calculation}, the above convolution only involves the variable $z$.
The renormalized semi-inclusive JFFs satisfy the following RG evolution equations:
\begin{equation}
\mu \frac{\dd{}}{\dd{\mu}}
\Delta _{\pqty{T}} \mathcal{G} _i^j \pqty{z, z_h, \omega _J, \mu}
=
\sum _k \int _z^1 \frac{\dd{z'}}{z'}
\Delta _{\pqty{T}}\gamma _{ik} \pqty{\frac{z}{z'}, \mu}
\Delta _{\pqty{T}} \mathcal{G} _i^j \pqty{z', z_h, \omega _J, \mu} ,
\end{equation}
where the anomalous dimension matrix is given by:
\begin{equation}
\Delta _{\pqty{T}}\gamma _{ik} = - \sum _k \int _z^1 \frac{\dd{z'}}{z'}
\pqty{\Delta _{\pqty{T}} Z}_{ik}^{-1} \pqty{\frac{z}{z'}, \mu}
\mu \frac{\dd{}}{\dd{\mu}} \Delta _{\pqty{T}} Z_{kj} \pqty{z', \mu} ,
\end{equation}
and $\pqty{\Delta _{\pqty{T}} Z}_{ik}^{-1}$ is the inverse of the renormalization matrix that is defined such that it satisfies:
\begin{equation}
\sum _k \int _z^1 \frac{\dd{z'}}{z'}
\pqty{\Delta _{\pqty{T}} Z}_{ik}^{-1} \pqty{\frac{z}{z'}, \mu}
\Delta _{\pqty{T}} Z_{kj} \pqty{z', \mu}
=
\delta _{ij} \delta \pqty{1 - z} .
\end{equation}
Up to $\order{\alpha _s}$, the renormalization matrix is
\begin{equation}
\Delta _{\pqty{T}} Z_{ij} \pqty{z, \mu}
=
\delta _{ij} \delta \pqty{1 - z}
+
\frac{\alpha _s \pqty{\mu}}{2 \pi} \frac{1}{\epsilon}
\Delta _{\pqty{T}} P_{ji} \pqty{z} ,
\end{equation}
and therefore the anomalous dimension matrix is given by:
\begin{equation}
\Delta _{\pqty{T}}\gamma _{ij} \pqty{z, \mu}
=
\frac{\alpha _s \pqty{\mu}}{\pi}
\Delta _{\pqty{T}} P_{ji} \pqty{z} ,
\end{equation}
this suggests that the evolution of the renormalized polarized semi-inclusive JFFs conforms to the timelike DGLAP equation for collinear polarized FFs \cite{Altarelli:1977zs}:
\begin{equation}
\mu \frac{\dd{}}{\dd{\mu}}
\Delta _{\pqty{T}} \mathcal{G} _i^h \pqty{z, z_h, \omega _J, \mu}
=
\frac{\alpha _s \pqty{\mu}}{\pi}
\sum _k \int _z^1 \frac{\dd{z'}}{z'}
\Delta _{\pqty{T}} P_{ji} \pqty{\frac{z}{z'}}
\Delta _{\pqty{T}} \mathcal{G} _i^h \pqty{z', z_h, \omega _J, \mu}\,.
\end{equation}
Notice that the hadronic JFFs have been reinstated, and the leading order splitting kernels are given in \cref{e.polarized_AP_splittings1}-\ref{e.polarized_AP_splittings5}.
An analogous finding was observed in the context of the semi-inclusive jet function in \cite{Kang:2016mcy}.

Now that we have eliminated the UV poles from renormalization, we will still have to deal with the IR poles, which will be addressed by matching onto the collinear polarized FFs.
Such matching can be done at a scale $\mu \gg \Lambda _{\mathrm{QCD}}$ as follows:
\begin{equation}
\Delta _{\pqty{T}} \mathcal{G}_i^h \pqty{z, z_h, \omega _J, \mu}
=
\sum _j \int _{z_h}^1 \frac{\dd{z_h'}}{z_h'}
\Delta_{\pqty{T}} \mathcal{J}_{ij} \pqty{z, z_h', \omega _J, \mu}
\Delta _{\pqty{T}} D^{h/j} \pqty{\frac{z_h}{z_h'}, \mu} ,
\end{equation}
where again, the hadronic semi-inclusive JFFs and collinear FFs are reinstated, and the relation is true up to a power correction of $\order{\Lambda _{\mathrm{QCD}}^2 / \pqty{\omega ^2 \tan[2](\mathcal{R} / 2)}}$ \cite{Jain:2011xz, Chien:2015ctp, Kang:2016ehg}.
It should be noted that in this case, the variable being convolved is $z_h$, while $z$ is merely a parameter.
This process is similar to the approach used for unpolarized JFFs described in \cite{Kang:2016ehg}, except that instead of the unpolarized FFs, in this instance, collinear polarized FFs are applied
\begin{equation}
\Delta _{\pqty{T}} D^{j/i} \pqty{z_h, \mu}
=
\delta _{ij} \delta \pqty{1-z_h}
+
\frac{\alpha _s}{2 \pi}
\Delta _{\pqty{T}} P_{ji} \pqty{z_h} 
\pqty{-\frac{1}{\epsilon}} .
\end{equation}
Finally, the matching coefficients $\Delta _{\pqty{T}} \mathcal{J}_{ij}$ for anti-$k_T$ jet algorithm can be presented as follows
\begingroup
\allowdisplaybreaks
\begin{align}
\Delta\mathcal{J}_{qq}(z,z_h,\omega_J,\mu)&=\delta(1-z)\delta(1-z_h)+\frac{\alpha_s}{2\pi}\bigg\{L\bigg[\Delta P_{qq}(z)\delta(1-z_h)-\Delta P_{qq}(z_h)\delta(1-z)\bigg]\nonumber\\
&+\delta(1-z)\left[2C_F(1+z_h^2)\left(\frac{\ln{(1-z_h)}}{1-z_h}\right)_++C_F(1-z_h)+\Delta\mathcal{I}_{qq}^{\text{anti}-k_T}(z_h) \right]\nonumber\\
&\left.-\delta(1-z_h)\left[2C_F(1+z^2)\left(\frac{\ln{(1-z)}}{1-z}\right)_++C_F(1-z)\right]\right\}\,,\label{eq:jqq}\\
\Delta\mathcal{J}_{qg}(z,z_h,\omega_J,\mu)&=\frac{\alpha_s}{2\pi}\bigg\{L\bigg[\Delta P_{gq}(z)\delta(1-z_h)-\Delta P_{gq}(z_h)\delta(1-z)\bigg]\nonumber\\
&+\delta(1-z)\left[2\Delta P_{gq}(z_h)\ln (1-z_h)-2C_F(1-z_h)+\Delta\mathcal{I}_{gq}^{\text{anti}-k_T}(z_h) \right]\nonumber\\
&-\delta(1-z_h)\bigg[2\Delta P_{gq}(z)\ln (1-z)-2C_F(1-z)\bigg]\bigg\}\\
\Delta\mathcal{J}_{gq}(z,z_h,\omega_J,\mu)&=\frac{\alpha_s}{2\pi}\bigg\{L\left[\Delta P_{qg}(z)\delta(1-z_h)-\Delta P_{qg}(z_h)\delta(1-z)\right]\nonumber\\
&+\delta(1-z)\left[2\Delta P_{qg}(z_h)\ln(1-z_h)+2T_F(1-z_h)+\Delta\mathcal{I}_{qg}^{\text{anti}-k_T}(z_h) \right]\nonumber\\
&-\delta(1-z_h)\bigg[2\Delta P_{qg}(z)\ln (1-z)+2T_F(1-z)\bigg]\bigg\}\,,\\
\Delta\mathcal{J}_{gg}(z,z_h,\omega_J,\mu)&=\delta(1-z)\delta(1-z_h)+\frac{\alpha_s}{2\pi}\bigg\{L\bigg[\Delta P_{gg}(z)\delta(1-z_h)-\Delta P_{gg}(z_h)\delta(1-z)\bigg]\nonumber\\
&+\delta(1-z)\bigg[4 C_A\left(2(1-z_h)^2+z_h\right)\left(\frac{\ln (1-z_h)}{1-z_h}\right)_+\nnu
&\hspace{2cm} -4C_A(1-z_h)+\Delta\mathcal{I}_{gg}^{\text{anti}-k_T}(z_h)\bigg]\nonumber\\
&-\delta(1-z_h)\left[4 C_A\left(2(1-z)^2+z\right) \left(\frac{\ln (1-z)}{1-z}\right)_+-4C_A(1-z)\right]\bigg\}\,,\\
\Delta_T\mathcal{J}_{qq}(z,z_h,\omega_J,\mu)&=\delta(1-z)\delta(1-z_h)+\frac{\alpha_s}{2\pi}\bigg\{L\bigg[\Delta_T P_{qq}(z)\delta(1-z_h)-\Delta_T P_{qq}(z_h)\delta(1-z)\bigg]\nonumber\\
&+\delta(1-z)\left[4C_Fz_h\left(\frac{\ln{(1-z_h)}}{1-z_h}\right)_++\Delta_T\mathcal{I}_{qq}^{\text{anti}-k_T}(z_h) \right]\nonumber\\
&-\delta(1-z_h)\left[4C_Fz\left(\frac{\ln{(1-z)}}{1-z}\right)_+\right]\bigg\}\,,\label{eq:jgg}
\end{align}
\endgroup
where
\begin{equation}
\begin{split}
\Delta \mathcal{I}_{ij}^{\mathrm{anti}-k_T} \pqty{z_h}
& =
2 \Delta P_{ji} \pqty{z_h} \ln{z_h} , \\
\Delta _T \mathcal{I}_{ij}^{\mathrm{anti}-k_T} \pqty{z_h}
& =
2 \Delta _T P_{ji} \pqty{z_h} \ln{z_h} . \\
\end{split}
\end{equation}


\subsection{Collinear JFFs in exclusive jet productions}
Exclusive jet production, such as proton-proton collisions leading to dijet events or electron-proton collisions producing electron-jet events, provides a valuable tool for understanding the fundamental dynamics of hadron structure and interactions. A key observable in these processes is the collinear jet fragmentation function, which describes the probability distribution for a parton in the jet to fragment into a particular hadron with a given momentum fraction.

The collinear jet fragmentation function plays a critical role in determining the properties of jets produced in these exclusive QCD processes, as it governs the hadronization of partons within the jet. This function is defined as the ratio of the differential cross-section for hadron-in-jet production to the differential cross-section for inclusive jet production. It depends on several variables, including the momentum fraction of the parton within the jet, the transverse momentum of the jet, and the factorization scale used to separate the hard and soft contributions to the jet production process.

Recent advances in perturbative QCD calculations and experimental measurements have led to a deeper understanding of the collinear jet fragmentation function and its role in exclusive QCD processes. In particular, studies have focused on the evolution of this function under renormalization group equations, which describe how it varies with changes in the factorization scale. Additionally, there has been increasing interest in the non-perturbative contributions to the collinear jet fragmentation function, which can be probed through experimental measurements of jet fragmentation functions.

In this section, we provide the derivation for the matching coefficients $\Delta_{(T)}\mathscr{J}_{i j}$ of the exclusive jet fragmentation functions $\Delta_{(T)}\mathscr{G}_i^h(\omega, R, z, \mu)$ with the collinear fragmentation function $\Delta_{(T)}D_i^h(z, \mu)$ for anti-${k}_{T}$ jets.
These results were first written down in the appendix of \cite{Waalewijn:2012sv}, with which our results are consistent.
We start by specifying the phase space constraint from the jet algorithm, which was nicely outlined in \cite{Ellis:2010rwa}.
Consider a parton splitting process, $i(\ell) \rightarrow j(q) + k(\ell-q)$, where an incoming parton $i$ with momentum $\ell$ splits into a parton $j$ with momentum $q$ and a parton $k$ with momentum $\ell-q$.
The four-vector $\ell^{\mu}$ can be decomposed in light-cone coordinates as $\ell^{\mu} = \pqty{\ell^+, \ell^- = \omega, 0_{\perp}}$ where $\ell^{\pm} = \ell^0 \mp \ell^z$.
The constraint for anti-$k_T$ algorithm with radius $R$ is given by:
\begin{align}
\Theta_{\text{anti-}k_T}
& =
\theta \pqty{\tan[2](\frac{R}{2}) - \frac{q^+ \omega^2}{q^- \pqty{\omega - \ell^-}^2}}\, .
\end{align}
For jet fragmentation functions, the above constraint lead to constraint on the jet invariant mass $m_J^2 = \omega \ell^{+}$ \cite{Procura:2011aq}, which is derived and listed as follows:
\begin{align}
\delta_{\text{anti-}k_T}
& =
\theta \pqty{z(1-z) \omega ^2 \tan[2](\frac{R}{2}) - m_J^2}
\theta \pqty{m_J^2} \,,
\end{align}
where $z = q^{-} / \omega$.
The exclusive JFF $\Delta_{(T)}\mathscr{G}_i^h(\omega, R, z, \mu)$ can be matched onto the fragmentation functions $\Delta_{(T)}D_i^h(z, \mu)$ as:
\begin{equation} \label{e.collinear_exclusive_JFF_matching}
\Delta _{(T)} \mathscr{G}_i^h (\omega, R, z, \mu)
=
\sum_j \int _z^1 \frac{\dd{x}}{x}
\Delta _{(T)} \mathscr{J}_{ij} (\omega, R, x, \mu)
\Delta _{(T)} D_j^h \pqty{\frac{z}{x}, \mu}
+
\order{\frac{\Lambda _{\mathrm{QCD}}^2}{\omega ^2 \tan[2](R / 2)}} \,,
\end{equation}
where $\Delta _{(T)} \mathscr{J}_{ij}$ are the matching coefficients.
The exclusive JFFs $\Delta _{(T)} \mathscr{G}_i^j \pqty{m_J^2, z, \mu}$ with $i, j \in \Bqty{q, g}$ has been extensively studied in \cite{Jain:2011xz, Ritzmann:2014mka}.
Using pure dimensional regularization with $4 - 2 \epsilon$ dimensions in the $\overline{\mathrm{MS}}$ scheme, the bare results at $\order{\alpha _s}$ can be written in the following compact form \cite{ Giele:1991vf,Ritzmann:2014mka,Chien:2015ctp}:
\begin{equation}
\Delta _{(T)} \mathscr{G}_{i, \text{bare}}^j \pqty{m_J^2, z}
=
\frac{\alpha _s}{2 \pi} \frac{\pqty{e^{\gamma _E} \mu^2}^\epsilon}{\Gamma (1 - \epsilon)}
\Delta _{(T)} \hat{P}_{ji}(z, \epsilon)
z^{-\epsilon} (1-z)^{-\epsilon} \pqty{m_J^2}^{-1-\epsilon}\, ,
\end{equation}
which is related to $\Delta _{(T)}\mathscr{G}_i^h (\omega, R, z, \mu)$ in \cref{e.collinear_exclusive_JFF_matching} by:
\begin{equation}
\Delta _{(T)} \mathscr{G}_i^h (\omega, R, z, \mu)
=
\int \dd{m_J^2} \Delta _{(T)} \mathscr{G}_i^h \pqty{m_J^2, z, \mu} \delta _{{\text{anti-}k_T}} \,,
\end{equation}
notice that we reinstated hadronic JFFs.
The splitting functions $\Delta _{(T)} \hat{P}_{ji} (z, \epsilon)$ are given in \cref{e.Delta_P_qq,e.Delta_P_gq,e.Delta_P_qg,e.Delta_P_gg,e.Delta_T_P_qq}.
By inserting them into \cref{e.collinear_exclusive_JFF_matching} and performing the integration over $m_J^2$ with the constraints imposed by the jet algorithm $\delta_{\text{anti-}k_T}$, one obtains the bare exclusive JFFs $\Delta_{(T)}\mathscr{G}_{i, \text { bare }}^j(\omega, R, z)$.
We present the results for anti-$k_T$ jets here, as their explicit expressions are not available in the literature:
\begingroup
\allowdisplaybreaks
\begin{align}
\Delta \mathscr{G}_{q, \text{bare}}^q (\omega, R, z)
& =
\frac{\alpha _s C_F}{2 \pi} \bigg\{\pqty{\frac{1}{\epsilon ^2} + \frac{3}{2 \epsilon} + \frac{L}{\epsilon}} \delta (1-z) - \frac{1}{\epsilon} \bqty{\Delta P_{qq} (z) + \frac{3}{2} \delta (1-z)} \nnu
& \quad + \delta (1-z) \pqty{\frac{L^2}{2}-\frac{\pi ^2}{12}} + \Delta P_{qq} (z) (-L + 2 \ln{z}) \nnu
& \quad + (1-z) + 2 \pqty{1+z^2} \pqty{\frac{\ln (1-z)}{1-z}}_+ \bigg\} \,, \label{e.bare_exclusive_JFFs_qq} \\
\Delta \mathscr{G}_{q, \text{bare}}^g (\omega, R, z)
& =
\frac{\alpha _s C_F}{2 \pi} \bigg\{\pqty{-\frac{1}{\epsilon}} \Delta P_{gq}(z) + \Delta P_{gq}(z) \Big[-L+2 \ln(z(1-z))\Big] - 2 (1-z) \bigg\}\, , \label{e.bare_exclusive_JFFs_qg} \\
\Delta \mathscr{G}_{g, \text{bare}}^q (\omega, R, z)
& =
\frac{\alpha _s T_F}{2 \pi} \bigg\{\pqty{-\frac{1}{\epsilon}} \Delta P_{qg}(z)+\Delta P_{qg}(z) \Big[-L+2\ln(z(1-z))\Big]+2(1-z) \bigg\}\, , \label{e.bare_exclusive_JFFs_gq} \\
\Delta \mathscr{G}_{g, \text{bare}}^g(\omega, R, z)
& =
\frac{\alpha _s C_A}{2 \pi} \bigg\{\pqty{\frac{1}{\epsilon ^2}+\frac{1}{\epsilon} \frac{\beta _0}{2 C_A}+\frac{L}{\epsilon}} \delta(1-z) + \pqty{-\frac{1}{\epsilon}} \bqty{\Delta P_{gg}(z)+\frac{\beta _0}{2 C_A} \delta (1-z)} \nnu
& \quad + \delta (1-z) \pqty{\frac{L^2}{2}-\frac{\pi^2}{12}} + \Delta P_{gg}(z)(-L+2\ln{z})-4(1-z) \nnu
& \quad + 4 \Big[2(1-z)^2+z\Big] \pqty{\frac{\ln (1-z)}{1-z}}_+ \bigg\} \,, \label{e.bare_exclusive_JFFs_gg} \\
\Delta _T \mathscr{G}_{q, \text{bare}}^q (\omega, R, z)
& =
\frac{\alpha _s C_F}{2 \pi} \bigg\{\pqty{\frac{1}{\epsilon ^2}+\frac{3}{2 \epsilon }+\frac{L}{\epsilon}} \delta (1-z) + \pqty{-\frac{1}{\epsilon}} \bqty{\Delta _T P_{qq}(z)+\frac{3}{2} \delta (1-z)} \nnu
& \quad + \delta (1-z) \pqty{\frac{L^2}{2}-\frac{\pi ^2}{12}} + \Delta _T P_{qq}(z)(-L+2\ln{z}) + 4 z \pqty{\frac{\ln (1-z)}{1-z}}_+ \bigg\} \,, \label{e.bare_exclusive_JFFs_qq_T}
\end{align}
where, as given in the main text, one has $\beta_0$ and $L$ defined:
\begin{equation}
\beta _0 \equiv \frac{11}{3} C_A - \frac{4}{3} T_F n_f,
\quad
L = \ln(\frac{\mu ^2}{\omega ^2 \tan[2](R/2)}) \,,
\end{equation}
\endgroup
and $\Delta_{(T)}{P}_{j i}(z)$ are given in \cref{e.polarized_AP_splittings1}-\cref{e.polarized_AP_splittings5}.
It is instructive to point out that the $\epsilon$ poles in the first term of \cref{e.bare_exclusive_JFFs_qq,e.bare_exclusive_JFFs_gg,e.bare_exclusive_JFFs_qq_T} correspond to ultraviolet (UV) divergences, and they are related to the renormalization of the JFF $\Delta_{(T)}\mathscr{G}_{i, \text{ bare}}^j (\omega, R, z)$.
All the remaining $\epsilon$ poles in \cref{e.bare_exclusive_JFFs_qq,e.bare_exclusive_JFFs_qg,e.bare_exclusive_JFFs_gq,e.bare_exclusive_JFFs_gg,e.bare_exclusive_JFFs_qq_T} are infrared (IR) poles, and they match exactly with those in the fragmentation functions $\Delta_{(T)}D_i^j(z, \mu)$, which we will show below.
$\Delta_{(T)}\mathscr{G}_{i, \text{bare}}^h (\omega, R, z)$ is renormalized by:
\begin{equation}
\Delta _{(T)} \mathscr{G}_{i, \text{bare}}^h (\omega, R, z)
=
\mathcal{Z}_{\mathscr{G}}^i(\mu)
\Delta _{(T)} \mathscr{G}_i^h (\omega, R, z, \mu)\,,
\end{equation}
where $i$ is not summed over in the above equation.
The corresponding renormalization group (RG) equation is given by:
\begin{equation}
\mu \frac{\dd{}}{\dd{\mu}} \Delta _{(T)} \mathscr{G}_i^h (\omega, R, z, \mu)
=
\gamma _{\mathscr{G}}^i(\mu) \Delta _{(T)} \mathscr{G}_i^h (\omega, R, z, \mu)\,,
\end{equation}
where the anomalous dimension $\gamma_{\mathscr{G}}^i(\mu)$ is:
\begin{equation} \label{e.exclusive_JFFs_gamma}
\gamma _{\mathscr{G}}^i(\mu)
=
- \pqty{\mathcal{Z}_{\mathscr{G}}^i(\mu)}^{-1}
\mu \frac{\dd{}}{\dd{\mu}} Z_{\mathscr{G}}^i(\mu)\, .
\end{equation}
The solution to \cref{e.exclusive_JFFs_gamma} is then:
\begin{equation}
\Delta _{(T)}\mathscr{G}_i^h (\omega, R, z, \mu)
=
\Delta _{(T)}\mathscr{G}_i^h \pqty{\omega, R, z, \mu_{\mathscr{G}}}
\exp(\int _{\mu_{\mathscr{G}}}^{\mu} \frac{\dd{\mu '}}{\mu '} \gamma _{\mathscr{G}}^i \pqty{\mu '}) \,,
\end{equation}
where the scale $\mu_{\mathscr{G}}$ should be the characteristic scale chosen such that large logarithms in the fixed-order calculation vanish.
The counter terms $\mathcal{Z}_{\mathscr{G}}^i(\mu)$ are given by
\footnote{Note here the counter terms for polarized quark and gluon JFFs are the same as those of the unpolarized ones as shown in \cite{Chien:2015ctp}.}
\begin{align}
\mathcal{Z}_{\mathscr{G}}^q (\mu)
& =
1 + \frac{\alpha _s}{2 \pi} C_F \bqty{\frac{1}{\epsilon^2}+\frac{3}{2 \epsilon}+\frac{L}{\epsilon}} , \label{e.exclusive_JFFs_counter_q} \\
\mathcal{Z}_{\mathscr{G}}^g (\mu)
& =
1 + \frac{\alpha _s}{2 \pi} C_A \bqty{\frac{1}{\epsilon^2} + \frac{1}{\epsilon} \frac{\beta _0}{2 C_A} + \frac{L}{\epsilon}}\, . \label{e.exclusive_JFFs_counter_g}
\end{align}
From these results we obtain the anomalous dimension $\gamma _{\mathscr{G}}^i (\mu)$ with the following form:
\begin{align}
\gamma _{\mathscr{G}}^i(\mu)
=
\Gamma _{\text{cusp}}^i \pqty{\alpha _s} \ln(\frac{\mu^2}{\omega^2 \tan[2](R/2)})
+
\gamma ^i \pqty{\alpha _s}\,,
\end{align}
where $\Gamma _{\text{cusp}}^i = \sum _n \Gamma _{n-1}^i \pqty{\frac{\alpha _s}{4 \pi}}^n$ and $\gamma ^i = \sum _n \gamma _{n-1}^i \pqty{\frac{\alpha _s}{4 \pi}}^n$.
The lowest-order coefficients can be extracted from the above calculations:
\begin{align}
\Gamma _0^q = 4 C_F, & \quad \gamma _0^q = 6 C_F , \\
\Gamma _0^g = 4 C_A, & \quad \gamma _0^g = 2 \beta _0 \,,
\end{align}
and higher-order results can be found in \cite{Jain:2011xz,Becher:2006mr,Becher:2009th,Echevarria:2012pw,Moch:2004pa}.
After the subtraction of the UV counter terms specified in \cref{e.exclusive_JFFs_counter_q,e.exclusive_JFFs_counter_g}, the renormalized JFF $\Delta _{(T)}\mathscr{G}_i^j (\omega, R, z, \mu)$ are given by:
\begingroup
\allowdisplaybreaks
\begin{align}
\Delta \mathscr{G}_q^q (\omega, R, z, \mu)
& =
\frac{\alpha _s C_F}{2 \pi} \bigg\{\pqty{-\frac{1}{\epsilon}} \bqty{\Delta P_{qq}(z)+\frac{3}{2} \delta (1-z)} + \delta (1-z) \pqty{\frac{L^2}{2}-\frac{\pi ^2}{12}} \nnu
& \quad + \Delta P_{qq}(z)(-L+2\ln{z}) + (1-z) + 2 \pqty{1+z^2} \pqty{\frac{\ln(1-z)}{1-z}}_+ \bigg\} \,, \\
\Delta \mathscr{G}_{q}^g (\omega, R, z, \mu)
& =
\frac{\alpha _s C_F}{2 \pi} \bigg\{\pqty{-\frac{1}{\epsilon}} \Delta P_{gq}(z)+\Delta P_{gq}(z) \Big[-L+2\ln(z(1-z))\Big]- 2(1-z) \bigg\} \,, \\
\Delta \mathscr{G}_g^q (\omega, R, z, \mu)
& =
\frac{\alpha _s T_F}{2 \pi} \bigg\{\pqty{-\frac{1}{\epsilon}} \Delta{P}_{q g}(z)+\Delta P_{qg}(z)\Big[-L+2\ln(z(1-z))\Big]+2(1-z) \bigg\} \,, \\
\Delta \mathscr{G}_g^g (\omega, R, z, \mu)
& =
\frac{\alpha _s C_A}{2 \pi} \bigg\{\pqty{-\frac{1}{\epsilon}} \bqty{\Delta P_{gg}(z)+\frac{\beta _0}{2 C_A} \delta (1-z)} \nnu
& \quad + \delta (1-z) \pqty{\frac{L^2}{2}-\frac{\pi^2}{12}}+\Delta P_{gg}(z)(-L+2\ln{z})-4(1-z) \nnu
& \quad + 4\Big[2(1-z)^2+z\Big] \pqty{\frac{\ln (1-z)}{1-z}}_+ \bigg\} \,, \\
\Delta _T \mathscr{G}_q^q (\omega, R, z, \mu)
& =
\frac{\alpha _s C_F}{2 \pi} \bigg\{\pqty{-\frac{1}{\epsilon}} \bqty{\Delta _T P_{qq}(z)+\frac{3}{2} \delta (1-z)} + \delta (1-z) \pqty{\frac{L^2}{2}-\frac{\pi^2}{12}} \nnu
& \quad + \Delta _T P_{qq}(z)(-L+2\ln{z})+4z \pqty{\frac{\ln (1-z)}{1-z}}_+ \bigg\}\, ,
\end{align}
\endgroup
where we can eliminate all large logarithms $L$ by choosing $\mu = \omega \tan(R / 2)$.
At the intermediate scale $\mu_{\mathscr{G}} \gg \Lambda_{\mathrm{QCD}}$, one can match the JFF $\Delta_{(T)}\mathscr{G}_i^h(\omega, R, z, \mu)$ onto the longitudinally (transversely) polarized fragmentation functions $\Delta_{(T)}D_j^h(x, \mu)$ as in \cref{e.collinear_exclusive_JFF_matching}.
In order to perform the matching calculation and determine the coefficients $\mathscr{J}_{i j}$, we simply need the perturbative results of the fragmentation functions $\Delta_{(T)} D_i^j(x, \mu)$ for a parton $i$ fragmenting into a parton $j$.
The renormalized $\Delta_{(T)}D_i^j(x, \mu)$ at $\order{\alpha _s}$ using pure dimensional regularization are given by:
\begingroup
\allowdisplaybreaks
\begin{align}
\Delta D_q^q (x, \mu)
& =
\delta (1-x) + \frac{\alpha _s C_F}{2 \pi} \pqty{-\frac{1}{\epsilon}} \bqty{\Delta P_{qq}(x)+\frac{3}{2} \delta (1-x)}\, , \\
\Delta D_q^g (x, \mu)
& =
\frac{\alpha _s C_F}{2 \pi} \pqty{-\frac{1}{\epsilon}} \Delta P_{gq}(x)\, , \\
\Delta D_g^q (x, \mu)
& =
\frac{\alpha _s T_F}{2 \pi} \pqty{-\frac{1}{\epsilon}} \Delta P_{qg}(x) \,, \\
\Delta D_g^g (x, \mu)
& =
\delta (1-x) + \frac{\alpha _s C_A}{2 \pi} \pqty{-\frac{1}{\epsilon}} \bqty{\Delta P_{gg}(x) + \frac{\beta _0}{2 C_A} \delta (1-x)} \,, \\
\Delta _T D_q^q (x, \mu)
& =
\delta (1-x) + \frac{\alpha _s C_F}{2 \pi} \pqty{-\frac{1}{\epsilon}} \bqty{\Delta _T P_{qq}(x) + \frac{3}{2} \delta (1-x)} \,.
\end{align}
\endgroup
Using the results for $\Delta _{(T)} \mathscr{G}_i^j (\omega, R, z, \mu)$ and $\Delta _{(T)} D_i^j (x, \mu)$, we obtain the following matching coefficients:
\begingroup
\allowdisplaybreaks
\begin{align}
\Delta \mathscr{J}_{q q} (\omega, R, z, \mu)
& =
\delta (1-z) + \frac{\alpha _s C_F}{2 \pi} \bigg[\delta (1-z) \pqty{\frac{L^2}{2}-\frac{\pi^2}{12}} - \Delta P_{qq}(z) L \nonumber \\
& \hspace{3.3 cm} + (1-z) + \Delta \hat{\mathscr{J}}_{qq}^{\text{anti-}k_T} \bigg]\, , \\
\Delta \mathscr{J}_{qg} (\omega, R, z, \mu)
& =
\frac{\alpha _s C_F}{2\pi} \bigg[-\Delta P_{gq}(z) L- 2(1-z) + \Delta \hat{\mathscr{J}}_{qg}^{\text{anti-}k_T}(z) \bigg] \,,\\
\Delta \mathscr{J}_{gq} (\omega, R, z, \mu)
& =
\frac{\alpha _s T_F}{2 \pi} \bigg[-\Delta P_{qg}(z) L + 2(1-z) + \Delta \hat{\mathscr{J}}_{gq}^{\text{anti-}k_T}(z)\bigg] \,, \\
\Delta \mathscr{J}_{gg} (\omega, R, z, \mu)
& =
\delta (1-z) + \frac{\alpha _s C_A}{2 \pi} \bigg[\delta (1-z) \pqty{\frac{L^2}{2}-\frac{\pi^2}{12}} - \Delta P_{gg}(z) L \nonumber \\
& \hspace{3.3 cm} - 4(1-z)+\Delta\hat{\mathscr{J}}_{g g}^{\text{anti-}k_T}(z)\bigg]\,,\\
\Delta _T \mathscr{J}_{qq} (\omega, R, z, \mu)
& =
\delta (1-z) + \frac{\alpha _s C_F}{2 \pi} \bigg[\delta (1-z) \pqty{\frac{L^2}{2}-\frac{\pi^2}{12}} - \Delta _T P_{qq}(z) L + \Delta _T \hat{\mathscr{J}}_q^{\text{anti-}k_T} \bigg] \,,
\end{align}
\endgroup
where $\Delta _{(T)} \hat{\mathscr{J}}_{i j}^{{\text{anti-}k_T}}(z)$ are jet-algorithm dependent.
For anti-$k_T$ jets, we have:
\begingroup
\allowdisplaybreaks
\begin{align}
\Delta \hat{\mathscr{J}}_{qq}^{\text{anti}-k_T}
& =
2 \Delta P_{qq}(z) \ln{z} + 2 \pqty{1+z^2} \pqty{\frac{\ln(1-z)}{1-z}}_+ \,, \\
\Delta \hat{\mathscr{J}}_{qg}^{\text{anti}-k_T}
& =
2 \Delta P_{gq}(z) \Big[\ln(z(1-z)) \Big] \,, \\
\Delta \hat{\mathscr{J}}_{gq}^{\text{anti}-k_T}
& =
2 \Delta P_{qg}(z) \Big[\ln(z(1-z)) \Big] \,, \\
\Delta \hat{\mathscr{J}}_{gg}^{\text{anti}-k_T}
& =
2 \Delta P_{gg}(z) \ln{z} + 4 \Big[2(1-z)^2 + z\Big] \pqty{\frac{\ln(1-z)}{1-z}}_+ \,, \\
\Delta _T \hat{\mathscr{J}}_{qq}^{\text{anti}-k_T}
& =
2 \Delta _T P_{qq}(z) \ln{z} + 4 z \pqty{\frac{\ln (1-z)}{1-z}}_+ \,,
\end{align}
\endgroup
where the functions $\Delta_{(T)}\hat{P}_{ji}$ have the following expressions
\begingroup
\allowdisplaybreaks
\begin{align}
\Delta\hat{P}_{qq}(z)&=\left[\frac{2}{(1-z)_+}-1-z\right]\,,\label{eq:dphqq}\\
\Delta\hat{P}_{gq}(z)&=\left[2-z\right]\,,\\
\Delta\hat{P}_{qg}(z)&=\left[2z-1\right]\,,\\
\Delta\hat{P}_{gg}(z)&=2\left[\frac{1}{(1-z)_+}-2z+1\right]\,,\\
\Delta_T\hat{P}_{qq}(z)&=\left[\frac{2z}{(1-z)_+}\right]\,,\label{eq:dphqqt}
\end{align}
\endgroup
The jet fragmentation function $\Delta_{(T)}\mathscr{G}_i^h(\omega, R, z, \mu)$ satisfies the following RG equation
\begin{align}
\mu \frac{\dd{}}{\dd{\mu}} \Delta_{(T)}\mathscr{G}_i^h(\omega, R, z, \mu)=\gamma_{\mathscr{G}}^i(\mu) \Delta_{(T)}\mathscr{G}_i^h(\omega, R, z, \mu)\,,
\end{align}
where the anomalous dimension $\gamma_{\mathscr{G}}^i(\mu)=\gamma_{J}^i(\mu)$ is the same as that of the unmeasured jet
function $J_i((\omega, R, \mu)$~\cite{Jain:2011xz,Waalewijn:2012sv,Ellis:2010rwa}. The solution to the RG equation is then
\begin{align}
\Delta_{(T)}\mathscr{G}_i^h(\omega, R, z, \mu)=\Delta_{(T)}\mathscr{G}_i^h\left(\omega, R, z, \mu_{\mathscr{G}}\right) \exp \left[\int_{\mu_{\mathscr{G}}}^\mu \frac{\dd{\mu^{\prime}}}{\mu^{\prime}} \gamma_{\mathscr{G}}^i\left(\mu^{\prime}\right)\right],\label{eq:exclu-rg}
\end{align}
where the scale $\mu_{\mathscr{G}}$ should be the characteristic scale that eliminates the large logarithms in the fixed-order perturbative calculations. In the large $z$ region, the scale choice $\mu_{\mathscr{G}}=$ $\omega \tan (R / 2)(1-z) \equiv p_{T R Z}$ resums~\cite{Waalewijn:2012sv} both $\ln R$ and $\ln (1-z)$. However, for consistency, this would require extracted fragmentation functions $\Delta_{(T)}D^{h/j}$ with a built-in resummation of logarithms in $(1-z)$, which is currently not available. It might be instructive to point out that with such a scale, the power corrections in \cref{e.collinear_exclusive_JFF_matching} will be of the order of $\Lambda_{\mathrm{QCD}}^2 /\left[\omega^2 \tan ^2(R / 2)(1-z)^2\right]$, similar to the usual threshold resummation, see, e.g. Ref.~\cite{Becher:2006mr}. For the numerical calculations presented in the next section, we will choose $\mu_{\mathscr{G}}=\omega \tan (R / 2)$ to resum $\ln R$ and comment on the effect of $\ln (1-z)$ resummation.

\section{Polarized transverse momentum dependent JFFs (TMDJFFs)}\label{sub:connection}

This section outlines the definition of TMD semi-inclusive jet fragmentation functions in SCET, which describe the distribution of hadron transverse momentum within a jet.
Both unpolarized and polarized fragmenting quarks are considered.
The operator definitions of these functions in SCET are presented, followed by their factorization formalism, which involves hard functions, soft functions, and TMDFFs.
The calculation of these functions is carried out to NLO at the parton level, and their RG evolution equations are derived and solved.

\subsection{TMD JFFs in semi-inclusive jet productions}
\label{ss.semi_inclusive_TMDJFFs}
\begin{figure}[htb]
\centering
\includegraphics[width = 0.5 \textwidth]{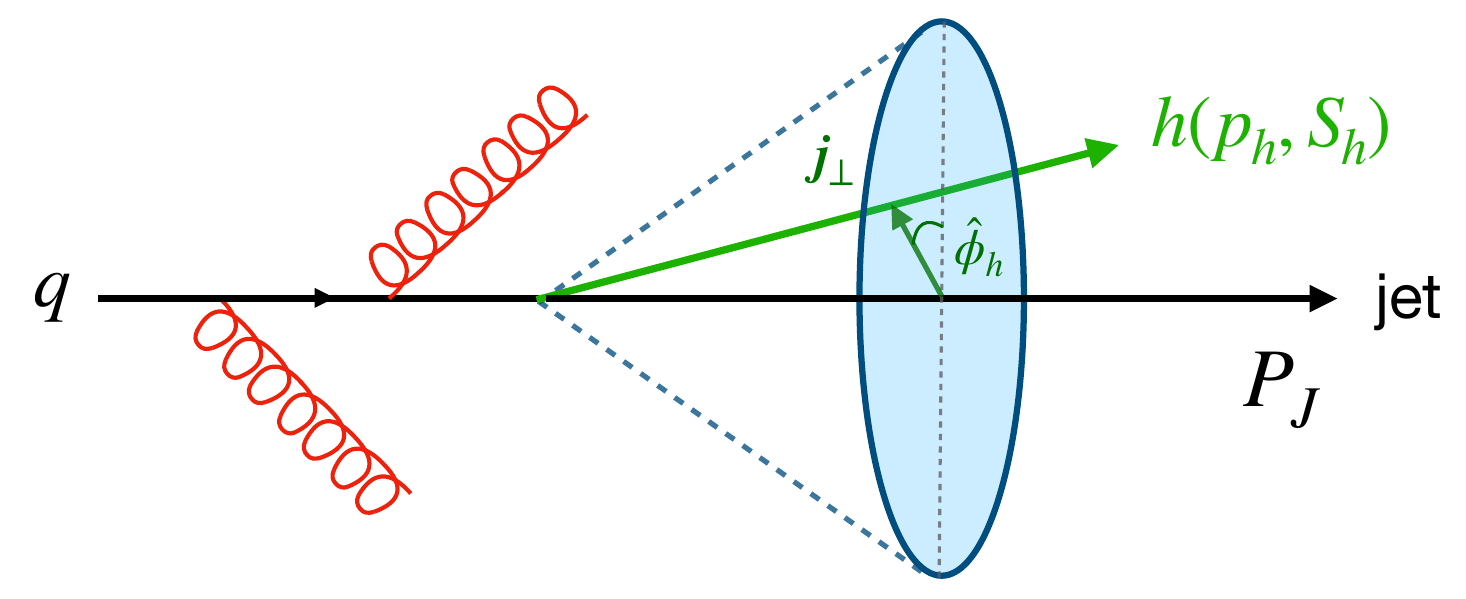}
\caption{Illustration for the distribution of hadrons inside a jet with transverse momentum ${\bm j}_\perp$ and azimuthal angle $\hat{\phi}_h$ with respect to the jet axis.}
\label{f.TMDJFF}
\end{figure}

In this section, we review the concept of polarized TMD jet fragmentation functions introduced in \cite{Kang:2020xyq}.
In \cref{f.TMDJFF}, where a quark-initiated jet is considered, a hadron $h$ is observed inside the jet, carrying a longitudinal momentum fraction $z_h$ of the jet and a transverse momentum $\boldsymbol{j}_{\perp}$ with respect to the jet axis. Both unpolarized and polarized fragmenting quarks are considered in our theoretical framework.
The operator definitions of these functions in SCET are presented, followed by their factorization formalism, which involves hard functions, soft functions, and TMD FFs. 

Based on the field and operator definitions introduced in \cref{ss.collinear_JFF}, we can define the general correlators for TMD JFFs initiated by quark or gluon as:
\begin{align}
\Delta^{h / q}\left(z, z_h, \boldsymbol{j}_{\perp}, S_h\right) &= \,\frac{z}{2 N_c} \delta\left(z_h-\frac{\omega_h}{\omega_J}\right)\label{e.TMD_JFF_collelator_q}\\
&\hspace{-1cm}\times\left\langle 0\left|\delta\left(\omega-n_a \cdot \mathcal{P}\right) \delta^2\left(\mathcal{P}_{\perp} / z_h+\boldsymbol{j}_{\perp}\right) \chi_n(0)\right|(J h) X\right\rangle\left\langle(J h) X\left|\bar{\chi}_n(0)\right| 0\right\rangle\,, \nnu
\Delta^{h / g, \mu v}\left(z, z_h, \boldsymbol{j}_{\perp}, S_h\right) &=\, \frac{z \omega}{(d-2)\left(N_c^2-1\right)} \delta\left(z_h-\frac{\omega_h}{\omega_J}\right)\label{e.TMD_JFF_collelator_g}\\
&\hspace{-1cm}\times\left\langle 0\left|\delta\left(p^{-}-n_a \cdot \mathcal{P}\right) \delta^2\left(\mathcal{P}_{\perp} / z_h+\boldsymbol{j}_{\perp}\right) \mathcal{B}_{n \perp}^\mu(0)\right|(J h) X\right\rangle\left\langle(J h) X\left|\mathcal{B}_{n \perp}^v(0)\right| 0\right\rangle\,.\nonumber
\end{align}
where the energy fractions $z$ and $z_h$ have been defined in \cref{sec:4.1}.

Next, one can parameterize the correlators in \cref{e.TMD_JFF_collelator_q,e.TMD_JFF_collelator_g} at the leading power:
\begin{align}
\Delta ^{h/q}(z, z_h, \boldsymbol{j}_{\perp}, S_h)
& =  
\Delta ^{h/q \, [\slashed{n}_b]} \frac{\slashed{n}_a}{2}
- \Delta^{h/q \, [\slashed{n}_b \gamma _5]} \frac{\slashed{n}_a \gamma _5}{2}
+ \Delta ^{h/q \, [i n_{b,\nu}\sigma ^{k \nu} \gamma _5]} \frac{i {n}_{a,\mu} \sigma^{k \mu} \gamma _5}{2} \,, \label{e.TMD_JFF_collelator_q_decomposition} \\
\Delta ^{h/g, ij} (z, z_h, \boldsymbol{j}_{\perp}, S_h)
& = \frac{1}{2} \delta _T^{ij} \Big(\delta _T^{kl} \Delta ^{h/g, kl} \Big) - \frac{i}{2} \epsilon _T^{ij} \Big(i \epsilon _T^{kl} \Delta ^{h/g, kl} \Big) + \hat{S} \Delta ^{h/g, ij} \label{e.TMD_JFF_collelator_g_decomposition}\, .
\end{align}
Here we have defined $\Delta^{h/q [\Gamma]} \equiv \frac{1}{4} \Tr(\Delta ^{h/q} \Gamma)$.
The three terms on the r.h.s. of \cref{e.TMD_JFF_collelator_q_decomposition} include TMD JFFs with unpolarized, longitudinally polarized, and transversely polarized initial quarks. More details have been presented in our previous work \cite{Kang:2020xyq}. In the following context, we use $n$ to represent light-cone vector $n_b$ for simplification and provide the parametrization of quark TMD JFFs
\begingroup
\allowdisplaybreaks
\begin{align}
\Delta ^{h/q [\slashed{n}]}
= & \,
\mathcal{D}_1^{h/q}(z,z_h,\boldsymbol{j}_{\perp})
- \frac{\epsilon ^{ij}_T j_{\perp}^i S_{h \perp}^j}{z_h M_h} \mathcal{D}_{1T}^{\perp h/q}(z,z_h,\boldsymbol{j}_{\perp}) \,, \\
\Delta ^{h/q [\slashed{n} \gamma _5]}
= & \,
\Lambda _h \mathcal{G}_{1L}^{h/q}(z,z_h,\boldsymbol{j}_{\perp})
- \frac{\boldsymbol{j}_{\perp} \cdot \boldsymbol{S}_{h \perp}}{z_h M_h} \mathcal{G}_{1T}^{h/q}(z,z_h,\boldsymbol{j}_{\perp}) \,, \\
\Delta ^{h/q [i n_\nu \sigma^{i\nu} \gamma _5]}
= & \,
S_{h\perp}^i \mathcal{H}_1^{h/q}(z,z_h,\boldsymbol{j}_{\perp})
- \frac{\epsilon^{ij}_T j_{\perp}^j}{z_h M_h} \mathcal{H}_{1}^{\perp h/q}(z,z_h,\boldsymbol{j}_{\perp})
- \frac{j_{\perp}^i}{z_h M_h} \Lambda _h {\mathcal{H}_{1L}^{\perp h/q}(z,z_h,\boldsymbol{j}_{\perp})} \nnu
& + \frac{j_{\perp}^i \boldsymbol{j}_{\perp} \cdot \boldsymbol{S}_{h \perp} - \frac{1}{2} \boldsymbol{j}_{\perp}^2 S_{h \perp}^i}{z_h^2 M_h^2} \mathcal{H}_{1T}^{\perp h/q} (z,z_h,\boldsymbol{j}_{\perp}) \,, 
\end{align}
\endgroup
\begin{table}
    \centering
\begin{tabular}{ |c|c|c|c| } 
 \hline
 \diagbox[width=4em]{$H$}{$q$} & $U$ & $L$ & $T$ \\ 
  \hline
 $U$ & $\mathcal{D}_1^{h/q} $&  &$\mathcal{H}_1^{\perp h/q}$  \\ 
  \hline
$L$ &  &  $\mathcal{G}^{h/q}_{1L}$& $\mathcal{H}_{1L}^{h/q}$\\ 
  \hline
$T$ & $\mathcal{D}_{1T}^{\perp h/q}$ & $\mathcal{G}^{h/q}_{1T}$ & $\mathcal{H}^{h/q}_{1}$, $\mathcal{H}_{1T}^{\perp h/q}$\\ 
  \hline
\end{tabular}
  \caption{Interpretation of TMDJFFs for quarks. The rows indicate the hadron polarization — unpolarized (U), longitudinally polarized (L), transversely polarized (T). And the columns indicate the quark polarization accordingly.}
  \label{tabTMDffq}
\end{table}
As for the gluon TMD JFFs given in \cref{e.TMD_JFF_collelator_g}, they are parametrized as
\begingroup
\allowdisplaybreaks
\begin{align}
\Delta ^{h/g, \alpha\beta} (z, z_h, \boldsymbol{j}_{\perp}, S_h)=&\frac{1}{2}\left[-g_T^{\alpha\beta}\mathcal{D}_1^{h/g}(z,z_h,\boldsymbol{j}_{\perp})+\frac{\boldsymbol{j}_\perp^2}{z_h^2M_h^2}\left(\frac{g_T^{\alpha\beta}}{2}+\frac{j_\perp^\alpha j_\perp^\beta}{\boldsymbol{j}_\perp^2}\right)\mathcal{H}_1^{\perp h/g}(z,z_h,\boldsymbol{j}_{\perp})\right]\nnu
&+\frac{\Lambda_h}{2}\left[-i\epsilon_T^{\alpha\beta}\mathcal{G}_{1L}^{h/g}(z,z_h,\boldsymbol{j}_{\perp})+\frac{\epsilon_T^{j_{\perp}\{\alpha}j_\perp^{\beta\}}}{2z_h^2M_h^2}\mathcal{H}_{1L}^{\perp h/g}(z,z_h,\boldsymbol{j}_{\perp})\right]\nnu
&+\frac{1}{2}\,\bigg[g_T^{\alpha\beta}\frac{\epsilon_T^{j_\perp S_{h\perp}}}{z_hM_h}\mathcal{D}_{1T}^{\perp h/g}(z,z_h,\boldsymbol{j}_{\perp})+i\epsilon_T^{\alpha\beta}\frac{\boldsymbol{j}_\perp\cdot\boldsymbol{S}_{h\perp}}{z_hM_h}\mathcal{G}_{1T}^{ h/g}(z,z_h,\boldsymbol{j}_{\perp})\nnu
&\qquad-\frac{\epsilon_T^{j_{\perp}\{\alpha}j_\perp^{\beta\}}}{2z_h^2M_h^2}\frac{\boldsymbol{j}_\perp\cdot\boldsymbol{S}_{h\perp}}{z_hM_h}\mathcal{H}_{1T}^{\perp h/g}(z,z_h,\boldsymbol{j}_{\perp})\nnu
&\qquad-\frac{\epsilon_T^{j_{\perp}\{\alpha}S_{h\perp}^{\beta\}}+\epsilon_T^{S_{h\perp}\{\alpha}j_\perp^{\beta\}}}{4z_hM_h}\mathcal{H}_{1T}^{ h/g}(z,z_h,\boldsymbol{j}_{\perp})\bigg]\,.\label{e.TMD_JFF_collelator_g_decomposition1}
\end{align}
\endgroup
Here functions $\mathcal{D}$, $\mathcal{G}$ and $\mathcal{H}$ on the r.h.s. of \cref{e.TMD_JFF_collelator_g_decomposition1} represent the TMD JFFs with unpolarized, circularly polarized, and linearly polarized initial gluons respectively and we have adopted the notation $v_T^{\{\alpha}w_T^{\beta\}}=v_T^\alpha w_T^\beta+v_T^\beta w_T^\alpha$ as applied in~\cite{Mulders:2000sh,Boussarie:2023izj}.
More specifically, one has each term in the r.h.s. of \cref{e.TMD_JFF_collelator_g_decomposition1} give as
\begin{align}
\delta _T^{\alpha\beta} \Delta ^{h/g, \alpha\beta}
= &\,\mathcal{D}_1^{h/g}(z,z_h,\boldsymbol{j}_{\perp})-\frac{\epsilon ^{ij}_T j_{\perp}^i S_{h \perp}^j}{z_hM_h}\mathcal{D}_1^{\perp h/g}(z,z_h,\boldsymbol{j}_{\perp})\,, \\
i \epsilon _T^{\alpha\beta} \Delta ^{h/g, \alpha\beta}
= & \,\Lambda_h \mathcal{G}_{1L}^{h/g}(z,z_h,\boldsymbol{j}_{\perp})-\frac{\boldsymbol{j}_\perp\cdot\boldsymbol{S}_{h\perp}}{z_hM_h}\mathcal{G}_{1T}^{ h/g}(z,z_h,\boldsymbol{j}_{\perp})\,,\\
\hat{S} \Delta ^{h/g, \alpha\beta}
= &\,\frac{j_{\perp}^\alpha j_{\perp}^\beta}{2 z_h^2 M_h^2} \mathcal{H}_1^{\perp h/g}(z,z_h,\boldsymbol{j}_{\perp})-\frac{\epsilon_T^{j_{\perp}\{\alpha}S_{h\perp}^{\beta\}}+\epsilon_T^{S_{h\perp}\{\alpha}j_\perp^{\beta\}}}{8z_hM_h}\mathcal{H}_{1T}^{ h/g}(z,z_h,\boldsymbol{j}_{\perp}) \\
& + \frac{\epsilon_T^{j_{\perp}\{\alpha}j_\perp^{\beta\}}}{4z_h^2M_h^2} \left(\Lambda _h \mathcal{H}_{1L}^{\perp , h/g}(z,z_h,\boldsymbol{j}_{\perp})-\frac{\boldsymbol{j}_\perp\cdot\boldsymbol{S}_{h\perp}}{z_hM_h}\mathcal{H}_{1T}^{\perp h/g}(z,z_h,\boldsymbol{j}_{\perp})\right)  \,,
\end{align}
where $\hat{S}O^{\alpha\beta}=\frac{1}{2}\left(O^{\alpha\beta}+O^{\beta\alpha}-\delta_T^{\alpha\beta}O^{\rho\rho}\right)$ and $\delta_T^{\alpha\beta}=-g_T^{\alpha\beta}$.
\begin{table}
    \centering
\begin{tabular}{ |c|c|c|c| } 
 \hline
 \diagbox[width=4em]{$H$}{$g$} & $U$ & $L$ & $T$ \\ 
  \hline
 $U$ & $\mathcal{D}_1^{h/g} $&  &$\mathcal{H}_1^{\perp h/g}$  \\ 
  \hline
$L$ &  &  $\mathcal{G}^{h/g}_{1L}$& $\mathcal{H}_{1L}^{h/g}$\\ 
  \hline
$T$ & $\mathcal{D}_{1T}^{\perp h/g}$ & $\mathcal{G}^{h/g}_{1T}$ & $\mathcal{H}^{h/g}_{1T}$, $\mathcal{H}_{1T}^{\perp h/g}$\\ 
  \hline
\end{tabular}
  \caption{Interpretation of TMDJFFs for gluons. The columns indicate the gluon polarization — unpolarized (U), circularly polarized (L), linearly polarized (T).}
  \label{tabTMDffg}
\end{table}
Since TMD JFFs represent the hadron fragmentation inside a fully reconstructed jet, their physical meaning is similar to that of standard TMD FFs as reviewed in \cite{Metz:2016swz}.
Consequently, we adopt the calligraphic font of the letters used for the TMD FFs as the notations of TMD JFFs with corresponding polarizations. 

In this study, following the unpolarized TMD JFFs calculation carried out in \cite{Kang:2017glf}, we will focus on the kinematic region where $\Lambda _{\mathrm{QCD}} \lesssim j_{\perp} \ll p_T R$, where the standard collinear factorization breaks down due to the presence of the large logarithms of the form $\ln(p_T R / j_{\perp})$.
This situation necessitates the adoption of the TMD factorization \cite{Collins:2011zzd}, which we will provide a detailed discussion of in the following sections.

\subsubsection{TMD Factorization}

In the kinematic region under consideration, the radiation relevant at leading power is restricted to collinear radiation within the jet, characterized by momentum that scales as $p_c = \pqty{p_c^- , p_c^+ , p_{c,T}} \sim p_T \pqty{1 , \lambda ^2 , \lambda}$, where $\lambda \sim j_{\perp} / p_T$.
Additionally, soft radiation of order $j_{\perp}$ is also relevant.
It is worth noting that harder emissions are only permitted outside the jet cone and will thus only impact the determination of the jet axis.
Consequently, the hadron transverse momentum $j_{\perp}$, which is defined with respect to the jet axis, remains intact from the radiations external to the jet.
A factorized formalism for the unpolarized TMD JFFs within SCET can thus be formulated as follows:
\begin{align} 
\mathcal{D}_1^{h/c} \pqty{z, z_h, \omega _J R, \boldsymbol{j}_{\perp}, \mu}
= & \, \hat{H}^U_{c \to i} \pqty{z, \omega _J R, \mu}
\int \dd[2]{\boldsymbol{k}_{\perp}} \dd[2]{\boldsymbol{\lambda}_{\perp}} \delta ^2 \pqty{z_h \boldsymbol{\lambda}_{\perp} + \boldsymbol{k}_{\perp} - \boldsymbol{j}_{\perp}} \label{eq:unp_JFF_FF1}\\
& \times D_1^{h/i} \pqty{z_h, \boldsymbol{k}_{\perp}, \mu, \nu} S_i \pqty{\boldsymbol{\lambda}_{\perp}, \mu, \nu R} \,, \nonumber
\end{align}
where $S_i\pqty{\boldsymbol{\lambda}_{\perp}, \mu, \nu R}$ denotes the soft radiation.
The $\delta$ function establishes a relationship between the hadron transverse momentum $\boldsymbol{j}_{\perp}$, relative to the jet axis, and two other momenta: the transverse component of soft radiation represented by $\boldsymbol{\lambda}_{\perp}$, and the hadron transverse momentum with respect to the soft radiation, denoted as $\boldsymbol{k}_{\perp}$.
Notice that $\boldsymbol{\lambda}_{\perp}$ is multiplied by $z_h$ to adjust for the dissimilarity between the fragmenting parton and the observed hadron.
As is common practice in TMD physics, we convert the aforementioned expression from the transverse momentum space to the coordinate $b$-space using the following transformation,
\begin{equation} \label{e.TMD_b_space}
\mathcal{D}_1^{h/c} \pqty{z, z_h, \omega _J R, \boldsymbol{j}_{\perp}, \mu}
=
\hat{H}^U_{c \to i} \pqty{z, \omega _J R, \mu}
\int \frac{\dd[2]{\boldsymbol{b}}}{(2 \pi)^2} e^{i \boldsymbol{j}_{\perp} \cdot \boldsymbol{b} / z_h} \widetilde{D}_1^{h/i} \pqty{z_h, \boldsymbol{b}, \mu, \nu} \widetilde{S}_i (\boldsymbol{b}, \mu, \nu R)\, ,
\end{equation}
where we have defined the Fourier transform for both $\widetilde{D}_1^{h/i} \pqty{z_h, \boldsymbol{b}, \mu, \nu}$ and $\widetilde{S}_i (\boldsymbol{b}, \mu, \nu R)$ as:
\begin{align}
\widetilde{D}_1^{h/i} \pqty{z_h, \boldsymbol{b}, \mu, \nu}
& =
\frac{1}{z_h^2} \int \dd[2]{\boldsymbol{k}_{\perp}} e^{-i \boldsymbol{k}_{\perp} \cdot \boldsymbol{b} / z_h} D_1^{h / i} \pqty{z_h, \boldsymbol{k}_{\perp}, \mu, \nu} \nnu
& =
\frac{1}{z_h^2}
\int \dd{k_{\perp}} k_{\perp} 2 \pi J_0 \pqty{\frac{b k_{\perp}}{z_h}} D_1^{h/i} \pqty{z_h, \boldsymbol{k}_\perp, \mu, \nu} \,, \label{e.TMD_FF_b_space} \\
\widetilde{S}_i(\boldsymbol{b}, \mu, \nu R)
& =
\int \dd[2]{\boldsymbol{\lambda _{\perp}}} e^{-i \boldsymbol{\lambda}_{\perp} \cdot \boldsymbol{b}} S_i \pqty{\boldsymbol{\lambda}_{\perp}, \mu, \nu R}\, .
\end{align}
Now let us define the operator $\mathscr{C}$:
\begin{align}
\mathscr{C} [\widetilde{D}^{h/i, (n)}]
=
\int \frac{b^{n+1} \dd{b}}{2 \pi n !} \pqty{\frac{z_h^2 M_h^2}{j_{\perp}}}^n J_n \pqty{\frac{j_{\perp} b}{z_h}} \widetilde{D}^{h/i, (n)} \pqty{z_h, \boldsymbol{b}, \mu, \nu} \widetilde{S}_i \pqty{\boldsymbol{b}, \mu, \nu R}\, ,
\end{align}
where by default $n = 0$, giving $\widetilde{D}^{h/i , (0)} \pqty{z_h, \boldsymbol{b}, \mu, \nu}$ the TMD FFs in the Fourier $b$-space.
This can be generalized to give the $n$-th moment:
\begin{equation}
\widetilde{D}^{h/i, (n)} \pqty{z_h, \boldsymbol{b}, \mu, \nu}
=
\frac{1}{z_h^2} \frac{2 \pi n !}{\pqty{z_h^2 M_h^2}^n}
\int \dd{k_{\perp}} k_{\perp} \pqty{\frac{k_{\perp}}{b}}^n J_n \pqty{\frac{b k_{\perp}}{z_h}} D^{h / i}\pqty{z_h, \boldsymbol{k}_{\perp}, \mu, \nu} \,. \label{e.TMD_FF_b_space_nth}
\end{equation}
One can easily verify that with $n = 0$ and $D = D_1$, \cref{e.TMD_FF_b_space_nth} gives \cref{e.TMD_FF_b_space}.
We can now write down the factorization for all the TMD JFFs as:
\begingroup
\allowdisplaybreaks
\begin{align}
\mathcal{D}_{1}^{h/c} \pqty{z, z_h, \omega _J R, \boldsymbol{j}_{\perp}, \mu}
& =
\hat{H}_{c \to i}^U (z, \omega _J R, \mu) \mathscr{C} \bqty{\widetilde{D}_{1}^{h/i}}\, , \\
\mathcal{D}_{1T}^{\perp, h/c} \pqty{z, z_h, \omega _J R, \boldsymbol{j}_{\perp}, \mu}
& =
\hat{H}_{c \to i}^U (z, \omega _J R, \mu) \mathscr{C} \bqty{\widetilde{D}_{1T}^{\perp, h/i, (1)}}\, , \\
\mathcal{G}_{1L}^{h/c} \pqty{z, z_h, \omega _J R, \boldsymbol{j}_{\perp}, \mu}
& =
\hat{H}_{c \to i}^L (z, \omega _J R, \mu) \mathscr{C} \bqty{\widetilde{G}_{1L}^{h/i}}\, ,\\
\mathcal{G}_{1T}^{h/c} \pqty{z, z_h, \omega _J R, \boldsymbol{j}_{\perp}, \mu}
& =
\hat{H}_{c \to i}^L (z, \omega _J R, \mu) \mathscr{C} \bqty{\widetilde{G}_{1T}^{h/i, (1)}}\, ,\\
\mathcal{H}_{1}^{h/c} \pqty{z, z_h, \omega _J R, \boldsymbol{j}_{\perp}, \mu}
& =
\hat{H}_{c \to i}^T (z, \omega _J R, \mu) \mathscr{C} \bqty{\widetilde{H}_{1}^{h/i}}\, , \\
\mathcal{H}_{1}^{\perp, h/c} \pqty{z, z_h, \omega _J R, \boldsymbol{j}_{\perp}, \mu}
& =
\hat{H}_{c \to i}^T (z, \omega _J R, \mu) \mathscr{C} \bqty{\widetilde{H}_{1}^{\perp, h/i, (1)}} \,,\\
\mathcal{H}_{1L}^{\perp, h/c} \pqty{z, z_h, \omega _J R, \boldsymbol{j}_{\perp}, \mu}
& =
\hat{H}_{c \to i}^T (z, \omega _J R, \mu) \mathscr{C} \bqty{\widetilde{H}_{1L}^{\perp, h/i, (1)}}\, , \\
\mathcal{H}_{1T}^{\perp, h/c} \pqty{z, z_h, \omega _J R, \boldsymbol{j}_{\perp}, \mu}
& =
\hat{H}_{c \to i}^T (z, \omega _J R, \mu) \mathscr{C} \bqty{\widetilde{H}_{1T}^{\perp, h/i, (2)}} \,.
\end{align}
\endgroup
Here the superscripts $U$, $L$ and $T$ of $\hat{H}_{c \to i}$ represent unpolarized, longitudinally polarized, or transversely polarized hard matching functions, and they will be provided in the next subsection.
The above equations also show how various TMD JFFs are matched onto their corresponding TMD FFs, with which the matching of the scenarios listed in \cref{t.JFF_and_FFs} can be performed.
The soft functions are identical for all scenarios and can be found in \cite{Kang:2017glf, Kang:2021ffh}.
The subsequent sections will provide detailed calculation of the hard matching functions and partonic TMD FFs, as well as their renormalization using RG evolutions.
\begin{table}
\begin{center}
\begin{tabular}{|c|c|c|c|}
\hline
\diagbox{$h$}{$q$} & U & L & T \\
\hline
U & $\mathcal{D}_1$ & ~ & $\mathcal{H}_1^{\perp}$ \\
\hline
L & ~ & $\mathcal{G}_{1L}$ & $\mathcal{H}_{1L}^{\perp}$ \\
\hline
T & $\mathcal{D}_{1T}^{\perp}$ & $\mathcal{G}_{1T}$ & $\mathcal{H}_1$, $\mathcal{H}_{1T}^{\perp}$ \\
\hline
\end{tabular}
\hfil
\begin{tabular}{ |c|c|c|c| }
\hline
\diagbox{$h$}{$q$} & U & L & T \\
\hline
U & $D_1$ & ~ & $H_1^{\perp}$ \\
\hline
L & ~ & $G_{1L}$ & $H_{1L}^{\perp}$ \\
\hline
T & $D_{1T}^{\perp}$ & $G_{1T}$ & $H_1$, $H_{1T}^{\perp}$ \\
\hline
\end{tabular}
\end{center}
\caption{
Summary of semi-inclusive TMD JFFs (left) and TMD FFs (right).
The header row represents the polarization of the fragmenting quarks while the header column indicates the corresponding polarizations of produced hadrons.}
\label{t.JFF_and_FFs}
\end{table}

\subsubsection{Hard matching functions}

The hard matching functions describe how the energetic parton $c$ produced in a hard scattering event fragments into the parton $i$ which initiates a jet with energy $\omega _J$ and radius $R$.
The out-of-jet diagrams for inclusive jet (substructure) observables \cite{Kang:2016ehg} can be used to calculate these functions up to NLO.
In \cite{Kang:2017mda}, the same hard matching functions were discovered in the context of central subjets that are measured in an inclusive jet sample.

The polarization of the final-state hadron does not affect the hard matching functions, as a result, the JFFs listed in the same column in \cref{t.JFF_and_FFs} share identical hard matching functions.

Our analysis incorporates three distinct hard matching functions, namely $\hat{H}_{c \to i}^U (z, \omega J R, \mu)$, $\hat{H}_{c \to i}^L (z, \omega J R, \mu)$, and $\hat{H}_{q \to q}^T (z, \omega J R, \mu)$.
The transversely polarized case only has the $q \to q$ type, due to the same reason as for the collinear semi-inclusive JFFs.
The unpolarized hard matching functions $\hat{H}_{c \to i}^U (z, \omega _J R, \mu)$ have been previously calculated in \cite{Kang:2017glf}.
Here we provide the expressions for the other types renormalized hard matching functions:
\begingroup
\allowdisplaybreaks
\begin{align}
\label{e.hard_matching_qq}
\hat{H}^L_{q\to q'}(z,\omega_J R,\mu) 
&= \delta_{qq'}\delta(1-z) +\delta_{qq'} \frac{\alpha_s}{2\pi} \bigg[C_F \delta(1-z) \pqty{-\frac{L^2}{2} - \frac{3}{2} L +\frac{\pi^2}{12}}
\nnu
& \quad
+ \Delta P_{qq}(z) L -2C_F(1+z^2)\left(\frac{\ln(1-z)}{1-z}\right)_+ -C_F(1-z)  \bigg] \,, 
\\[.2cm]
\hat{H}^L_{q\to g}(z,\omega_J R,\mu) 
 &=\frac{\alpha_s}{2\pi}\bigg[\Big(L - 2 \ln(1-z) \Big) \Delta P_{gq}(z) + 2 C_F (1-z) \bigg]\,, 
 \\[.2cm]
\label{e.hard_matching_gg}
\hat{H}^L_{g\to g}(z, \omega_J R, \mu) 
& = \delta(1-z) + \frac{\alpha_s}{2\pi}\bigg[ \delta (1-z) \pqty{-C_A\frac{L^2}{2} - \frac{\beta_0}{2} L + C_A\frac{\pi^2}{12}}
\nnu
& \quad
+ \Delta P_{gg}(z) L + 4C_A (1-z) - 4C_A (2(1-z)^2+z) \left(\frac{\ln(1-z)}{1-z}\right)_{+} \bigg]\,,
\\[.2cm]
\hat{H}^L_{g\to q}(z,\omega_J R, \mu)
& =  \frac{\alpha_s}{2\pi}\bigg[\Big(L - 2\ln(1-z) \Big)  \Delta P_{qg}(z) - 2 T_F (1-z) \bigg]\,, \\
\label{e.hard_matching_qq_T}
\hat{H}^T_{q\to q'} (z,\omega_JR,\mu)&=\delta_{qq'} \delta(1-z) + \delta_{qq'}\frac{\alpha_s}{2\pi}\bigg\{\Delta_T P_{qq}(z)L+C_F\bigg[-4z\left(\frac{\ln(1-z)}{1-z}\right)_+\nonumber\\
& \quad
+\left(-\frac{3}{2}L-\frac{L^2}{2}+\frac{\pi^2}{12}\right)\delta(1-z)\bigg]\bigg\} ,
\end{align}
\endgroup
where the leading order splitting kernels are given in \cref{e.polarized_AP_splittings1}-\ref{e.polarized_AP_splittings5}.
The RG equations take the form:
\begingroup
\allowdisplaybreaks
\begin{align}
\mu \frac{\dd}{\dd{\mu}} \hat{H}_{i \to j}^L \pqty{z, \omega _J R , \mu}&=\sum _k \int _z^1 \frac{\dd{z'}}{z'} \gamma _{i k} \pqty{\frac{z}{z'} , \omega _J R , \mu}
\hat{H}_{k \to j}^L \pqty{z', \omega _J R , \mu}\, , \\
\mu \frac{\dd}{\dd{\mu}} \hat{H}_{q \to q'}^T \pqty{z, \omega _J R , \mu}&=\sum _k \int _z^1 \frac{\dd{z'}}{z'} \gamma _{i k} \pqty{\frac{z}{z'} , \omega _J R , \mu}
\hat{H}_{q \to q'}^T \pqty{z', \omega _J R , \mu}\, .
\end{align}
\endgroup
The anomalous dimensions $\gamma _{i j} \pqty{\frac{z}{z'} , \omega _J R , \mu}$ are given by
\begingroup
\allowdisplaybreaks
\begin{align}
\gamma _{q q}^L &=\frac{\alpha _s}{\pi} \pqty{\Delta P_{q q} \pqty{z} - C_F L \delta \pqty{1 - z} - \frac{3 C_F}{2} \delta \pqty{1 - z}}\, , \\
\gamma _{g q}^L &=\frac{\alpha _s}{\pi} \Delta P_{g q} \pqty{z}\, , \\
\gamma _{g g}^L &=\frac{\alpha _s}{\pi} \pqty{\Delta P_{g g} \pqty{z} - C_A L \delta \pqty{1 - z} - \frac{\beta _0}{2} \delta \pqty{1 - z}} \,, \\
\gamma _{q g}^L &=\frac{\alpha _s}{\pi} \Delta P_{q g} \pqty{z}\, , \\
\gamma _{q q}^T &=\frac{\alpha _s}{\pi} \pqty{\Delta _T P_{q q} \pqty{z} - C_F L \delta \pqty{1 - z} - \frac{3 C_F}{2} \delta \pqty{1 - z}}\, . 
\end{align}
\endgroup
The natural scale here is $\mu \sim \omega _J \tan(\mathcal{R}/2)$, followed from the definition that $L \equiv \ln(\frac{\mu ^2}{\omega _J^2 \tan[2](\mathcal{R}/2)})$ as in \cref{sss.collinear_siJFFs_calculation}.
Thus by solving the RG equations and evolving the hard matching functions from scale $\mu \sim \omega _J \tan(\mathcal{R}/2)$ to the hard scattering scale $\mu \sim p_T$, we resummed the large logarithms of jet radius $\ln R$.

\subsubsection{Transverse momentum dependent FFs}
\label{ss.TMDFFs}
Following the calculation in \cite{Kang:2017glf}, we can obtain the perturbative result for bare TMDFFs of longitudinally polarized and transversely polarized hadrons.
(For brevity, we will only list their matching coefficients later.)
Additionally, we study the TMDFFs in Fourier space or $b$-space, following the convention in \cite{Collins:2011zzd, Echevarria:2014xaa}, we define the TMDFFs in $b$-space as:
\begingroup
\allowdisplaybreaks
\begin{align}
D_1^{h/i} \pqty{z_h, \boldsymbol{b}, \mu , \nu}
& =
\frac{1}{z_h^2} \int \dd[2]{\boldsymbol{k}_{\bot}}
e^{- i \boldsymbol{k}_{\bot} \cdot \boldsymbol{b} / z_h}
D_1^{h/i} \pqty{z_h, \boldsymbol{k}_{\bot}, \mu , \nu} , \\
G_{1L}^{h/i} \pqty{z_h, \boldsymbol{b}, \mu , \nu}
& =
\frac{1}{z_h^2} \int \dd[2]{\boldsymbol{k}_{\bot}}
e^{- i \boldsymbol{k}_{\bot} \cdot \boldsymbol{b} / z_h}
G_{1L}^{h/i} \pqty{z_h, \boldsymbol{k}_{\bot}, \mu , \nu} , \\
H_1^{h/i} \pqty{z_h, \boldsymbol{b}, \mu , \nu}
& =
\frac{1}{z_h^2} \int \dd[2]{\boldsymbol{k}_{\bot}}
e^{- i \boldsymbol{k}_{\bot} \cdot \boldsymbol{b} / z_h}
H_1^{h/i} \pqty{z_h, \boldsymbol{k}_{\bot}, \mu , \nu} .
\end{align}
\endgroup
In the perturbative region $1/b \gg \Lambda_{\mathrm{QCD}}$, these TMDFFs can be matched onto the standard collinear FFs, with the matching relation is given by:
\begingroup
\allowdisplaybreaks
\begin{align}
D_1^{h/i}(z_h, {\bm b}, \mu, \nu)
& = \frac{1}{z_h^2}\int_{z_h}^1 \frac{d\hat z_h}{\hat z_h} \C_{j \to i}\left(\frac{z_h}{\hat z_h}, {\bm b}, \mu, \nu \right) D^{h/j} (\hat z_h, \mu) \nonumber \\
& \equiv \frac{1}{z_h^2} \C_{j \to i}\otimes D^{h/j}(z_h, \mu) \, , \\
G_{1L}^{h/i}(z_h, {\bm b}, \mu, \nu)
& = \frac{1}{z_h^2}\int_{z_h}^1 \frac{d\hat z_h}{\hat z_h} \Delta\C_{j \to i}\left(\frac{z_h}{\hat z_h}, {\bm b}, \mu, \nu \right) \Delta D^{h/j} (\hat z_h, \mu) \nonumber \\
& \equiv \frac{1}{z_h^2} \Delta\C_{j \to i}\otimes \Delta D^{h/j}(z_h, \mu) \, , \label{e.TMDFF_G_1L} \\
H_1^{h/q}(z_h, {\bm b}, \mu, \nu) & = \frac{1}{z_h^2}\int_{z_h}^1 \frac{d\hat z_h}{\hat z_h} \Delta_T\C_{j \to i}\left(\frac{z_h}{\hat z_h}, {\bm b}, \mu, \nu \right)  \Delta_T D^{h/j}(\hat z_h, \mu) \nonumber \\
& \equiv \frac{1}{z_h^2} \Delta_T\C_{j \to i}\otimes \Delta_T D^{h/j} (z_h, \mu) \, , \label{e.TMDFF_H_1}
\end{align}
\endgroup
where the matching coefficients are denoted by $(\Delta)\C_{j \to i}$.
The unpolarized $\C_{j\leftarrow i}$ are already given in \cite{Kang:2017glf}, while for the polarized TMDFFs, we have
\begingroup
\allowdisplaybreaks
\begin{align}
\label{e.Delta_C_qq}
\Delta \C_{q'\leftarrow q}(z_h, {\bm b}, \mu, \nu) = & 
\delta_{qq'}\Bigg\{\delta(1-z_h)   
- \frac{\alpha_s}{2\pi} \ln\left(\frac{\mu^2}{z_h^2\mu_b^2}\right) \Delta P_{qq}(z_h)
\\
&\hspace{-0.5cm}+\frac{\alpha_s}{2\pi} C_F \bigg[\ln\left(\frac{\mu^2}{\mu_b^2}\right) \left(2 \ln\left(\frac{\nu}{\omega_J}\right) + \frac{3}{2}\right) \delta(1-z_h) + (1-z_h)\bigg]\Bigg\}\,,
\nnu[.3cm]
\label{e.Delta_C_gq}
\Delta \C_{g\leftarrow q}(z_h, {\bm b}, \mu, \nu) =& \frac{\alpha_s}{2 \pi}\Big[ -\ln\left(\frac{\mu^2}{z_h^2\mu_b^2}\right)\Delta P_{gq}(z_h) - 2 C_F (1-z_h)\Big]\,,
\\[.3cm]
\label{e.Delta_C_gg}
\Delta \C_{g\leftarrow g}(z_h, {\bm b}, \mu, \nu) =& \delta(1-z_h) - \frac{\alpha_s}{2\pi} \ln\left(\frac{\mu^2}{z_h^2\mu_b^2}\right)\Delta P_{gg}(z_h)
\\
&\hspace{-0.5cm}+\frac{\alpha_s}{2\pi} C_A \bigg[\ln\left(\frac{\mu^2}{\mu_b^2}\right) \left(2 \ln\left(\frac{\nu}{\omega_J}\right) + \frac{\beta_0}{2C_A}\right) \delta(1-z_h) - 2 \pqty{1 - z_h} \bigg]\,,
\nnu[.3cm]
\label{e.Delta_C_qg}
\Delta \C_{q\leftarrow g}(z_h, {\bm b}, \mu, \nu) =&  \frac{\alpha_s}{2\pi}\Big[ -\ln\left(\frac{\mu^2}{z_h^2\mu_b^2}\right) \Delta P_{qg}(z_h) +  2 T_F (1-z_h) \Big]\,.\\
\label{e.Delta_C_qq_T}
\Delta_T \tilde{{C}}_{q'\leftarrow q}(z_h,\textbf{b},\mu,\nu)=&\delta_{qq'}\bigg\{\delta(1-z_h)-\frac{\alpha_s}{2\pi}\ln\left(\frac{\mu^2}{z_h^2\mu_b^2}\right)\Delta_T P_{qq}(z_h)\\
&+\frac{\alpha_sC_F}{2\pi}\left[\ln\left(\frac{\mu^2}{\mu_b^2}\right)\left(2\ln\left(\frac{\nu}{\omega_J}\right)+\frac{3}{2}\right)\delta(1-z_h)\right]\bigg\} ,\nonumber
\end{align}
\endgroup
where $\mu _b \equiv 2 e^{- \gamma _E} / b$.
The UV divergences in polarized TMDFFs \cref{e.TMDFF_G_1L,e.TMDFF_H_1} can be renormlalized by the RG equations:
\begin{align}
\mu \frac{\dd}{\dd{\mu}} G_{1L}^{j/i} \pqty{z, \omega _J R , \mu , \nu}&=\sum _k \int _z^1 \frac{\dd{z'}}{z'} \gamma _{i\to k}^{\mu , L} \pqty{\frac{z}{z'} , \omega _J R , \mu , \nu}
G_{1L}^{j/k} \pqty{z', \omega _J R , \mu , \nu} , \\
\mu \frac{\dd}{\dd{\mu}} H_1^{j/i } \pqty{z, \omega _J R , \mu , \nu}&=\sum _k \int _z^1 \frac{\dd{z'}}{z'} \gamma _{i \to k}^{\mu , T} \pqty{\frac{z}{z'} , \omega _J R , \mu , \nu}
H_1^{j/k} \pqty{z', \omega _J R , \mu , \nu} ,
\end{align}

with anomalous dimensions:
\begingroup
\allowdisplaybreaks
\begin{align}
\gamma _{q \to q}^{\mu , L}
& =
\frac{\alpha _s}{\pi} C_F
\pqty{\frac{3}{2} + 2 \ln(\frac{\nu}{\omega _J})}
\delta \pqty{1 - z_h}
- \frac{\alpha _s}{\pi}
  \Delta P_{q q} \pqty{z_h} \,, \\
\gamma _{q \to g}^{\mu , L}
& =
\frac{- \alpha _s}{\pi} \Delta P_{g q} \pqty{z_h}\, , \\
\gamma _{g \to g}^{\mu , L}
& =
\frac{\alpha _s}{\pi}
\pqty{\frac{\beta _0}{2} + 2 C_A \ln(\frac{\nu}{\omega _J})}
\delta \pqty{1 - z_h}
- \frac{\alpha _s}{\pi}
  \Delta P_{g g} \pqty{z_h} \,, \\
\gamma _{g \to q}^{\mu , L}
& =
\frac{- \alpha _s}{\pi} \Delta P_{q g} \pqty{z_h} \,, \\
\gamma _{q \to q}^{\mu , T}
& =
\frac{\alpha _s}{\pi} C_F\pqty{\frac{3}{2} + 2 \ln(\frac{\nu}{\omega _J})}
\delta \pqty{1 - z_h}
- \frac{\alpha _s}{\pi}\Delta _T P_{q q} \pqty{z_h}\, .
\end{align}
\endgroup
The rapidity divergences in \cref{e.TMDFF_G_1L,e.TMDFF_H_1} can be renormalized by the rapidity renormalization group (RRG) equations:
\begin{align}
\nu \frac{\dd}{\dd{\nu}} G_{1L}^{j/i} \pqty{z, \omega _J R , \mu , \nu}&=\sum _k \int _z^1 \frac{\dd{z'}}{z'} \gamma _{i\to k}^{\nu , L} \pqty{\frac{z}{z'} , \omega _J R , \mu , \nu}
G_{1L}^{j/k} \pqty{z', \omega _J R , \mu , \nu} \,, \\
\nu \frac{\dd}{\dd{\nu}} H_1^{j/i} \pqty{z, \omega _J R , \mu , \nu}&=\sum _k \int _z^1 \frac{\dd{z'}}{z'} \gamma _{i \to k}^{\nu , T} \pqty{\frac{z}{z'} , \omega _J R , \mu , \nu}
H_1^{j/k} \pqty{z', \omega _J R , \mu , \nu} \,, 
\end{align}

with anomalous dimensions:
\begingroup
\allowdisplaybreaks
\begin{align}
\gamma _{q \to q}^{\nu , L}
&=
\frac{\alpha _s}{\pi} C_F
\ln(\frac{\mu ^2}{\mu _b^2})
\delta \pqty{1 - z_h} \,, \\
\gamma _{q \to g}^{\nu , L}
&=
\gamma _{g \to q}^{\nu , L}
= 0 \,, \\
\gamma _{g \to g}^{\nu , L}
&=
\frac{\alpha _s}{\pi} C_A
\ln(\frac{\mu ^2}{\mu _b^2})
\delta \pqty{1 - z_h}\, , \\
\gamma _{q \to q}^{\nu , T}
&=
\frac{\alpha _s}{\pi} C_F
\ln(\frac{\mu ^2}{\mu _b^2})
\delta \pqty{1 - z_h} \,. 
\end{align}
\endgroup
It is worth pointing out that so far we have been working in the $\overline{\mathrm{MS}}$ scheme, which is different from the simplest minimal subtraction scheme by insertion of a factor $S_{\epsilon} = \pqty{4 \pi e^{- \gamma _E}}^{\epsilon}$.

\subsection{TMD JFFs in exclusive jet productions}
\label{ss.exclusive_TMDJFFs}
\subsubsection{Factorization}
Within this section, we study the distribution of the energy fraction $z_h$ and transverse momentum $j_{\perp}$ of hadrons found within an exclusive jet. In the region of small $j_{\perp}$, where $j_{\perp}$ is much less than the product of the transverse momentum of the jet and its size, i.e., $j_{\perp} \ll p_{J T} R$, the exclusive transverse momentum dependent (TMD) jet fragmentation function $\tilde{\mathscr{G}}_c^h\left(z_h, p_{J T} R, {j}_{\perp}, \mu\right)$ encompasses contributions from both collinear and collinear-soft modes, as was previously demonstrated by~\cite{Bain:2016rrv}. Through applying the factorization introduced in~\cite{Kang:2017glf}, the factorized formalism of the exclusive TMD jet fragmentation function $\tilde{\mathscr{G}}_c^h\left(z_h, p_{J T} R, j_{\perp}, \mu\right)$ can be obtained. Without measuring any out-of-jet radiation, one has
\begin{align}
\tilde{\mathscr{G}}_i^h\left(z_h, p_{J T} R, {j}_{\perp}, \mu\right)=\int \dd[2]{\boldsymbol{k}_{\perp}} \dd[2]{\boldsymbol{\lambda}_{\perp}} \delta^2\left(z_h \boldsymbol{\lambda}_{\perp}+\boldsymbol{k}_{\perp}-\boldsymbol{j}_{\perp}\right) D^{h / i}\left(z_h, \boldsymbol{k}_{\perp}, \mu, v\right) S_i\left(\boldsymbol{\lambda}_{\perp}, \mu, v R\right)\,.\label{eq:exclusive_TMDJFFs}
\end{align}
Here the transverse component of the hadron momentum relative to the jet direction is denoted by $\boldsymbol{j}_{\perp}$. The collinear mode is described using the conventional transverse momentum dependent fragmentation functions (TMDFFs) $D^{h / i}\left(z_h, \boldsymbol{k}_{\perp}, \mu, \nu\right)$, whereas the collinear-soft mode is captured by the soft function $S_i\left(\boldsymbol{\lambda}_{\perp}, \mu, \nu R\right)$. In addition to the usual renormalization scale $\mu$, the scale $\nu$ is again related to the rapidity divergence. It is worth noting the difference between the above refactorization and those for TMD hadron distribution inside a single inclusive jet produced in proton-proton collisions, $p+p \rightarrow \left(\mathrm{jet}  h\right)+X$, in~\cite{Kang:2017glf}, where an additional hard factor emerges that captures out-of-jet radiation with a characteristic scale of around $\sim p_{J T} R$. In this study of exclusive jet production in the back-to-back region, any out-of-jet radiation is prohibited.

As is common practice in TMD physics, we convert the aforementioned expression from transverse momentum space to coordinate $b$-space using the following transformation.
\begin{align}
\tilde{\mathscr{G}}_i^h\left(z_h, p_{J T} {R}, {j}_{\perp}, \mu\right)=\int \frac{\dd[2]{\boldsymbol{b}}}{(2 \pi)^2} e^{i \boldsymbol{j}_{\perp} \cdot \boldsymbol{b} / z_h} D^{h / i}\left(z_h, \boldsymbol{b}, \mu, \nu\right) S_i(\boldsymbol{b}, \mu, \nu R),\label{eq:eq19}
\end{align}
where we have defined the Fourier transform for both $D^{h / i}\left(z_h, \boldsymbol{b}, \mu, \nu\right)$ and $S_i(\boldsymbol{b}, \mu, \nu R)$ as
\begin{align}
D^{h / i}\left(z_h, \boldsymbol{b}, \mu, \nu\right) & =\frac{1}{z_h^2} \int \dd[2]{\boldsymbol{k}_{\perp}} e^{-i \boldsymbol{k}_{\perp} \cdot \boldsymbol{b} / z_h} D^{h / i}\left(z_h, \boldsymbol{k}_{\perp}, \mu, \nu\right) \\
S_i(\boldsymbol{b}, \mu, \nu R) & =\int \dd[2]{\bm\lambda}_{\perp} e^{-i \boldsymbol{\lambda}_{\perp} \cdot \boldsymbol{b}^{\prime}} S_i\left({\bm \lambda}_{\perp}, \mu, \nu R\right)
\end{align}
In~\cite{Kang:2017glf}, the perturbative results up to next-to-leading order and the renormalization for both $D^{h / i}\left(z_h, \boldsymbol{b}, \mu, \nu\right)$ and $S_i(\boldsymbol{b}, \mu, \nu R)$ were thoroughly investigated. The ``proper'' in-jet TMD fragmentation function $\mathcal{D}^{h / i,R}$ was defined in that study as well:
\begin{align}
\mathcal{D}^{h / i,R}\left(z_h, \boldsymbol{b}, \mu\right)=D^{h / i}\left(z_h, \boldsymbol{b}, \mu, \nu\right) S_i(\boldsymbol{b}, \mu, \nu R)\,,
\end{align}
where rapidity divergence $\nu$ between $D^{h / i}\left(z_h, \boldsymbol{b}, \mu, \nu\right)$ and $S_i(\boldsymbol{b}, \mu, \nu R)$ cancels out, resulting in no rapidity divergence and, consequently, no $\nu$-dependence on the left-hand side. It was also discovered that $\mathcal{D}^{h / i, R}$ evolves according to the following equation,
\begin{align}
\mathcal{D}^{h / i,R}\left(z_h, \boldsymbol{b}, \mu\right) & =\hat{D}^{h / i}\left(z_h, \boldsymbol{b}, \mu_J\right) \exp \left[\int_{\mu_j}^\mu \frac{\dd{\mu^{\prime}}}{\mu^{\prime}}\left(-2 \Gamma_{\text {cusp }}^i\left(\alpha_s\right) \ln\left(\frac{p_{JT}R}{\mu}\right)+\gamma^i\left(\alpha_s\right)\right)\right] \nnu
& =\hat{D}^{h / i}\left(z_h, \boldsymbol{b}, \mu_J\right) \exp \left[\int_{\mu_J}^\mu \frac{\dd{\mu^{\prime}}}{\mu^{\prime}} \gamma_J^i\left(\mu^{\prime}\right)\right]
\end{align}
The aforementioned equation is valid when $\mu_J=p_{J T} R$, and $\hat{D}^{h / i}\left(z_h, \boldsymbol{b}, \mu_J\right)$ refers to the ``properly"-defined transverse momentum dependent fragmentation functions (TMDFFs), which are typically measured in semi-inclusive deep inelastic scattering and/or back-to-back hadron pair production in $e^{+} e^{-}$ collisions~\cite{Collins:2011zzd}. By substituting this result into \cref{eq:eq19}, we obtain ,
\begin{align}
\tilde{\mathscr{G}}_i^h\left(z_h, p_{J T} R, {j}_{\perp}, \mu\right) & =\left[\int \frac{\dd[2]{\boldsymbol{b}}}{(2 \pi)^2} e^{i\frac{  \boldsymbol{j}_{\perp}\cdot \boldsymbol{b}}{z_h}} \hat{D}^{h / i}\left(z_h, \boldsymbol{b}, \mu_J\right)\right] \exp \left[\int_{\mu_J}^\mu \frac{\dd{\mu^{\prime}}}{\mu^{\prime}} \gamma_J^i\left(\mu^{\prime}\right)\right], \nnu
& \equiv {D}^{h / i}\left(z_h, {j}_{\perp}, \mu_J\right) \exp \left[\int_{\mu_J}^\mu \frac{\dd{\mu^{\prime}}}{\mu^{\prime}} \gamma_J^i\left(\mu^{\prime}\right)\right]\, .\label{eq:exclu-rg2}
\end{align}
An important observation is that the exponential factor in the evolution equation is the same for the jet function $J_i\left(p_{J T} R, \mu\right)$ described in previous studies~\cite{Jain:2011xz,Waalewijn:2012sv,Ellis:2010rwa,Kang:2019ahe}, the collinear jet fragmentation function $\mathscr{G}_i^h\left(z_h, p_{J T} R, \mu\right)$ presented in Eq.\cref{eq:exclu-rg}, and the TMD jet fragmentation function $\tilde{\mathscr{G}}_i^h\left(z_h, p_{J T} R, j_{\perp}, \mu\right)$ in Eq.\cref{eq:exclu-rg2}. This implies that the renormalization group equation is the same for all of them\footnote{It is worth noting that this factor differs from that of the hadron distribution inside jets for single inclusive jet production, which follows time-like DGLAP equations and has been extensively studied in previous works~\cite{Kang:2017yde,Kang:2016ehg,Kang:2017glf,Kang:2017mda,Cal:2019hjc}.}. We employ the same parametrization as presented in~\cite{Kang:2017glf} the proper TMD fragmentation functions ${D}^{h / i}\left(z_h, j_{\perp}, \mu_J\right)$,
\begin{align}
{{D}}^{h / i}\left(z_h, j_{\perp}, \mu_J\right)=\frac{1}{z_h^2} \int \frac{b\dd{b}}{2 \pi} J_0\left(\frac{j_{\perp} b}{z_h}\right) C_{j \leftarrow i} \otimes D^{h / j}\left(z_h, \mu_{b_*}\right) e^{-S_{\mathrm{pert}}^i\left(b_*, \mu_s\right)-S_{\mathrm{NP}}^{\mathrm{i}}(b, \mu J)}\,.\label{eq:eq25}
\end{align}
In order to avoid the Landau pole of the strong coupling $\alpha_s$~\cite{Collins:1984kg}, we have employed the $b_*$-prescription. The coefficient functions $C_{j \leftarrow i}$, the perturbative Sudakov factor $S_{\mathrm{pert}}^i\left(b_*, \mu_J\right)$, and the non-perturbative Sudakov factor $S_{\mathrm{NP}}^i\left(b, \mu_J\right)$ are defined in~\cite{Kang:2017glf}. The expressions for these factors are computed at next-to-leading order for $C_{j \leftarrow i}$ and at the next-to-leading logarithmic level for $S_{\mathrm{pert}}^i\left(b_*, \mu_J\right)$. The integration in~\cref{eq:eq25} involves an oscillating Bessel function $J_0$, which we handle using an optimized Ogata quadrature method developed in~\cite{Kang:2019ctl} to ensure better numerical convergence and reliability.

\subsubsection{Connecting TMD JFFs to TMD FFs}\label{sec:appTMDFFs}
In \cref{eq:exclusive_TMDJFFs}, we wrote down the relation between the $\tilde{\mathscr{G}}_i^h\left(z_h, p_{ T} R, {j}_{\perp}, \mu\right)$ and the unpolarized TMDFF in $j_\perp \ll p_T R$ region. To write down explicit relations between other TMDJFFs and TMDFFs, we start by recalling the parametrization of the TMDFF correlator~\cite{Metz:2016swz} in the momentum space,
\begin{align}
{\Delta}\left(z_h, \boldsymbol{k}_\perp,S_h\right) =& \sum_X \int \frac{d \xi^{+}d^2{\bm \xi}_T}{(2 \pi)^3} e^{i (k^{-} \xi^{+}+{\bm k}_{\perp}\cdot{\bm \xi}_T)/z_h}\left\langle 0\left| \psi_{q}\left(\xi^{+}, 0^{-}, {\bm \xi}_{T}\right)\right| p_h, S_{h} ; X\right\rangle \nnu
& \times\left\langle p_h, S_{h} ; X\left|\bar{\psi}_{q}\left(0^{+}, 0^{-}, {\bm 0}_{T}\right) \right| 0\right\rangle \,,
\end{align}
where $\boldsymbol{k}_\perp$ is the transverse momentum of the final hadron $h$ with respect to the fragmenting quark $q$ and we suppress the Wilson lines that make the correlator gauge invariant. To the leading twist accuracy, the parametrization has been given in~\cref{eq:TMDFF} and the TMDFF correlator in $\bm{b}$-space is given in~\cref{eq:TMDFFb}

Working with an assumption that soft function is independent of the polarization, we can now write down the relations between all of the TMDJFFs and TMDFFs. We find for a general TMDJFF $\tilde{\mathscr{F}}$ that
\begin{align}
\tilde{\mathscr{F}}^{h/q}(z_h,j_\perp,\mu, \zeta_J) &= \int \frac{b^{n+1}\,db}{2\pi n!}\left(\frac{z_h^2M_h^2}{j_\perp}\right)^n J_n\left(\frac{j_\perp b}{z_h}\right) \tilde{F}^{h/q(n),{\rm unsub}}(z_h,b^2,\mu,\zeta'/\nu^2)\tilde{S}_q(b^2,\mu,\nu \mathcal{R})\nonumber\\
&=\int \frac{b^{n+1}\,db}{2\pi n!}\left(\frac{z_h^2M_h^2}{j_\perp}\right)^n J_n\left(\frac{j_\perp b}{z_h}\right) \tilde{F}^{h/q(n)}(z_h,b^2,\mu,\zeta'\mathcal{R}^2) 
\nnu
&= F^{h/q}(z_h,j_\perp^2,\mu,\zeta_J)\,.
\label{eq:TMDJFFrel}
\end{align}
The values of $n$ on the right-hand-side of \cref{eq:TMDJFFrel} follows the $n$ values of the parametrization given in \cref{eq:TMDFFb}. Therefore, all of the TMDJFFs are equal to their corresponding TMDFF at the scale $\zeta_J$. As the TMD evolutions are assumed to be polarization independent, we follow the same parametrization as that of the unpolarized TMDFF to include evolution effects for the other TMDFFs.


\chapter{Applications of JFFs}\label{sec:ajff}
\begin{quote}
\rule{0.875\textwidth}{0.5pt}\\
The Polarized Jet Fragmentation Functions (JFFs) introduced in the previous chapter has rich applications in high-energy physics. Specifically, we present the phenomenological example of polarized JFFs in two important processes: Inclusive jet production in proton-proton collisions at RHIC kinematics and back-to-back electron-jet production in electron-proton collisions at the future Electron-Ion Collider. We show how the polarized JFFs can be used to understand the spin structure of hadrons in these processes. This chapter intends to elucidate the applications of polarized Jet Fragmentation Functions (JFFs) to these pivotal processes and underscore their significance for forthcoming research in spin and TMD physics.\\
\rule{0.875\textwidth}{0.5pt}
\end{quote}

\section{Inclusive jet in $pp$ collision}
\subsection{Theoretical framework}
\begin{figure}[htb]
    \centering
    \includegraphics[width=3.8in]{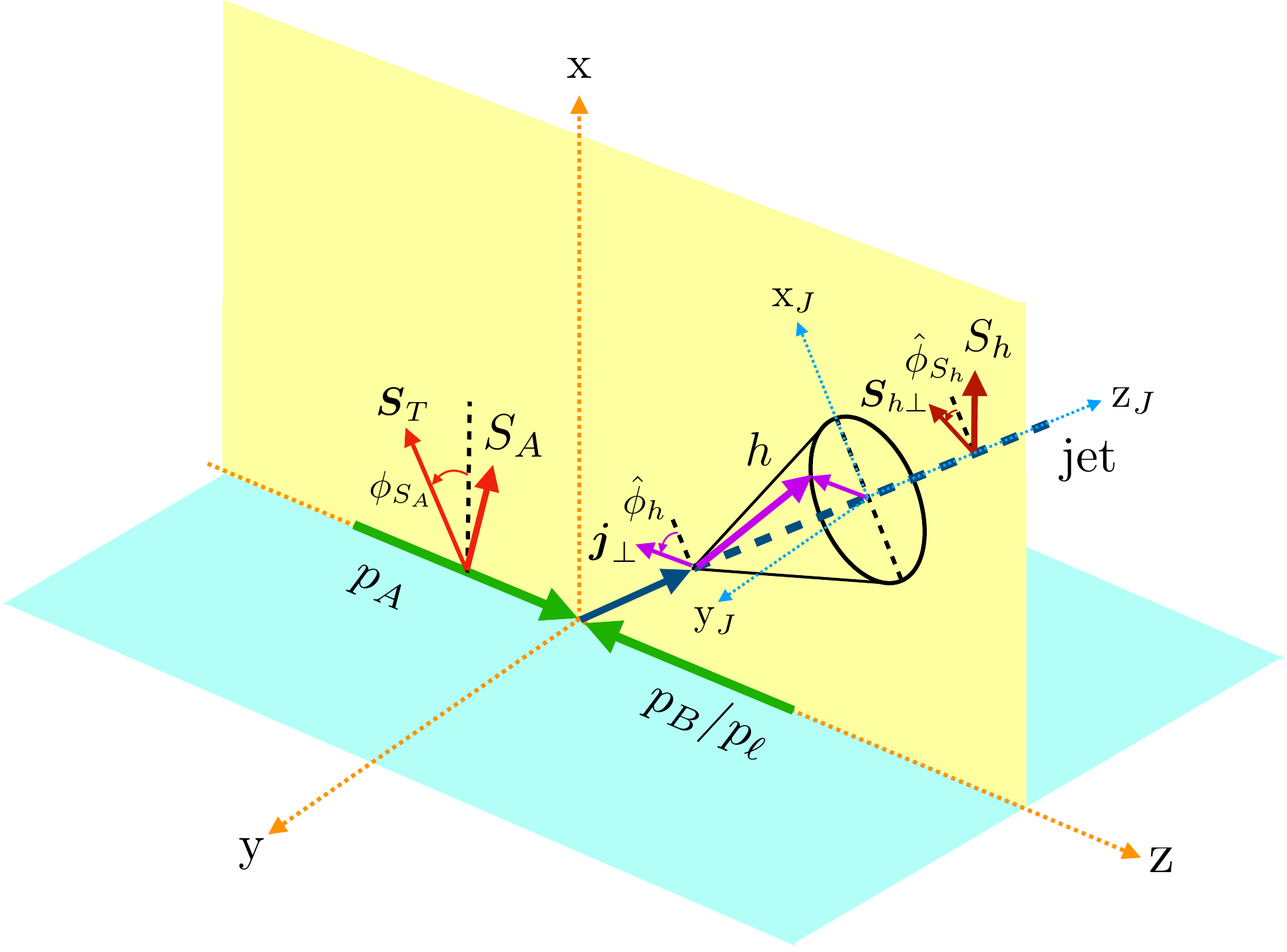}   
    \caption{Illustration for the distribution of hadrons inside jets in the collisions of a polarized proton and an unpolarized proton or lepton, where $S_A$ indicates the spin of the incoming proton, $S_{h}$ is the spin of the produced hadron in a jet. The jet axis and colliding direction define the $xz$-plane.}
    \label{fig:illu}
\end{figure}
We first consider hadron distribution inside jets in lepton-proton or proton-proton collisions as illustrated in \cref{fig:illu},
\begin{align}
p(p_A,{S}_{A}) + \Big(p(p_B) / e(p_\ell)\Big) \rightarrow \left(\text{jet}(\eta_J, p_{JT}, R)\ h(z_h,{\bm j}_\perp,S_h)\right)+X\,.
\end{align}
In this scenario, we consider a collision involving a polarized proton with spin $S_A$ and momentum $p_A$ moving in the positive $z$ direction. This polarized proton scatters off an unpolarized proton (or lepton) with momentum $p_B$ ($p_\ell$), which is moving in the negative $z$ direction. During this collision, a jet is reconstructed using the standard anti-$k_T$ algorithm~\cite{Cacciari:2008gp} with specific parameters such as the jet radius $R$, rapidity $\eta_J$, and transverse momentum $p_{JT}$. Within this jet, we observe a hadron with spin $S_h$, which carries a longitudinal momentum fraction $z_h$ of the jet and has a transverse momentum $\bm{j}_\perp$ with respect to the jet axis. Our investigation focuses on studying this process in the center-of-mass (CM) frame of the $ep$ collision. In this frame, the incoming momenta $p{A,B}$ and the proton spin vector $S_A$ can be expressed as follows: 
\bea
{p}_A^\mu=&\sqrt{\frac{s}{2}} n_a^\mu + \mathcal{O}(M)\,,\label{eq:light-cone_pA}\\ 
{p}_B^\mu=&\sqrt{\frac{s}{2}} n_b^\mu + \mathcal{O}(m_e)\,,\label{eq:light-cone_pB}\\
S_A^\mu =&\left[\lambda_p \frac{p_A^+}{M},-\lambda_p \frac{M}{2p_A^+},{\bm S}_T\right]\,.
\label{eq:light-cone_SA}
\eea
Here the center-of-mass energy is denoted by $s=\left(p_A+p_B\right)^2$, and we use $\lambda_p$ and $\bm S_T$ to represent the helicity and transverse spin vector of the incoming proton, respectively. The masses of the proton and electron are denoted by $M$ and $m_e$, respectively. We also define the standard quantities $Q^2 = -(p_B-p_D)^2$, representing the virtuality of the exchanged photon, and the event inelasticity $y= Q^2/\left(x_B s\right)$, where $x_B$ is the standard Bjorken-$x$~\cite{Bacchetta:2006tn}.

In this chapter, we focus on the measurement of the jet fragmentation function in an inclusive jet sample in proton-proton (or lepton-proton) collisions. This measurement involves the summation over all particles in the final state $X$ besides the observed jet. Hence, we are probing the semi-inclusive Jet Fragmentation Functions (JFFs) discussed in the previous section. However, for exclusive jet production, such as the back-to-back lepton-jet in lepton-proton collisions or $Z$-jet correlations in proton-proton collisions, one should utilize the exclusive version of the JFFs. More detailed information can be found in references~\cite{Kang:2019ahe,Liu:2018trl,Buffing:2018ggv}.

For single inclusive jet production, the most general azimuthal dependence of the hadron distribution inside the jet, differential in both $z_h$ and ${\bm j}_\perp$, can be expressed as follows:
\begingroup
\allowdisplaybreaks
\begin{align}
\label{eq:gen_str}
\frac{d\sigma^{p(S_A)+p/e\to (\text{jet}\,h(S_h))X}}{dp_{JT} d\eta_{J} dz_h d^2{\bm j}_\perp}
=&F_{UU,U}+|{\bm S}_T|\sin(\phi_{S_A}-\hat{\phi}_h)F^{\sin(\phi_{S_A}-\hat{\phi}_h)}_{TU,U}\nnu
&+\Lambda_h\left[\lambda_p F_{LU,L}+|{\bm S}_T|\cos(\phi_{S_A}-\hat{\phi}_h)F^{\cos(\phi_{S_A}-\hat{\phi}_h)}_{TU,L}\right]\nnu
&+|{\bm S}_{h\perp}|\bigg[\sin(\hat{\phi}_h-\hat{\phi}_{S_h})F^{\sin(\hat{\phi}_h-\hat{\phi}_{S_h})}_{UU,T}+\lambda_p\cos(\hat{\phi}_h-\hat{\phi}_{S_h}) F^{\cos(\hat{\phi}_h-\hat{\phi}_{S_h})}_{LU,T}\nnu
&\hspace{1.6cm}+|{\bm S}_T|\cos(\phi_{S_A}-\hat{\phi}_{S_h})F^{\cos(\phi_{S_A}-\hat{\phi}_{S_h})}_{TU,T}\nnu
&\hspace{1.6cm}+|{\bm S}_T|\cos(2\hat{\phi}_h-\hat{\phi}_{S_h}-\phi_{S_A})F^{\cos(2\hat{\phi}_h-\hat{\phi}_{S_h}-\phi_{S_A})}_{TU,T}\bigg]\,.
\end{align}
\endgroup
Here $F_{AB, C}$ represents the spin-dependent structure functions, where $A$, $B$, and $C$ indicate the polarizations of the incoming proton $A$, incoming proton $B$ (or electron), and the fragmented hadron inside the jet, respectively. The parameters $\lambda_p$ and $\abs{\boldsymbol{S}_{T}}$ denote the longitudinal and transverse spin of the initial polarized hadron, while $\Lambda_h$ and $\abs{\boldsymbol{S}_{h\perp}}$ represent the longitudinal and transverse polarization of the hadron inside the jet, measured in the fragmenting parton helicity frame.

In \cref{eq:gen_str}, the structure function $F_{UU, U}(z_h, j_{\perp})$ is defined by the expression:
\begin{align}
F_{UU, U} (z_h, j_{\perp}) & =
\frac{\alpha _s^2}{s} \sum _{a, b, c}
\int _{x_1^{\min}}^1 \frac{\dd{x_1}}{x_1} f_1^{a/A}(x_1, \mu)
\int _{x_2^{\min}}^1 \frac{\dd{x_2}}{x_2} f_2^{b/B}(x_2, \mu)
\nonumber \\
& \quad \times
\int _{z^{\min}}^1 \frac{\dd{z}}{z^2}
\hat{H}_{ab}^c \pqty{\hat{s}, \hat{p}_T, \hat{\eta}, \mu}
\mathcal{D}_1^{h/c} \pqty{z, z_h, j_{\perp}^2, Q}
\nonumber \\
& \equiv \mathcal{C} \bqty{f f \mathcal{D}_1 \hat{H}}\, , \label{e.F_UUU}
\end{align}
where $f_1^{a/A} (x_1, \mu)$ and $f_2^{b/B} (x_2, \mu)$ are the collinear unpolarized PDFs, $\mathcal{D}_1^{h/q (g)} (z_h, j_{\perp}^2, Q)$ is the unpolarized TMD JFF and $\hat{H}_{ab}^c$ is the hard function for unpolarized parton $a$, $b$ to unpolarized parton $c$.
The lower integration limits $x_1^{\min}$, $x_2^{\min}$ and $z^{\min}$ can be found in \cite{Kang:2016ehg, Kaufmann:2015hma}.
The variables $\hat{s}$, $\hat{p}_T$ and $\hat{\eta}$ are the squared parton center of mass energy, transverse momentum and rapidity of parton $c$, respectively, and are related to their hadron analogues as:
\begin{equation}
\hat{s} = x_1 x_2 s \,,
\quad
\hat{p}_T = p_T / z\, ,
\quad
\hat{\eta} = \eta - \frac{1}{2} \ln(\frac{x_1}{x_2})\, .
\end{equation}
In \cref{e.F_UUU}, we have also defined the notation $\mathcal{C} \bqty{f f \mathcal{D}_1 \hat{H}}$, where parton flavors are summed for PDFs and JFFs along with their corresponding unpolarized hard functions. Similarly to the notation for unpolarized case given by \cref{e.F_UUU}, we write down the rest structure functions in \cref{eq:gen_str}~\cite{KangLeeZhao:inprogress}:
\begingroup
\allowdisplaybreaks
\begin{align}
F^{\sin(\phi _S - \phi _h)}_{TU, U} (z_h, j_{\perp})
& =
\mathcal{C} \bqty{\frac{j_{\perp}}{z_h M_h} h_{1} f_1 \mathcal{H}_{1}^{\perp} \Delta_T \hat{H}}
\,, \\
F_{LU, L} (z_h, j_{\perp})
& =
\mathcal{C} \Big[g_{1 L} f_1 \mathcal{G}_{1 L} \Delta _L \hat{H} \Big]
\,, \\
F^{\cos(\phi _S - \phi _h)}_{TU, L} (z_h, j_{\perp})
& =
- \mathcal{C} \bqty{\frac{j_{\perp}}{z_h M_h} h_1 f_1 \mathcal{H}_{1 L}^{\perp} \Delta _T \hat{H}}
\,, \\
F^{\sin(\phi _h - \phi _{S_h})}_{UU, T} (z_h, j_{\perp})
& =
- \mathcal{C} \bqty{\frac{j_{\perp}}{z_h M_h} f_1 f_1 \mathcal{D}_{1 T}^{\perp} \hat{H}}
\,, \\
F^{\cos(\phi _h - \phi _{S_h})}_{LU, T} (z_h, j_{\perp})
& =
- \mathcal{C} \bqty{\frac{j_{\perp}}{z_h M_h} g_{1 L} f_1 \mathcal{G}_{1 T} \Delta _L \hat{H}}
\,, \\
F^{\cos(\phi _S - \phi _{S_h})}_{TU, T} (z_h, j_{\perp})
& =
\mathcal{C} \Big[h_1 f_1 \mathcal{H}_{1} \Delta _T \hat{H} \Big]
\,, \label{e.F_TUT} \\
F^{\cos(2 \phi _h - \phi _S - \phi _{S_h})}_{TU, T} (z_h, j_{\perp})
& =
- \mathcal{C} \bqty{\frac{j^2_{\perp}}{2 z_h^2 M_h^2} h_{1} f_1 \mathcal{H}_{1 T}^{\perp} \Delta_T \hat{H}}\, .
\end{align}
\endgroup
The polarization relations for hard functions $\hat{H}_{ab}^c$, $\Delta _L\hat{H}_{ab}^c$ and $\Delta _T\hat{H}_{ab}^c$ are given in \cref{t.hard_functions}.
\begin{table}
\centering
\begin{tabular}{|c|c|c|c|}
\hline
\diagbox{$c$}{$a$} & U & L & T \\
\hline
U & $\hat{H}_{ab}^c$ & & \\
\hline
L &  & $\Delta _L \hat{H}_{ab}^c$ & \\
\hline
T & & & $\Delta _T \hat{H}_{ab}^c$ \\
\hline
\end{tabular}
\caption{Hard functions for parton $a$, $b$ to parton $c$. Parton $b$ is always unpolarized.
}
\label{t.hard_functions}
\end{table}. 

In the center-of-mass frame, as depicted in \cref{fig:illu}, the azimuthal angle $\phi_{S_A}$ corresponds to the transverse spin ${\bm S}_T$ of the incoming proton. Here, we align the jet momentum in the $x$-$z$ plane. Simultaneously, we select another reference frame where the jet momentum aligns in the $+z_J$ direction, and the $x_J$ axis is in the $x$-$z$ plane. In this frame, we measure the transverse momentum ${\bm j}_\perp$ and transverse spin vector ${\bm S}_{h\perp}$ of the produced hadron inside the jet using the $x_J$-$y_J$-$z_J$ reference frame. The corresponding azimuthal angles in this frame are defined as $\hat{\phi}_h$ and $\hat{\phi}_{S_h}$.

\subsection{Collinear JFFs}
\subsubsection{General structure of the observables}
\label{sub:general}
To study collinear JFFs, one integrates over ${\bm j}_\perp$ and measures only the $z_h$-distribution of hadrons inside the jet, namely one has
\begin{align}
\label{eq:long_str}
\frac{d\sigma^{p(S_A)+p/e\to (\text{jet}\,h(S_h))X}}{dp_{JT} d\eta_{J} dz_h}=&F_{UU,U}+\Lambda_h\,\lambda_p F_{LU,L}+|{\bm S}_{h\perp}||{\bm S}_T|\cos(\phi_{S_A}-\hat{\phi}_{S_h})F^{\cos(\phi_{S_A}-\hat{\phi}_{S_h})}_{TU,T}\,.
\end{align}
The structure functions $F_{AB,C}$ in \cref{eq:gen_str} represent the collinear versions of the structure functions. When considering polarization for both incoming particles, additional terms emerge in both \cref{eq:gen_str} and \cref{eq:long_str}. In the following section, we present key practical applications, focusing on two specific investigations. Firstly, we study the longitudinal spin transfer, which is captured by the ${\bm j}_\perp$-integrated version of $F_{LU,L}$. Secondly, we analyze the transverse polarization of $\Lambda$ production, characterized by the structure function $F^{\sin(\hat{\phi}_h-\hat{\phi}_{S_h})}_{UU,T}$. Prior research has already explored the unpolarized hadron distribution in jets, described by $F_{UU,U}$\cite{Kang:2016ehg,Kang:2017glf}, and the hadron Collins asymmetry, represented by $F^{\sin(\phi_{S_A}-\hat{\phi}{h})}_{TU,U}$\cite{Kang:2017btw}.

\subsubsection{Example: Longitudinally polarized $\Lambda$}
\label{sec:helicity}
It is widely recognized in the scientific community that $\Lambda (\bar{\Lambda})$-hyperons offer a valuable platform for investigating spin-dependent fragmentation, owing to their polarizations being determinable through the dominant weak decay channel $\Lambda \to p,\pi,(\bar{\Lambda} \to \bar{p},\pi)$. Over time, numerous diverse measurements of polarized $\Lambda (\bar{\Lambda})$-hyperons have been conducted, and in this section, we present predictions concerning the longitudinal polarization of $\Lambda$ particles within jets.

\begin{figure}
    \centering
    \includegraphics[width=6in]{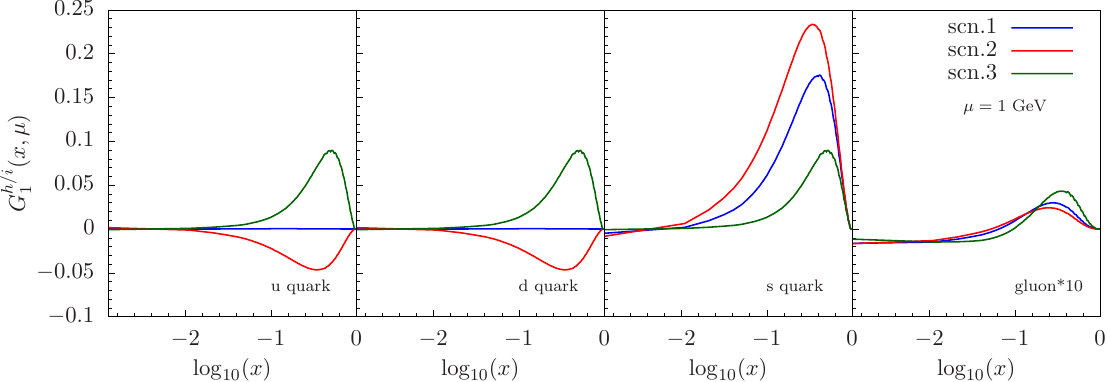}   
    \caption{The longitudinally polarized fragmentation functions plotted at $\mu = 1~\text{GeV}$ for various scenarios \cite{deFlorian:1997zj} consistent with the LEP data. Scenario 1 has only polarized $s$ quark nonvanishing, scenario 2 has $u$ and $d$ quark equal, but with opposite sign from $s$ quark, scenario 3 has $u,~d$, and $s$ quark equal to each other. In all scenarios, gluon vanishes at the input scale and is generated entirely from QCD evolution.}
    \label{fig:polFF}
\end{figure}

The longitudinal polarized $\Lambda/\bar{\Lambda}$ fragmentation functions were first determined by analyzing LEP data to NLO \cite{deFlorian:1997zj, Buskulic:1996vb} (an earlier study can be found in \cite{Burkardt:1993zh}). The analysis, however, was not able to constrain the valence fragmentation functions for all flavors, but different helicity FFs motivated by various scenarios (as shown in \cref{fig:polFF}) were found to equally describe the LEP data. Subsequently, studies of the longitudinal spin transfer $D_{LL}$ in longitudinally polarized proton-proton collisions, $\vec{p}p \to \vec{\Lambda}X$, as a function of rapidity, have been proposed to discriminate between the different scenarios of valence spin-dependent fragmentation functions. The positive rapidity region (forward region of the polarized proton) corresponds to the valence region of the polarized proton, and thus spin transfers are dominated by $u$ and $d$ quarks inside the polarized proton. Analyzing $D_{LL}$ as a function of rapidity can, therefore, distinguish the different scenarios of helicity FFs of $u$ and $d$ quarks.

In the past decade, the STAR collaboration at RHIC has conducted several measurements and analyses for such longitudinal spin transfer to $\Lambda$ and $\bar{\Lambda}$ hyperons \cite{Xu:2006my, Abelev:2009xg, Adam:2018kzl}. However, the measurements have been binned in only two rapidity bins, the negative ($-1.2 < \eta_J < 0$) and positive ($0 < \eta_J < 1.2$) bins (at the most recent measurement \cite{Adam:2018kzl}), and have not yet been able to discriminate between the different scenarios proposed in \cite{deFlorian:1997zj}.

To achieve more continuous binning in the transverse momentum $p_{\Lambda T}$ of the $\Lambda$ particle, we propose the use of the relevant collinear helicity JFF to directly scan through the helicity FFs. This can be accomplished by studying the $z_\Lambda = p_{\Lambda T}/p_{J T}$ distribution of longitudinally polarized $\Lambda$ particles inside a jet in longitudinally polarized proton-proton collisions, $\vec{p}+p\to (\text{jet}\vec{\Lambda})+X$. We can define an analogous longitudinal spin transfer $D_{LL}^{\text{jet}\Lambda}$ for $\Lambda$ polarization in the jet as follows
\begin{align}
\label{eq:assyjet}
D_{LL}^{\text{jet}\Lambda} = \frac{d\Delta\sigma^{\vec{p}p\to  (\mathrm{jet}\vec{\Lambda})X}}{dp_{JT}d\eta_J dz_\Lambda}   \Bigg/ \frac{d\sigma^{pp\to  (\mathrm{jet}\Lambda)X}}{dp_{JT}d\eta_J dz_\Lambda} = \frac{F_{LU,L}}{F_{UU,U}}  \,,
\end{align}
where the structure functions ${F_{LU,L}}$ and $F_{UU,U}$ given in \cref{eq:long_str} have been applied in the above expression. For the unpolarized cross section in the denominator, factorization is given by~\cite{Kang:2016ehg}
\begin{align}
\frac{d\sigma^{pp\to  (\mathrm{jet}\Lambda)X}}{dp_{JT}d\eta_{J} dz_\Lambda} =\sum_{a,b,c}\int_{x_a^{\text{min}}}^1\frac{dx_a}{x_a}f_a(x_a,\mu)\int_{x_b^{\text{min}}}^1\frac{dx_b}{x_b}f_b(x_b,\mu)\int_{z_c^{\text{min}}}^1\frac{dz_c}{z_c^2}H_{ab}^c \,\mathcal{G}_c^{h}(z_c,z_\Lambda,p_{JT} R,\mu)\,.
\end{align}
On the other hand, the numerator of \cref{eq:assyjet} is defined as $d\Delta\sigma = [d\sigma(+,+) - d\sigma(+,-)]/2$ with the first and second index denoting the helicities $\lambda_p$ and $\Lambda_\Lambda$ respectively. Then the numerator of \cref{eq:assyjet} can be written in the following factorized form
\begin{align}
\label{eq:longfact}
\frac{d\Delta\sigma^{\vec{p}p\to  (\mathrm{jet}\vec{h})X}}{dp_{JT}d\eta_J dz_\Lambda}= \sum_{a,b,c}\int_{x_a^{\text{min}}}^1\frac{dx_a}{x_a}g_a(x_a,\mu)\int_{x_b^{\text{min}}}^1\frac{dx_b}{x_b}f_b(x_b,\mu)\int_{z_c^{\text{min}}}^1\frac{dz_c}{z_c^2}\,\Delta_{LL}H_{ab}^{c} \,\Delta\mathcal{G}_c^{h}(z_c,z_\Lambda,p_{JT}R,\mu)\,.
\end{align}
Here $f_{a}(x_a, \mu)$ and $g_a(x_a, \mu)$ are the unpolarized PDFs and helicity parton distribution functions, and $H_{ab}^c$ (or $\Delta_{LL} H_{ab}^c$) are the corresponding hard functions, respectively. Finally, $\mathcal{G}_c^{h}$ and $\Delta\mathcal{G}_c^{h}$ are the relevant unpolarized collinear JFFs and helicity JFFs defined in the previous section. Also note that $\mathcal{G}_1^{h/c}$ can be matched onto the standard helicity FFs as 
\begin{align}
\label{eq:longmatch}
\Delta\mathcal{G}_i^{h}(z,z_h,p_{JT} R,\mu) = \sum_j\int_{z_h}^1 \frac{dz_h'}{z_h'}\Delta\mathcal{J}_{ij}(z,z_h',p_{JT} R,\mu)\,\Delta D_j^{h}\left(\frac{z_h}{z_h'},\mu\right)\,,
\end{align}
where the matching coefficients $\Delta\mathcal{J}_{ij}$ has been provided in \cref{eq:jqq}-\cref{eq:jgg} at the next-to-leading order (NLO).

In the following section, we present numerical predictions for the production of longitudinally polarized $\Lambda$ particles within jets, considering the kinematic conditions of both the RHIC and EIC experiments. To describe the collinear helicity parton distributions and unpolarized PDFs, we make use of the polarized and unpolarized NNPDF sets~\cite{ball2013unbiased, ball2013parton}. As for the fragmentation functions for $\Lambda$, including both unpolarized and longitudinally polarized cases, we adopt the leading-order set provided in \cite{deFlorian:1997zj}. This set includes three distinct scenarios, visually represented in \cref{fig:polFF}.

\begin{figure}[htb]
\centering
\begin{subfigure}{.5\textwidth}
  \centering
  \includegraphics[width=.85\linewidth]{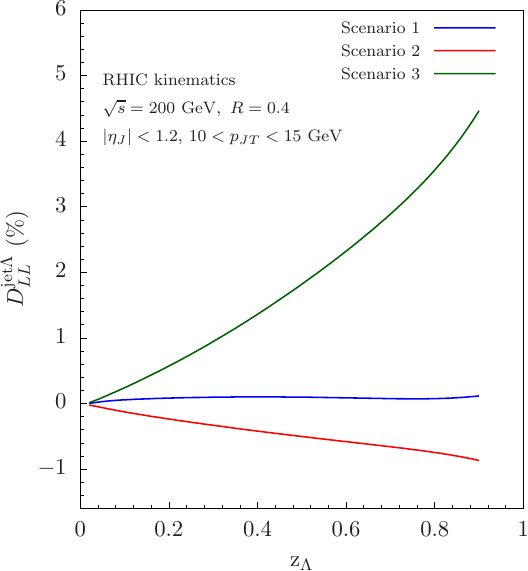} 
  \caption{}
    \label{fig:DLLpp}
\end{subfigure}%
\begin{subfigure}{.5\textwidth}
  \centering
  \includegraphics[width=.85\linewidth]{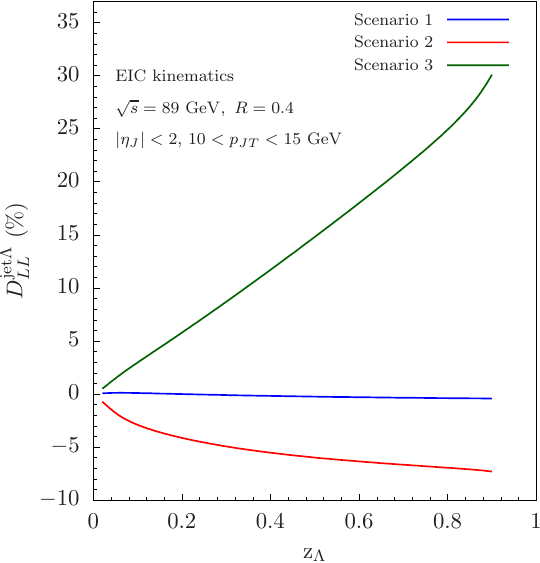} 
  \caption{}
  \label{fig:DLLep}
\end{subfigure}
\caption{Calculated results for the asymmetry $D_{LL}^{\text{jet}\Lambda}$ at both the RHIC (left) and EIC (right) experimental setups, considering various scenarios proposed in~\cite{deFlorian:1997zj} and depicted in \cref{fig:polFF}. The predictions we have obtained demonstrate the discriminative power of $D_{LL}^{\text{jet}\Lambda}$ in distinguishing between the different proposed scenarios.}
\end{figure}

In line with the recent measurement by STAR at RHIC~\cite{Adam:2018kzl}, our initial investigation focuses on studying the polarization of $\Lambda$ particles within jets. The jets are reclustered using the anti-$k_T$ algorithm with a jet radius $R=0.4$ in proton-proton collisions at a center-of-mass energy of $\sqrt{s}=200$ GeV. The predictions for $D_{LL}^{\text{jet}\Lambda}$ are presented in \cref{fig:DLLpp}, where the jet transverse momentum and rapidity are integrated over the ranges $10<p_{JT}<15$ GeV and $|\eta_J|<1.2$, respectively. The observed asymmetry is found to be on the order of a few percent, and the distinct behavior in the spin asymmetry arises from polarized Fragmentation Functions (FFs) within different scenarios. This behavior can be attributed to the dominant contributions of $u$ and $d$ quarks in this kinematic region, leading $D_{LL}^{\text{jet}\Lambda}$ to follow the signs of the corresponding helicity FFs $G_1^{\Lambda/u,d}$. Interestingly, as $z_\Lambda$ directly probes the polarized FFs in \cref{fig:polFF}, there is no need for further differential information to discriminate between the three scenarios. As a result, this newly proposed observable has the capability to differentiate among the scenarios without requiring binning in $\eta_J$, as seen in the case of single inclusive $\Lambda$ production~\cite{deFlorian:1998ba}.

Expanding our investigation, we now study longitudinally polarized $\Lambda$ production at the EIC with a center-of-mass energy of $\sqrt{s}=89$ GeV. We consider $R=0.4$ anti-$k_T$ jets with transverse momentum and rapidity in the range $10<p_{JT}<15$ GeV and $|\eta_J|<2$, respectively. Notably, due to the enhancement from the leading-order $eq \to eq$ process, $D_{LL}^{\text{jet}\Lambda}$ reaches magnitudes on the order of tens of percent at the EIC, as demonstrated in \cref{fig:DLLep}. This significant enhancement ensures the clear discrimination of different scenarios for helicity FFs under the experimental conditions of the EIC.

\subsection{TMD JFFs}
\subsubsection{Example: Transversely polarized $\Lambda$ from unpolarized scatterings}
\label{sec:PFFs}
To study the distribution of hadrons inside the jet with respect to their transverse momentum ${\bm j}_\perp$ and any spin-dependent correlations, it is crucial to use Transverse Momentum Dependent Jet Fragmentation Functions (TMDJFFs). For instance, if the goal is to measure both unpolarized and transversely polarized $\Lambda$ production within the jet in the context of unpolarized proton-proton or proton-lepton collisions, such as $p+\left(p/e\right)\to (\text{jet}{\Lambda^\uparrow})+X$, the TMDJFFs $\mathcal{D}_{1}^{h/i}(z, z_h, {\bm j}_\perp, p{JT}R, \mu)$ and $\mathcal{D}_{1T}^{\perp, h/i}(z, z_h, {\bm j}_\perp, p_{JT}R, \mu)$ are required.

Recent experimental efforts by the Belle Collaboration have successfully measured transversely polarized $\Lambda$ particles in the back-to-back production of $\Lambda$ and a light hadron during $e^+ e^-$ collisions, i.e., $e^+ e^- \to \Lambda^\uparrow + h + X$ \cite{Guan:2018ckx}. Subsequently, the corresponding TMD Parton Fragmentation Functions have been extracted from these measurements in \cite{Callos:2020qtu,DAlesio:2020wjq}.

In a previous work \cite{Kang:2017glf}, it was demonstrated that for the scattering illustrated in \cref{fig:illu}, the unpolarized TMDJFF $\mathcal{D}_{1}^{h/i}$ is related to the standard TMDFF $D_{1}^{h/i}$. Expanding on the same approach, we present only the final results here, while leaving the detailed analysis for a future publication. Notably, the evolution of $\mathcal{D}_{1T}^{\perp h/i}$ involves an unpolarized parton that initiates the jet, implying that its evolution follows the same DGLAP evolution equations as the unpolarized case.
\begin{align}
\mu\frac{d}{d\mu}\mathcal{D}_{1T}^{\perp\,h/i}(z,z_h,{\bm j}_\perp, p_{JT} R, \mu) = \frac{\alpha_s(\mu)}{\pi} 
 \sum_j \int_z^1  \frac{dz'}{z'} P_{ji}\left(\frac{z}{z'}, \mu \right) \mathcal{D}_{1T}^{\perp\,h/j}(z',z_h, {\bm j}_\perp, p_{JT} R, \mu)\,,
\end{align}
The evolution of $\mathcal{D}_{1T}^{\perp,h/i}$ from the typical jet scale $\mu_J\sim p_{JT} R$ to the hard scale $\mu\sim p_{JT}$ is accomplished through resummation, effectively addressing the logarithm of the jet radius $R$. At the scale $\mu_J$, the expression for $\mathcal{D}_{1T}^{\perp,h/i}$ reads as follows:
\begin{align}
\mathcal{D}_{1T}^{\perp\, h/i}(z,z_h,{\bm j}_\perp,p_{JT} R,\mu_J) = \mathcal{C}_{i\to j}(z,p_{JT} R,\mu_J)\,D_{1T}^{\perp\, h/j}(z_h,{\bm j}_\perp;\mu_J)\,.
\label{eq:PFFs}
\end{align}
Here $D_{1T}^{\perp, h/j}$ represents the Transverse Momentum Dependent Parton Fragmentation Function (TMD PFF) that characterizes the fragmentation of an unpolarized parton into a transversely polarized hadron. The coefficient functions $\mathcal{C}_{i\to j}$ remain the same as those in the unpolarized scenario and can be found in~\cite{Kang:2017glf}.

In this phenomenological example, we concentrate on the production of transversely polarized $\Lambda$ particles within a jet, considering unpolarized proton-proton and lepton-proton collisions, i.e., $p+e/p\to (\text{jet},\Lambda^\uparrow)+ X$. This process has been proposed as a promising measurement at the LHC~\cite{Boer:2010yp,Boer:2007nh}. Our analysis involves measuring three key observables: the longitudinal momentum fraction $z_\Lambda$ carried by the $\Lambda$ particle within the jet, the transverse momentum ${\bm j}_\perp$ relative to the jet direction, and the transverse spin ${\bm S}_{h\perp}$ of the $\Lambda$ particle. As emphasized earlier in \cref{sub:general}, the transverse momentum ${\bm j}_\perp$ and the transverse spin ${\bm S}_{h\perp}$ of the $\Lambda$ particle exhibit a correlation, giving rise to an azimuthal dependence of the form $\sin(\hat{\phi}_h - \hat{\phi}_{S_h})$,
\begin{align}
\frac{d\sigma}{dp_{JT} d\eta_{J} dz_\Lambda d^2{\bm j}_\perp}=&F_{UU,U}+|{\bm S}_{h\perp}| \sin(\hat{\phi}_h-\hat{\phi}_{S_h})F^{\sin(\hat{\phi}_h-\hat{\phi}_{S_h})}_{UU,T}+\cdots\,,
\end{align}
where ``$\cdots$'' includes all the other terms in \cref{eq:gen_str}. And we have also defined the $\Lambda$ transverse polarization inside the jet as
\begin{align}
P_\Lambda = \frac{F^{\sin(\hat{\phi}_h-\hat{\phi}_{S_h})}_{UU,T}}{F_{UU,U}}\,,
\end{align}
where one can write the following factorized form for the structure functions in the numerator and denominator above
\begin{align}
F_{UU,U}&=\sum_{a,b,c}\int_{x_a^{\text{min}}}^1\frac{dx_a}{x_a}f_a(x_a,\mu)\int_{x_b^{\text{min}}}^1\frac{dx_b}{x_b}f_b(x_b,\mu)\int_{z_c^{\text{min}}}^1\frac{dz_c}{z_c^2}H_{ab}^c \,\mathcal{D}_1^{\Lambda/c}(z_c,z_\Lambda,{\bm j}_\perp,p_{JT}R,\mu)\,,
\\
\label{eq:PFFTMDJFF}
F_{UU,T}^{\sin(\hat{\phi}_h-\hat{\phi}_{S_h})}
&=\sum_{a,b,c}\int_{x_a^{\text{min}}}^1\frac{dx_a}{x_a}f_a(x_a,\mu)\int_{x_b^{\text{min}}}^1\frac{dx_b}{x_b}f_b(x_b,\mu)\int_{z_c^{\text{min}}}^1\frac{dz_c}{z_c^2}H_{ab}^c \,\mathcal{D}_{1T}^{\perp\, \Lambda/c}(z_c,z_\Lambda,{\bm j}_\perp,p_{JT}R,\mu)\,.
\end{align}
It is worth noting that the expression for the unpolarized structure function $F_{UU,U}$ has been previously established in~\cite{Kang:2017glf}. Building upon this, we can derive the formalism for the spin-dependent structure function $F_{UU,T}^{\sin(\hat{\phi}_h-\hat{\phi}_{S_h})}$. These structure functions are sensitive to both the unpolarized TMDJFFs $\mathcal{D}_1^{\Lambda/c}$ and the spin-dependent TMDJFFs $\mathcal{D}_{1T}^{\perp \Lambda/c}$. The Renormalization Group (RG) evolution equations for these functions and their connections to the standard TMD Fragmentation Functions (TMDFFs) have been provided in \cref{sub:connection}. In particular, the TMDJFF $\mathcal{D}_{1T}^{\perp \Lambda/c}$ is related to the TMD Parton Fragmentation Function (TMD PFF) $D_{1T}^{\perp \Lambda/c}(z_h, {\bm j}_\perp; \mu_J)$ as presented in \cref{eq:PFFs}.

To estimate the transverse polarization of $\Lambda$ particles inside the jet, we adopt a specific model for the TMD PFFs. Utilizing the so-called $b_*$ prescription \cite{Collins:1984kg}, we combine TMD evolution with the recent Gaussian fit of the Belle data to parametrize $D_{1T}^{\perp \Lambda/c}$ at the jet scale $\mu_J\sim p_{JT}R$ as follows:
\begin{align}
D_{1T}^{\perp \Lambda/c}(z_\Lambda,{\bm j}_\perp;\mu_J)=\frac{1}{z^2_\Lambda}\left(\frac{1}{2z_\Lambda}\right)\int_0^{\infty}\frac{b^2\text{d}b}{2\pi}J_1\left(\frac{j_\perp b}{z_\Lambda}\right)F_c(z_\Lambda,\mu_{b_*})\ e^{-S_{\text{pert}}^i(b_*,\ \mu_J)-S_{\text{NP}}^i(b,\ \mu_J)},
\end{align}
where $S_{\text{pert}}^i$ are the usual perturbative Sudakov factors~\cite{Collins:2011zzd}, and  $F_c(z_\Lambda,\mu_{b_*})$ is fitted from the recent Belle data and has the following functional form 
\begin{align}
\label{eq:Fc}
F_c(z_\Lambda,\mu_b^*) \equiv \mathcal{N}_c(z_\Lambda)\,D_1^{\Lambda/c}(z_\Lambda,\mu_b^*)\,.
\end{align}
In the present context, the unpolarized collinear $c\to \Lambda$ fragmentation functions are denoted as $D_{1}^{\Lambda/c}(z_\Lambda, \mu_b^*)$, and the parametrization of $\mathcal{N}_c(z_\Lambda)$ for different quark flavors can be found in~\cite{Callos:2020qtu}. It is noteworthy that, as of now, there is no available extraction for the gluon Transverse Momentum Dependent Parton Fragmentation Function (TMD PFF) denoted as $D_{1T}^{\perp \Lambda/g}$. Consequently, we do not include this quantity in the numerical study below. However, we do incorporate the unpolarized gluon Transverse Momentum Dependent Fragmentation Function (TMDFF) denoted as $D_{1}^{\Lambda/g}$ in the calculation of $F_{UU,U}$. Also we set the scale $\mu_{b_*} = 2e^{-\gamma_E}/{b_*}$, where $b_* = b/\sqrt{1+b^2/b^2_{\text{max}}}$. Subsequently, we choose to parametrize the non-perturbative Sudakov factor for quark TMD PFFs using the prescription presented in~\cite{Kang:2017glf,Kang:2015msa,Su:2014wpa}.
\begin{align}
S^q_{\text{NP}}(b,\mu_J)&=\frac{g_2}{2}\ln\left(\frac{b}{b_*}\right)\ln\left(\frac{p_{JT} R}{Q_0}\right)+\frac{\langle M_D^2\rangle}{4z_\Lambda^2}b^2\,,
\end{align}
with $Q_0^2 = 2.4~\text{GeV}^2,\ b_{\text{max}} = 1.5~\text{GeV}^{-1},\ g_2 = 0.84$, and $\langle M_D^2\rangle=0.118$~GeV$^2$~\cite{Callos:2020qtu}. 

\begin{figure}
\centering
\begin{subfigure}{.33\textwidth}
  \centering
  \includegraphics[width=.99\linewidth]{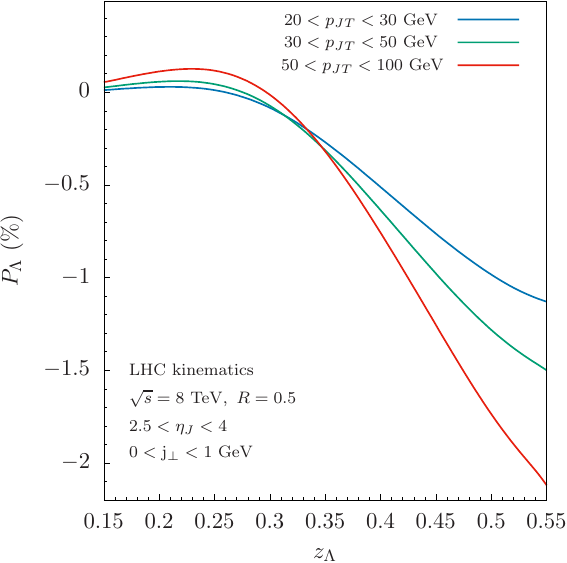} 
  \caption{}
  \label{fig:Plhcb}
\end{subfigure}
\begin{subfigure}{.33\textwidth}
  \centering
  \includegraphics[width=.99\linewidth]{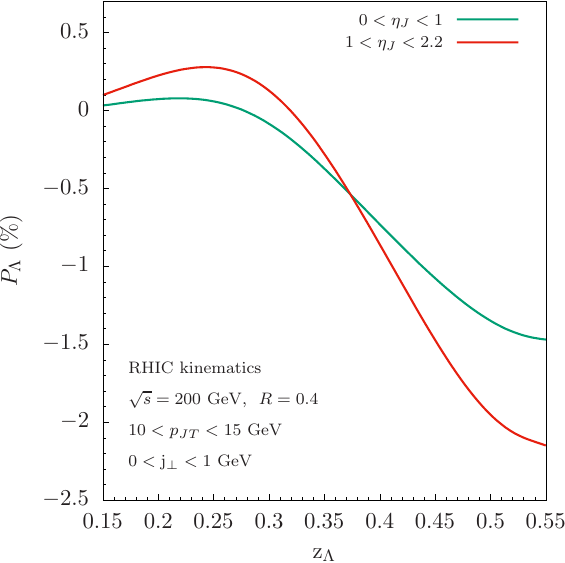} 
  \caption{}
    \label{fig:Ppp}
\end{subfigure}%
\begin{subfigure}{.33\textwidth}
  \centering
  \includegraphics[width=.99\linewidth]{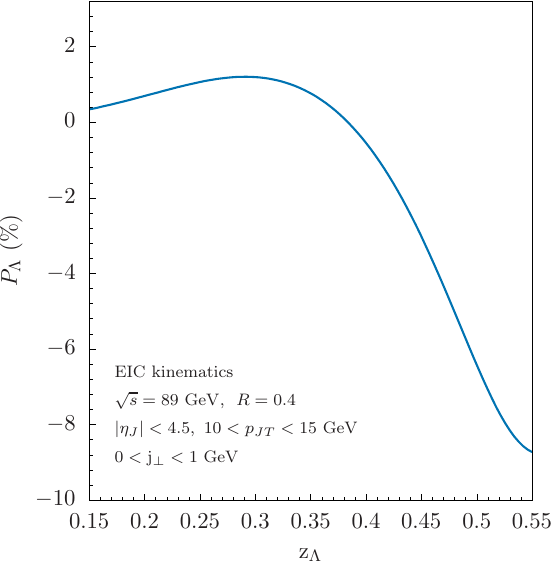} 
  \caption{}
  \label{fig:Pep}
\end{subfigure}
\caption{Predictions for the asymmetry $P_\Lambda$ for the LHC (left), RHIC (middle) and EIC (right) kinematics.}
\end{figure}

Here we present our predictions for $\Lambda$ polarization $P_\Lambda$ at various colliders, including the LHC, RHIC, and the future EIC. For the LHC, we utilize anti-$k_T$ jets with $R = 0.5$, while for the RHIC and EIC, we use $R = 0.4$. We consistently employ LO NNPDF \cite{ball2013parton} as the PDF sets and the AKK08 \cite{Albino:2008fy} parametrization of the $\Lambda$ fragmentations to remain consistent with the extraction of TMD Parton Fragmentation Functions from~\cite{Callos:2020qtu}. For the LHC, we adopt the kinematics used in the recent LHCb measurements for the distribution of charged hadrons in $Z$-tagged jets~\cite{Aaij:2019ctd} in proton-proton collisions at a center-of-mass energy of $\sqrt{s}=8~$TeV, focusing on the forward rapidity regions $2.5 < \eta_J < 4$. \cref{fig:Plhcb} illustrates the $z_\Lambda$ distribution of $P_\Lambda$ with $j_\perp$ integrated over $0 < j_\perp < 1 \text{GeV}$ for three distinct ranges of jet transverse momenta: $20~<p_{JT} <30$ GeV, $30<p_{JT} <50$ GeV, and $50<p_{JT} <100$ GeV. The asymmetry $P_\Lambda$ exhibits a positive value initially and then becomes negative around $z_\Lambda \sim 0.3$. This behavior can be attributed to the positive PFF of the $u$ quark extracted from\cite{Callos:2020qtu}, which dominates in the small $z_\Lambda$ region. As $z_\Lambda \gtrsim 0.3$, the negative PFF of the $d$ quark takes over. Additionally, we observe that the asymmetry $P_\Lambda$ is enhanced with increasing jet transverse momenta, as the quark jet fraction increases with $p_{JT}$.

For the RHIC kinematics, we consider the transverse momentum of the jets to be $10~<p_{JT} <15$ GeV, and investigate two different ranges of rapidity, namely, $0 < \eta_J < 1$ and $1 < \eta_J < 2.2$. The latter range may be accessible once a forward detector upgrade is made available at sPHENIX\cite{Adare:2015kwa}. \cref{fig:Ppp} presents our results, differential in $z_\Lambda$, again with $j_\perp$ integrated over $0 < j_\perp < 1 \text{GeV}$. Similar to the case at the LHC, the $u$ quark PFF dominates when $z_\Lambda \lesssim 0.3$, but is overtaken by the negative $d$ quark PFF at larger values of $z_\Lambda$. Furthermore, we observe that valence contributions are enhanced when looking at the more forward rapidity region of $1 < \eta_J < 2.2$. Since valence quarks, especially $u$ and $d$, have the largest PFFs, the size of $P_\Lambda$ is enhanced for the more forward rapidity region. Nevertheless, the $\Lambda$ polarization $P_{\Lambda}$ in proton-proton collisions at both LHC and RHIC is at the level of about $2\%$, similar in magnitude to the recent ATLAS measurement for the transverse polarization of single inclusive $\Lambda$ production in $pp\to \Lambda^\uparrow+X$~\cite{ATLAS:2014ona}.

Finally, we present our predictions for $P_\Lambda$ as a function of $z_\Lambda$ for the future Electron-Ion Collider (EIC) operating at a center-of-mass energy of $\sqrt{s}=89~$GeV, as illustrated in \cref{fig:Pep}. In this case, the transverse momentum and rapidity of the jets are confined to the intervals $10<p_{JT} <15$ GeV and $|\eta_J| < 2$, respectively. As before, we integrate $j_\perp$ over $0 < j_\perp < 1\ \text{GeV}$. While the overall qualitative behavior of the results is similar to those obtained for the LHC and RHIC kinematics, there are some notable distinctions. Specifically, the $u$ quark PFF dominates over a larger region of $z_\Lambda$, and as a consequence, $P_\Lambda$ remains positive until $z_\Lambda \lesssim 0.4$. Due to the prevalence of quark PFFs in lepton-proton collisions, the polarization size is larger, with $P_\Lambda\sim 10\%$ at $z_\Lambda\sim 0.5$. Consequently, this jet substructure observable holds potential as a feasible measurement candidate at the future EIC.

\section{Back-to-back $e$ + jet in $ep$ collision}\label{sec:ep-epoljet0}
In this section, we study the back-to-back electron-jet production with unpolarized hadron observed inside jets. Recently, a novel avenue for investigating Transverse Momentum Dependent Parton Distribution Functions (TMDPDFs) and Transverse Momentum Dependent Fragmentation Functions (TMDFFs) has emerged through the study of back-to-back electron+jet production and the corresponding jet substructure at the Electron-Ion Collider (EIC)\cite{Liu:2018trl,Arratia:2020nxw,Liu:2020dct}. This work introduces a comprehensive theoretical framework for analyzing the distribution of hadrons within a jet in the context of back-to-back electron-jet production at electron-proton($ep$) colliders, described by the process:
\bea
e^- + p \rightarrow e^- + \left(\text{jet}\left({\bm q}_T\right) h\left(z_h, {\bm j}_\perp\right)\right)+X\,,
\eea
where both the incoming particles (electron and proton) and the outgoing hadrons inside the jet can possess arbitrary polarizations. Here, ${\bm q}_T$ represents the transverse momentum imbalance between the final-state electron and the jet, measured with respect to the beam direction of the electron and proton. Meanwhile, $z_h$ denotes the momentum fraction of the jet carried by the produced hadron, and ${\bm j}_\perp$ corresponds to the transverse momentum of the hadron within the jet relative to the jet axis.

In particular, besides the electron-jet transverse momentum imbalance~$\bm{q}_T$, we also observe transverse momentum~$\bm{j}_\perp$ distribution of hadrons inside the jet with respect to the jet axis. Observation of a hadron inside a jet makes the process sensitive to a TMDPDF and a TMDJFF simultaneously. Unlike the counterpart process involving a hadron without observation of a jet, such as SIDIS~\cite{Bacchetta:2006tn}, further dependence in $\bm{j}_\perp$ allows the two TMDs to be separately constrained. In this section, we consider only unpolarized hadron (such as pions) inside the jet and we write down the complete azimuthal modulations for the cross section. The well-known Collins asymmetry for hadrons in a jet in $ep$ collisions is one of such azimuthal modulations~\cite{Arratia:2020nxw}. 

\subsection{Theoretical formalism}
We provide the theoretical framework where one measures distribution of unpolarized hadrons inside the jet described by the following process,
\begin{align}
p({p}_A,{S}_{A})+e({p}_B, \lambda_e)\rightarrow \Big[\text{jet}({p}_C)\, h\left(z_h, \bm{j}_\perp\right)\Big]+e({p}_D)+X\,.\label{eq:epjeth}
\end{align}
In this scattering process, an electron with momentum ${p}_B$, moving along the ``negative $z$" direction, interacts with a polarized proton with momentum ${p}_A$ and polarization $S_A$, which is moving along the ``positive $z$" direction. This collision results in the production of a jet with momentum ${p}_C$ and an electron with momentum ${p}_D$ in the final state. The electron can be either unpolarized or longitudinally polarized, with a helicity denoted as $\lambda_e$.

To study this process, we utilize the center-of-mass frame of the $ep$ collision, where the incoming momenta $p_{A,B}$ and the proton spin vector $S_A$ have been given in \cref{eq:light-cone_pA}-\ref{eq:light-cone_SA}. For our analysis, we consider the final observed jet to be produced in the $xz$-plane, as shown in \cref{fig:illu}, with its four-momentum $p_C$ expressed as
\bea
{p}_C^\mu=&E_J(1,\sin\theta_J,0,\cos\theta_J)\,,
\label{eq:jetmom}
\eea
The angle $\theta_J$ is measured concerning the beam direction. The $xz$-plane, formed by the jet momentum and the directions of the incoming electron-proton beams, is referred to as the scattering plane. It's important to mention that we represent $p_C^\mu$ using the $(t,x,y,z)$ momentum notation to distinguish it from the light-cone component representation, which is denoted by brackets.

In equation \cref{eq:epjeth}, we also introduce the definitions of $z_h$ and $\bm{j}_\perp$. Here, $z_h$ represents the longitudinal momentum fraction of the jet carried by the hadron $h$, and $\bm{j}_\perp$ denotes the hadron's transverse momentum concerning the jet axis. The specifics of this scattering process are depicted in figure \cref{fig:illu}. In this context, the symbol ``$\perp$" is used to indicate the transverse vector relative to the jet axis. Additionally, $\bm{j}_\perp$ is parametrized in the $ep$ center-of-mass frame as
\begin{align}
\bm{j}_{\perp}=&j_\perp(\cos\hat{\phi}_h\cos\theta_J,\sin\hat{\phi}_h,-\cos\hat{\phi}_h\sin\theta_J)\,.
\label{eq:bmj}
\end{align} 
In the context of the previous section (refer to equation \cref{eq:jetmom}), the angle $\theta_J$ is defined. We adopt a slight abuse of notation, as discussed earlier below equation \cref{eq:sT}, using $j_\perp = |\bm{j}_{\perp}|$ to represent the magnitude of the transverse vector $\bm{j}_{\perp}$. Additionally, $\hat{\phi}_{h}$ corresponds to the azimuthal angle of the produced hadron transverse momentum $\bm{j}_\perp$ in the jet frame denoted by $x_J y_J z_J$ (as illustrated in \cref{fig:illu}). It is important to recall that the scattering plane is defined as the $xz$-plane, formed by the jet momentum and the incoming electron-proton beam directions. Also, note that we differentiate the azimuthal angle measured in this jet frame (as shown in \cref{fig:illu}) with a hat symbol.

We focus on the kinematic region where both the electron and the jet are produced back-to-back, and the transverse momentum imbalance $q_T = |\bm q_T|$ is much smaller than the average transverse momentum $p_T = |\bm p_T|$ of the electron and the jet, such that $q_T \ll p_T$. In this context, $\bm q_T$ and $\bm p_T$ are given by:
\bea
\bm{q}_T &= \bm{p}_{C,T} + \bm{p}_{D,T}\,,
\\
\bm{p}_T &=  \left(\bm{p}_{C,T} - \bm{p}_{D,T}\right)/2\,.
\label{eq:pT}
\eea
In the region where $q_T \lesssim p_T$, one must carry out the fixed-order matching (so-called $Y$ term~\cite{Collins:1984kg}). We focus on the resummation region in this paper. 

Next we parametrize $\bm{q}_T$ and transverse spin vector $\bm{S}_T$ in terms of their azimuthal angles as
\bea
\label{eq:qT}
\bm{q}_T  &= q_T(\cos\phi_{q},\sin\phi_{q})\,,\\
\label{eq:sT}
\bm{S}_T &= S_T(\cos\phi_{S_A},\sin\phi_{S_A})\,.
\eea
The vectors denoted by the subscript $T$ indicate that they are transverse concerning the beam direction. The azimuthal angles are measured in a frame where the incoming beams and the outgoing jet define the $xz$-plane.

It is important to be cautious regarding a slight abuse of notation, where we use $S_T = |\bm{S}_T|$ and $q_T=|\bm{q}_T|$ to represent the magnitudes of the transverse vectors in equations \cref{eq:qT} and \cref{eq:sT}, respectively. However, using the representation of the four-vector as $q_T^\mu = (0,0,\bm{q}_T)$ (similarly for $S_T^\mu$) would lead to a contradiction, namely $q_T^2 = - \bm{q}_T^2$, if one interprets $q_T^2$ as $q_T^\mu ,q_{T,\mu}$. To avoid confusion, we consistently use $q_T$ and $S_T$ to represent the magnitudes of the transverse momentum and spin, explicitly writing indices, for instance $q_T^\mu ,q_{T,\mu}$, to denote the invariant mass of a four-momentum. Additionally, unbolded text with Latin indices, such as ${k}_T^i$ or ${S}_T^i$, is used to represent the components of transverse vectors.
\subsection{With unpolarized hadron-in-jet}
The differential cross section of the back-to-back electron-jet production with unpolarized hadron observed inside jets is given by 
\begin{align}
\label{eq:unpjeth}
&\frac{d\sigma^{p(S_A)+e(\lambda_e)\to e+(\text{jet}\,h)+X}}{d{p}^2_Tdy_Jd^2{\bm q}_Tdz_h d^2{\bm j}_\perp}=F_{UU,U}+\cos(\phi_{q}-\hat{\phi}_h)F^{\cos(\phi_{q}-\hat{\phi}_h)}_{UU,U}\nnu
&\hspace{1.5cm}+\lambda_p\bigg\{\lambda_e F_{LL,U}+\sin(\phi_{q}-\hat{\phi}_h)F_{LU,U}^{\sin(\phi_{q}-\hat{\phi}_h)}\bigg\}\nnu
&\hspace{1.5cm}+S_T\bigg\{\sin(\phi_{q}-{\phi}_{S_A})F^{\sin(\phi_{q}-{\phi}_{S_A})}_{TU,U}+\sin(\phi_{S_A}-\hat{\phi}_h)F^{\sin(\phi_{S_A}-\hat{\phi}_h)}_{TU,U}\nnu
&\hspace{3cm}+\lambda_e\cos(\phi_{q}-{\phi}_{S_A})F^{\cos(\phi_{q}-{\phi}_{S_A})}_{TL,U}\nnu
&\hspace{3cm}+\sin(2\phi_{q}-\hat{\phi}_h-\phi_{S_A})F^{\sin(2\phi_{q}-\hat{\phi}_h-\phi_{S_A})}_{TU,U}\bigg\}\,,
\end{align}
where $F_{AB,C}$ denote the spin-dependent structure functions, with $A$, $B$, and $C$ respectively indicating the polarization of the incoming proton, of the incoming electron, and the outgoing hadron inside a jet. In this section we only consider the distribution of unpolarized hadrons inside the jet, thus we always have $C = U$. The polarization of the hadrons in jets will be discussed in the next section. 


For the unpolarized hadron, the correlator is parametrized by TMDJFFs at the leading twist accuracy as
\begin{align}\label{eq:unpjffdef} 
\Delta_{\rm jet}^{h/q}\left(z_h,{\bm j}_\perp,S_h\right)=\ &\frac{1}{2}\Bigg[\mathcal{D}_1^{h/q}(z_h,j_\perp^2)\sla{n}_J+\mathcal{H}_{1}^{\perp h/q}(z_h,j_\perp^2)\frac{\sigma^{\mu\nu} n_{J,\mu} j_{\perp\nu}}{z_h\,M_h}\Bigg] \nonumber\\&+ \text{spin dependent terms}\,,
\end{align}
Here we introduce the jet light-cone vectors $n_J=\frac{1}{\sqrt{2}}(1,0,0,1)$ and $\bar n_J=\frac{1}{\sqrt{2}}(1,0,0,-1)$ in the $x_Jy_Jz_J$ coordinate system, and they are defined with the jet momentum along the $n_J$ direction  as shown in \cref{fig:illu}. In the first row of \cref{tabTMDffq} we have summarized the physical interpretations of the unpolarized quark TMDJFFs. That is to say,, $\mathcal{D}_1^{h/q}$ describes an unpolarized quark initiating a jet in which an unpolarized hadron is observed, while $\mathcal{H}_{1}^{\perp h/q}$ describes a transversely polarized quark initiating a jet in which an unpolarized hadron is observed. Thus one have TMDJFFs $\mathcal{D}_1^{h/q}$ closely related to the unpolarized TMDFFs $D_1^{h/q}$~\cite{Kang:2020xyq,Kang:2017glf,Kang:2017btw}, and accordingly TMDJFFs $\mathcal{H}_{1}^{\perp h/q}$ are closely related to the Collins TMDFFs $H_{1}^{\perp h/q}$. 

We now illustrate the factorization of the structure functions in the region $q_T\sim j_\perp \ll p_T R$. The factorization formula for $F_{UU,U}$ in \cref{eq:unpjeth} is given as follows
\begin{align}
\label{eq:FUUUbefore}
F_{UU,U} =&\hat{\sigma}_0\,H(Q,\mu)\sum_qe_q^2\, {\mathcal{D}}_{1}^{h/q}(z_h,j_\perp^2,\mu, \zeta_J)
\int\frac{b \,db}{2\pi}J_0(q_Tb)\,x\,\tilde{f}^{q}_{1}(x,b^2, \mu,\zeta)
\nonumber\\
&\times \bar{S}_{\rm global}(b^2,\mu)\bar{S}_{cs}(b^2,R,\mu)\,,
\end{align}
where we include renormalization scale $\mu$ and Collins-Soper parameter $\zeta_J$ for the TMDJFFs. As we will demonstrate below, $\sqrt{\zeta_J} = p_T R$. 

In the kinematic region $j_\perp\ll p_TR$, the unpolarized TMDJFF ${\mathcal{D}}_{1}^{h/q}(z_h,j_\perp^2,\mu, \zeta_J)$ can be further factorized in terms of the corresponding unpolarized TMDFF and a collinear-soft function as~\cite{Kang:2017glf,Kang:2020xyq}
\begin{align}
{\mathcal{D}}_{1}^{h/q}(z_h,j_\perp^2,\mu, \zeta_J) &=  \int_{\bm{k}_\perp,\, \bm{\lambda}_\perp}\, D_1^{h/q,\, 
\rm unsub}(z_h,k_\perp^2,\mu,\zeta'/\nu^2) S_q(\lambda_\perp^2,\mu,\nu \mathcal{R})\nonumber\\
& =\int \frac{b\,db}{2\pi} J_0\left(\frac{j_\perp b}{z_h}\right) \tilde{D}_1^{h/q,\,\rm unsub}(z_h,b^2,\mu,\zeta'/\nu^2)S_q(b^2,\mu,\nu \mathcal{R})\,,
\label{unp_JFF_FF}
\end{align}
where we use the short-hand notation of
$\int_{\bm{k}_\perp,\, \bm{\lambda}_\perp}=\int d^2\bm{k}_\perp d^2\bm{\lambda}_\perp \delta^2(z_h\bm{\lambda}_\perp + \bm{k}_\perp-\bm{j}_\perp)$ in the first line, and $\sqrt{\zeta'} = \sqrt{2}n_J\cdot p_J$ is the Collins-Soper parameter for the TMDFFs. On the other hand, $S_q(b^2, \mu, \nu \mathcal{R})$ is the collinear-soft function with the following expressions~\cite{Kang:2017glf,Kang:2017mda}
\begin{align}
S_q(b^2,\mu,\nu \mathcal{R})=1 - \frac{\alpha_s C_F}{4\pi} &\Bigg[2\left(\frac{2}{\eta}+\ln\frac{\nu^2\mathcal{R}^2}{4\mu^2}\right)\left(\frac{1}{\epsilon}+\ln\frac{\mu^2}{\mu_b^2}\right)
+\ln^2\frac{\mu^2}{\mu_b^2}-\frac{2}{\epsilon^2}+\frac{\pi^2}{6}
\Bigg]\,,
\label{eq:S_q}
\end{align}
where $\displaystyle\mathcal{R} = \frac{R}{\cosh{y_J}}$. To proceed, comparing the standard soft function $\sqrt{S_{ab}(b^2,\mu,\nu)}$ where
\bea
{S}_{ab}(b^2,\mu,\nu)&=1 - \frac{\alpha_s C_F}{2\pi} \Bigg[2\left(\frac{2}{\eta}+\ln\frac{\nu^2}{\mu^2}\right)\left(\frac{1}{\epsilon}+\ln\frac{\mu^2}{\mu_b^2}\right)+
\ln^2\frac{\mu^2}{\mu_b^2}
-\frac{2}{\epsilon^2}+\frac{\pi^2}{6}
\Bigg]\,.
\label{eq:S-ab}
\eea
with \cref{eq:S_q} we realize $S_q(b^2,\mu,\nu\mathcal{R}) = \sqrt{S_{ab}(b^2,\mu,\nu)}|_{\nu \to\nu\mathcal{R}/2}$ at the NLO and thus
\begin{align}
\tilde{D}_1^{h/q,\,\rm unsub}(z_h,b^2,\mu,\zeta'/\nu^2)S_q(b^2,\mu,\nu \mathcal{R})
= & \tilde{D}_1^{h/q,\,\rm unsub}(z_h,b^2,\mu, 
\zeta'/\nu^2) \sqrt{S_{ab}(b^2,\mu,\nu\mathcal{R}/2)} 
\nnu 
= & \tilde{D}_1^{h/q}(z_h,b^2,\mu, 
\zeta' \mathcal{R}^2/4)\,.
\label{eq:Dunpb}
\end{align}
Finally using the fact that $\displaystyle\sqrt{\zeta'} \mathcal{R}/2 = \sqrt{2}n_J\cdot p_J {R}/({2\cosh y_J}) = p_T R  \equiv \sqrt{\zeta_J}$, we obtain the following relation between TMDJFF ${\mathcal{D}}_{1}^{h/q}$ and TMDFF $D_{1}^{h/q}$
\begin{align}
{\mathcal{D}}_{1}^{h/q}(z_h,j_\perp^2,\mu, \zeta_J) = \int \frac{b\,db}{2\pi} J_0\left(\frac{j_\perp b}{z_h}\right) \tilde{D}_1^{h/q}(z_h,b^2,\mu,\zeta_J) = D_1^{h/q}(z_h,j_\perp^2,\mu,\zeta_J)\,.
\label{unp_JFF_FF2}
\end{align}
Alternatively, the TMDJFF is equal to the TMDFF at the scale $\zeta_J$. The parametrization of TMDFF follows the similar form as that of the TMDPDF. Using the CSS formalism, the $b$-space unpolarized TMDFF can be expressed as 
\begin{align}
\tilde{D}_1^{h/q}(z_h,b^2,\mu, 
\zeta_J) &=\frac{1}{z_h^2}\left[\hat{C}_{i\leftarrow q}\otimes D_{1}^{h/i}\right]\left(z_h,\mu_{b_*}\right)
\exp\left[-S_{\rm pert}\left(\mu, \mu_{b_*} \right) - S_{\rm NP}^D\left(z_h, b, Q_0, \zeta_J\right)\right]\,,
\label{eq:D1param}
\end{align}
where we have performed an operator product expansion in terms of the unpolarized collinear FFs $D_1^{h/i}(x, \mu_{b_*})$ with the convolution defined as follows 
\begin{align}
\left[\hat{C}_{i\leftarrow q}\otimes D_{1}^{h/i}\right]\left(z_h,\mu_{b_*}\right) = \int_{z_h}^{1} \frac{d\hat{z}_h}{\hat{z}_h} \hat{C}_{i\leftarrow q}\left(\frac{z_h}{\hat{z}_h}, \mu_{b_*} \right) D_1^{h/i}\left(\hat{z}_h, \mu_{b_*}\right)\,.
\end{align}
We take into account the summation over repeated indices, and we adopt the same $b_*$ prescription as utilized in TMDPDFs. The coefficient functions $\hat{C}_{i\leftarrow q}$ at the NLO can be looked up in~\cite{Kang:2015msa}, and results for even higher orders are also accessible in~\cite{Echevarria:2016scs, Luo:2019hmp, Luo:2020epw, Ebert:2020qef}. The perturbative Sudakov factor is equivalent to that of the TMDPDFs, as provided by
\bea
\label{eq:Sud-pert}
S_{\rm pert}\left(\mu, \mu_{b_*} \right)  = -\tilde K(b_*,\mu_{b_*})\ln\left(\frac{\sqrt{\zeta}}{\mu_{b_*}}\right) -\int_{\mu_{b_{*}}}^{\mu} \frac{d\mu'}{\mu'}\left[\gamma_F\left(\alpha_s(\mu'), \frac{\zeta}{\mu'^2}\right) \right]\,.
\eea
On the other hand, for the non-perturbative Sudakov factor $S_{\rm NP}^D\left(z_h, b, Q_0, \zeta_J\right)$, we use the parametrization~\cite{Sun:2014dqm,Echevarria:2020hpy}
\begin{align}
\label{eq:Sud-NPD}
S_{\rm NP}^D(z_h, b, Q_0,\zeta_J) = \frac{g_2}{2}\ln{\frac{\sqrt{\zeta_J}}{Q_0}}\ln{\frac{b}{b_*}}+g_1^D \frac{b^2}{z_h^2}\,.
\end{align}
The values of $Q_0$, $g_2$, and $g_1^f$ are provided as $Q_0=\sqrt{2.4}$~GeV, $g_2=0.84$, $g_1^f=0.106$ GeV$^2$, and $g_1^D=0.042$ GeV$^2$.

Applying similar arguments, one can establish analogous relationships between other TMDJFFs and TMDFFs. The explicit expressions of the remaining structure functions in terms of TMDJFFs and TMDPDFs are presented in the subsequent subsection, and a summary of the azimuthal asymmetries associated with them can be found in \cref{scn2tab}.

\begin{table}
\centering
\begin{tabular}{ |c|c|c|c|c| } 
 \hline
 \diagbox[width=1.8cm]{JFF}{PDF} & $f_1$ & $g_{1L}$ & $f_{1T}^\perp$  & $g_{1T}$ \\ 
  \hline
  $\mathcal{D}_1$ & 1 & 1  & $\sin(\phi_{q}-{\phi}_{S_A})$ & $\cos(\phi_{q}-{\phi}_{S_A})$     \\ 
  \hline
  \hline
   \diagbox[width=1.8cm]{JFF}{PDF} & $h_1^\perp$ & $h_{1L}^\perp$ & $h_1$  & $h_{1T}^\perp$  \\ 
  \hline 
  $\mathcal{H}^\perp_1$ & $\cos(\phi_{q}-\hat{\phi}_h)$ & $\sin(\phi_{q}-\hat{\phi}_h)$ & $\sin({\phi}_{S_A}-\hat{\phi}_h)$ & $\sin(2\phi_{q}-\hat{\phi}_{h}-\phi_{S_A})$\\
   \hline 
\end{tabular}
  \caption{Summary of the characteristic azimuthal asymmetry with which different TMDPDFs and TMDJFFs arise for back-to-back electron-jet production, with unpolarized hadrons observed inside the jet. See \cref{eq:unpjeth} and~\cref{eq:strhlep1}-\cref{eq:strhlep8} for parametrizations of structure functions.} \label{scn2tab}
\end{table}

\subsubsection{Soft function}\label{sec:th-soft}
In \cref{sec:th-fac}, we presented the factorization formula for the process involving an unpolarized hadron inside a jet, using the narrow jet cone limit where $R\ll 1$. However, in this subsection, we will derive the factorization formula without making the narrow cone approximation.

As the jet radius $R$ is not a small parameter, the narrow cone approximation is not suitable for constructing the factorization formula. Generally, the factorized cross section can be expressed as the product of the hard, soft, and TMD collinear functions, denoted as $d\sigma\sim H f_{1}^q\otimes D_{1}^{h/q}\otimes S$, where the soft function $S$ depends on both $\bm q_T$ and $\bm j_\perp$, and it also accounts for the jet algorithm dependence within $S$.

Specifically, the physics scale inside the jet is represented by $j_\perp$, whereas the scale outside the jet is represented by $q_T$. Since we assume $q_T \sim j_\perp$, there are no large logarithms present inside the soft function. However, if there is a scale hierarchy between $q_T$ and $j_\perp$, then it becomes necessary to consider the refactorization of the soft function, as demonstrated in~\cite{Becher:2016omr}. The factorized cross section takes the form of 
\begin{align}\label{eq:fac_Lr}
    \frac{d\sigma^{p(S_A)+e(\lambda_e)\to e+(\text{jet}\,h)+X}}{d{p}^2_Tdy_Jd^2{\bm q}_Tdz_h d^2{\bm j}_\perp}=&\hat\sigma_0 H(Q,\mu) \int\frac{d^2\bm b}{(2\pi)^2} e^{i \bm{b} \cdot \bm{q}_T} \int\frac{d^2 \bm b'}{(2\pi)^2} e^{i \bm{b}' \cdot \bm{j}_\perp}   \\
    &\hspace{-2.5cm} \times \sum_q e_q^2 \tilde{D}_{1}^{h/q,{\rm unsub}}(z_h,b'^2,\mu,\zeta'/\nu^2) x \tilde{f}^{q,{\rm unsub}}_{1}(x,b^2,\mu,\zeta/\nu^2) S(\bm b,\bm b',y_J,R,\mu,\nu), \notag
\end{align}
where $\bm b$ and $\bm b'$ are conjugate variables of $\bm q_T$ and $\bm j_\perp$, separately. At one-loop order,  we consider only one soft gluon emission, which is either inside or outside the jet cone. Then the soft function can be factorized as
\begin{align}\label{eq:sbpsb}
    S(\bm b,\bm b',y_J,R,\mu,\nu) = S_{\rm in}(\bm b',y_J,R,\mu,\nu) S_{\rm out}(\bm b,y_J,R,\mu,\nu). 
\end{align}
Using the above relation, we find that $q_T$ and $j_\perp$ dependence in the cross section are fully factorized. It is noted that the above factorization is an approximation, and beyond the one-loop order the expressions depending on both $b$ and $b'$ can show up. Explicitly, the one-loop soft function $S(\bm b,\bm b',y_J,R,\mu,\nu)$ is given by
\begin{align}\label{snlo}
S^{\rm NLO}(\bm b,\bm b',y_J,R,\mu,\nu) = & \,C_F \frac{\alpha_s \mu^{2\epsilon} \pi^\epsilon e^{\gamma_E\epsilon}}{\pi^2} \int d^d k \, \delta(k^2)\theta(k^0) \frac{n\cdot n_J}{n\cdot k \, k \cdot n_J} \left( \frac{\nu}{2k^0}\right)^\eta \notag \\
& \hspace{1.5cm} \times \left[ \theta(\Delta R - R) e^{i \bm k_T \cdot \bm b}  + \theta(R-\Delta R) e^{i \bm k_\perp \cdot \bm b'} \right]\,.
\end{align}
Here the jet light-cone vector $n_J$ has been defined below \cref{eq:unpjffdef}. In \cref{snlo}, $\Delta R$ denotes the distance between the jet and the soft emission in rapidity and azimuthal angle plane, which is defined by $\Delta R=\sqrt{(y-y_J)^2+(\phi-\phi_J)^2}$. Therefore, $\theta(\Delta R - R)$ and $\theta(R - \Delta R)$ indicate the soft gluon with momentum $k$ is radiated outside and inside the jet, respectively. In the CM frame of incoming beams the vectors $\bm b$ and $\bm b'$ are defined as
\begin{align}
\bm b&=b(\cos\phi_1,\sin \phi_1,0), \\
\bm b'&=b'(\cos\hat\phi_2\cos\theta_J,\sin\hat\phi_2,-\cos\hat\phi_2\sin\theta_J),
\end{align}
respectively. Here, without loss of generality, we have chosen $\phi_J=0$. It is noted that the vector $\bm b'$ is the conjugate variable of $\bm k_\perp$, which describes the transverse momentum perpendicular to the jet axis.  In the jet frame, it is given as 
\begin{align}
\bm b'= b'(\cos \hat\phi_2,\sin\hat\phi_2)_J,
\end{align}
After performing the rotation transformation in the $xz$ plane, we obtain its expression in the CM frame of incoming beams.

In \cref{snlo} the contribution outside the jet region can be rewritten as
\begin{align}
\theta(\Delta R-R) = 1 - \theta(R-\Delta R),
\end{align}
where the first term on the right side indicates that the soft radiation is independent on the jet definition, so it is also called the global soft function $S_{\rm global}$. Then $S_{\rm out}(b,y_J,R,\mu,\nu)$ in \cref{eq:sbpsb} is given by
\begin{align}
    S_{\rm out}(\bm b,y_J,R,\mu,\nu)=S_{\rm global}(\bm b,y_J,R,\mu,\nu)+S^{\rm in}_{\rm I}(\bm b,R,\mu)
\end{align}
Next we define the contribution from the second term as $S^{\rm in}_{\rm I}$, which is 
\begin{align}
S^{\rm in}_{\rm I}(\bm b,R,\mu) = -C _F \frac{\alpha_s \mu^{2\epsilon} \pi^\epsilon e^{\gamma_E\epsilon}}{\pi^2} \int d^d k \, \delta(k^2)\theta(k^0) \frac{n\cdot n_J}{n\cdot k \, k \cdot n_J}  \theta(R-\Delta R) e^{i \bm k_T \cdot \bm b}\,.
\end{align}
Note that since the integral does not contain the rapidity divergence anymore after constraining the angular integration only inside the jet, one ignores the rapidity regulator here. Then the phase space integration can be expressed as
\begin{align}
\int d^{d} k \delta\left(k^{2}\right) \theta(k^{0})=\frac{\pi^{\frac{1}{2}-\epsilon}}{\Gamma\left(\frac{1}{2}-\epsilon\right)} \int_{0}^{\pi} d \phi \sin ^{-2 \epsilon} \phi \int d y \int d k_{T}\, k_{T}^{1-2 \epsilon}.
\end{align}
After integrating $k_T$ with the Fourier transformation factor, we have
\begin{align}
\int d k_{T}\, k_{T}^{-1-2 \epsilon} e^{i k_{T} b \cos (\phi-\phi_1)}=\Gamma(-2 \epsilon)\left[-i b \cos \left(\phi-\phi_{1}\right)\right]^{2 \epsilon}.
\end{align}
The integration region of $y$ and $\phi$ are constrained by the jet cone as $\theta [R-\phi^2-(y-y_J)^2]$, and we express the integration variables as
\begin{align}
y=r\cos\chi+y_J, ~~~~~~\phi=r\sin\chi.
\end{align}
Then we obtain
\begin{align}
\int d y \int_{0}^{\pi} d \phi \,\theta\left[R^{2}-\left(y-y_{J}\right)^{2}-\phi^{2}\right]=\int_{0}^{R} d r \, r \int_{0}^{\pi} d \chi.
\end{align}
In the small $R$ limit, after taking the leading contribution of the integrand in the $r\ll 1$ region, we have
\begin{align}
S^{\rm in}_{\rm I}(\bm b,R,\mu) =&\, -\frac{\alpha_s}{2\pi}C_F \Bigg[  \frac{1}{\epsilon^{2}}+\frac{2}{\epsilon}\ln \left(\frac{-2i \cos \phi_{1}\mu}{\mu_{b}R} \right)+ 2\ln^2\left( \frac{-2i \cos \phi_{1}\mu}{\mu_{b}R}\right)+ \frac{\pi^2}{4}\Bigg] \,.
\end{align}  
which is also know as the collinear-soft function $S_{ cs}$. 

Similarly, we define $S_{\rm in}(\bm b',y_J,R,\mu,\nu)$ in \cref{eq:sbpsb}, namely the contribution from $\theta(R-\Delta R)$ in \cref{snlo} as $S^{\rm in}_{\rm II}$, where $k_\perp^\mu$ is defined as
\begin{align}
    k_\perp^\mu = k^\mu - \frac{\bar n_J \cdot k}{2} n_J^\mu - \frac{n_J \cdot k}{2} \bar n_J^\mu.
\end{align}
In the small $R$ limit, we have
\begin{align}
S^{\rm in}_{\rm II}(b',\mu,\nu\mathcal{R}) = & \, -\frac{\alpha_s}{2 \pi} C_{F}\Bigg[-\frac{1}{\epsilon^{2}}+\frac{2}{\eta} \left(\frac{1}{\epsilon }+\ln\frac{\mu^2}{\mu_b'^2}\right)+\frac{1}{\epsilon} \ln \left(\frac{\nu^2 \mathcal{R}^2}{4\mu^2}\right) \notag \\
& \hspace{2cm} + \ln\frac{\mu^2}{\mu_b'^2} \ln \left(\frac{\nu^2 \mathcal{R}^2}{4\mu^2}\right) +  \frac{1}{2}\ln^2\frac{\mu^2}{\mu_b'^2} + \frac{\pi^2}{12} \Bigg],
\end{align}
which is the same as the one-loop expression of $S_q$ given in \cref{eq:S_q}. Therefore we show that in the small $R$ approximation, the one-loop soft function $S^{\rm NLO}$ is
\begin{align}
S^{\rm NLO} &= S_{\rm global}(\bm b,\mu,\nu)+S^{\rm in}_{\rm I}(\bm b,\mu)+S^{\rm in}_{\rm II}(b',\mu,\nu\mathcal{R}) + \mathcal{O}(R^2),\notag \\
&=S_{\rm global}(\bm b,\mu,\nu)+S_{ cs}(\bm b,\mu)+S_{q}(b',\mu,\nu\mathcal{R})+ \mathcal{O}(R^2)\,.
\end{align}
In other words the soft function $S$ in the factorization formula \cref{eq:fac_Lr} can be expressed as
\begin{align}
S=S_{\rm global}\,S_{ cs}\,S_{q} + \mathcal{O}(R^2). 
\end{align}
Finally we obtain \cref{eq:FUUUbefore}, the factorization formula for the process of the unpolarized hardon production inside jet in the narrow cone approximation. 


\subsection{With polarized hadron-in-jet}\label{sec:ep-epoljet1}
Continue with the theoretical background for $ep\rightarrow e+{\rm jet}(h)+X$ provided in~\cref{sec:ep-epoljet0}, we obtain the differential cross section of the back-to-back electron-jet production with polarized hadron observed inside jets is given by
\begingroup
\allowdisplaybreaks
\bea
\label{eq:poljeth}
&\frac{d\sigma^{p(S_A)+e(\lambda_e)\to e+(\text{jet}\,h(S_h))+X}}{d{p}^2_Tdy_Jd^2{\bm q}_Tdz_h d^2{\bm j}_\perp}=F_{UU,U}+\cos({\phi}_{q}-\hat{\phi}_{h})F_{UU,U}^{\cos({\phi}_{q}-\hat{\phi}_{h})}\nnu
&+\lambda_p\bigg\{\lambda_eF_{LL,U}+\sin(\phi_{q}-\hat{\phi}_h)F_{LU,U}^{\sin(\phi_{q}-\hat{\phi}_h)}\bigg\}\nnu
&+S_T\bigg\{\sin(\phi_{q}-{\phi}_{S_A})F^{\sin(\phi_{q}-{\phi}_{S_A})}_{TU,U}+\lambda_e\cos(\phi_{q}-{\phi}_{S_A})F^{\cos(\phi_{q}-{\phi}_{S_A})}_{TL,U}\nnu
&\hspace{1.3cm}+\sin(\phi_{S_A}-\hat{\phi}_h)F^{\sin(\phi_{S_A}-\hat{\phi}_h)}_{TU,U}+\sin(2\phi_{q}-\hat{\phi}_h-\phi_{S_A})F^{\sin(2\phi_{q}-\hat{\phi}_h-\phi_{S_A})}_{TU,U}\bigg\}\nnu
&+\lambda_h\bigg\{\lambda_e F_{UL,L}+\sin(\hat{\phi}_h-\phi_{q})F_{UU,L}^{\sin(\hat{\phi}_h-\phi_{q})}+\lambda_p \bigg[F_{LU,L}+{\cos(\hat{\phi}_h-\phi_{q})F_{LU,L}^{\cos(\hat{\phi}_h-\phi_{q})}}\bigg]\nnu
&\hspace{1.3cm}+S_T\bigg[\cos(\phi_{q}-{\phi}_{S_A})F^{\cos(\phi_{q}-{\phi}_{S_A})}_{TU,L}+\lambda_e\sin(\phi_{q}-{\phi}_{S_A})F^{\sin(\phi_{q}-{\phi}_{S_A})}_{TL,L}\nnu
&\hspace{2.5cm}+{\cos(\phi_{S_A}-\hat{\phi}_h)F^{\cos(\phi_{S_A}-\hat{\phi}_h)}_{TU,L}}+{\cos(2\phi_{q}-\phi_{S_A}-\hat{\phi}_h)F^{\cos(2\phi_{q}-\phi_{S_A}-\hat{\phi}_h)}_{TU,L}}\bigg]\bigg\}\nnu
&+S_{h\perp}\bigg\{\sin(\hat{\phi}_h-\hat{\phi}_{S_h})F^{\sin(\hat{\phi}_h-\hat{\phi}_{S_h})}_{UU,T}+\lambda_e{\cos(\hat{\phi}_h-\hat{\phi}_{S_h})F_{UL,T}^{\cos(\hat{\phi}_h-\hat{\phi}_{S_h})}}\nnu
&\hspace{1.3cm}+\sin(\hat{\phi}_{S_h}-\phi_{q})F_{UU,T}^{\sin(\hat{\phi}_{S_h}-\phi_{q})}+\sin(2\hat{\phi}_h-\hat{\phi}_{S_h}-\phi_{q})F_{UU,T}^{\sin(2\hat{\phi}_h-\hat{\phi}_{S_h}-\phi_{q})}\nnu
&\hspace{1.3cm}+\lambda_p\bigg[{\cos(\hat{\phi}_h-\hat{\phi}_{S_h}) F^{\cos(\hat{\phi}_h-\hat{\phi}_{S_h})}_{LU,T}}+\cos(\phi_{q}-\hat{\phi}_{S_h}) F^{\cos(\phi_{q}-\hat{\phi}_{S_h})}_{LU,T}\nnu
&\hspace{2.5cm}+\cos(2\hat{\phi}_h-\hat{\phi}_{S_h}-\phi_{q}) F^{\cos(2\hat{\phi}_h-\hat{\phi}_{S_h}-\phi_{q})}_{LU,T}+\lambda_e\sin(\hat{\phi}_h-\hat{\phi}_{S_h})F_{LL,T}^{\sin(\hat{\phi}_h-\hat{\phi}_{S_h})}\bigg]\nnu
&\hspace{1.3cm}+S_T\bigg[\cos(\phi_{S_A}-\hat{\phi}_{S_h})F^{\cos(\phi_{S_A}-\hat{\phi}_{S_h})}_{TU,T}+\cos(2\hat{\phi}_h-\hat{\phi}_{S_h}-\phi_{S_A})F^{\cos(2\hat{\phi}_h-\hat{\phi}_{S_h}-\phi_{S_A})}_{TU,T}\nnu
&\hspace{2.5cm}+\sin(\hat{\phi}_h-\hat{\phi}_{S_h})\sin(\phi_{q}-{\phi}_{S_A})F_{TU,T}^{\sin(\hat{\phi}_h-\hat{\phi}_{S_h})\sin(\phi_{q}-{\phi}_{S_A})}\nnu
&\hspace{2.5cm}+{\cos(\hat{\phi}_h-\hat{\phi}_{S_h})\cos(\phi_{q}-\phi_{S_A}))F_{TU,T}^{\cos(\hat{\phi}_h-\hat{\phi}_{S_h})\cos(\phi_{q}-\phi_{S_A})}}\nnu
&\hspace{2.5cm}+\cos(2\phi_{q}-\phi_{S_A}-\hat{\phi}_{S_h})F_{TU,T}^{\cos(2\phi_{q}-\phi_{S_A}-\hat{\phi}_{S_h})}\nnu
&\hspace{2.5cm}+\cos(2\hat{\phi}_h-\hat{\phi}_{S_h}+2\phi_{q}-\phi_{S_A})F_{TU,T}^{\cos(2\hat{\phi}_h-\hat{\phi}_{S_h}+2\phi_{q}-\phi_{S_A})}\nnu
&\hspace{2.5cm}+\lambda_e\cos(\hat{\phi}_h-\hat{\phi}_{S_h})\sin(\phi_{S_A}-\phi_{q})F_{TL,T}^{\cos(\hat{\phi}_h-\hat{\phi}_{S_h})\sin(\phi_{S_A}-\phi_{q})}\nnu
&\hspace{2.5cm}+\lambda_e\sin(\hat{\phi}_h-\hat{\phi}_{S_h})\cos(\phi_{S_A}-\phi_{q}))F_{TL,T}^{\sin(\hat{\phi}_h-\hat{\phi}_{S_h})\cos(\phi_{S_A}-\phi_{q})}\bigg]\bigg\}\,,
\eea
\endgroup
where the structure functions with unpolarized hadron in the final state $C=U$ already made appearances in \cref{eq:unpjeth} in last section. More explicitly, the expressions of the structure functions in Eqs.\ \cref{eq:unpjeth} and\ \cref{eq:poljeth}. To give a compact presentation, we define
\bea
\mathcal{C}_{mnk}[ {\mathcal{D}}_q(z_h,j_\perp^2;\mu)\tilde{A}^{(n)}(x,b^2)]=&\hat{\sigma}_{k}H(Q, \mu)\sum_qe_q^2\left(\frac{ j_\perp}{z_hM_h}\right)^m{\mathcal{D}}_q(z_h,j_\perp^2,\mu, \zeta_J)\nnu
&\times M^n\int\frac{b^{n+1}db}{2\pi n!}J_n(q_T b)x\tilde{A}^{(n)}(x,b^2)\,,
\eea
where $m$, $n$ and $k$ can be $m=0,\ 1,\ 2$, $n=0,\ 1,\ 2$ and $k=0,\ L,\ T$.

Partonic cross section $\hat{\sigma}_k$ describes scattering of electron-quark with different polarizations depending on the value of $k$. 
The $k=0$ corresponds to the partonic scattering $eq\rightarrow eq$ or $eq_L\rightarrow eq_L$, $k=L$ corresponds to the partonic scattering $e_Lq_L\rightarrow eq$ or $eq\rightarrow e_Lq_L$, and $k=T$ corresponds to the partonic scattering $eq_T\rightarrow eq_T$. Their expressions are given as
\bea
\hat{\sigma}_0=\frac{\alpha_{\rm em}\alpha_s}{sQ^2}\frac{2(\hat{u}^2+\hat{s}^2)}{\hat{t}^2}\,,\label{eq:sigma0}\\
\hat{\sigma}_L=\frac{\alpha_{\rm em}\alpha_s}{sQ^2}\frac{2(\hat{u}^2-\hat{s}^2)}{\hat{t}^2}\,,\label{eq:sigmaL}\\
\hat{\sigma}_T=\frac{\alpha_{\rm em}\alpha_s}{sQ^2}\left(\frac{-4\hat{u}\hat{s}}{\hat{t}^2}\right)\,.
\label{eq:sigma_T}
\eea
Then, we find
\begingroup
\allowdisplaybreaks
\begin{align}
F_{UU,U}(q_T,j_\perp) &= \mathcal{C}_{000}[{\mathcal{D}}_{1,q}\tilde{f}_1\bar{S}_{\rm global}\bar{S}_{cs}]\,,
\label{eq:strhlep1}
\\
F_{UU,U}^{\cos(\phi_q-\hat{\phi}_h)}(q_T,j_\perp) &= \mathcal{C}_{11T}[{\mathcal{H}}_{1,q}^{\perp}\tilde{h}_1^{\perp(1)}\bar{S}_{\rm global}\bar{S}_{cs}]\,,
\label{eq:strhlep2}
\\
F_{LL,U}(q_T,j_\perp)&=\mathcal{C}_{00L}[{\mathcal{D}}_{1,q}\tilde{g}_{1L}\bar{S}_{\rm global}\bar{S}_{cs}]\,,
\label{eq:strhlep3}
\\
F_{LU,U}^{\sin(\phi_q-\hat{\phi}_h)}(q_T,j_\perp)&=\mathcal{C}_{11T}[{\mathcal{H}}_{1,q}^{\perp}\tilde{h}_{1L}^{\perp(1)}\bar{S}_{\rm global}\bar{S}_{cs}]\,,
\label{eq:strhlep4}
\\
F_{TU,U}^{\sin({\phi}_{q}-{\phi}_{S_A})}(q_T,j_\perp)&= \mathcal{C}_{010}\left[{\mathcal{D}}_{1,q}\tilde{f}_{1T}^{\perp(1)}\bar{S}_{\rm global}\bar{S}_{cs}\right]\,,
\label{eq:strhlep5}
\\
F_{TU,U}^{\sin({\phi}_{S_A}-\hat{\phi}_{h})}(q_T,j_\perp)&= \mathcal{C}_{10T}\left[{\mathcal{H}}_{1,q}^{\perp}\tilde{h}_{1}\bar{S}_{\rm global}\bar{S}_{cs}\right]\,,
\label{eq:strhlep6}
\\
F_{TL,U}^{\cos({\phi}_{q}-{\phi}_{S_A})}(q_T,j_\perp)&=\mathcal{C}_{01L}\left[{\mathcal{D}}_{1,q}\tilde{g}_{1T}^{(1)}\bar{S}_{\rm global}\bar{S}_{cs}\right]\,,
\label{eq:strhlep7}
\\
F_{TU,U}^{\cos(2{\phi}_{q}-\hat{\phi}_h-{\phi}_{S_A})}(q_T,j_\perp)&=\mathcal{C}_{12T}\left[{\mathcal{H}}_{1,q}^{\perp}\tilde{h}_{1T}^{\perp(2)}\bar{S}_{\rm global}\bar{S}_{cs}\right]\,,
\label{eq:strhlep8}
\\
F_{UL,L}(q_T,j_\perp) &= \mathcal{C}_{00L}[{\mathcal{G}}_{1L,q}\tilde{f}_1\bar{S}_{\rm global}\bar{S}_{cs}]\,,
\label{eq:strh1}
\\
F_{UU,L}^{\sin(\hat{\phi}_h-\phi_q)}(q_T,j_\perp) &= \mathcal{C}_{11T}[{\mathcal{H}}_{1L,q}^{\perp}\tilde{h}_1^{\perp(1)}\bar{S}_{\rm global}\bar{S}_{cs}]\,,
\label{eq:strh2}
\\
F_{LU,L}(q_T,j_\perp)&=\mathcal{C}_{000}[{\mathcal{G}}_{1L,q}\tilde{g}_{1L}\bar{S}_{\rm global}\bar{S}_{cs}]\,,
\label{eq:strh3}
\\
F_{LU,L}^{\cos(\hat{\phi}_h-\phi_q)}(q_T,j_\perp)&=-\mathcal{C}_{11T}[{\mathcal{H}}_{1L,q}^{\perp}\tilde{h}_{1L}^{\perp(1)}\bar{S}_{\rm global}\bar{S}_{cs}]\,,
\label{eq:strh4}
\\
F_{TU,L}^{\cos({\phi}_{q}-{\phi}_{S_A})}(q_T,j_\perp)&= \mathcal{C}_{010}\left[{\mathcal{G}}_{1L,q}\tilde{g}_{1T}^{(1)}\bar{S}_{\rm global}\bar{S}_{cs}\right]\,,
\label{eq:strh5}
\\
F_{TL,L}^{\sin({\phi}_{q}-{\phi}_{S_A})}(q_T,j_\perp)&= \mathcal{C}_{01L}\left[{\mathcal{G}}_{1L,q}\tilde{f}_{1T}^{\perp(1)}\bar{S}_{\rm global}\bar{S}_{cs}\right]\,,
\label{eq:strh6}
\\
F_{TU,L}^{\cos({\phi}_{S_A}-\hat{\phi}_{h})}(q_T,j_\perp)&=- \mathcal{C}_{10T}\left[{\mathcal{H}}_{1L,q}^{\perp}\tilde{h}_{1}\bar{S}_{\rm global}\bar{S}_{cs}\right]\,,
\label{eq:strh7}
\\
F_{TU,L}^{\cos(2{\phi}_{q}-{\phi}_{S_A}-\hat{\phi}_{h})}(q_T,j_\perp)&=- \mathcal{C}_{12T}\left[{\mathcal{H}}_{1L,q}^{\perp}\tilde{h}_{1T}^{\perp(2)}\bar{S}_{\rm global}\bar{S}_{cs}\right]\,,
\label{eq:strh8}
\\
F_{UU,T}^{\sin(\hat{\phi}_{h}-\hat{\phi}_{S_h})}(q_T,j_\perp)&= \mathcal{C}_{100}\left[{\mathcal{D}}_{1T,q}^{\perp}\tilde{f}_{1}\bar{S}_{\rm global}\bar{S}_{cs}\right]\,,
\label{eq:strh9}
\\
F_{UL,T}^{\cos(\hat{\phi}_{h}-\hat{\phi}_{S_h})}(q_T,j_\perp)&=- \mathcal{C}_{00L}\left[{\mathcal{G}}_{1T,q}\tilde{f}_{1}\bar{S}_{\rm global}\bar{S}_{cs}\right]\,,
\label{eq:strh10}
\\
F_{UU,T}^{\sin(\hat{\phi}_{S_h}-{\phi}_{q})}(q_T,j_\perp)&= \mathcal{C}_{01T}\left[{\mathcal{H}}_{1,q}\tilde{h}_{1}^{\perp(1)}\bar{S}_{\rm global}\bar{S}_{cs}\right]\,,
\label{eq:strh11}
\\
F_{UU,T}^{\cos(2\hat{\phi}_{h}-{\phi}_{q}-\hat{\phi}_{S_h})}(q_T,j_\perp)&= \mathcal{C}_{21T}\left[{\mathcal{H}}_{1T,q}^{\perp}\tilde{h}_{1}^{\perp(1)}\bar{S}_{\rm global}\bar{S}_{cs}\right]\,,
\label{eq:strh12}
\\
F_{LU,T}^{\cos(\hat{\phi}_{h}-\hat{\phi}_{S_h})}(q_T,j_\perp)&=- \mathcal{C}_{100}\left[{\mathcal{G}}_{1T,q}\tilde{g}_{1L}\bar{S}_{\rm global}\bar{S}_{cs}\right]\,,
\label{eq:strh13}
\\
F_{LU,T}^{\cos(\hat{\phi}_{S_h}-{\phi}_{q})}(q_T,j_\perp)&= \mathcal{C}_{01T}\left[{\mathcal{H}}_{1,q}\tilde{h}_{1L}^{\perp(1)}\bar{S}_{\rm global}\bar{S}_{cs}\right]\,,
\label{eq:strh14}
\\
F_{LU,T}^{\cos(2\hat{\phi}_{h}-{\phi}_{q}-\hat{\phi}_{S_h})}(q_T,j_\perp)&= \mathcal{C}_{21T}\left[{\mathcal{H}}_{1T,q}^{\perp}\tilde{h}_{1L}^{\perp(1)}\bar{S}_{\rm global}\bar{S}_{cs}\right]\,,
\label{eq:strh15}
\\
F_{LL,T}^{\sin(\hat{\phi}_{h}-\hat{\phi}_{S_h})}(q_T,j_\perp)&= \mathcal{C}_{10L}\left[{\mathcal{D}}_{1T,q}^{\perp}\tilde{g}_{1L}\bar{S}_{\rm global}\bar{S}_{cs}\right]\,,
\label{eq:strh16}
\\
F_{TU,T}^{\cos({\phi}_{S_A}-\hat{\phi}_{S_h})}(q_T,j_\perp)&= \mathcal{C}_{00T}\left[{\mathcal{H}}_{1,q}\tilde{h}_{1}\bar{S}_{\rm global}\bar{S}_{cs}\right]\,,
\label{eq:strh17}
\\
F_{TU,T}^{\cos(2\hat{\phi}_{h}-\hat{\phi}_{S_h}-{\phi}_{S_A})}(q_T,j_\perp)&= \mathcal{C}_{10T}\left[{\mathcal{H}}_{1T,q}^{\perp}\tilde{h}_{1}\bar{S}_{\rm global}\bar{S}_{cs}\right]\,,
\label{eq:strh18}
\\
F_{TU,T}^{\sin(\hat{\phi}_{h}-\hat{\phi}_{S_h})\sin({\phi}_q-{\phi}_{S_A})}(q_T,j_\perp)&= \mathcal{C}_{110}\left[{\mathcal{D}}_{1T,q}^{\perp}\tilde{f}_{1T}^{\perp(1)}\bar{S}_{\rm global}\bar{S}_{cs}\right]\,,
\label{eq:strh19}
\\
F_{TU,T}^{\cos(\hat{\phi}_{h}-\hat{\phi}_{S_h})\cos({\phi}_q-{\phi}_{S_A})}(q_T,j_\perp)&=- \mathcal{C}_{110}\left[{\mathcal{G}}_{1T,q}\tilde{g}_{1T}^{(1)}\bar{S}_{\rm global}\bar{S}_{cs}\right]\,,
\label{eq:strh20}
\\
F_{TU,T}^{\cos(2{\phi}_{q}-\hat{\phi}_{S_h}-{\phi}_{S_A})}(q_T,j_\perp)&= \mathcal{C}_{02T}\left[{\mathcal{H}}_{1,q}\tilde{h}_{1T}^{\perp(2)}\bar{S}_{\rm global}\bar{S}_{cs}\right]\,,
\label{eq:strh21}
\\
F_{TU,T}^{\cos(2{\phi}_{h}-\hat{\phi}_{S_h}+2{\phi}_{q}-{\phi}_{S_A})}(q_T,j_\perp)&= \mathcal{C}_{12T}\left[{\mathcal{H}}_{1T,q}^{\perp}\tilde{h}_{1T}^{\perp(2)}\bar{S}_{\rm global}\bar{S}_{cs}\right]\,,
\label{eq:strh22}
\\
F_{TL,T}^{\cos(\hat{\phi}_{h}-\hat{\phi}_{S_h})\sin({\phi}_{S_A}-{\phi}_q)}(q_T,j_\perp)&=\mathcal{C}_{11L}\left[{\mathcal{G}}_{1T,q}\tilde{f}_{1T}^{\perp(1)}\bar{S}_{\rm global}\bar{S}_{cs}\right]\,,
\label{eq:strh23}
\\
F_{TL,T}^{\sin(\hat{\phi}_{h}-\hat{\phi}_{S_h})\cos({\phi}_{S_A}-{\phi}_q)}(q_T,j_\perp)&=\mathcal{C}_{11L}\left[{\mathcal{D}}_{1T,q}^{\perp}\tilde{g}_{1T}^{(1)}\bar{S}_{\rm global}\bar{S}_{cs}\right]\,,
\label{eq:strh24}
\end{align}
\endgroup
where TMDJFFs found in the above equations can also be simplified in terms of TMDFFs and collinear-soft function in the region $j_\perp \ll p_T R$.

\begin{figure}
\centering
\includegraphics[width = 0.43\textwidth]{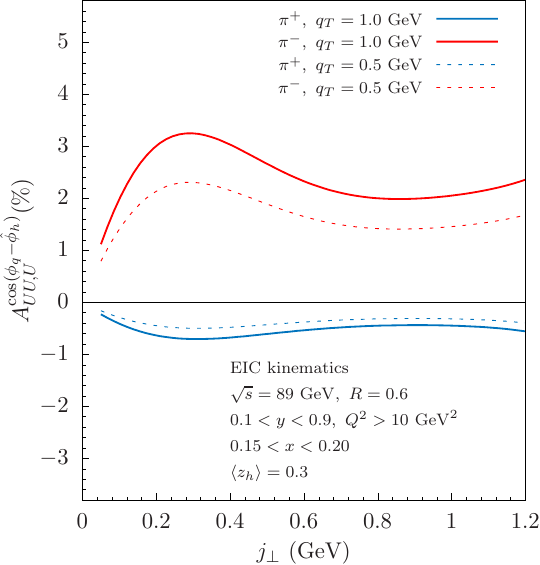}
\hspace{1cm}\includegraphics[width = 0.43\textwidth]{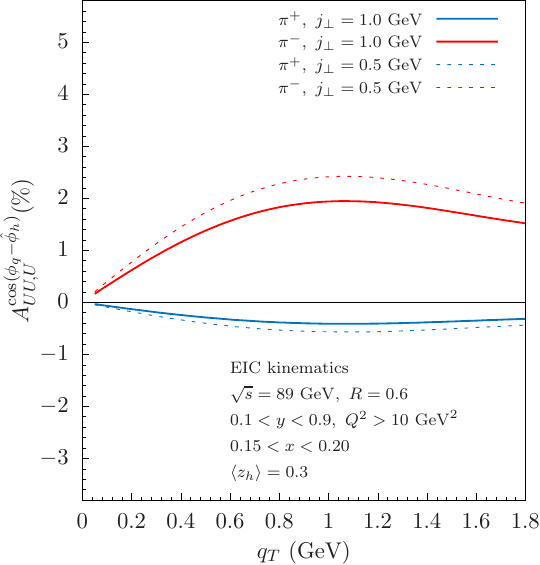}
\caption{The horizontal ($j_\perp$-dependent) and vertical ($q_T$-dependent) slices of $A_{UU,U}^{\cos(\phi_{q}-\hat{\phi}_{h})}$ for unpolarized $\pi^\pm$ in jet production with electrons in unpolarized $ep$ collisions are depicted. The left panel shows the variation as a function of the transverse momentum $j_\perp$, with fixed values of $q_T=1.0$ GeV and $0.5$ GeV for $\pi^\pm$. The right panel illustrates the dependence on the jet imbalance $q_T$, while $j_\perp$ is fixed at $1.0$ GeV and $0.5$ GeV for $\pi^\pm$, respectively. These results are obtained using EIC (Electron-Ion Collider) kinematics with a center-of-mass energy of $\sqrt{s}=89$ GeV, jet radius $R=0.6$, inelasticity $y$ in the range $[0.1,0.9]$, $Q^2>10$ GeV$^2$, Bjorken-$x$ within the interval $[0.15,0.20]$, and an average momentum fraction of $\langle z_h\rangle=0.3$.}
    \label{fig:scn2_2d}
\end{figure}

\subsection{Example 1: Boer-Mulders correlation with Collins function}
\label{sec3:pheno}
In this section, we demonstrate the application of this process in constraining TMD functions through a new phenomenological study as an example. The azimuthal modulation under investigation is denoted as $A_{UU,U}^{\cos(\phi_{q}-\hat{\phi}_h)}$, which depends on the azimuthal angles $\phi_q$ and $\hat \phi_h$. This modulation allows us to study the Boer-Mulders TMDPDFs $h_{1}^\perp$ and the Collins TMDFFs $H_{1}^\perp$.

Previously, the well-known Collins asymmetry for hadrons in a jet, resulting from the collisions between an unpolarized electron and a transversely polarized proton, and represented as a $\sin(\phi_{S_A}-\hat{\phi}_h)$ modulation in \cref{eq:unpjeth}, has been investigated~\cite{Arratia:2020nxw}. In this section, we conduct a new phenomenological study, focusing on the azimuthal modulation $\cos(\phi_{q}-\hat{\phi}_h)$ in \cref{eq:unpjeth}. This azimuthal dependence arises in the distribution of unpolarized hadrons in collisions involving unpolarized electrons and protons. The relevant structure function, $F_{UU,U}^{\cos(\phi_{q}-\hat{\phi}_h)}$, probes the Boer-Mulders function $h_1^\perp$ in the unpolarized proton, coupled with the Collins fragmentation function $H_1^\perp$. The advantage of this asymmetry is that it does not require any polarization of either the beams or the final-state hadron, making it accessible even to the HERA experiment. Therefore, we present numerical results for both HERA and EIC kinematics.

To proceed, we define the new azimuthal asymmetry by normalizing the structure function $F_{UU,U}^{\cos(\phi_{q}-\hat{\phi}_h)}$ with respect to the unpolarized and azimuthal-independent structure function $F_{UU,U}$, as follows:
\begin{align}
\label{eq:A-UUU}
A_{UU,U}^{\cos(\phi_{q}-\hat{\phi}_h)}=\frac{F_{UU,U}^{\cos(\phi_{q}-\hat{\phi}_h)}}{F_{UU,U}}\,.
\end{align}
Namely the azimuthal asymmetry is defined as the ratio of two structure functions: the denominator $F_{UU,U}$ and the numerator $F_{UU,U}^{\cos(\phi_{q}-\hat{\phi}_h)}$ in \cref{eq:unpjeth}. The factorization formula and the parametrization of the unpolarized Transverse Momentum Dependent (TMD) parton distribution functions (PDFs) for the denominator $F_{UU,U}$ were presented in \cref{eq:FUUUbefore} and discussed subsequently. On the other hand, the structure function $F_{UU,U}^{\cos(\phi_{q}-\hat{\phi}_h)}$ depend on the Boer-Mulders TMDPDF $h_1^\perp$ and the Collins Transverse Momentum Dependent Fragmentation Function (TMDJFF) $\mathcal{H}_1^\perp$.

The Boer-Mulders function describes the distribution of transversely polarized quarks inside an unpolarized proton. When such a transversely polarized quark scatters with an unpolarized electron, it undergoes transverse spin transfer, resulting in a transversely polarized quark initiating a jet with a distribution of unpolarized hadrons measured inside the jet. On the other hand, the Collins function describes the correlation of a transversely polarized quark fragmenting into an unpolarized hadron. Thus, the structure function $F_{UU,U}^{\cos(\phi_{q}-\hat{\phi}_h)}$ is related to the Collins function.

The factorization formula for $F_{UU,U}^{\cos(\phi_{q}-\hat{\phi}_h)}$ is given in \cref{eq:strhlep2}, and for convenience, it is explicitly expressed here as follows:
\begin{align}
F_{UU,U}^{\cos(\phi_{q}-\hat{\phi}_h)} =&\hat{\sigma}_T \,H(Q,\mu)\sum_q e_q^2\, \frac{ j_\perp}{z_hM_h} H_1^{\perp\,h/q}(z_h,j_\perp^2,\mu, \zeta_J)
\nnu
&\times 
M \int\frac{b^2 \,db}{2\pi}J_1(q_Tb)\,x\,\tilde{h}_1^{\perp\, q(1)}(x,b^2, \mu,\zeta)\bar{S}_{\rm global}(b^2,\mu)\bar{S}_{cs}(b^2,R,\mu)\,,
\label{eq:FUUU-spin}
\end{align}
The transverse spin-transfer cross section $\hat\sigma_T$ is given by \cref{eq:sigma_T}. To establish the relationship between the TMDJFF $\mathcal{H}_1^{\perp,h/q}$ and the Collins TMDFF $H_1^{\perp,h/q}$, we have followed the same procedure as we did for the case of the unpolarized TMDJFF $\mathcal{D}_1^{h/q}$ from \cref{unp_JFF_FF} to \cref{unp_JFF_FF2}.

It is important to emphasize that $F_{UU,U}^{\cos(\phi_{q}-\hat{\phi}_h)}$ is differential in both $q_T$ and $j_\perp$, enabling us to separately constrain the Boer-Mulders function $\tilde{h}_1^{\perp}$ and the Collins function $H_1^{\perp,h/q}$. This characteristic is evident in \cref{eq:FUUU-spin}, where all the $q_T$-dependence is contained in the Fourier transform $b$-integral, while the $j_\perp$-dependence is outside this integration. This physical distinction is expected since $q_T$ and $j_\perp$ are measured with respect to two different directions, namely the beam direction and the jet direction, respectively. This advantageous feature contrasts with the usual TMD measurements, such as in Drell-Yan production, where all transverse momenta are measured with respect to the beam direction.

For the phenomenology in the subsequent analysis, we employ the Collins TMDFFs extracted from~\cite{Kang:2015msa}, which possess proper TMD evolution. However, to proceed, we still require the parametrization for the Boer-Mulders function $h_1^{\perp}$. For the numerical studies below, we adopt the Boer-Mulder functions extracted from~\cite{Barone:2009hw}, which are based on the usual Gaussian model. Subsequently, we establish a parametrization for $\tilde{h}_1^{\perp, q(1)}(x,b^2, \mu,\zeta)$ with TMD evolution:
\begin{align}
\tilde{h}_1^{\perp\, q(1)}(x,b^2, \mu,\zeta) = 
2\pi h_{1}^{\perp\, q(1)}\left(x,\mu_{b_*}\right)
\exp\left[-S_{\rm pert}\left(\mu, \mu_{b_*} \right) - S_{\rm NP}^{h_{1}^\perp}\left(x, b, Q_0, \zeta\right)\right]\,,
\end{align}
where the collinear function $h_{1}^{\perp\, q(1)}\left(x,\mu_{b_*}\right)$ is constructed in the Gaussian model from~\cite{Barone:2009hw} via
\begin{align}
h_{1}^{\perp\, q(1)}(x)=\int d^2\boldsymbol{k}_T\frac{k_T^2}{2M^2}h_{1}^{\perp\,q}(x,k_T^2)\,.
\end{align}
On the other hand, we have the non-perturbative Sudakov factor $S_{\rm NP}^{h_{1}^\perp}$ given as
\begin{align}
S_{\rm NP}^{h_{1}^\perp}\left(x, b, Q_0, \zeta\right) = \frac{g_2}{2}\ln{\frac{\sqrt{\zeta}}{Q_0}}\ln{\frac{b}{b_*}}+g_1^{h_{1}^\perp} b^2\,.
\end{align}
Here $g_1^{h_{1}^\perp}$ is again related to the intrinsic Gaussian width for the TMDPDF $h_1^\perp$ in the transverse momentum space
\begin{align}
g_1^{h_{1}^\perp} = \frac{\langle k_T^2\rangle_{h_1^\perp}}{4} = 0.036{\rm ~GeV}^2\,,
\end{align}
where we used $\langle k_T^2\rangle_{h_1^\perp}=\frac{M_1^2\langle k_T^2\rangle}{M_1^2+\langle k_T^2\rangle}$  with $\langle k_T^2\rangle=0.25$ GeV$^2$ and $M_1^2=0.34$ GeV$^2$ from~\cite{Barone:2009hw}. 

\begin{figure}[hbt!]
\hspace{1.3cm}\includegraphics[width = 0.8\textwidth]{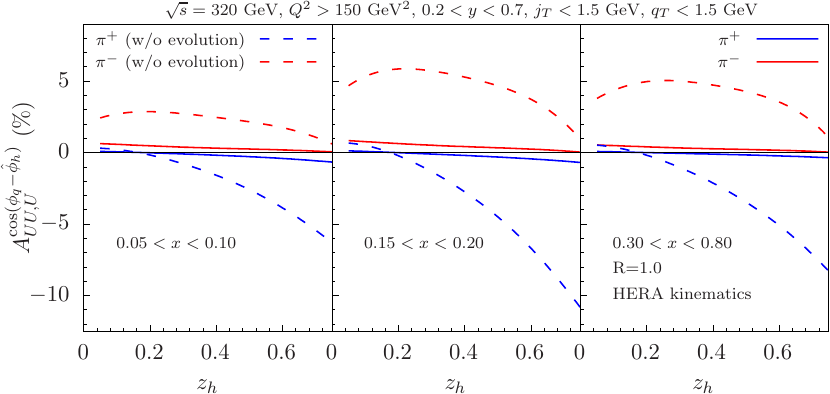}
\caption{Numerical results of $A_{UU,U}^{\cos(\phi_{q}-\hat{\phi}_{h})}$ as a function of hadron momentum fraction $z_h$ for unpolarized $\pi^\pm$ in jet production with electron in unpolarized $ep$ collision predicted for HERA using three different bins of $x$:  $[0.05,0.1],\ [0.15,0.2]$, and $[0.3,0.8]$. The solid (dashed) curves are the calculations with (without) TMD evolution. We apply the center-of-mass energy $\sqrt{s}=320$ GeV of HERA kinematics, jet radius $R=1.0$, $Q^2>150$ GeV$^2$, inelasticity $y$ in $[0.2,0.7]$ with transverse momentum imbalance $q_T$ and final hadron transverse momentum in jet are both smaller than $1.5$ GeV.}
    \label{fig:scn2_hera}
\end{figure}

The azimuthal asymmetry $A_{UU,U}^{\cos(\phi_{q}-\hat{\phi}_{h})}$ involves only unpolarized proton and electron beams, making it suitable for study in the HERA experiment at DESY. We present numerical results for both HERA and future Electron-Ion Collider (EIC) kinematics. In \cref{fig:scn2_hera}, we plot the azimuthal asymmetry $A_{UU,U}^{\cos(\phi_{q}-\hat{\phi}_{h})}$ for unpolarized $\pi^\pm$ production inside a jet with a radius $R=1$ using HERA kinematics~\cite{DiStalk}. Specifically, we choose an electron-proton center-of-mass energy of $\sqrt{s}=320$ GeV and apply the cuts $Q^2 > 150$ GeV$^2$ and $0.2< y< 0.7$. Furthermore, we integrate over the hadron transverse momentum $j_\perp$ and the imbalance $q_T$ with $0<j_\perp<1.5$ GeV and $0<q_T<1.5$ GeV. Our analysis is performed as a function of the hadron momentum fraction $z_h$ using three different bins of $x$: $[0.05,0.1]$, $[0.15,0.2]$, and $[0.3,0.8]$.

It is important to note that the cuts on $Q^2$, $y$, and $x$ are chosen to constrain the jet $p_T$ directly, keeping us in the TMD factorization regime discussed above. The numerical results are presented with and without TMD evolution between scales, represented by solid and dashed lines in the figures, respectively. Without TMD evolution, the azimuthal correlations are assumed to be purely Gaussian, as shown by the dashed lines. We find that the azimuthal asymmetry is negative for $\pi^+$ production inside the jet and positive for $\pi^-$ production inside the jet, with a magnitude of up to around $1\%$ for HERA when TMD evolution is considered. However, without TMD evolution, the size of the azimuthal asymmetry can be much larger, reaching around $\sim 5\%$. This is consistent with the expectation that TMD evolution suppresses the asymmetry as radiation broadens the distribution. Thus, the azimuthal asymmetry $A_{UU,U}^{\cos(\phi_{q}-\hat{\phi}_{h})}$ serves a dual purpose. On one hand, it allows us to extract information about the Boer-Mulders TMDPDFs and Collins TMDFFs. On the other hand, this asymmetry provides valuable constraints for the TMD evolution of these TMD functions.

\begin{figure}[hbt!]
\centering
\includegraphics[width = 0.9\textwidth]{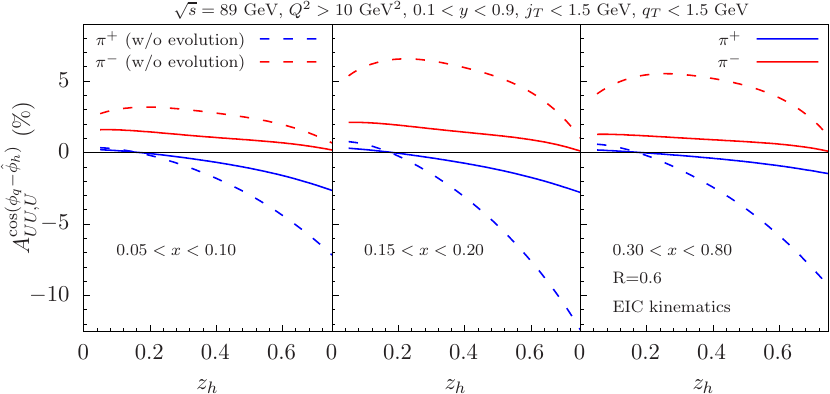}
\caption{Numerical results of $A_{UU,U}^{\cos(\phi_{q}-\hat{\phi}_{h})}$ as a function of hadron momentum fraction $z_h$ for unpolarized $\pi^\pm$ in jet production with electron in unpolarized $ep$ collision predicted for EIC  using three different bins of $x$:  $[0.05,0.1],\ [0.15,0.2]$, and $[0.3,0.8]$. The solid (dashed) curves are the calculations with (without) TMD evolution. We apply the center-of-mass energy $\sqrt{s}=89$ GeV of EIC kinematics, jet radius $R=0.6$, $Q^2>10$ GeV$^2$, inelasticity $y$ in $[0.1,0.9]$ and both transverse momentum imbalance $q_T$ and final hadron transverse momentum in jet smaller than $1.5$ GeV.}
    \label{fig:scn2_eic}
\end{figure}
We also present the same asymmetry using EIC kinematics with a jet radius of $R=0.6$ in \cref{fig:scn2_eic}. The asymmetry is shown as a function of $z_h$ in three different bins of $x$: $[0.05,0.1]$, $[0.15,0.2]$, and $[0.3,0.8]$. For these calculations, we use a center-of-mass energy of $\sqrt{s}=89$ GeV and apply the following cuts: $Q^2>10$ GeV$^2$, $0.1<y<0.9$, and $0<j_\perp, , q_T < 1.5$ GeV. The azimuthal asymmetry exhibits similar trends as with HERA kinematics, but with larger magnitudes, i.e., $\sim 2-3\%$ ($5-10\%$) with (without) TMD evolution, in comparison to the asymmetry expected with HERA kinematics. These findings indicate that experimental measurements of $A_{UU,U}^{\cos(\phi_{q}-\hat{\phi}_{h})}$ at EIC could be quite promising and can be used to constrain TMD evolution for Boer-Mulders and Collins functions. For the remainder of the paper, we will only present the numerical results with TMD evolution.

Alternatively, instead of integrating over $q_T$ and $j_\perp$, we can create plots that are simultaneously differential in $q_T$ and $j_\perp$. As discussed earlier, this approach is useful as TMDPDFs and TMDFFs are separately sensitive to $q_T$ and $j_\perp$, respectively. For the EIC kinematics with jet radius $R=0.6$, an inelasticity cut of $0.1<y<0.9$, momentum fraction $\langle z_h\rangle=0.3$, and $Q^2>10$ GeV$^2$, we generate three-dimensional and contour plots of the azimuthal asymmetry $A_{UU,U}^{\cos(\phi_{q}-\hat{\phi}_{h})}$ in \cref{fig:scn2_pip} and\ref{fig:scn2_pim}. To gain better insights into the unpolarized and azimuthal-dependent structure function, we present the three-dimensional and contour plots of $F_{UU,U}$, $F_{UU,U}^{\cos(\phi_{q}-\hat{\phi}_{h})}$, and their ratio $A_{UU,U}^{\cos(\phi_{q}-\hat{\phi}_{h})}$ in these figures. In the first row of both figures, we observe the Sudakov peak from the unpolarized TMDPDF and TMDFF for constant $j_\perp$ and $q_T$ slices, respectively. In the second row, the shape of the constant $j_\perp$ slices, i.e., the $q_T$-dependence at a constant $j_\perp$, is determined by the Boer-Mulders function. On the other hand, the shape of the constant $q_T$ slices, i.e., the $j_\perp$-dependence at a constant $q_T$, is determined by the Collins function. Finally, the ratio of these plots, which defines the asymmetry, is given in the third row. We find that the spin asymmetry for $\pi^+$ production inside the jet tends to be negative, around $\sim 1\%$, while for $\pi^-$ production inside the jet, it is positive with a magnitude of approximately $\sim 3\%$.

\begin{figure}
\includegraphics[width = 0.4\textwidth]{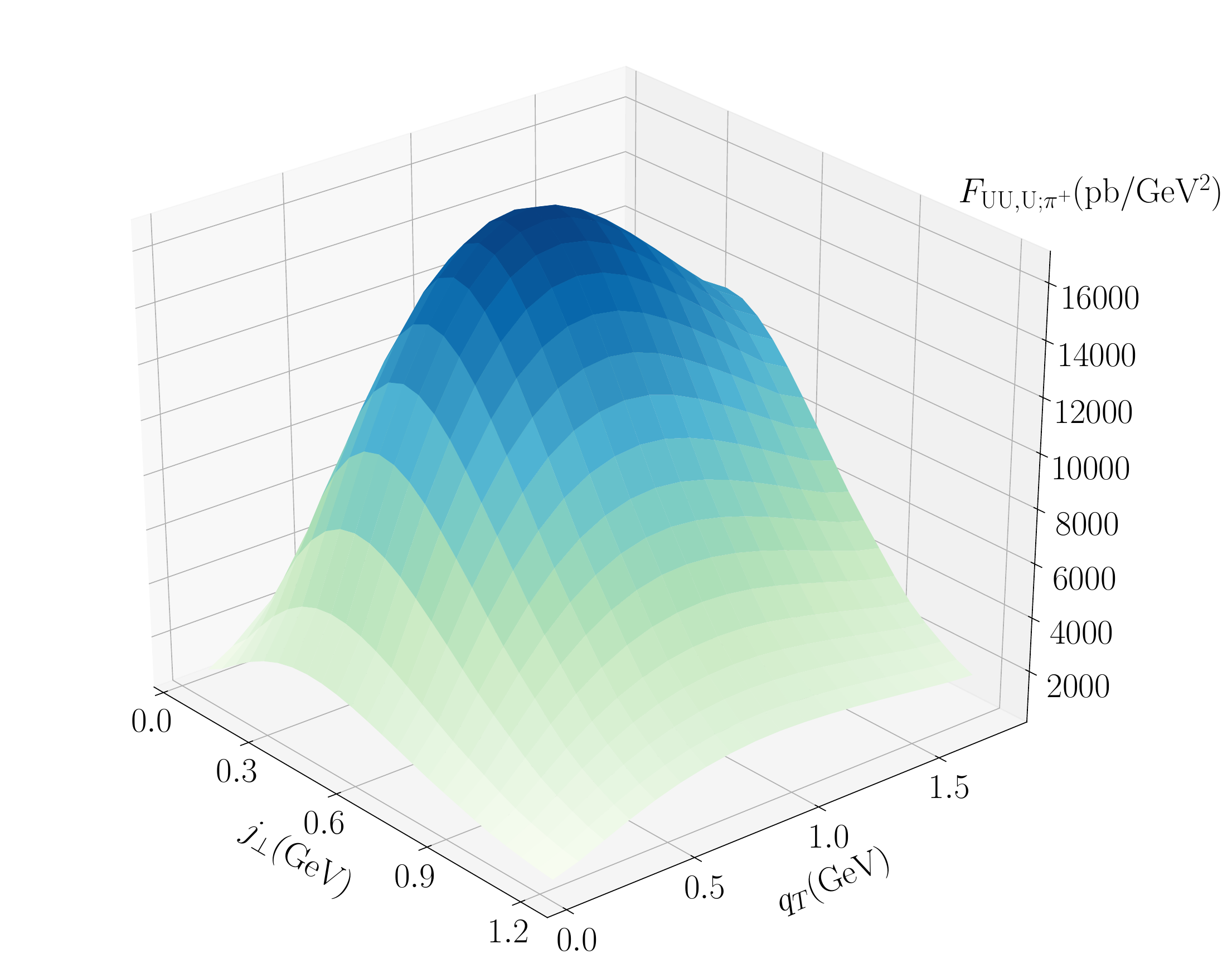}\hspace{-1.5cm}
\includegraphics[width = 0.37\textwidth]{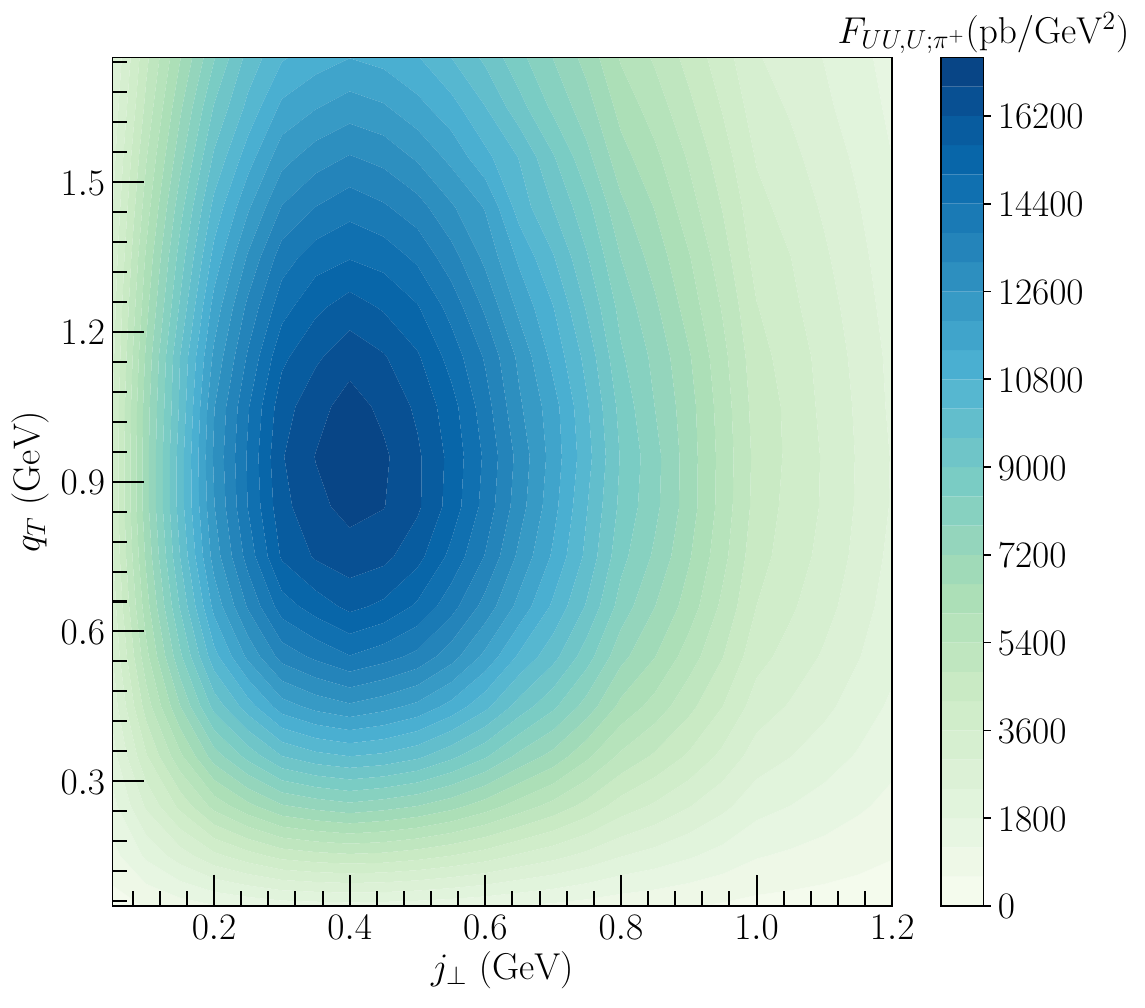}
\includegraphics[width = 0.4\textwidth]{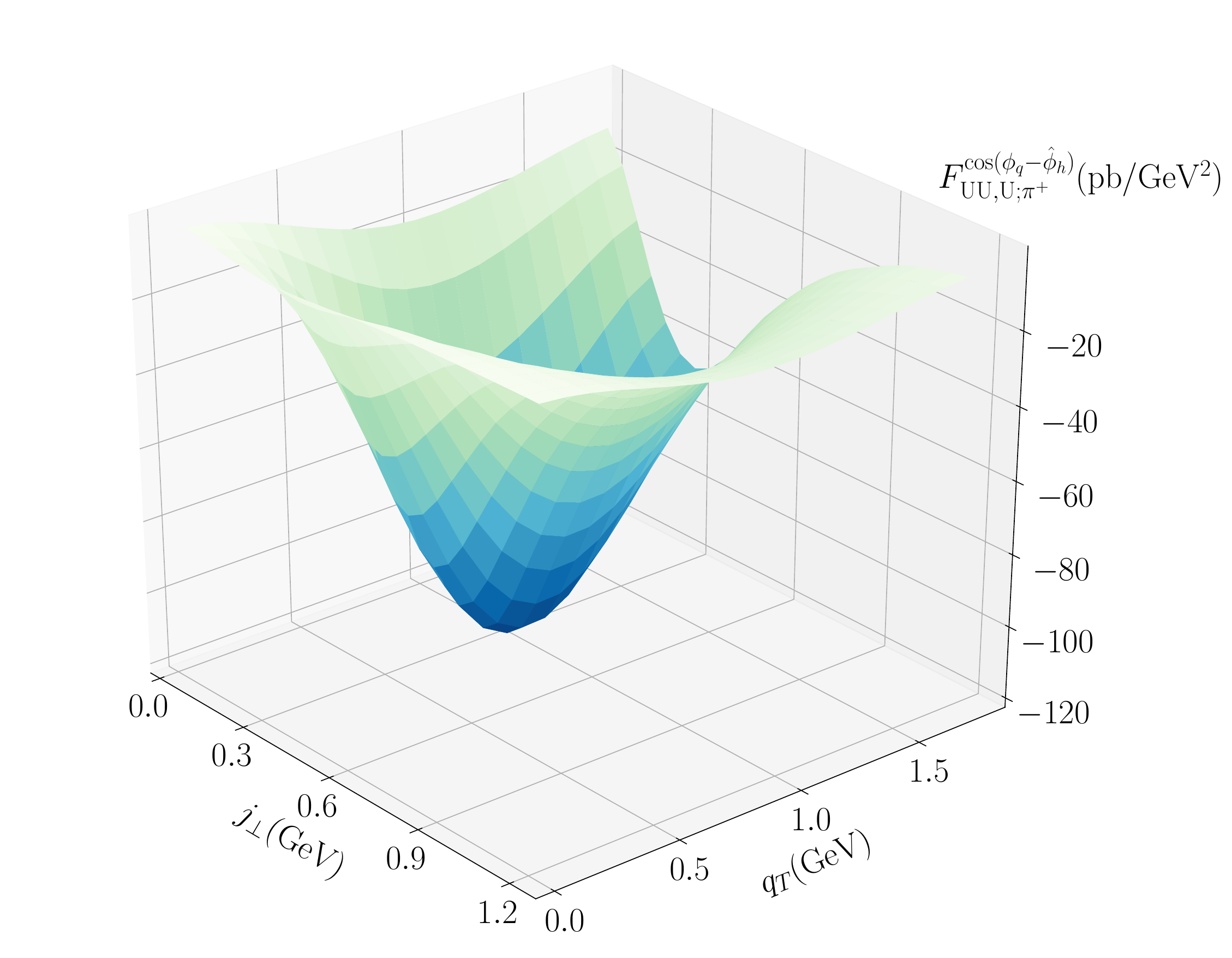}
\includegraphics[width = 0.38\textwidth]{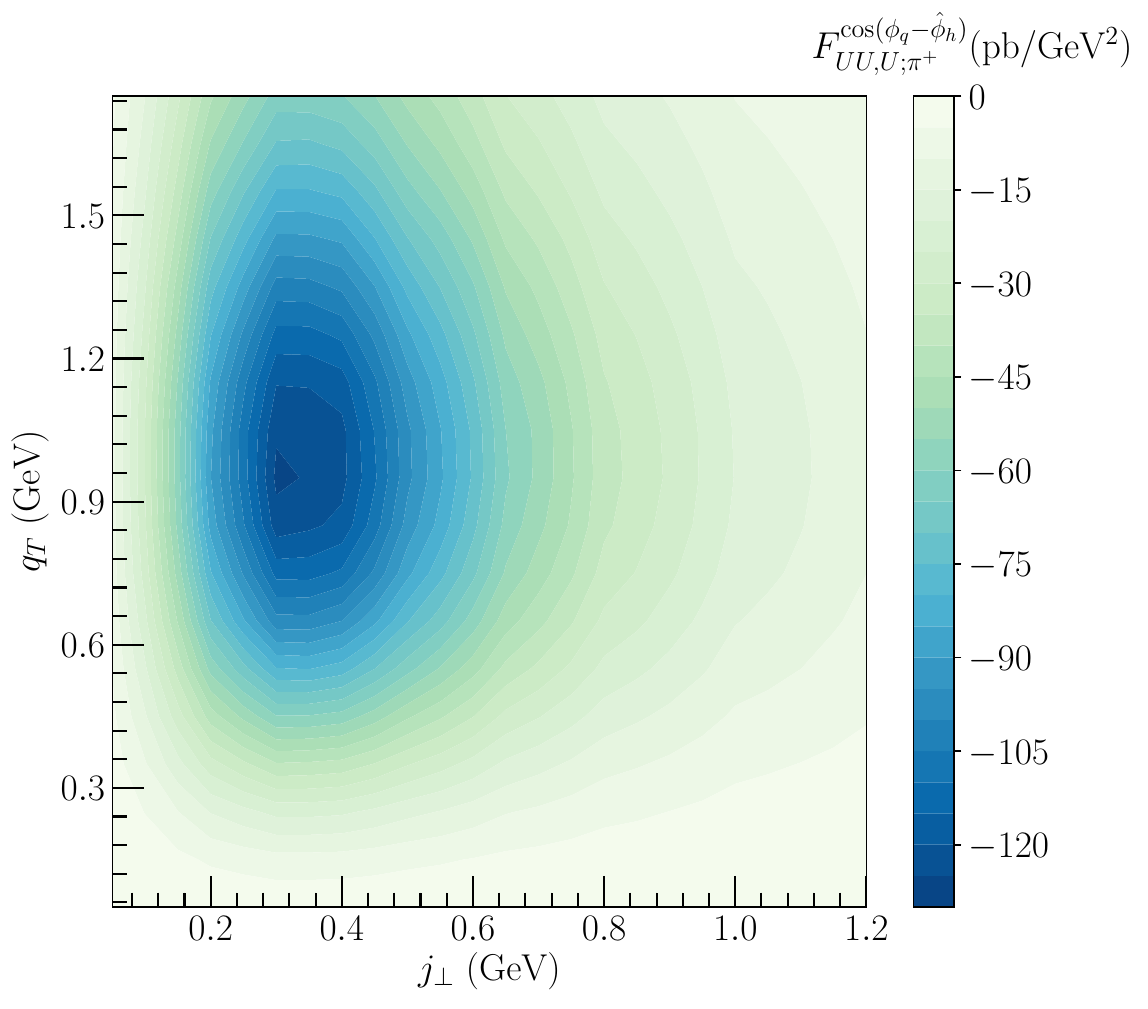}
\includegraphics[width = 0.4\textwidth]{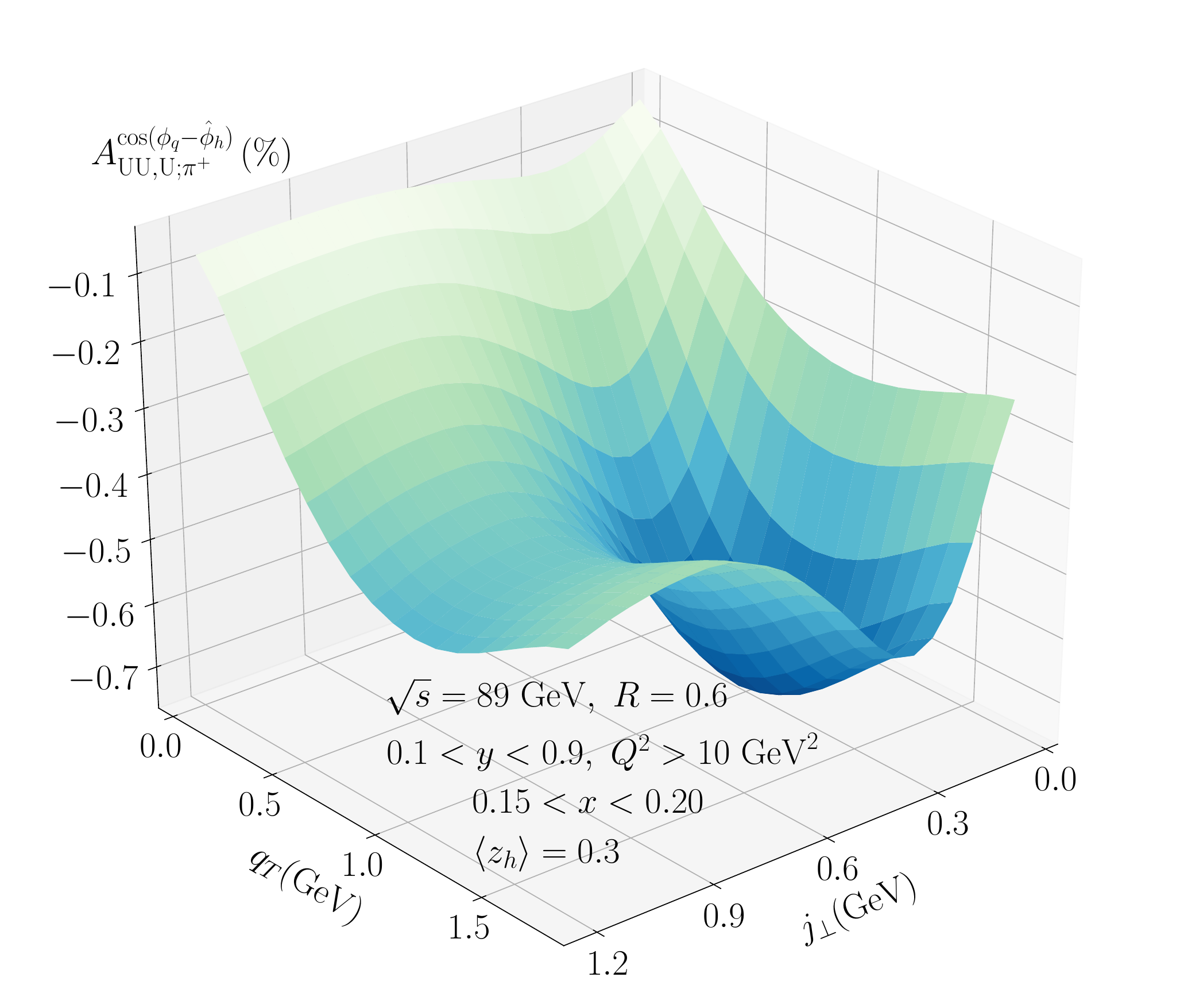}
\hspace{3.5cm}\includegraphics[width = 0.37\textwidth]{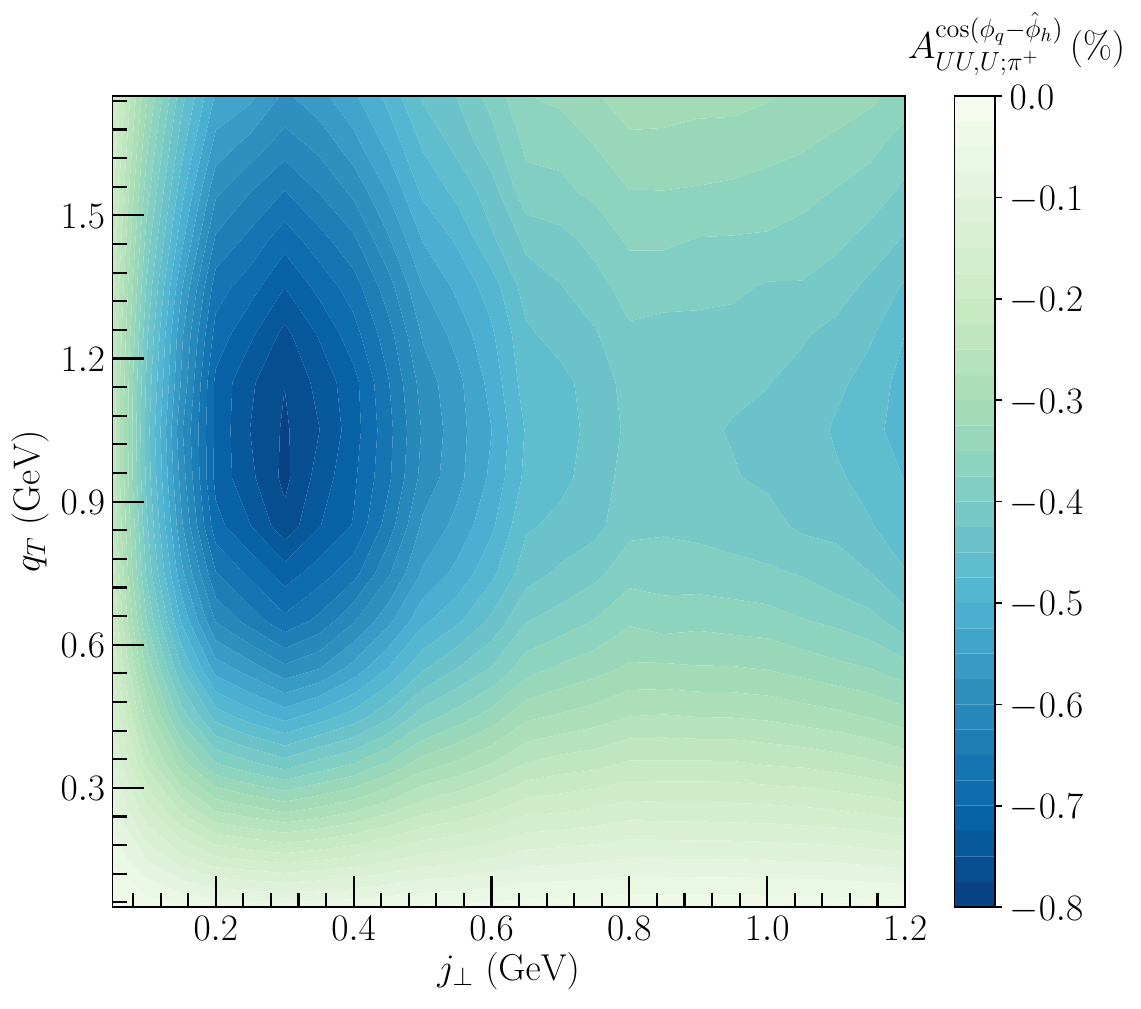}
\caption{$F_{UU,U}$ (first row), $F_{UU,U}^{\cos(\phi_{q}-\hat{\phi}_{h})}$ (second row) and $A_{UU,U}^{\cos(\phi_{q}-\hat{\phi}_{h})}$ (third row) as a function of jet imbalance $q_T$ and $j_\perp$ for unpolarized $\pi^+$ in jet production with electron in unpolarized $ep$ collision with EIC kinematics, where we have applied $\sqrt{s}=89$ GeV, jet radius $R=0.6$, inelasticity $y$ in range $[0.1,0.9]$, $Q^2>10$ GeV$^2$, Bjorken-$x$ in $[0.15,0.20]$ and average momentum fraction $\langle z_h\rangle=0.3$. Left column: Three dimensional plots of the structure functions and their ratio in $q_T$ and $j_\perp$. Right column: Contour plots of the structure functions and their ratio.}
    \label{fig:scn2_pip}
\end{figure}

\begin{figure}
\includegraphics[width = 0.4\textwidth]{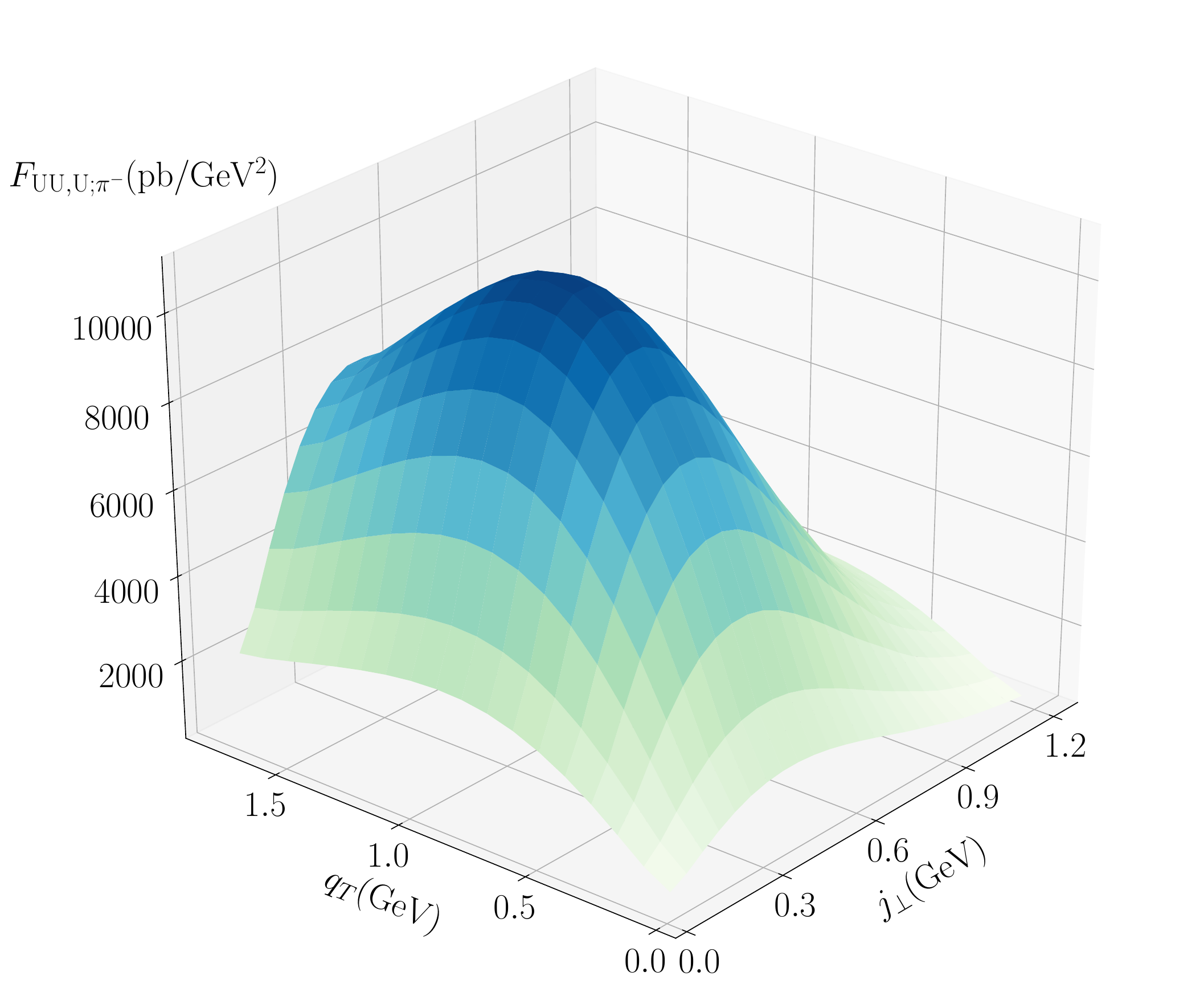}\hspace{-1.7cm}
\includegraphics[width = 0.37\textwidth]{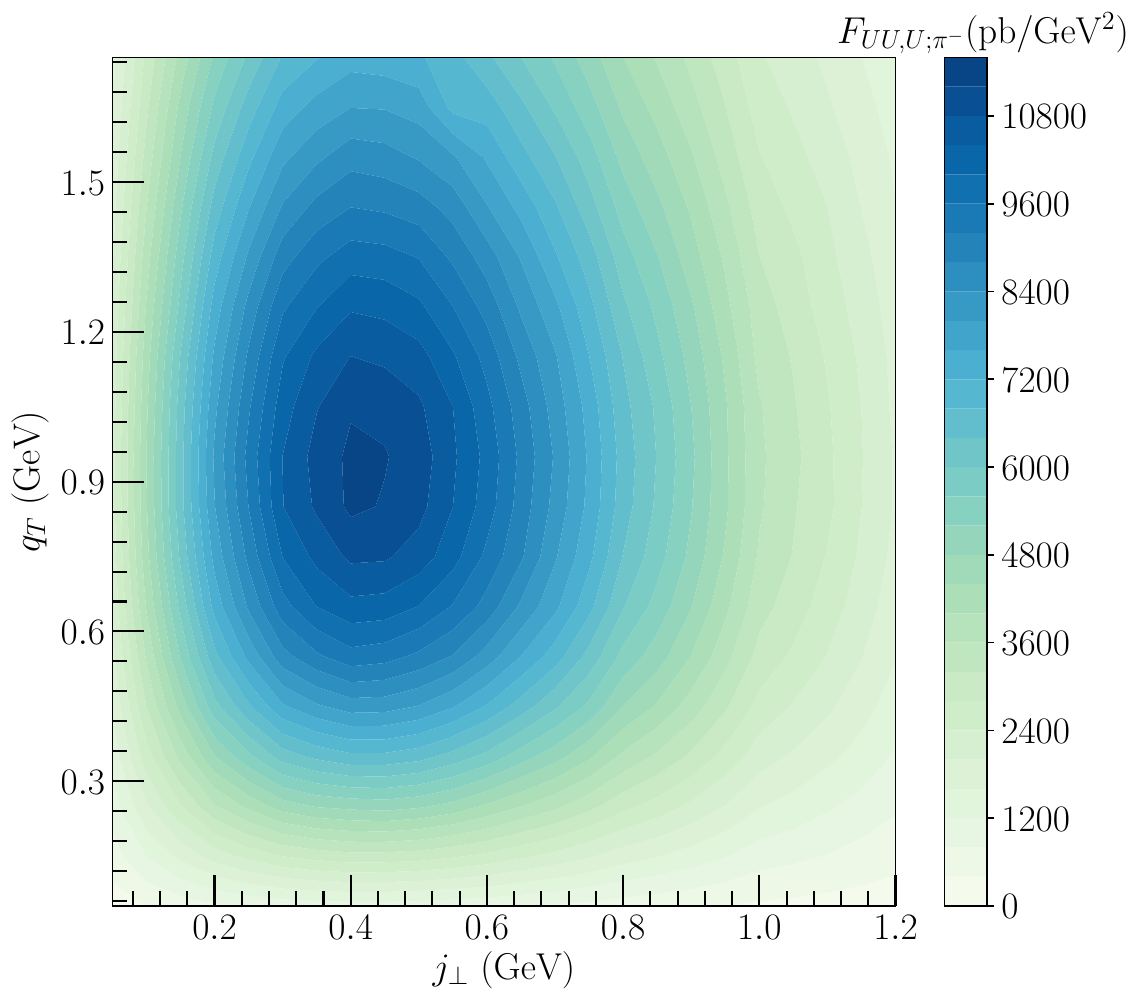}
\includegraphics[width = 0.4\textwidth]{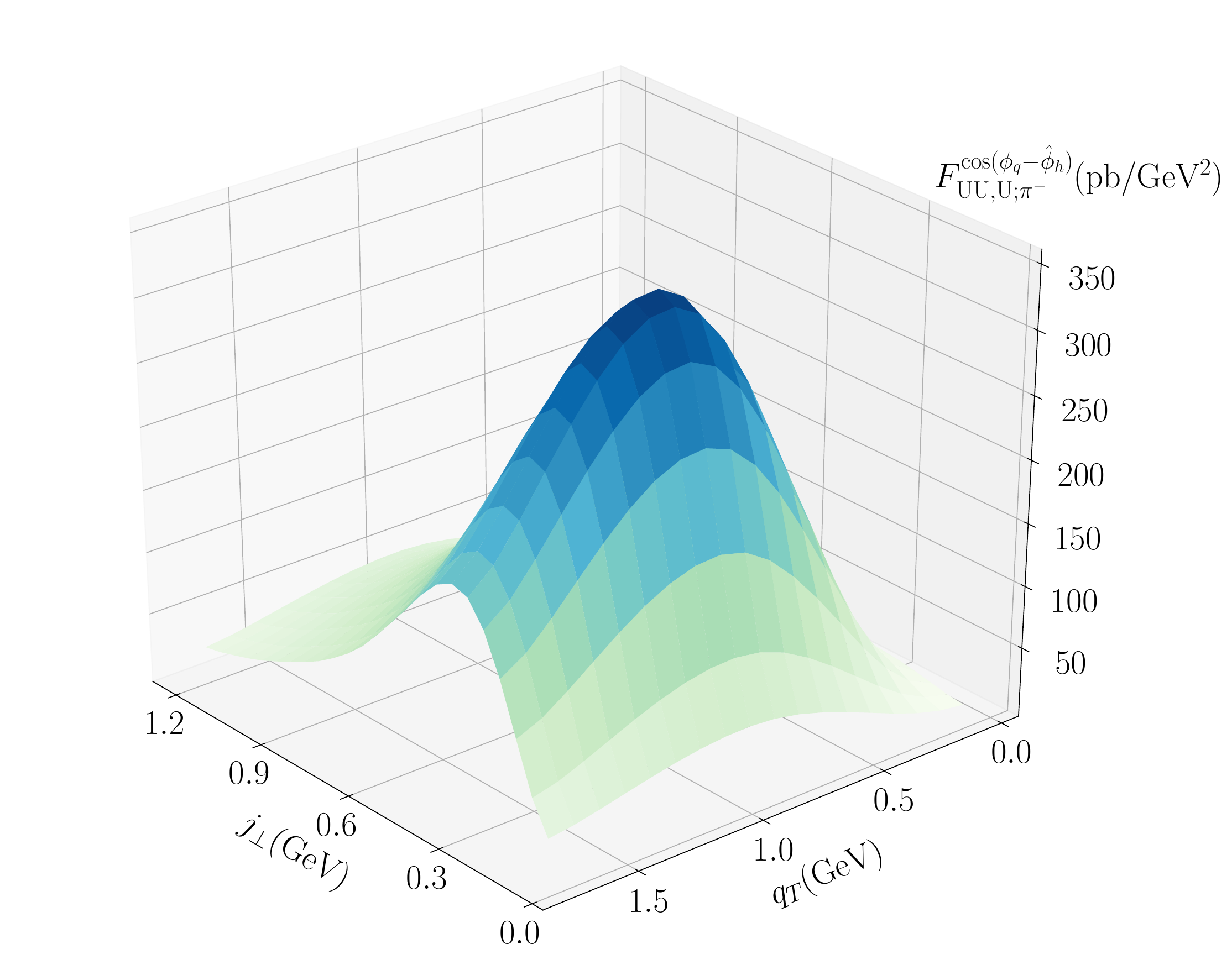}
\includegraphics[width = 0.38\textwidth]{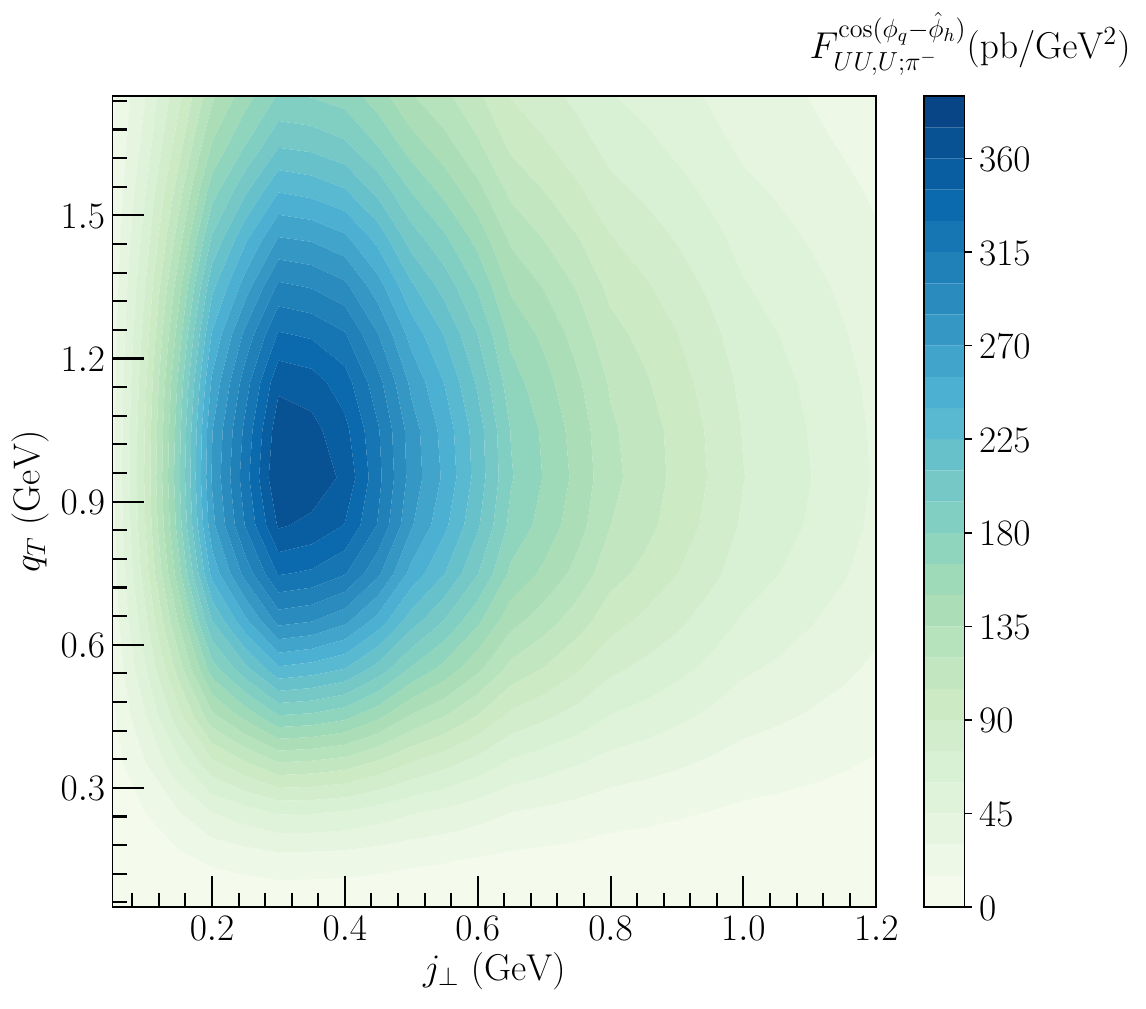}
\includegraphics[width = 0.4\textwidth]{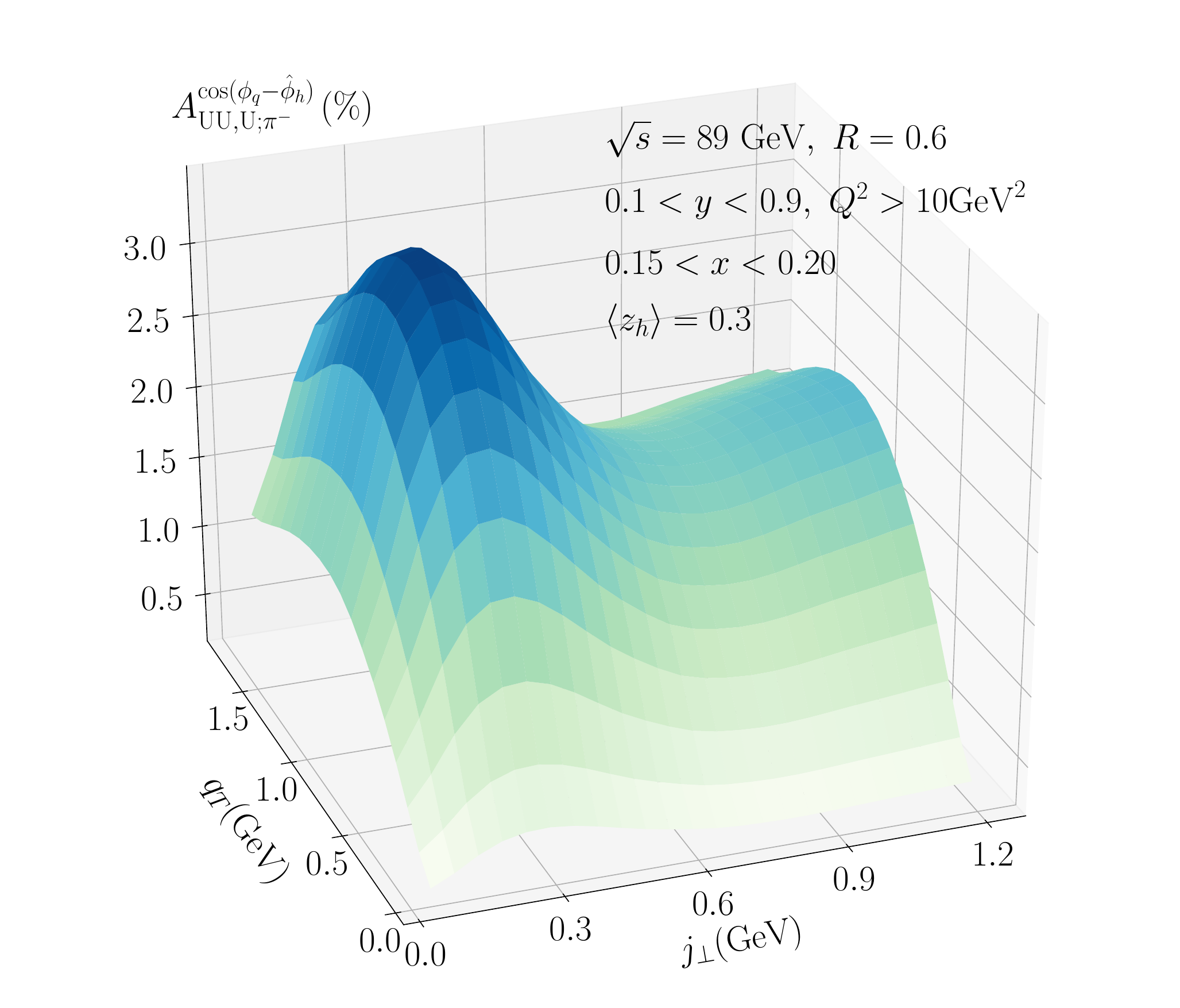}
\hspace{3.5cm}\includegraphics[width = 0.37\textwidth]{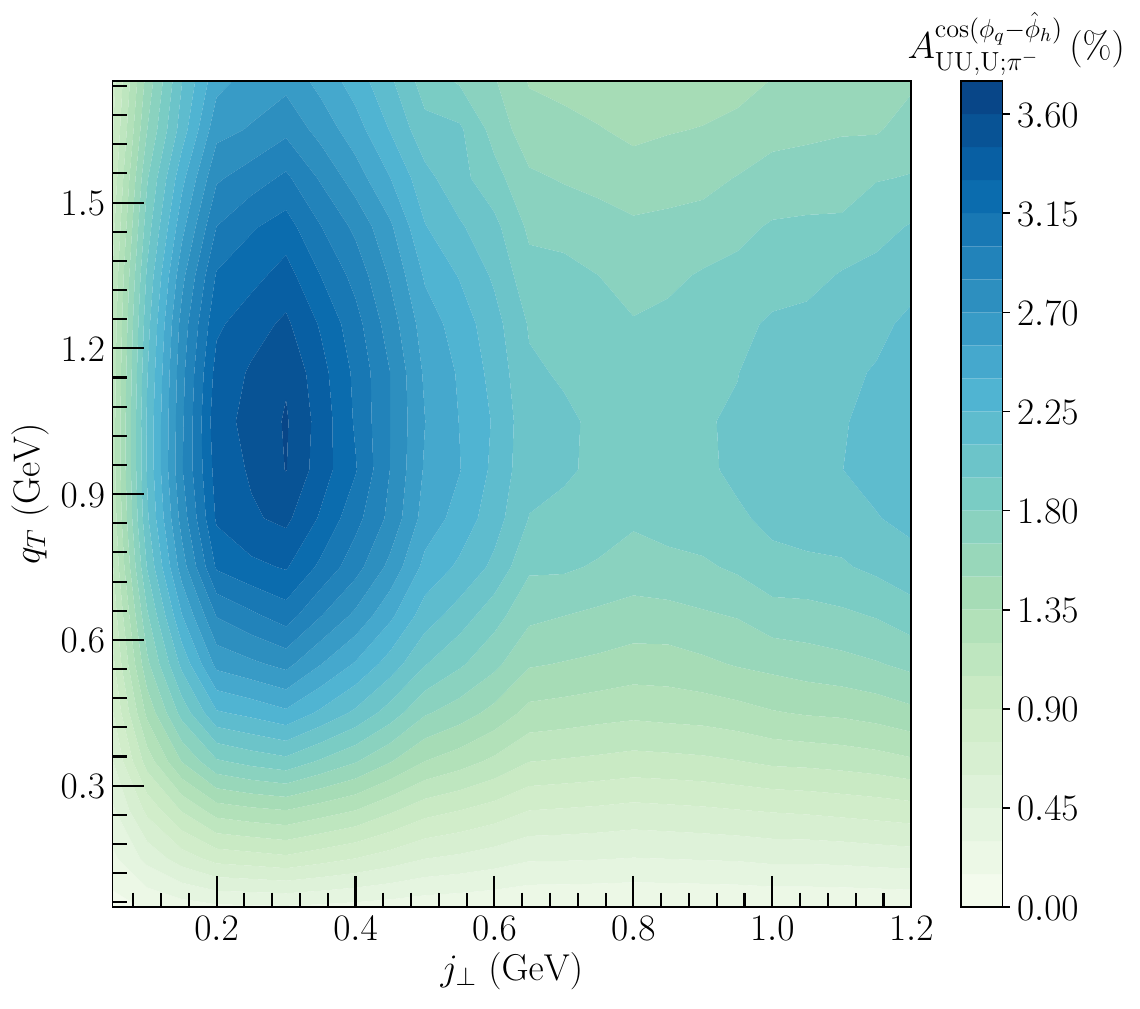}
\caption{$F_{UU,U}$ (first row), $F_{UU,U}^{\cos(\phi_{q}-\hat{\phi}_{h})}$ (second row) and $A_{UU,U}^{\cos(\phi_{q}-\hat{\phi}_{h})}$ (third row) as a function of jet imbalance $q_T$ and $j_\perp$ for unpolarized $\pi^-$ in jet production with electron in unpolarized $ep$ collision with EIC kinematics, where we have applied $\sqrt{s}=89$ GeV, jet radius $R=0.6$, inelasticity $y$ in range $[0.1,0.9]$, $Q^2>10$ GeV$^2$, Bjorken-$x$ in $[0.15,0.20]$ and average momentum fraction $\langle z_h\rangle=0.3$. Left column: Three dimensional plots of the structure functions and their ratio in $q_T$ and $j_\perp$. Right column: Contour plots of the structure functions and their ratio.}
    \label{fig:scn2_pim}
\end{figure}

To give a more straightforward interpretation, in \cref{fig:scn2_2d} we show the horizontal ($j_\perp$-dependent) slices for $q_T=1.0$ GeV (solid curves) and $q_T=0.5$ GeV (dashed curves) of the third row of \cref{fig:scn2_pip} and \ref{fig:scn2_pim} in the left plot with blue curves representing $\pi^+$ and red curves representing $\pi^-$ productions in jet. As for the right plot of \cref{fig:scn2_2d}, we provide the vertical ($q_T$-dependent) slices for $j_\perp=1.0$ GeV (solid curves) and $j_\perp=0.5$ GeV (dashed curves) of the third row of \cref{fig:scn2_pip} and \ref{fig:scn2_pim} with blue curves representing $\pi^+$ and red curves representing $\pi^-$ productions in jet. With the reasonable asymmetry of order negative  $\sim 1\%$ for $\pi^+$ and positive $\sim 3\%$ for $\pi^-$ with the TMD evolution turned on, this is a promising observable at the EIC to study the Boer-Mulders functions and Collins fragmentation functions.

\subsection{Example 2: $\Lambda$ transverse polarization inside the jet}
\label{sec4:pheno}
As an example of application of studying the back-to-back electron-jet production with a polarized hadron inside the jet, we study transverse spin asymmetry of a $\Lambda$ particle inside the jet, $A^{\sin(\hat{\phi}_\Lambda-\hat{\phi}_{S_\Lambda})}_{UU,T}$, which arises from the structure function $F^{\sin(\hat{\phi}_\Lambda-\hat{\phi}_{S_\Lambda})}_{UU,T}$. The spin asymmetry is defined as
\bea\label{eq:auut}
A_{UU,T}^{\sin(\hat{\phi}_{\Lambda}-\hat{\phi}_{S_\Lambda})} = \frac{F^{\sin(\hat{\phi}_\Lambda-\hat{\phi}_{S_\Lambda})}_{UU,T}}{F_{UU,U}}\,.
\eea
The asymmetry can be measured in the unpolarized electron-proton collisions by observing the distribution of transversely polarized $\Lambda$s inside the jet. The $\Lambda$ transverse spin vector $\bm{S}_{\Lambda\perp}$ and the transverse momentum $\bm{j}_\perp$ with respect to the jet axis can correlate with each other, and leads to the $\sin(\hat{\phi}_{\Lambda}-\hat{\phi}_{S_\Lambda})$ correlation between their azimuthal angles. In practice, this is the mechanism which can describe the transverse polarization of $\Lambda$ particles inside the jet. As can be seen from \cref{eq:strh9}, the structure function $F^{\sin(\hat{\phi}_\Lambda-\hat{\phi}_{S_\Lambda})}_{UU,T}$ depends on the unpolarized TMDPDF $f_1^q(x,k_T^2)$ and TMDJFF $\mathcal{D}_{1T}^{\perp\ \Lambda/q}(z_\Lambda,j_\perp^2)$. The TMDJFF $\mathcal{D}_{1T}^{\perp\ \Lambda/q}(z_\Lambda,j_\perp^2)$ describes distribution of transversely polarized $\Lambda$ inside the jet initiated by an unpolarized quark. This is reminiscent of the polarizimg TMDFF $D_{1T}^{\perp\ \Lambda/q}(z_\Lambda,j_\perp^2)$, which describes distribution of transversely polarized $\Lambda$ fragmented from an unpolarized quark. For this reason, we will also refer to the TMDJFF $\mathcal{D}_{1T}^{\perp\ \Lambda/q}(z_\Lambda,j_\perp^2)$ as polarizing TMDJFF.

The factorization formula of the denominator $F_{UU,U}$ was presented in \cref{eq:FUUUbefore}, which is expressed in terms of the unpolarized TMDPDF and TMDFF, and was extensively discussed there. On the other hand, the factorization formula for $F^{\sin(\hat{\phi}_\Lambda-\hat{\phi}_{S_\Lambda})}_{UU,T}$ is given in \cref{eq:strh9}, which is explicitly expressed as
\bea
\label{eq:FUUT-lambda}
F^{\sin(\hat{\phi}_\Lambda-\hat{\phi}_{S_\Lambda})}_{UU,T} =&\hat{\sigma}_0 \,H(Q,\mu)\sum_q e_q^2\, \frac{ j_\perp}{z_hM_h} D_{1T}^{\perp\,\Lambda/q}(z_\Lambda,j_\perp^2,\mu, \zeta_J)
\nnu
&\times 
\int\frac{b \,db}{2\pi}J_0(q_Tb)\,x\,\tilde{f}_1^{q}(x,b^2, \mu,\zeta)\bar{S}_{\rm global}(b^2,\mu)\bar{S}_{cs}(b^2,R,\mu)\,,
\eea
where we also used \cref{eq:TMDJFFrel} to express the polarizing TMDJFF $\mathcal{D}_{1T}^{\perp\,\Lambda/q}$ in terms of polarizing TMDFF $D_{1T}^{\perp\,\Lambda/q}$. The derivation is again similar to that for the case of the unpolarized TMDJFF $\mathcal{D}_1^{h/q}$ and the corresponding unpolarized TMDFF $D_1^{h/q}$, as shown from \cref{unp_JFF_FF} to \cref{unp_JFF_FF2}.

We perform a numerical analysis to make predictions for the transverse polarization of $\Lambda$ inside the jet in back-to-back electron-jet production at the EIC. To achieve this, we incorporate TMD evolution into the unpolarized TMDFF $D_1^{\Lambda/q}$ and the polarizing TMDFF $D_{1T}^{\perp,\Lambda/q}$, which were previously extracted in~\cite{Callos:2020qtu}. The extraction of the polarizing TMDFF $D_{1T}^{\perp,\Lambda/q}$ in~\cite{Callos:2020qtu} is based on a Gaussian model, and we extend the parametrization to include TMD evolution. This is done using data from the recent measurement of back-to-back $\Lambda$ and a light hadron production in $e^+ e^-$ collisions, $e^+e^-\to \Lambda + h + X$, conducted by the Belle Collaboration~\cite{Guan:2018ckx}. With these modifications, we obtain the following expression for $D_{1T}^{\perp,\Lambda/q}$, which is required in the TMD factorization formula in \cref{eq:FUUT-lambda}:
\bea
D_{1T}^{\perp\,\Lambda/q}(z_\Lambda,j_\perp^2,\mu, \zeta_J)
=\int \frac{b^{2}\,db}{2\pi}\left(\frac{z_{\Lambda}^2M_{\Lambda}^2}{j_\perp}\right) J_1\left(\frac{j_\perp b}{z_{\Lambda}}\right) \tilde{D}_{1T}^{\perp(1)\,\Lambda/q}(z_\Lambda,b^2,\mu, 
\zeta_J)\,,
\eea
and $\tilde{D}_{1T}^{\perp(1)\,\Lambda/q}$ on the right-hand side takes the following form
\bea
\tilde{D}_{1T}^{\perp(1)\,\Lambda/q}(z_\Lambda,b^2,\mu, 
\zeta_J) = \,&\frac{\langle M_D^2\rangle}{2z_{\Lambda}^5M_\Lambda^2}\mathcal{N}_q(z_{\Lambda})D_1^{\Lambda/q}(z_\Lambda,\mu_{b_*})
\nnu
&\times \exp\left[-S_{\rm pert}\left(\mu, \mu_{b_*} \right) - S_{\rm NP}^{D_{1T}^\perp}\left(z_{\Lambda}, b, Q_0, \zeta_J\right)\right]\,,
\eea
where $\langle M_D^2\rangle=0.118$~GeV$^2$ and $\mathcal{N}_q(z_{\Lambda})=N_qz_{\Lambda}^{\alpha_q}(1-z_{\Lambda})^{\beta_q}\frac{(\alpha_q+\beta_q)^{(\alpha_q+\beta_q)}}{\alpha_q^{\alpha_q}\beta_q^{\beta_q}}$ with parameters $N_q,\ \alpha_q$ and $\beta_q$ determined in~\cite{Callos:2020qtu}. The non-perturbative Sudakov factor $S_{\rm NP}^{D_{1T}^\perp}$ is given by
\bea
S_{\rm NP}^{D_{1T}^\perp}(z_{\Lambda}, b, Q_0,\zeta_J) = \frac{g_2}{2}\ln{\frac{\sqrt{\zeta_J}}{Q_0}}\ln{\frac{b}{b_*}}+g_1^{D_{1T}^\perp} \frac{b^2}{z_{\Lambda}^2}\,,
\eea
with the parameter $g_1^{D_{1T}^\perp} = \langle M_D^2\rangle/4 = 0.0295$ GeV$^2$. We similarly include TMD evolution to the Gaussian model extraction of the unpolarized Lambda TMDFF of~\cite{Callos:2020qtu} to arrive at the same form as \cref{eq:D1param} and~\cref{eq:Sud-NPD}, except that we use the AKK08 parametrizations~\cite{Albino:2008fy} for the collinear $q\to \Lambda$ FFs $D_1^{\Lambda/q}(z_\Lambda, \mu_{b_*})$.

\begin{figure}[t]
\includegraphics[width = 0.48\textwidth]{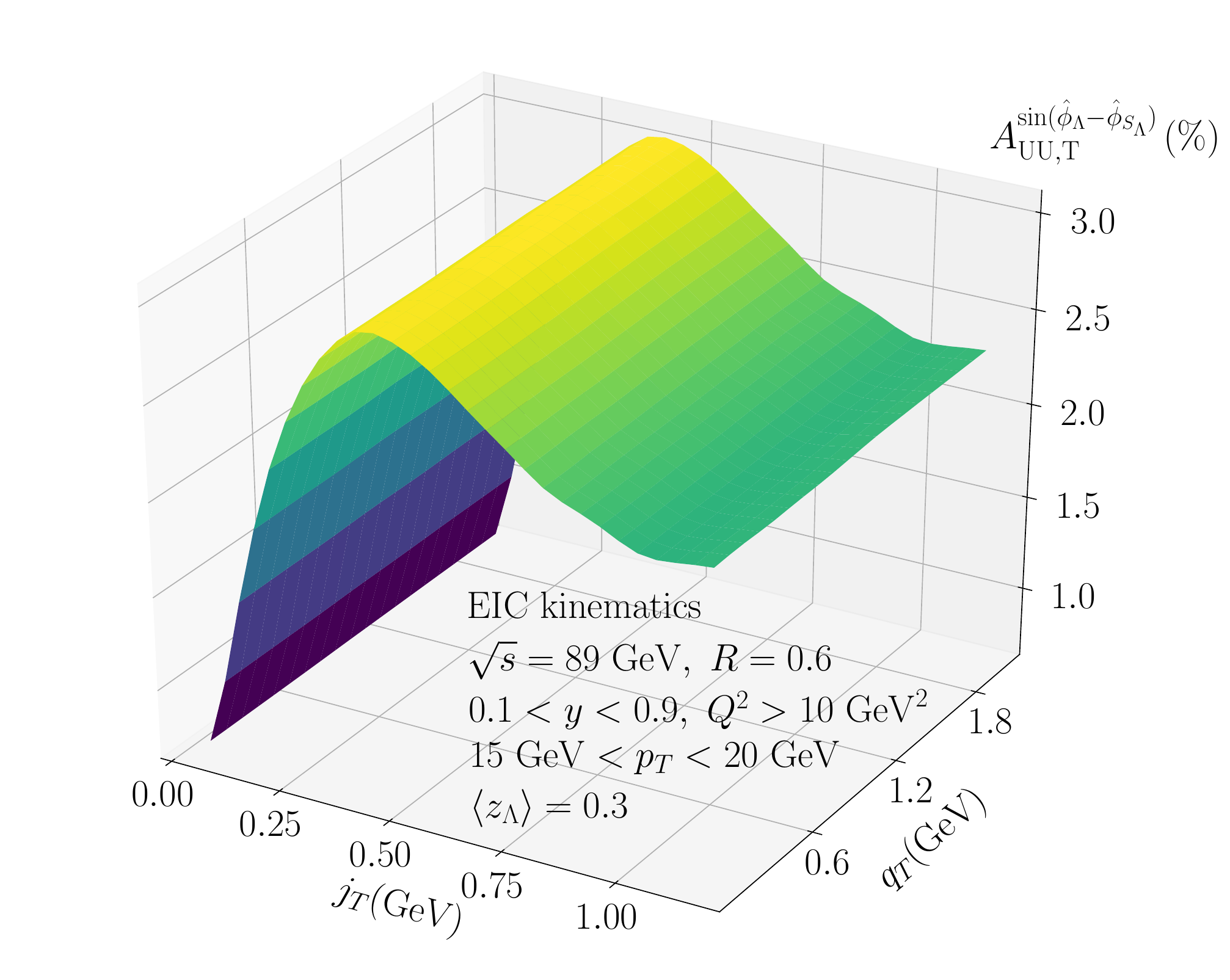}
    \includegraphics[width = 0.45\textwidth]{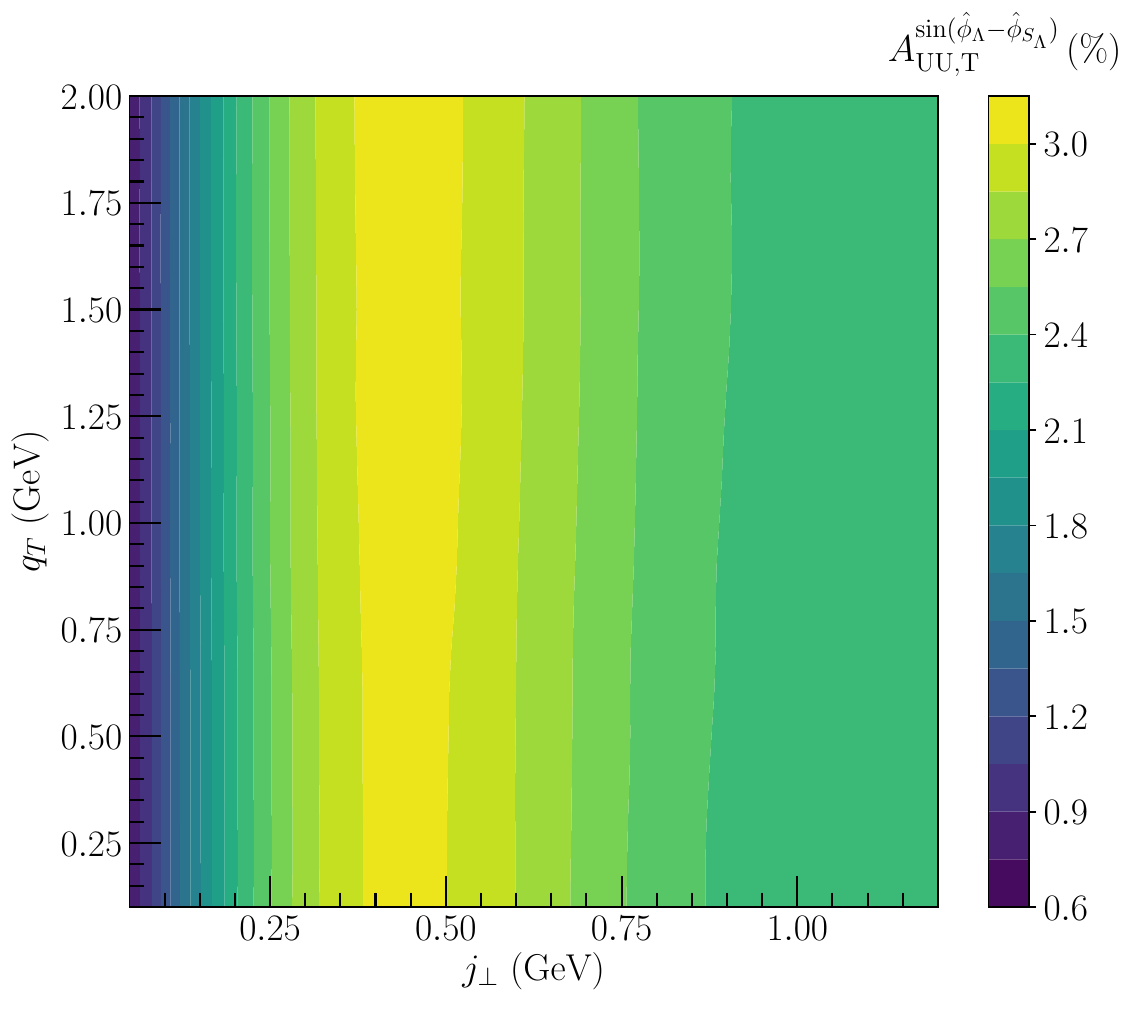}
\caption{$A_{UU,T}^{\sin(\hat{\phi}_{\Lambda}-\hat{\phi}_{S_\Lambda})}$ as a function of jet imbalance $q_T$ and $j_\perp$ for transversely polarized $\Lambda$ in jet production with electron in unpolarized $ep$ collision with EIC kinematics, where we have applied $\sqrt{s}=89$ GeV, jet radius $R=0.6$, inelasticity $y$ in range $[0.1,0.9]$, $Q^2>10$ GeV$^2$, jet transverse momentum $p_T$ in range $[15,20]$ GeV and average momentum fraction $\langle z_\Lambda\rangle=0.3$. Left: Three-dimensional plot of the spin asymmetry in $q_T$ and $j_\perp$. Right: Contour plot of the same quantity.}
    \label{fig:scn3}
\end{figure}

Let's make some predictions for the transverse polarization of $\Lambda$ at the future EIC. In \cref{fig:scn3}, we present the asymmetry $A_{UU,T}^{\sin(\hat{\phi}_{\Lambda}-\hat{\phi}_{\Lambda})}$ as a function of both the imbalance $q_T$ and the transverse momentum $j_\perp$, using EIC kinematics. For this analysis, we set the center-of-mass energy to $\sqrt{s}=89$ GeV, and integrate over the inelasticity $y$ and Bjorken $x$ in the ranges $0.1<y<0.9$ and $0.15<x<0.20$, respectively.

The asymmetry in \cref{eq:auut} depends on the unpolarized TMDPDF $f_1$, which results in constant values in constant $j_\perp$ slices, as expected. Since the dependence on TMDPDFs cancels in the ratio, this asymmetry is particularly useful for extracting the polarizing TMDFF $D_{1T}^{\perp,\Lambda/q}$. This advantage becomes apparent when comparing it to standard SIDIS measurements where the polarizing TMDFF $D_{1T}^{\perp,\Lambda/q}$ would still be convolved with the unpolarized TMDPDF $f_1$.

Additionally, in \cref{fig:scn3_2d}, we provide horizontal slices of the contour plots $A_{UU,T}^{\sin(\hat{\phi}_{\Lambda}-\hat{\phi}_{S_\Lambda})}$ shown in \cref{fig:scn3}, focusing on the variation with transverse momentum $j_\perp$. The left panel represents the EIC kinematics, while the right panel corresponds to the HERA kinematics. For the EIC case, as $j_\perp$ increases, the asymmetry rises up to $3\%$ at $j_\perp=0.4$ GeV and then gradually decreases to approximately $2.5\%$, indicating the feasibility of measurements at the future EIC. On the other hand, for the HERA kinematics, the spin asymmetry is smaller, approximately $\sim 1\%$, and it remains hopefully measurable.

\begin{figure}[t]
\centering
\includegraphics[width = 0.45\textwidth]{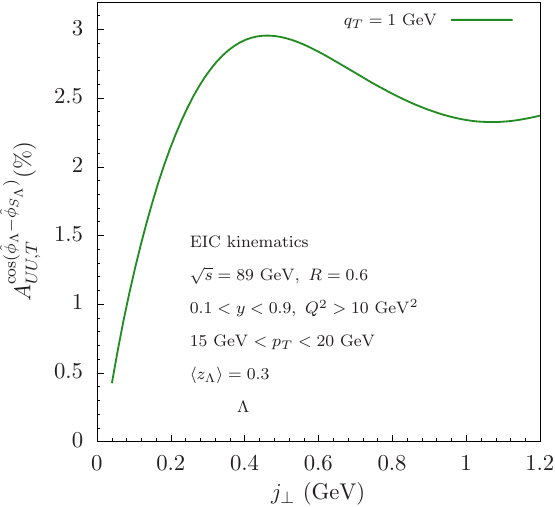}\hspace{0.5cm}
\includegraphics[width = 0.44\textwidth]{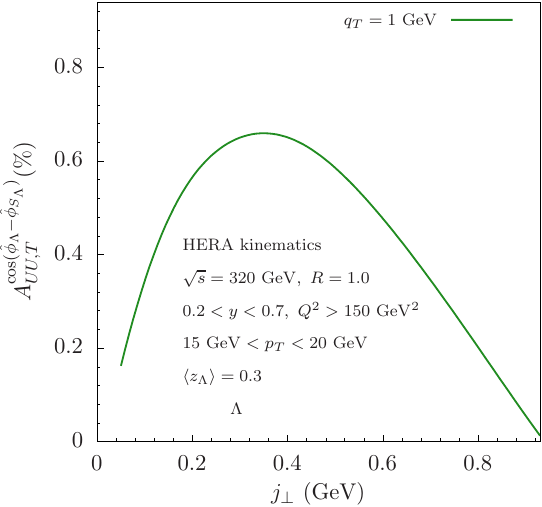}
\caption{Horizontal slices of $A_{UU,T}^{\sin(\hat{\phi}_{\Lambda}-\hat{\phi}_{S_\Lambda})}$ in \cref{fig:scn3} as a function of transverse momentum $j_\perp$ for transversely polarized $\Lambda$ in jet production with electron in unpolarized $ep$ collision. We have jet transverse momentum $p_T$ in range $[15,20]$ GeV and average momentum fraction $\langle z_\Lambda\rangle=0.3$. Left panel: EIC kinematics, we have applied $\sqrt{s}=89$ GeV, jet radius $R=0.6$, inelasticity $y$ in range $[0.1,0.9]$, $Q^2>10$ GeV$^2$. Right panel: HERA kinematics, where we have applied $\sqrt{s}=320$ GeV, jet radius $R=1.0$, inelasticity $y$ in range $[0.2,0.7]$, $Q^2>150$ GeV$^2$.}
    \label{fig:scn3_2d}
\end{figure}
\end{part}

\begin{part}{Quantum Simulation for the QCD Phase Diagram}
\chapter{Quantum Computing algorithms for studying QCD}\label{sec:npqcd}
\begin{quote}
\rule{0.875\textwidth}{0.5pt}\\
The application of quantum computing in QCD is a promising area of research that may lead to significant breakthroughs in our understanding of fundamental physics. In this chapter, we illustrate how one can apply quantum algorithms to study chiral phase diagram and chirality imbalance of a low energy model of QCD, which offers a unique perspective on studying QCD phase transition and other thermal behaviors. \\
\rule{0.875\textwidth}{0.5pt}
\end{quote}

\section{Introduction}
Quantum computing is an emerging field that deals with the study and development of computer technology based on quantum mechanics. Unlike classical computing, which uses classical bits to store and manipulate information, quantum computing uses quantum bits (qubits), which can exist in multiple states simultaneously, enabling more efficient and powerful computation. 

A range of quantum computing applications have arisen in high-energy particle and nuclear physics in recent years~\cite{Preskill_2018,Jordan1130,kaplan2017ground,Preskill:2019WY,Klco_2019,PhysRevLett.120.210501,PhysRevA.98.032331,PhysRevD.101.074038,PhysRevD.101.074512,Roggero_2020,Chang2019,cloet2019opportunities,Nachman_2021,PhysRevD.102.016007,Wei_2020,Holland_2020,PhysRevLett.124.080501,Shaw_2020,Liu_2020,kreshchuk2020quantum,PhysRevResearch.2.023342,Klco_2020,Di_Matteo_2021,davoudi2020search,Bepari_2021}. In this chapter, we will provide a new approach of studying and understanding QCD, especially the QCD phase diagram using quantum algorithms.


It is well known that quantum computers can in principle simulate the time evolution of quantum field theories such as QCD and provide deeper insights into the behavior of quarks and gluons and help us understand the properties of strongly interacting matter at extreme temperatures and densities~\cite{Jordan:2011ne}. In 2012, Jordan, Lee, and Preskill developed a quantum algorithm to compute scattering amplitudes for scalar relativistic quantum field theory with self-interactions ($\phi^4$ theory)~\cite{Jordan:2011ne,Jordan:2011ci}. Since then, quantum simulations for QCD and nuclear physics, in general, have received a lot of attention, especially in recent years~\cite{Lamm:2019uyc,Mueller:2019qqj,Perez-Salinas:2020nem,Echevarria:2020wct,Li:2021kcs,Shaw:2020udc,deJong:2020tvx,deJong:2021tvx,Wei:2019rqy,Kreshchuk:2020kcz}. It is believed that quantum computing has the potential to revolutionize the study of QCD by enabling simulations of larger systems than classical computers can handle.  

More specifically, here we focus on quantum simulation of chiral phase transition, which is a phenomenon where the chirality of particles is imbalanced due to temperature or density. The studies presented in this thesis include quantum simulation of chiral phase transitions and studying chirality imbalance with quantum algorithms for a low energy model of QCD, \textcolor{red}{the Nambu-Jona-Lasinio (NJL) model in 1+1 dimensions.} 
These studies have potential implications for future high-energy experiments and the development of new theoretical frameworks in QCD.

\section{Preliminaries of quantum computation}
The basic unit of information in a quantum computer is the qubit, which can exist in a superposition of two states, typically denoted as $|0\rangle$ and $|1\rangle$ as depicted in \cref{fig:bloch} shown on a bloch sphere. This means that a qubit can exist in both states simultaneously, with a probability of being in each state given by the amplitude of the corresponding state vector. Furthermore, two or more qubits can be entangled, which means that the state of one qubit is dependent on the state of the other, even if they are physically separated. This property of entanglement allows quantum computers to perform certain tasks exponentially faster than classical computers.

\begin{figure}
\centering
\includegraphics[width=0.64\textwidth,height=0.29\textwidth]{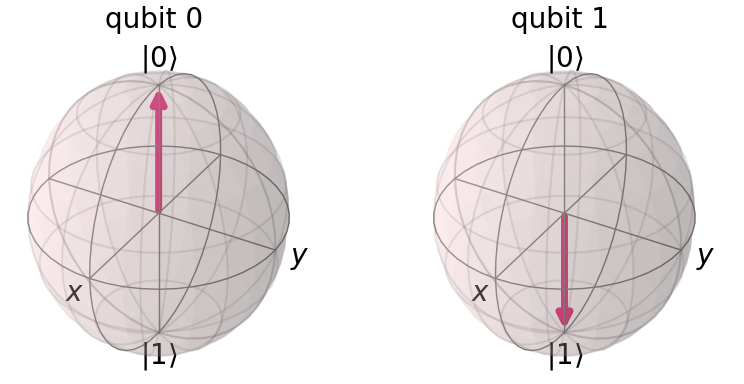}
\caption{Quantum states $|0\rangle$ and $|1\rangle$ on the bloch sphere. Plot generated from \texttt{QISKIT}~\cite{gadi_aleksandrowicz_2019_2562111} (IBM).}\label{fig:bloch}
\end{figure}

In the rest of this subsection, we introduce the four main postulates of quantum mechanics related to Quantum Computing~\cite{nielsen00}. All postulates concern closed quantum systems (i.e., systems isolated from environments) only.
\subsection{State space postulate.} The set of all quantum states of a quantum system forms a complex vector space with inner product structure called the state space $\mathcal{H}$. If the $\mathcal{H}$ is finite dimensional state space, it is isomorphic to some $\mathbb{C}^N$, i.e. $\mathcal{H} \cong \mathbb{C}^N$. And one may naively take $\mathcal{H}=\mathbb{C}^N$ without loss of generality. Always we assume $N=2^n$ for some non-negative integer $n$ which is called the number of quantum bits (qubits). In the Dirac notation, a quantum state $|\psi\rangle \in \mathbb{C}^N$ and its Hermitian conjugate $\langle\psi|$ can be expressed in terms of 
\begin{equation}
|\psi\rangle=\left(\begin{array}{c}
\psi_0 \\
\psi_1 \\
\vdots \\
\psi_{N-1}
\end{array}\right)\,,\quad \langle\psi|=|\psi\rangle^{\dagger}=\left(\begin{array}{llll}
{\psi}_0^* & {\psi}_1^* & \cdots & {\psi}_{N-1}^*
\end{array}\right)\,.
\end{equation}
The inner product is thus
\begin{equation}
\langle\psi | \varphi\rangle:=\langle\psi, \varphi\rangle=\sum_{i \in[N]} {\psi}_i^* \varphi_i \,,
\end{equation}
with $[N]=\{0, \ldots, N-1\}$. Let $\{|i\rangle\}$ be the standard basis of $\mathbb{C}^N$. The $i$-th entry of $\psi$ can be written as an inner product $\psi_i=\langle i |\psi\rangle$. Then $|\psi\rangle\langle\varphi|$ should be interpreted as an outer product, with $(i, j)$-th matrix element given by
\begin{equation}
\langle i|(|\psi\rangle\langle\varphi|)| j\rangle=\langle i | \psi\rangle\langle\varphi | j\rangle=\psi_i \bar{\varphi}_j
\end{equation}
Two state vectors $|\psi\rangle$ and $c|\psi\rangle$ for some $0 \neq c \in \mathbb{C}$ always denote to the same physical state. In other words, the complex scalar $c$ has no observable effects. Therefore, for simplicity, we can always assume that $|\psi\rangle$ is normalized to be a unit vector, meaning $\langle\psi | \psi\rangle=1$. 

Sometimes, for convenience, we may prefer to work with unnormalized states, denoted by $\psi$ without the ket notation $|\cdot\rangle$. However, when considering normalized state vectors, the complex number $c=e^{\mathrm{i} \theta}$, where $\theta \in [0,2 \pi)$, is referred to as the global phase factor.

As an example, one can define in a single qubit system corresponds to a state space $\mathcal{H}\cong\mathbb{C}^2$ that 
\begin{equation}
|0\rangle=\left(\begin{array}{c}
1 \\
0 
\end{array}\right),\quad |1\rangle=\left(\begin{array}{c}
0 \\
1 
\end{array}\right)\,.
\end{equation}
Since the state space of the spin- $1/2$ system is also isomorphic to $\mathbb{C}^2$, this is also called the single spin system, where $|0\rangle$, $|1\rangle$ are referred to as the spin-up and spin-down state, respectively. A general
state vector in $\mathcal{H}$ takes the form
\begin{equation}
|\psi\rangle=a|0\rangle+b|1\rangle=\left(\begin{array}{c}
a \\
b 
\end{array}\right)\,,\quad a,\ b\in\mathbb{C}\,,
\end{equation}
with the normalization condition $|a|^2+|b|^2=1$.
\subsection{Quantum operator postulate.} The evolution of a quantum state from $|\psi\rangle \rightarrow\left|\psi^{\prime}\right\rangle \in$ $\mathbb{C}^N$ is always achieved via a unitary operator $U \in \mathbb{C}^{N \times N}$, i.e.,
\begin{equation}
\left|\psi^{\prime}\right\rangle=U|\psi\rangle, \quad U^{\dagger} U=I_N
\end{equation}
Here $U^{\dagger}$ is the Hermitian conjugate of a matrix $U$, and $I_N$ is the $N$-dimensional identity matrix. When the dimension is apparent, we may also simply write $I \equiv I_N$. In quantum computation, a unitary matrix is often referred to as a gate.

For example, for a single qubit, the Pauli matrices are
\begin{equation}
\sigma_x=X=\left(\begin{array}{ll}
0 & 1 \\
1 & 0
\end{array}\right), \quad \sigma_y=Y=\left(\begin{array}{cc}
0 & -\mathrm{i} \\
\mathrm{i} & 0
\end{array}\right), \quad \sigma_z=Z=\left(\begin{array}{cc}
1 & 0 \\
0 & -1
\end{array}\right) .
\end{equation}
Together with the two-dimensional identity matrix, they form a basis of all linear operators on $\mathbb{C}^2$
Some other commonly used single qubit operators include, to name a few:
\begin{itemize}
    \item Hadamard gate
\begin{equation}
H=\frac{1}{\sqrt{2}}\left(\begin{array}{cc}
1 & 1 \\
1 & -1
\end{array}\right)\,.
\end{equation}
    \item Phase gate
\begin{equation}
S=\left(\begin{array}{ll}
1 & 0 \\
0 & \mathrm{i}
\end{array}\right)\,.
\end{equation}
    \item T gate:
\begin{equation}
T=\left(\begin{array}{cc}
1 & 0 \\
0 & e^{\mathrm{i} \pi / 4}
\end{array}\right)\,.
\end{equation}
\end{itemize}
To avoid notation conflicts, we will use the typewriter font such as $\mathtt{H}, \mathtt{X}$ for these single-qubit gates (one common scenario is to distinguish the Hadamard gate $\mathtt{H}$ from a Hamiltonian $H$ ). An operator acting on an $n$-qubit quantum state space is called an $n$-qubit operator.

In the quantum circuit language, time flows from the left to right, i.e., the input quantum state appears on the left, and the quantum operator appears on the right, and each “wire” represents a qubit as shown in \cref{fig:qc-circ}. Note here $\displaystyle|+\rangle=\frac{|0\rangle+|1\rangle}{\sqrt{2}}$ represents the eigenstate of Pauli-$X$ matrix.
\begin{figure}
\centering
\includegraphics[width=0.74\textwidth]{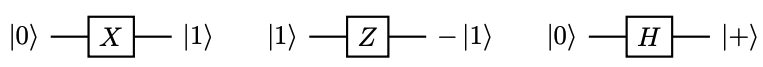}
\caption{A few examples of a quantum circuit.}\label{fig:qc-circ}
\end{figure}

Starting from an initial quantum state $|\psi(0)\rangle$, the quantum state can evolve in time, which gives a single parameter family of quantum states denoted by $\{|\psi(t)\rangle\}$. These quantum states are related to each other via a quantum evolution operator $U$ :
\begin{equation}
\psi\left(t_2\right)=U\left(t_2, t_1\right) \psi\left(t_1\right),
\end{equation}
where $U\left(t_2, t_1\right)$ is unitary for any given $t_1, t_2$. Here $t_2>t_1$ refers to quantum evolution forward in ime, $t_2<t_1$ refers to quantum evolution backward in time, and $U\left(t_1, t_1\right)=I$ for any $t_1$.

The quantum evolution under a time-independent Hamiltonian $H$ satisfies the time-independent Schr\"{o}dinger equation
\begin{equation}
\mathrm{i} \partial_t|\psi(t)\rangle=H|\psi(t)\rangle
\end{equation}
Here $H=H^{\dagger}$ is a Hermitian matrix. The corresponding time evolution operator is
\begin{equation}
U\left(t_2, t_1\right)=e^{-\mathrm{i} H\left(t_2-t_1\right)}, \quad \forall t_1, t_2
\end{equation}
In particular, $U\left(t_2, t_1\right)=U\left(t_2-t_1, 0\right)$. On the other hand, for any unitary matrix $U$, we can always find a Hermitian matrix $H$ such hat $U=e^{\mathrm{i} H}$.
\subsection{Quantum measurement postulate.} 
Next we shall focus our discussion solely on a specific type of quantum measurements known as projective measurements, without losing generality. It is worth noting that all quantum measurements, which can be described as positive operator-valued measures (POVMs), can be equivalently represented using projective measurements in an enlarged Hilbert space through the Naimark dilation theorem.

In a finite-dimensional context, a quantum observable can always be associated with a Hermitian matrix $M$, which possesses the spectral decomposition as follows:
\begin{equation}
M=\sum_m \lambda_m P_m
\end{equation}
Here $\lambda_m \in \mathbb{R}$ are the eigenvalues of $M$, and $P_m$ is the projection operator onto the eigenspace associated with $\lambda_m$, i.e., $P_m^2=P_m$.

When a quantum state $|\psi\rangle$ is measured by a quantum observable $M$, the outcome of the measurement is always an eigenvalue $\lambda_m$, with probability
\begin{equation}
p_m=\left\langle\psi\left|P_m\right| \psi\right\rangle
\end{equation}
After the measurement, the quantum state becomes
\begin{equation}
|\psi\rangle \rightarrow \frac{P_m|\psi\rangle}{\sqrt{p_m}}
\end{equation}
Note that this is not a unitary process!
In order to evaluate the expectation value of a quantum observable $M$, we first use the resolution of identity:
\begin{equation}
\sum_m P_m=I
\end{equation}
This implies the normalization condition,
\begin{equation}
\sum_m p_m=\sum_m\left\langle\psi\left|P_m\right| \psi\right\rangle=\langle\psi | \psi\rangle=1 .
\end{equation}
Together with $p_m \geq 0$, we find that $\left\{p_m\right\}$ is indeed a probability distribution.
The expectation value of the measurement outcome is
\begin{equation}
\mathbb{E}_\psi(M)=\sum_m \lambda_m p_m=\sum_m \lambda_m\left\langle\psi\left|P_m\right| \psi\right\rangle=\langle\psi|\left(\sum_m \lambda_m P_m\right)| \psi\rangle=\langle\psi|M| \psi\rangle
\end{equation}
\subsection{Tensor product postulate.} For a quantum state consists of $m$ components with state spaces $\left\{\mathcal{H}_i\right\}_{i=0}^{m-1}$, the state space is their tensor products denoted by $\mathcal{H}=\otimes_{i=0}^{m-1} \mathcal{H}_i$. Let $\left|\psi_i\right\rangle$ be a state vector in $\mathcal{H}_i$, then
\begin{equation}
|\psi\rangle=\left|\psi_0\right\rangle \otimes \cdots \otimes\left|\psi_{m-1}\right\rangle
\end{equation}
in $\mathcal{H}$. However, not all quantum states in $\mathcal{H}$ can be written in the tensor product form above. Let $\left\{\left|e_j^{(i)}\right\rangle\right\}_{j \in\left[N_i\right]}$ be the basis of $\mathcal{H}_i$, then a general state vector in $\mathcal{H}$ takes the form
\begin{equation}
|\psi\rangle=\sum_{j_0 \in\left[N_0\right], \ldots, j_{m-1} \in\left[N_{m-1}\right]} \psi_{j_0 \cdots j_{m-1}}\left|e_{j_0}^{(0)}\right\rangle \otimes \cdots \otimes\left|e_{j_{m-1}^{(m-1)}}^{(m)}\right\rangle .
\end{equation}
Here $\psi_{j_0 \cdots j_{m-1}} \in \mathbb{C}$ is an entry of a $m$-way tensor, and the dimension of $\mathcal{H}$ is therefore $\prod_{i \in[m]} N_i$. The state space of $n$-qubits is $\mathcal{H}=\left(\mathbb{C}^2\right)^{\otimes n} \cong \mathbb{C}^{2^n}$, rather than $\mathbb{C}^{2 n}$. We also use the notation
\begin{equation}
|01\rangle \equiv|0,1\rangle \equiv|0\rangle|1\rangle \equiv|0\rangle \otimes|1\rangle, \quad\left|0^{\otimes n}\right\rangle=|0\rangle^{\otimes n}
\end{equation}
Furthermore, $x \in\{0,1\}^n$ is called a classical bit-string, and $\left\{|x\rangle | x \in\{0,1\}^n\right\}$ is called the computational basis of $\mathbb{C}^{2^n}$.

As an example, in a two-qubit system corresponds to a state space $\mathcal{H}=\left(\mathbb{C}^2\right)^{\times 2}\cong\mathbb{C}^4$, one can define the standard basis
\begin{equation}
|00\rangle=\left(\begin{array}{c}
1 \\
0 \\
0 \\
0
\end{array}\right),\quad |01\rangle=\left(\begin{array}{c}
0 \\
1 \\
0 \\
0 
\end{array}\right)\,,\quad |10\rangle=\left(\begin{array}{c}
0 \\
0 \\
1 \\
0 
\end{array}\right)\,,\quad |11\rangle=\left(\begin{array}{c}
0 \\
0 \\
0 \\
1 
\end{array}\right)\,.
\end{equation}
The Bell state (also called the EPR pair) is defined to be
\begin{equation}
|\psi\rangle=\frac{1}{2}\left(|00\rangle+|11\rangle\right)=\frac{1}{2}\left(\begin{array}{c}
1 \\
0 \\
0 \\
1 
\end{array}\right)\,,
\end{equation}
which cannot be written as any product state $|a\rangle\otimes|b\rangle$. There are many important quantum operators on the two-qubit quantum system. One of them is the CNOT gate, with matrix representation
\begin{equation}
\mathrm{CNOT}=\left(\begin{array}{llll}
1 & 0 & 0 & 0 \\
0 & 1 & 0 & 0 \\
0 & 0 & 0 & 1 \\
0 & 0 & 1 & 0
\end{array}\right) .
\end{equation}
The quantum circuit for the CNOT gate is shown in \cref{fig:cnot}.
\begin{figure}
\centering
\includegraphics[width=0.32\textwidth]{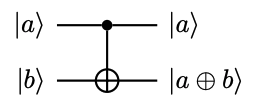}
\caption{The quantum circuit for the CNOT gate}\label{fig:cnot}
\end{figure}
In other words, when acting on the standard basis, we have
\begin{equation}
\operatorname{CNOT} \begin{cases}|00\rangle & =|00\rangle \\ |01\rangle & =|01\rangle \\ |10\rangle & =|11\rangle \\ |11\rangle & =|10\rangle\end{cases}
\end{equation}
This can be compactly written as
\begin{equation}
\mathrm{CNOT}|a\rangle|b\rangle=|a\rangle|a \oplus b\rangle\,.
\end{equation}
Here $a \oplus b=(a+b) \bmod 2$ is the ``exclusive or" (XOR) operation.

\section{Applications in quantum field theory}
Noisy Intermediate-Scale Quantum (NISQ) technology has demonstrated its ability to solve complex problems such as real-time dynamics~\cite{Chiesa2019,Smith2019,Zhang2017,doi:10.1126/science.1232296,PhysRevB.101.014411,Feynman,doi:10.1126/science.273.5278.1073,DeJong:2020riy,deJong:2021wsd}, relativistic behaviors, many-body systems~\cite{wallraff_strong_2004,majer_coupling_2007,jordan_quantum_2012,zohar_simulating_2012,zohar_cold-atom_2013,banerjee_atomic_2013,banerjee_atomic_2012,wiese_ultracold_2013,wiese_towards_2014,jordan_quantum_2014,garcia-alvarez_fermion-fermion_2015,marcos_two-dimensional_2014,bazavov_gauge-invariant_2015,zohar_quantum_2015,mezzacapo_non-abelian_2015,dalmonte_lattice_2016,zohar_digital_2017,martinez_real-time_2016,bermudez_quantum_2017,gambetta_building_2017,krinner_spontaneous_2018,macridin_electron-phonon_2018,zache_quantum_2018,zhang_quantum_2018,klco_quantum-classical_2018,klco_digitization_2019,gustafson_quantum_2019,nuqs_collaboration_ensuremathsigma_2019,magnifico_real_2020,jordan_quantum_2019,lu_simulations_2019,klco_minimally-entangled_2020,lamm_simulation_2018,klco_su2_2020,alexandru_gluon_2019,mueller_deeply_2020,lamm_parton_2020,chakraborty_digital_2020,Bermudez:2018eyh,Ziegler:2020zkq,Ziegler:2021yua}, ground state estimations~\cite{Arute2020,Ma2020,Kandala2019,PhysRevX.6.031007,Kandala2017,Peruzzo2014,PhysRevX.8.011021}, and finite-temperature properties of various systems~\cite{Bauer,PhysRevA.61.022301,PhysRevLett.103.220502,PhysRevLett.108.080402,Temme2011,Yung754,Li:2021kcs,Zhang:2020uqo}. While digital quantum simulations of thermal physical systems have been studied previously, finite-temperature physics on quantum computers remains a challenging area with limited understanding~\cite{PRXQuantum.2.010317}. Recently, various algorithms for imaginary time evolution on quantum computers have been introduced, including the Quantum Imaginary Time Evolution (\texttt{QITE}) algorithm~\cite{Motta,McArdle,Verein,PhysRevB.100.094434,Nishi,Gomes,Yeter-Aydeniz21,Yeter-Aydeniz}, which uses a unitary operation to simulate imaginary time evolution. \texttt{QITE} has been applied to calculate finite-temperature observables such as energy~\cite{PRXQuantum.2.010317,Ville:2021hrl}, magnetization in the Transverse Field Ising Model (TFIM)~\cite{Ville:2021hrl}, and more. These advancements open up new opportunities for quantum simulation in various fields, including materials science, condensed matter physics, and quantum chemistry.

Quantum field theory is the branch of theoretical physics that deals with the study of quantum mechanics applied to fields, such as the electromagnetic field and the Higgs field. It is the framework used to describe the behavior of subatomic particles, including quarks and leptons. Quantum field theory plays a crucial role in the study of high-energy physics and hadron physics.

Quantum computing has the potential to provide significant improvements in the field of quantum field theory, especially in the study of high-energy physics and hadron physics. One area where quantum computing can be useful is in simulating the behavior of subatomic particles, which is a computationally intensive task that is currently beyond the reach of classical computers. By leveraging the power of quantum computing, researchers can simulate larger and more complex systems, enabling more accurate predictions of the behavior of particles.

Another area where quantum computing can be useful is in solving certain optimization problems that arise in quantum field theory. For example, the problem of finding the ground state of a system of interacting particles is an important problem in quantum field theory. This problem can be reduced to an optimization problem, which can be solved using quantum annealing, a specific form of quantum computing that is well-suited for solving optimization problems.

In theoretical physics, the complex phase structure of strongly interacting matter, as described by Quantum Chromodynamics, holds great significance~\cite{PhysRevD.77.114028,COSTA2007431,PhysRevD.77.014006,PhysRevD.91.056003,Jiang2011,PhysRevD.90.114031,Shi2014}. In the infinite quark limit, the deconfinement phase transition is expected as a result of spontaneous $\mathbb{Z}_N$ symmetry breaking, with the transition occurring as a function of temperature~\cite{Ratti:2021ubw,Holland:2000uj}. Alternatively, in the massless quark limit, the chiral phase transition has been extensively studied and has an order parameter~\cite{Rajagopal:1995bc,Borsanyi:2010cj,HotQCD:2018pds,Borsanyi:2020fev,HotQCD:2019xnw,Ding:2020xlj}. QCD is dominated by non-perturbative effects at low energies, which makes it challenging to study the chiral phase transition of strongly interacting matter. Furthermore, other phases of strongly interacting matter are thought to exist, such as the Color-Flavor Locked (CFL) phase, which occupy the high density, low temperature region of the phase diagram.

In conclusion, quantum computing has the potential to revolutionize many fields of science, including physics, chemistry, and cryptography. In the field of quantum field theory, quantum computing can provide significant improvements in the study of high-energy physics and hadron physics by enabling more accurate predictions of the behavior of subatomic particles and solving optimization problems. As quantum computing technology continues to evolve, it is likely that we will see more applications in quantum field theory and other areas of physics.

\subsection{Discretization of the Continuous Theory}
In order to address the non-perturbative nature of lattice Quantum Chromodynamics, a transformation is employed to promote all SU(3)-valued gauge fields $A_\mu(x)$ to elements in the SU(3) group, denoted as $U_\mu(x)$. This is accomplished through the expression:
\begin{equation}
U_\mu(x)=\exp \left[\mathrm{iag}_s A_\mu(x)\right]\,,
\end{equation}
where $a$ and $g_s$ correspond to the lattice spacing and the strong coupling constant, respectively. Throughout the remainder of this document, the focus will primarily be on the treatment of $U_\mu(x)$.

To facilitate further analysis, a Wick rotation is performed, resulting in the substitution:
\begin{equation}
x_0 \rightarrow -i x_4\,.
\end{equation}
As a consequence, the metric takes on a Euclidean form:
\begin{equation}
x^2 = x_\mu x_\mu = x_1^2 + x_2^2 + x_3^2 + x_4^2\,,
\end{equation}
and the weight factor in the path integral becomes real as well,
\begin{equation}
\left\langle 0\left|T\left\{\psi\left(x_1\right) \bar{\psi}\left(x_2\right)\right\}\right| 0\right\rangle=\frac{1}{Z} \int[d \bar{\psi}][d \psi]\left[d U_\mu\right] \psi\left(x_1\right) \bar{\psi}\left(x_2\right) \exp \left[-S\left[U_\mu, \bar{\psi}, \psi\right]\right]\,.\label{eq:(20)}
\end{equation}
Here the symbol $S$ is still used to denote the Euclidian action.

The fermionic component of Lagrangian Equation~\cref{eq:(1.10)} can also be expressed in a straightforward manner as
\begin{equation}
S_F=a^3 a_0 \sum_n \bar{\psi}(n)\left(\sum_{\mu=1}^4 i \gamma_\mu \frac{U_\mu(n) \psi(n+\hat{\mu})-U_{-\mu}(n) \psi(n-\hat{\mu})}{2 a}+m \psi(n)\right)
\end{equation}
Consequently, the lattice Dirac matrix $D(n \mid m)$ can be written as
\begin{equation}
D(n \mid m)_{\alpha i, \beta j}=\sum_{\mu=1}^4 i\left(\gamma_\mu\right)_{\alpha \beta} \frac{U_\mu(n)_{i j} \delta_{n+\hat{\mu}, m}-U_{-\mu}(n)_{i j} \delta_{n-\hat{\mu}, m}}{2 a}+m \delta_{\alpha \beta} \delta_{i j} \delta_{n m}\,,
\end{equation}
where we explicitly indicate the lattice sites $m$ and $n$, Dirac indices $\alpha$ and $\beta$, as well as color indices $a$ and $b$. However, this naive form of the Dirac matrix exhibits the fermion doubling problem~\cite{Nielsen:1981hk}, characterized by the presence of unphysical poles. Various well-established methods have been developed to resolve this issue, including Wilson fermions~\cite{Wilson:1974sk}, Kogut-Susskind staggered fermions~\cite{Kogut:1974ag}, and domain wall fermions~\cite{Kaplan:1992bt}.

After addressing the fermion doubling problem, it becomes advantageous to integrate out the fermionic degrees of freedom in the path integral by employing Gaussian integration over Grassmann numbers. This allows us to obtain a path integral solely in terms of gauge fields. For instance, the partition function can be written as:
\begin{equation}
Z=\int \mathcal{D}[U] \operatorname{det}[D] e^{-S_W[U]}
\end{equation}
Here, the matrix $D$ represents the Dirac matrix with a chosen method for handling the fermion doubling problem. The size of the Dirac matrix scales linearly with the lattice size, resulting in a computational cost significantly higher than that of pure Yang-Mills theory.

With the path integral expressed solely in terms of gauge fields, we can now proceed to numerically evaluate it. Given the high dimensionality of the integral, it is practical to employ Monte Carlo methods with importance sampling. In the case of thermal calculations without real-time evolution, the weight factor $e^{-S}$ is always real and positive, enabling it to be interpreted as the probability distribution function for the configuration of gauge fields. Through Markov chain Monte Carlo (MCMC) methods, we sample $N_s$ configurations of gauge links across the lattice according to the distribution function $e^{-S}$. 

 Given that $S\left[U_\mu, \bar{\psi}, \psi\right]$ is a real valued number, the path integral formulation defines a probability that can be used in a Monte Carlo algorithm. Samples of $U_\mu, \bar{\psi}$ and $\psi$ are drawn according to the probability exp $\left[-S\left[U_\mu, \bar{\psi}, \psi\right]\right]$ and the desired quantity $\psi\left(x_1\right) \bar{\psi}\left(x_2\right)$ is measured on these samples. By taking the average of the values we acquire an approximation of the true correlation function
\begin{equation}
\langle\mathcal{O}\rangle \simeq \frac{1}{n} \sum_j[\mathcal{O}]_j\,.
\end{equation}
with the notation $[\mathcal{O}]_j$ measured on the $j$-th sample. According to the strong law of large numbers, the right-hand side of the expression serves as an unbiased estimator for the true expectation, and its variance decreases as $\mathcal{O}(1/n)$ as the sample size $n$ grows larger.

\subsection{Staggered fermions}
Staggered fermions are a specific type of fermion discretization used in lattice QCD. They involve a clever arrangement of fermion fields on the lattice that exploits the taste symmetry to represent multiple fermion flavors. In staggered fermions, each lattice site is associated with a single fermion flavor, reducing the computational cost compared to other discretization schemes. This approach retains chiral symmetry at the level of the lattice action, making staggered fermions particularly advantageous for large-scale QCD simulations.

The transformation from a Dirac fermion field to a staggered fermion field is achieved through the process known as rooting or taking the nth root of the fermion determinant. By taking the nth root, where n is the number of fermion flavors, each flavor of the original Dirac fermion field is associated with a distinct staggered fermion field. This procedure effectively eliminates unwanted degrees of freedom known as doublers that arise due to the lattice discretization, resulting in a reduced number of fermion degrees of freedom. The resulting staggered fermion formulation retains essential properties of QCD and enables efficient numerical simulations on the lattice grid.

More specifically, the method of staggered fermion fields can be described using the following steps:
\begin{enumerate}
    \item The transformation from a Dirac fermion field to a staggered fermion field involves rooting:
\begin{equation}
    \Psi(x) \rightarrow \sqrt[n]{\det(D)} \Psi_s(x)\,.
\end{equation}
Here, $\Psi(x)$ represents the original Dirac fermion field, $\Psi_s(x)$ denotes the staggered fermion field, and $D$ represents the fermion operator on the lattice.
    \item The staggered fermion fields can be expressed in terms of staggered spinors:
\begin{equation}
\Psi_s(x) = \sum_{\alpha} \chi_{\alpha}(x) \eta_{\alpha}\,.
\end{equation}
In this equation, $\chi_{\alpha}(x)$ represents the staggered spinor associated with the lattice site $x$ and flavor $\alpha$, while $\eta_{\alpha}$ denotes the staggered spinor phase factor.
    \item The staggered fermion action, incorporating the hopping and mass terms, can be written as:
\begin{equation}
S = \sum_{x,y,\alpha} \bar{\chi}_{\alpha}(x) D_{xy} \chi_{\alpha}(y)\,.
\end{equation}
Here, $\bar{\chi}_{\alpha}(x)$ represents the staggered spinor conjugate, and $D_{xy}$ represents the staggered fermion hopping matrix on the lattice.
\end{enumerate}
The rooting transformation reduces the number of fermion degrees of freedom, and the staggered fermion fields are represented using staggered spinors, enabling the formulation of the staggered fermion action for lattice QCD simulations.

In summary, staggered fermion fields are primarily employed in lattice QCD simulations as a discretization scheme for representing fermions on a lattice. They possess taste symmetry and are computationally efficient. Staggered fermions provide a way to describe multiple fermion flavors on the lattice and retain certain symmetries of the continuum QCD theory. While staggered fermions are widely used in lattice QCD, they are not directly related to the Jordan-Wigner transformation.

\subsection{Jordan-Wigner transformation}
The Jordan-Wigner transformation~\cite{Jordan1928} is a powerful mathematical technique used to establish a connection between lattice fermions and quantum spin systems. By employing this transformation, fermion creation and annihilation operators are mapped onto spin operators, enabling the study of lattice QCD and other quantum field theories through the analogy of condensed matter systems. This transformation plays a crucial role in bridging different physical systems, allowing researchers to explore QCD phenomena by leveraging techniques and insights from condensed matter physics. The Jordan-Wigner transformation provides a valuable tool for investigating the behavior of lattice fermions and advancing our understanding of fundamental interactions in high-energy physics.

In practical terms, the Jordan-Wigner transformation involves expressing fermion operators in terms of spin operators by introducing string operators that keep track of the fermion parity. This transformation enables the formulation of lattice fermion systems as spin models, which can be analyzed using powerful techniques from condensed matter physics. Through the application of the Jordan-Wigner transformation, researchers gain a deeper understanding of lattice QCD dynamics and can explore various phenomena by drawing analogies to condensed matter systems. This transformative technique broadens the scope of investigation and facilitates the study of complex quantum field theories, ultimately contributing to advancements in our understanding of fundamental particles and their interactions.

In the Jordan-Wigner transformation, one first maps fermion creation and annihilation operators onto spin operators:
\begin{equation}
c_j^{\dagger} = \prod_{k<j} \sigma_k^z \sigma_j^+\,,\quad
c_j = \prod_{k<j} \sigma_k^z \sigma_j^-\,,
\end{equation}
where $c_j^{\dagger}$ and $c_j$ represent the creation and annihilation operators for fermions at lattice site $j$, respectively, and $\sigma_j^+$ and $\sigma_j^-$ are the spin raising and lowering operators at the corresponding site.

Next, the fermionic number operator can be expressed in terms of spin operators:
\begin{equation}
n_j = \frac{1}{2}\left(1 - \sigma_j^z\right)\,,
\end{equation}
with $n_j$ representing the fermionic number operator at site $j$, and $\sigma_j^z$ is the spin operator corresponding to the $z$-component of spin at that site.

Then the fermionic hopping term can be written in terms of spin operators:
\begin{equation}
c_i^{\dagger} c_j = \frac{1}{4}\left(\sigma_i^x - i\sigma_i^y\right)\left(\sigma_j^x + i\sigma_j^y\right)\,.
\end{equation}
Here, $c_i^{\dagger}$ and $c_j$ are the creation and annihilation operators for fermions at lattice sites $i$ and $j$, respectively, while $\sigma_i^x$, $\sigma_i^y$, $\sigma_j^x$, and $\sigma_j^y$ are spin operators corresponding to the $x$ and $y$ components of spin at the respective sites.

These equations illustrate the mapping of fermion operators to spin operators through the Jordan-Wigner transformation. By expressing fermionic creation, annihilation, and number operators in terms of spin operators, researchers can analyze lattice fermion systems using the formalism of spin models, facilitating the exploration of quantum field theories and condensed matter physics phenomena.

The Jordan-Wigner transformation establishes a connection between lattice fermions and quantum spin systems. It is a mathematical mapping that enables the representation of fermion operators in terms of spin operators. The Jordan-Wigner transformation is used in various areas of theoretical physics, particularly in the study of quantum field theories and condensed matter physics. In the context of lattice QCD, the Jordan-Wigner transformation is not typically employed directly, as other discretization schemes, such as staggered fermions, are more commonly utilized.


To discretize a fermion field, a two-step process involving the transformation into staggered fermion fields and subsequent application of the Jordan-Wigner transformation can be employed. Initially, the fermion field is transformed into staggered fermion fields using a rooting procedure. This involves taking the $n$th root of the fermion determinant, where $n$ corresponds to the number of fermion flavors. By doing so, each flavor of the original fermion field is associated with a distinct staggered fermion field, effectively reducing the number of fermion degrees of freedom.

Once the staggered fermion fields are obtained, the Jordan-Wigner transformation is applied to each of these fields individually. This transformation maps the staggered fermion fields to quantum spin systems, allowing for the representation of the staggered fermion operators in terms of spin operators. This enables the study of the fermionic lattice system through the analogy of condensed matter spin models. By applying the Jordan-Wigner transformation to each staggered fermion field, researchers can explore the behavior of lattice fermions using the language and techniques of condensed matter physics, gaining insights into the underlying physics of the system.

\section{Study QCD chiral phase diagram with Quantum Simulations}

\subsection{QCD phase diagram}
The study of the QCD phase diagram has experienced tremendous progress with lattice simulations in the last few years. The phase transition line is typically determined by extrapolating chiral observables to finite chemical potential $\mu_B$, and finding the temperature at which the chiral condensate has an inflection point, or the chiral susceptibility has a peak.
The transition temperature as a function of $\mu_B$ can be written as ${T_c(\mu_B)}/{ T_c(\mu_B=0)}=1- \kappa_2 \left({\mu_B}/{T_c(\mu_B)}\right)^2- \kappa_4 \left({\mu_B}/{T_c(\mu_B)}\right)^4+\cdots$.
A high-precision result for the crossover temperature $T_c(\mu_B=0)$ has become available~\cite{Borsanyi:2020fev,HotQCD:2018pds,Borsanyi:2020fev}.
Similar coefficients for the extrapolation of the transition temperature to finite strangeness, electric charge and isospin chemical potentials were obtained in Ref. \cite{HotQCD:2018pds}.
No sign of criticality is observed from lattice QCD simulations up to $\mu_B\simeq300$ MeV \cite{Borsanyi:2020fev,HotQCD:2019xnw}.
Future challenges include the extrapolation of the phase transition line to larger values of chemical potential and more stringent constraints on the location of the critical point.

Based on lattice QCD calculations, the QGP-hadron gas transition at vanishing net-baryon density is understood to be a smooth crossover with the transition temperature $T_c = 156\pm1.5$ MeV~\cite{Aoki:2006we}. 
Model studies indicate a first-order phase boundary at large net-baryon density (baryon chemical potential  $\mu_B$) \cite{Fukushima:2013rx}.
If there is a crossover and a first order transition line, they will be joined at the QCD critical point \cite{Stephanov:1998dy,Stephanov:1999zu,Bzdak:2019pkr}. 
State-of-the-art lattice calculations further predicted that the chiral crossover region extends into the finite chemical potential region $\mu_B/T \le 2$~\cite{Bazavov:2020bjn}, see \cref{fig:phasestructure}. Precise calculations in the higher $\mu_B$ region become more difficult and experimental measurements are essential to determine if a QCD critical point exists.  

\begin{figure}[ht]
  \begin{center}
  \vspace{-1cm} 
    \includegraphics[width=0.8\textwidth]{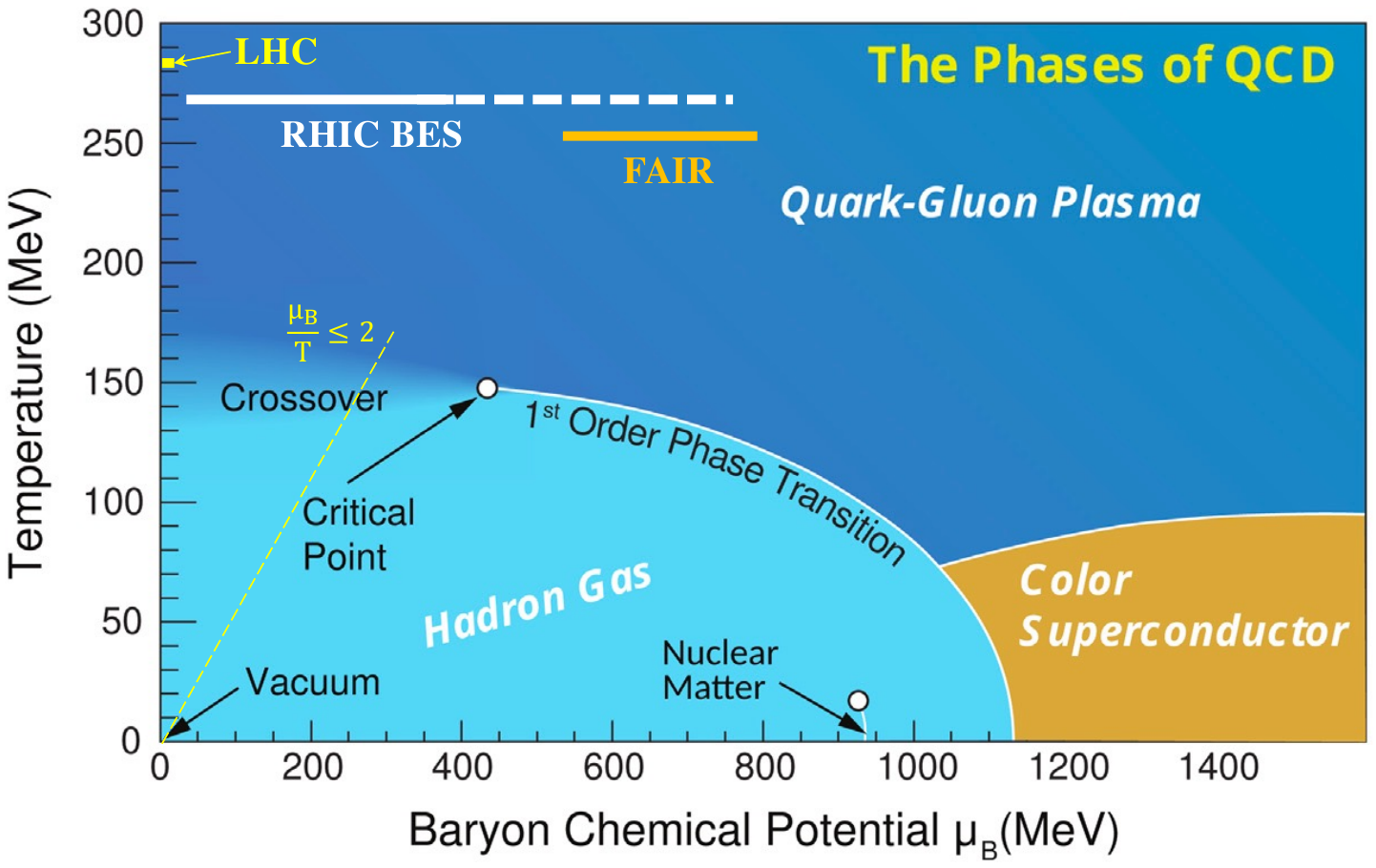}
  \end{center}
    \caption{Sketch of the QCD phase diagram, incorporating a conjectured critical end point and first order transition. 
    The yellow line indicates the region of the phase diagram where lattice QCD can reliably predict the smooth crossover region of the hadron-QGP transition, up to $\mu_B/T \leq 2$. Figure adapted from \cite{Geesaman:2015fha}. 
    }
\label{fig:phasestructure} 
\end{figure}

\subsection{Quantum Simulations}
In order to study the structure of the QCD chiral phase diagram, which divides the temperature-chemical potential plane into regions based on the physical properties of strongly interacting matter or ``phases", researchers typically employ lattice QCD calculations. However, this method has limitations due to the fermion sign problem. At nonzero baryochemical potential, the QCD action is no longer necessarily real, making the use of a Monte Carlo evaluation of thermal expectation values based on $e^{-S}$ impossible~\cite{Ratti:2021ubw, Philipsen:2007rj}. To address this issue, simulations on imaginary chemical potentials are conducted~\cite{deForcrand:2002hgr,DElia:2002tig,Wu:2006su,DElia:2007bkz,Conradi:2007be,deForcrand:2008vr,DElia:2009pdy,Moscicki:2009id}, and the expectation value of a given observable is expanded in powers of $\mu/T$~\cite{Allton:2002zi,Allton:2005gk,Gavai:2008zr,MILC:2008reg,Kaczmarek:2011zz}: $\langle\mathcal O\rangle = \sum_k \mathcal{O}_k(T)(\mu/T)^k$, then the coefficients $\mathcal{O}_k(T)$ are evaluated on the lattice ~\cite{Ratti:2021ubw, Allton:2005gk,Philipsen:2007rj}. This approach is limited to a potentially small range of baryochemical potentials and assumes that the expectation values are analytic in $\mu$~\cite{Ratti:2021ubw}. This assumption is unlikely to hold for some observable in the vicinity of a finite-order phase transition. Despite this obstacle, researchers can use quantum computers to model the lattice system, eliminating the need for a Monte Carlo analysis by taking advantage of the statistical properties of the quantum computer. This approach has been suggested by Feynman~\cite{Feynman} and has been shown in~\cite{Kharzeev:2020kgc} to avoid the sign problem.

\new{In the so-called near-term noisy intermediate quantum (NISQ) era~\cite{Preskill_2018}, directly simulating QCD is not possible. Instead, we will study chiral phase transition via a low energy model of QCD, the Nambu-Jona-Lasinio (NJL) model in 1+1 dimension.} The NJL model has proven to be a practical and convenient tool to investigate the QCD chiral phase transition, owing to its ability to provide insight into the mechanism of chiral symmetry breaking and access to the dynamical mass~\cite{PhysRevD.77.114028,COSTA2007431,Lu2015,doi:10.1142/S0217751X15501997,CUI2015172,PhysRevD.88.114019,PhysRevD.91.036006,PhysRevC.80.065805,PhysRevC.79.035807,PhysRevC.75.015805,PhysRevD.86.071502}. This effective model, tailored for low-energy two-flavor QCD, can also be utilized for analytical calculations at finite temperature and chemical potential. An even simpler version of the NJL model is the Gross-Neveu (GN) model~\cite{Gross:1974jv}, a renormalizable and asymptotically free $(1+1)$-dimensional theory that is composed of $N$ fermion species interacting via four-fermion contact interaction, and it exhibits chiral symmetry breaking at low energy by means of the generation of a nonvanishing chiral condensate $\langle\bar\psi\psi\rangle$. 

In our work~\cite{Czajka:2021yll,Czajka:2022plx}, we aim to examine the chiral condensate of the GN model at finite temperature and chemical potential~\cite{Thies:2019ejd} and compare our findings with quantum simulations. In particular, we illustrate how the quantum algorithm can be used to study the chiral phase transition of the GN model in 1+1 dimension. \new{In this section, we provide our study that leverages a 4-qubit system with each staggered fermion field located in one qubit 
to simulate the $(1+1)$ dimensional NJL Hamiltonian by implementing single-qubit gates and the CNOT gate via the Jordan-Wigner transformation~\cite{Jordan1928}. This Hamiltonian, when approached through its continuum limit, corresponds to the NJL model. We will provide detailed quantum simulations in the next subsections, here let us emphasize the following important points:
\begin{itemize}
\item 
Ultimately, to make contact with the continuum field theory, any such simulation will have to be performed on a series of increasing lattices, and the result extrapolated to that where the lattice sites $N\to \infty$ and the lattice spacing $a\to 0$. Any parameters of the theory present in the continuum must be suitably matched for this procedure to yield meaningful results. Standard procedures for how this should be done derived from the lattice QCD community can be accessed in reputable sources~\cite{MILC:1998ras,Fukushima:2010fe,Philipsen:2021qji}, and for the NJL model in 3+1 dimensions in~\cite{Walters:2003fxa}. In this chapter, we work at fixed lattice size $a$ with a small number of lattice sites $N$ and we leave the detailed investigation of these issues to future work. In this regard, our work~\cite{Czajka:2021yll,Czajka:2022plx} is primarily centered around the formulation and development of the quantum algorithm to illustrate how this can be done for a system with a small number of qubits. 
\item 
In the community, there has been a lot of important studies that use relatively small number of qubits, e.g.~\cite{QuBiPF:2020iiz,Xie:2022jgj,Bauer:2019qxa}, which allow further development on the actual quantum algorithms and directly use it in the current simulations. While its contributions have yet to yield novel insights into new realms of physics, the concerted efforts in this direction have nurtured a fertile ground for the exploration of new research possibilities.
\end{itemize}
}

The rest of this subsection is organized as follows: In \cref{sec:theory}, we furnish analytical calculations to explore the finite temperature behavior of the NJL model with consideration of chemical potentials. Sec.~\ref{sec:njlQITE} provides an overview of the discretization of the field theory on a lattice and \cref{sec:QITE} shows the quantum algorithm used, namely the \texttt{QITE} algorithm. Subsequently, we present a comparison between the results obtained by analytical calculations, exact diagonalization, and quantum simulation at various chemical potentials in \cref{sec:pheno}, where we find a remarkable consistency between them. 

\subsection{Investigated model: Nambu-Jona-Lasinio model}\label{sec:theory}
This section provides a comprehensive exposition on the Nambu-Jona-Lasinio (NJL) model, focusing on its phase transition in the context of finite temperature and chemical potential. 

The NJL model is governed by a Lagrangian density in $(1+1)$-dimensional Minkowski space, which takes the form of the following expression:~\cite{Nambu:1961tp,Nambu:1961fr}
\begin{align}
\label{eq:Lagrangian1}
\mathcal{L}_{\textrm NJL}&=\bar{\psi}(i\slashed{\partial}-m)\psi+g\left[(\bar{\psi}\psi)^2+\bar{\psi}i\gamma_5\tau_a\psi)^2\right]\,,
\end{align}
where the operator $\slashed{\partial}=\gamma_\mu\partial^\mu$, the $\tau_a$ are the Pauli matrices in isospin space, $m$ denotes the bare quark mass, and $g$ is the dimensionless coupling constant. Notably, the gamma matrices $\gamma_0$, $\gamma_1$, and $\gamma_5$ in $(1+1)$ dimensions are given by the following expressions:
\begin{align}
\gamma_0=Z,\quad\gamma_1=-iY,\quad\gamma_5=\gamma_0\gamma_1=-X\,,
\end{align}
where, $X$, $Y$, and $Z$ represent the Pauli matrices:
\begin{align}
X = \begin{pmatrix}
  0 & 1\\
  1 & 0
\end{pmatrix}, \ 
Y = \begin{pmatrix}
  0 & -i \\
  i & 0
\end{pmatrix}, \ 
Z = \begin{pmatrix}
  1 & 0 \\
  0 & -1
\end{pmatrix}
\end{align}
A simplified version of the NJL model is the Gross-Neveu (GN) model~\cite{Gross:1974jv}, which is governed by the following Lagrangian expression:
\begin{align}
\label{eq:Lagrangian2}
\mathcal{L_{\textrm GN}}&=\bar{\psi}(i\slashed{\partial}-m)\psi+g(\bar{\psi}\psi)^2\,.
\end{align}
One can also include the chiral chemical potential $\mu_5$ that simulates the chiral asymmetry between right- and left-handed quarks coupled with the chirality charge density operator $n_5=\bar{\psi}\gamma_0\gamma_5\psi$ and obtain:
\begin{align}\label{eq:Lagrangian05}
\mathcal{L}=&\bar{\psi}(i\slashed{\partial}-m)\psi+g(\bar{\psi}\psi)^2+\mu\bar{\psi}\gamma_0\psi+\mu_5\bar{\psi}\gamma_0\gamma_5\psi\,.
\end{align}
We will study the QCD phase diagram at both $\mu_5=0$ and $\mu_5\ne 0$ in this section. 

\subsubsection{QCD phase diagram at finite chemical potential}
To begin with, we study the dependence on chemical potential and provide an analytical calculation with zero chiral potential. In the limit where the bare quark mass is zero, the Lagrangian satisfies a discrete symmetry group known as $\mathbb{Z}_{2, L}\times\mathbb{Z}_{2, R}\equiv{\pm 1, \pm \gamma_5}$, under which the fields transform by a left-action, i.e. $\psi \mapsto G\cdot \psi$ for any group element $G\in\mathbb{Z}_{2, L}\times\mathbb{Z}_{2, R}$. In this work, we investigate the chiral phase transition of the Gross-Neveu (GN) model with a non-zero chemical potential $\mu$ ~\cite{Thies:2019ejd}:
\begin{align}
\label{eq:Lagrangian}
\mathcal{L}&=\bar{\psi}(i\slashed{\partial}-m)\psi+g(\bar{\psi}\psi)^2+\mu\bar{\psi}\gamma_0\psi\,,
\end{align}
This model, which is a simplified version of the Nambu-Jona-Lasinio (NJL) model, has been previously utilized for studying the chiral phase transition at finite values of both temperature $T$ and chemical potential $\mu$. For the sake of consistency throughout this section, we will refer to the Lagrangian in \cref{eq:Lagrangian} as the NJL model.

In order to discern between different phases, we investigate the chiral condensate $\langle\bar\psi\psi\rangle$ in the vacuum, which is widely recognized as an order parameter~\cite{RevModPhys.53.43,Nambu:1961tp,PhysRevD.24.450,POLYAKOV1978477,Fang:2018vkp} in the chiral limit with zero quark mass. As $\bar\psi\psi$ transforms nontrivially under $\mathbb Z_{2,L}\times\mathbb Z_{2,R}$, a nonzero value of $\langle\bar\psi\psi\rangle$ requires spontaneous breaking of this symmetry, thus a transition from non-vanishing to vanishing values of the condensate indicates a chiral phase transition. Hence, with non-zero quark mass, the chiral condensate $\langle\bar\psi\psi\rangle$ can be considered as a quasi-order parameter~\cite{PhysRevD.98.094501}.

The transition between the chirally symmetric and broken phases can be examined in the \textit{mean field} approximation. \new{This part is widely studied and its results are also well-known in the literature. Below, we review the important results in details in order to validate our quantum simulation results shown in the next section. It is important to emphasize again that the purpose is to illustrate how specific quantum algorithms can be achieved to simulate the NJL model in 1+1 dimensions in quantum circuits, and that it is not meant to provide new physics insights.} In the mean field approximation, the quantity $\bar\psi\psi$ can be expressed as $\bar\psi\psi = \langle\bar\psi\psi\rangle + \sigma$~\cite{Gross:1974jv,WALECKA1974491}, where $\sigma$ is a small, real scalar field (``fluctuations''), i.e., $|\sigma / \langle\bar\psi\psi\rangle| \ll 1$, and $\langle\bar\psi\psi\rangle$ is a constant that is independent of coordinates. $\langle\bar\psi\psi\rangle$ depends on both the chemical potential $\mu$ and the temperature $T$. Thus 
\begin{align}
\left(\bar\psi\psi\right)^2 &= \left(\langle\bar\psi\psi\rangle + \sigma\right)^2= \langle\bar\psi\psi\rangle^2+2\sigma \langle\bar\psi\psi\rangle+ \mathcal O(\sigma^2)\nnu
&=\langle\bar\psi\psi\rangle^2+2\left(\bar\psi\psi-\langle\bar\psi\psi\rangle\right)\langle\bar\psi\psi\rangle+ \mathcal O(\sigma^2)\nnu
&=-\langle\bar\psi\psi\rangle^2+2\bar\psi\psi\langle\bar\psi\psi\rangle+ \mathcal O(\sigma^2)\,,
\end{align}
and the Lagrangian is expressed as follows:
\begin{align}
\mathcal L = \bar\psi(i\slashed\partial - m + 2g\langle\bar\psi\psi\rangle + \mu\gamma_0)\psi - g\langle\bar\psi\psi\rangle^2 + \mathcal O(\sigma^2)\,.
\end{align}
By disregarding the terms of order $\mathcal O(\sigma^2)$, we obtain the Lagrangian $\mathcal L_\mathrm{Dirac}$ that describes a free Dirac fermion (subject to finite chemical potential) possessing a mass $M = m - 2g\langle\bar\psi\psi\rangle$, along with a constant potential $\mathcal V = g\langle\bar\psi\psi\rangle^2 = (M - m)^2/4g$. This yields an efficient, linearized Lagrangian that reads as follows:
\begin{align}
\label{Leff}
\mathcal L_\mathrm{eff} = \bar\psi(i\slashed\partial - M + \mu\gamma_0)\psi - \frac{(M - m)^2}{4g} = \mathcal L_\mathrm{Dirac} - \mathcal V\,,
\end{align}
In consequence, the Lagrangian is now an effective one, denoted as $\mathcal L_\mathrm{eff}$, with the small fluctuation terms represented by $\mathcal O(\sigma^2)$, and the mass of the fermion field $M$ is now a function of the chemical potential $\mu$ and temperature $T$, namely $M = M(\mu, T)$.

The determination of the chiral condensate $\langle\bar\psi\psi\rangle$, or equivalently the effective mass $M$, can be achieved by minimizing the Grand Canonical Potential $\Omega(\mu, T; M) = -\frac{T}{L}\log\mathcal Z$ under the condition of thermal equilibrium. It is worth noting that in this context, the ``volume" refers to a one-dimensional space that can be represented by a length $L$. The partition function $\mathcal Z$, expressed in the path integral formulation, is given by~\cite{Kapusta}
\begin{equation}
    \mathcal Z = \int \mathcal{D}\psi\int \mathcal{D}\bar{\psi}\exp\left[\int_0^\beta d\tau\int dx \mathcal L_E \right]\,.
\end{equation}
Here Euclidean Lagrangian $\mathcal L_E$ is derived through a transformation of variables to imaginary time, with $\tau = it$, while the coldness $\beta = 1/T$. Notably, the effective Lagrangian (\cref{Leff}) leads to $\mathcal L_{\mathrm{eff}, E} = \mathcal L_{\mathrm{Dirac}, E} - \mathcal V$, which implies...
\begin{align}
    \mathcal Z &= \int \mathcal{D}\psi\int \mathcal{D}\bar{\psi}\exp\left[\int_0^\beta d\tau\int dx \,\mathcal L_{\mathrm{eff}, E} \right] \nonumber\\
    &= \int \mathcal{D}\psi\int \mathcal{D}\bar{\psi}\exp\left[\int_0^\beta d\tau\int dx\, (\mathcal L_{\mathrm{Dirac}, E} - \mathcal V) \right] \nonumber\\
    &= e^{-\beta L\mathcal V}\int \mathcal{D}\psi\int \mathcal{D}\bar{\psi}\exp\left[\int_0^\beta d\tau\int dx\, \mathcal L_{\mathrm{Dirac}, E} \right] \nonumber\\
    &= e^{-L\mathcal V/T}\mathcal Z_\mathrm{Dirac} \nonumber\\
    &= e^{-\frac{L}{T}(\Omega_\mathrm{Dirac} + \mathcal V)}\,.
\end{align}
Given by the above derivation, one obtains the Grand Canonical Potential $\Omega = \Omega_\mathrm{Dirac} + \mathcal V$, with $\Omega_\mathrm{Dirac}$ representing the Grand Canonical Potential of a free Dirac field of mass $M$ in $(1+1)$ dimensions~\cite{Kapusta,Buballa:2003qv}:
\begin{align}
\Omega_\mathrm{Dirac}(\mu, T; M) = -\frac{2}{\pi}\int_0^\infty dk\left[\omega_k + T\log(1 + e^{-\beta(\omega_k + \mu)})  + T\log(1 + e^{-\beta(\omega_k - \mu)})\right]\,.
\end{align}
Here $\omega_k = \sqrt{k^2 + M^2}$. Next, by defining the potential $\mathcal V = (M - m)^2/4g$, one has
\begin{align}
\label{GC}
\Omega(\mu, T; M) = \frac{(M - m)^2}{4g}-\frac{2}{\pi}\int_0^\infty dk\left[\omega_k + T\log(1 + e^{-\beta(\omega_k + \mu)})  + T\log(1 + e^{-\beta(\omega_k - \mu)})\right]\,.
\end{align}
This quantity displays a divergent behavior, which requires the implementation of a regularization procedure. \new{To facilitate a comparative analysis with the Lattice model detailed in \cref{sec:njlQITE} below, we naturally introduce a hard cutoff $\Lambda = \pi/a$ for the upper bound in the momentum integration, where $a$ represents the lattice spacing. With this explicit finite cutoff introduced, we left the results un-removed by renormalization and dependent on the lattice spacing.} As a result, one can compute the effective mass $M$ for fixed values of $\mu$ and $T$ by minimizing the Grand Canonical Potential $\Omega(\mu, T; M)$ given by \cref{GC} with respect to $M$. Subsequently, the chiral condensate can be obtained from $\langle\bar\psi\psi\rangle = (m - M)/2g$. For our simulation, we adopted the parameter values $m = 100$ MeV and $g = a = 1$ MeV$^{-1}$, which are well-suited for our calculations. 

In particular, we analyzed $\mu$ and $T$ over the interval $[0 \ \mathrm{MeV}, 300 \ \mathrm{MeV}]$ and present the resulting mass surface in \cref{fig:3dplot} below.
\begin{figure}[h]
\centering
\includegraphics[width=0.48\textwidth,trim={0 0.6cm 0 0},clip]{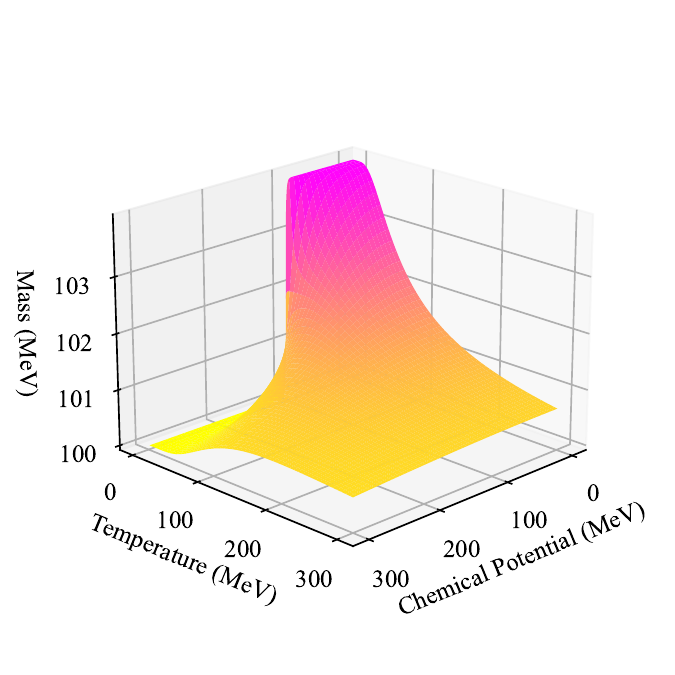}
\includegraphics[width=0.48\textwidth]{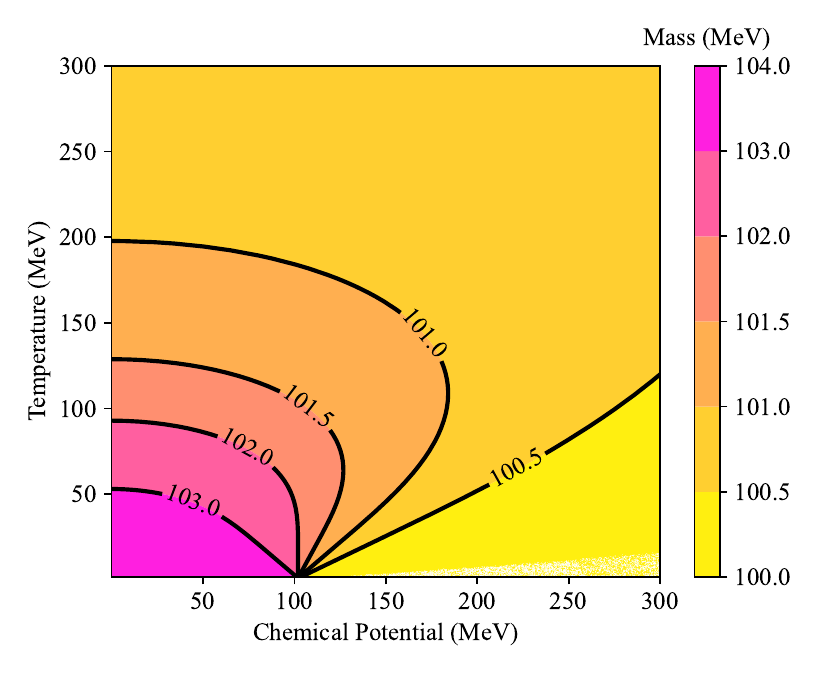}
\caption{The mass of the fermion field, expressed as an effective quantity $M$, is depicted as a three-dimensional surface that varies with the baryochemical potential $\mu$ and temperature $T$. In the accompanying contour plot, the contours of constant $M$ are shown to provide a more detailed representation of the mass landscape.}\label{fig:3dplot}
\end{figure}
The findings indicate that when $\mu^2 + T^2$ is sufficiently low, a dynamically induced mass of approximately $\Delta m \equiv M-m = 4$ MeV emerges in addition to the mass $m = 100$ MeV present in the Lagrangian. Importantly, $M$ approaches $m$ asymptotically, as expected.

It is worth mentioning that historically, the GN model has primarily been studied due to its property of being \textit{asymptotically free}~\cite{Gross:1974jv}. Therefore, at extremely high temperatures or chemical potentials, it is anticipated that a free field theory will be restored.

\subsubsection{Study Chirality Imbalance with Quantum Simulations}
Another very interesting feature for phase diagram is the significance of a chiral chemical potential $\mu_5$. The existence of $\mu_5$ has shown important consequences at the so-called chiral magnetic effect~\cite{Kharzeev:2007tn,Kharzeev:2007jp,Fukushima:2008xe} in heavy ion collisions, which has been under intensive experimental investigation~\cite{STAR:2021mii,Zhang:2021kxj}. In this section, we focus on studying the effects of phase transition at finite $T$, $\mu$, and $\mu_5$.



The examination of the chiral chemical potential $\mu_5$ effects on the QCD chiral phase transition involves studying the QCD phase diagram and the behavior of the total chirality charge $N_5$ under the influence of an external magnetic field at finite temperature $T$ and baryon chemical potential $\mu$. To capture the effects of topological charge changing transitions, a finite chiral chemical potential $\mu_5$ is introduced. In the high temperature/density regime, a chirality charge may arise due to fermion helicity flipping during the deconfinement phase transition from hadronic matter to quark-gluon plasma. Experimental observations indicate that the chirality charge reaches an equilibrium value shortly after a heavy-ion collision, making it essential to explore the chiral imbalance in the QCD phase diagrams for an accurate description of heavy-ion collisions.

The Nambu-Jona-Lasinio (NJL) model has been a prominent choice for studying the chiral magnetic effect and the QCD chiral phase transition~\cite{Nambu:1961tp,Nambu:1961fr,PhysRevD.77.114028,COSTA2007431,Lu2015,doi:10.1142/S0217751X15501997,CUI2015172,PhysRevD.88.114019,PhysRevD.91.036006,PhysRevC.80.065805,PhysRevC.79.035807,PhysRevC.75.015805,PhysRevD.86.071502}. As an effective model for QCD, it allows for analytical calculations at finite temperature $T$, chemical potentials $\mu$, and $\mu_5$.

\subsubsection{Effective mass}
Next we present theoretical calculations for the vacuum chiral condensate $\langle\bar\psi\psi\rangle$ at various temperatures $T$, chemical potentials $\mu$, and $\mu_5$. The chiral condensate $\langle\bar\psi\psi\rangle$, which is a well-known order parameter~\cite{RevModPhys.53.43,PhysRev.122.345,PhysRevD.24.450,POLYAKOV1978477,Fang:2018vkp} for the chiral phase transition in the chiral limit ($m\sim 0$), has been studied in the \textit{mean field} approximation. To achieve this, as first introduced in~\cite{Gross:1974jv,WALECKA1974491}, we define $\bar\psi\psi = \langle\bar\psi\psi\rangle + \sigma$ with a constant term $\langle\bar\psi\psi\rangle$ and a small real scalar field $\sigma$, which corresponds to fluctuations about the vacuum value. Then, we drop terms that are $\mathcal O (\sigma^2)$. By doing this, the four-fermion contact interaction $g(\bar\psi\psi)^2$ can be expressed as
\bea
g(\bar\psi\psi)^2=&g\left(\langle\bar\psi\psi\rangle + \sigma\right)^2= 2g\bar\psi\psi\langle\bar\psi\psi\rangle-g\langle\bar\psi\psi\rangle^2+ \mathcal O(\sigma^2)\,.
\eea
Moreover, one can define the effective mass $M(\mu,\mu_5,T)$ through the following expression:
\bea
M(\mu,\mu_5,T) = m - 2g\langle\bar\psi\psi\rangle(\mu,\mu_5,T),
\label{eq:relation}
\eea
with the Lagrangian is defined as $\mathcal L = \mathcal L_\mathrm{eff} + \mathcal O(\sigma^2)$, and
\bea
\label{Leff}
\mathcal L_\mathrm{eff} =& \bar\psi(i\slashed\partial - M + \mu\gamma_0+\mu_5\gamma_0\gamma_5)\psi - \frac{(M - m)^2}{4g}\nnu 
=& \mathcal L_\mathrm{Dirac} - \mathcal V\,.
\eea

In accordance with previous studies~\cite{Kapusta,Buballa:2003qv}, we express the Grand Canonical Potential $\Omega_\mathrm{Dirac}$ of $\mathcal L_\mathrm{Dirac}$, which has a mass $M$, as follows:
\bea
&\Omega_\mathrm{Dirac}(\mu, \mu_5, T; M)= -\frac{2}{\pi}\sum_{s=\pm 1}\int_0^\infty \bigg[T\ln(1 + e^{-\beta(\omega_{k,s} + \mu)}) + T\ln(1 + e^{-\beta(\omega_{k,s} - \mu)})+\omega_{k,s}\bigg]dk\,,
\eea
Here, the energy spectrum of the free fermions $\omega_{k,s} = \sqrt{(k+s\mu_5)^2 + M^2}$ with $s=\pm 1$. To obtain the grand canonical potential for the NJL model, we add the potential $\mathcal V = (M - m)^2/4g$ to the expression.
\bea
\label{GC2}
&\Omega(\mu, \mu_5, T; M)= \mathcal{V}-\frac{2}{\pi}\sum_s\int_0^\infty \bigg[T\ln(1 + e^{-\beta(\omega_{k,s} + \mu)}) + T\ln(1 + e^{-\beta(\omega_{k,s} - \mu)})+\omega_{k,s}\bigg]dk\,.
\eea

To obtain the value of the effective mass, the regularization of the grand canonical potential $\Omega(\mu, \mu_5, T; M)$ is necessary to account for its divergent behavior. For consistency with numerical simulations using a lattice spacing $a$, a natural hard momentum cutoff $\Lambda=\pi/a$ is adopted for the integral in \cref{GC2}. Subsequently, the effective mass $M$ is determined at fixed $\mu,\mu_5$, and $T$ through numerical minimization of $\Omega(\mu, \mu_5, T; M)$ with respect to $M$, which amounts to solving the gap equation. 
\bea
\frac{\partial \Omega(\mu, \mu_5, T; M)}{\partial M} = 0\,.
\eea
This leads to the expression for the chiral condensate as $\langle\bar\psi\psi\rangle = (m - M)/2g$ as given in \cref{eq:relation}.

In the presence of a magnetic field and under an imbalance of right/left-handed chirality, a finite induced current is generated along the magnetic field direction. This phenomenon is known as the Chiral Magnetic Effect~\cite{Kharzeev:2007tn,Kharzeev:2007jp,Fukushima:2008xe}. Specifically, when the number of right-handed quarks ($N_R$) is not equal to that of left-handed quarks ($N_L$), a separation of positive and negative charges takes place along the magnetic field. The CME arises from the axial anomaly and topological objects in QCD and can produce observable effects that are useful for studying topological $\mathcal{P}$- and $\mathcal{CP}$-odd excitations~\cite{Witten:1979vv,Veneziano:1979ec,Schafer:1996wv,Vicari:2008jw}.

\subsubsection{Chirality charge density}
In the Lagrangian equation \cref{eq:Lagrangian}, we incorporate a novel term featuring the chiral chemical potential $\mu_5$ coupled with the chirality charge density operator $n_5=\bar{\psi}\gamma_0\gamma_5\psi$. This operator is a distinguishing attribute of hot and dense QCD matter since it violates the conservation law owing to the chiral anomaly. 
Although $\mu_5$ is introduced to examine topological charge fluctuations, it is considered a time-independent variable representing the chirality imbalance. Similar to the chiral condensate $\langle\bar{\psi}\psi\rangle$, the chirality charge density $n_5=\langle\bar{\psi}\gamma_0\gamma_5\psi\rangle$ is also a global quantity that does not depend on the coordinates. To relate the induced electric current density with the chirality density, the correlation between $n_5$ and $\mu_5$ is crucial, and the chirality charge density $n_5$ can be computed using the method proposed in~\cite{Fukushima:2010fe}.
\bea
n_5=&-\frac{\partial\Omega(\mu, \mu_5, T; M)}{\partial\mu_5}\,,\label{eq:mu5_calc}
\eea
The expression for the grand potential $\Omega(\mu, \mu_5, T; M)$ is provided in Eqn.~\cref{GC2}. In the subsequent section, we will furnish the analytical computations of $n_5$, which will be contrasted with the results obtained using \texttt{QITE} and exact diagonalization.

\subsection{Discritization of NJL Hamiltonian}\label{sec:njlQITE}
In this subsection, we present the constituent parts of the quantum algorithms employed for the NJL model. Specifically, we begin by outlining the discretization of the NJL Hamiltonian using a lattice grid, before introducing the quantum imaginary time evolution algorithm.

In order to investigate the chiral phase transition and chirality imbalance in the NJL model, we incorporate terms associated with non-zero chemical potential $\mu$ and chiral chemical potential $\mu_5$. These terms simulate the chiral asymmetry between right- and left-handed quarks coupled with the chirality charge density operator $n_5=\bar{\psi}\gamma_0\gamma_5\psi$. Hence, the revised Lagrangian is given by\footnote{In our work~\cite{Czajka:2021yll}, we explored the characteristics of the chiral condensate $\langle\bar\psi\psi\rangle$ under the conditions of finite and non-zero temperature $T$ and chemical potential $\mu$, while setting $\mu_5=0$ and in~\cite{Czajka:2022plx} we included finite chiral potential in the study.}:
\begin{align}\label{eq:Lagrangian5}
\mathcal{L}=&\bar{\psi}(i\slashed{\partial}-m)\psi+g(\bar{\psi}\psi)^2+\mu\bar{\psi}\gamma_0\psi+\mu_5\bar{\psi}\gamma_0\gamma_5\psi\,.
\end{align}
Then the corresponding Hamiltonian $\mathcal H = i\psi^\dagger\partial_0\psi - \mathcal L$ can be expressed as a function of \cref{eq:Lagrangian5}.
 \begin{align}
\label{eq:Hamiltonian05}
\mathcal{H}=&\bar{\psi}(i\gamma_1{\partial_1}+m)\psi-g(\bar{\psi}\psi)^2-\mu\bar{\psi}\gamma_0\psi-\mu_5\bar{\psi}\gamma_0\gamma_5\psi\,.
\end{align}
To avoid ambiguity, we specify that in this section, the term ``NJL model" is employed to refer to the Hamiltonian as stated in Equation~\cref{eq:Hamiltonian05}.

For the discretization of a theory of the Dirac fermion field $\psi(x)$ with components $\rho(x)$ and $\eta(x)$, we set a staggered fermion field $\chi_n$ on a spatial lattice of spacing $a$. Using an even integer $N$ and integer values $n=0,1,\cdots N/2$, the sites with even indices $2n$ are assigned to the staggered fermion field $\chi_{2n}$ with spinor $\rho(x=n)/\sqrt{a}$, while sites with odd indices $2n+1$ are assigned to the staggered fermion field $\chi_{2n+1}$ with spinor $\eta(x=n)/\sqrt{a}$. The Dirac fermion fields $\psi(x)$ are then represented by these staggered fermion fields~\cite{Borsanyi:2010cj,Borsanyi:2013bia,Aoki:2005vt,HotQCD:2014kol,Aubin:2019usy},
\begin{align}
\psi(x=n)=\left(\begin{array}{cc}
     &\rho(x=n)/\sqrt{a}\  \\
     &\eta(x=n)/\sqrt{a}\
\end{array}\right)=\left(\begin{array}{cc}
     &\chi_{2n}\  \\
     &\chi_{2n+1}\
\end{array}\right)\,.\label{eq:staggered}
\end{align}
Once one converts the Dirac fermions to staggered fermions, by applying the Jordan-Wigner transformation ~\cite{Jordan1928}, 
\begin{align}
\chi_n=\frac{X_n-iY_n}{2}\prod_{i=0}^{n-1}\left(-iZ_i\right)\,,\label{eq:jw-trans}
\end{align}
After the application of the Jordan-Wigner transformation to convert Dirac fermions into staggered fermions, the spin representation of the operators is obtained for quantum simulation. In \cref{eq:jw-trans}, the symbol $X_n$ represents the Pauli-$X$ matrix operating on the $n$-th grid on the lattice, and so on. Consequently, the following equivalence up to a constant term can be readily confirmed, 
\begingroup
\allowdisplaybreaks
\begin{align}
&\int dx\bar{\psi}\psi=\sum_{n=0}^{N-1}(-1)^n\chi_n^\dagger\chi_n=\sum_{n=0}^{N-1}(-1)^n\frac{Z_n}{2}\,,\label{eq:stagg1}\\
&\int dx\bar{\psi}\gamma_0\psi=\sum_{n=0}^{N-1}\chi_n^\dagger\chi_n=\sum_{n=0}^{N-1}\frac{Z_n}{2}\,,\\
&\int dx\bar{\psi}i\gamma_1\partial_1\psi=-\frac{i}{2a}\sum_{n=0}^{N-1}\left[\chi_n^\dagger\chi_{n+1}-\chi_{n+1}^\dagger\chi_n\right]=\sum_{n=0}^{N-1}\frac{1}{4a}\left(X_{n}X_{n+1}+Y_{n}Y_{n+1}\right)\,,\\
&\int dx(\bar{\psi}\psi)^2=\frac{1}{a}\left[\sum_{n=0}^{N-1}(-1)^n\chi_n^\dagger\chi_n\right]^2=\sum_{n,m=0}^{N-1}(-1)^{n+m}\frac{Z_nZ_m}{4a}\,,\label{eq:stagg2}\\
&\int dx\bar{\psi}\gamma_0\gamma_5\psi=-\sum_{n=0}^{N/2-1}\left(\chi_{2n}^\dagger\chi_{2n+1}+\chi_{2n+1}^\dagger\chi_{2n}\right)=\frac{1}{2}\sum_{n=0}^{N/2-1}\left(X_{2n}Y_{2n+1}-Y_{2n}X_{2n+1}\right)\,.\label{eq:stagg3}
\end{align}
\endgroup
where the subscripts denote the index of qubit upon which the single-qubit gate is applied. Subsequently, the Hamiltonian in spin representation takes the form of
\begin{align}
H=&\int dx \left[\bar{\psi}(i\gamma_1\partial_1+m)\psi-g(\bar{\psi}\psi)^2-\mu\bar{\psi}\gamma_0\psi]\right]\nonumber\\
=&-\frac{i}{2a}\sum_{n=0}^{N-1}\left[\chi_n^\dagger\chi_{n+1}-\chi_{n+1}^\dagger\chi_n\right]+m\sum_{n=0}^{N-1}(-1)^n\chi_n^\dagger\chi_n-\frac{g}{a}\left[\sum_{n=0}^{N-1}(-1)^n\chi_n^\dagger\chi_n\right]^2\nnu
&-\mu\sum_{n=0}^{N-1}\chi_n^\dagger\chi_n+\mu_5\sum_{n=0}^{N/2-1}\left[\chi_{2n}^\dagger\chi_{2n+1}+\chi_{2n+1}^\dagger\chi_{2n}\right]\,.\label{Hamiltonia_all}
\end{align}
\new{Note that hewe we choose a periodic boundary condition, namely $\chi_N \to \chi_0$, and thus the
$\chi_{N-1}$ field would be coupled with the $\chi_0$ field.
It is important to emphasize that in the continuum limit, the true physics should, of course not depend, on the boundary condition except for quantities that specifically pertain to the boundary. However, since we choose a very small number of lattice sites $N=4$, it would be important to study the dependence on the boundary condition. We leave such a study in the future where we would also increase the number of lattice sites in the simulation.}
Applying the Jordan-Wigner transformation to the Hamiltonian given in \cref{Hamiltonia_all} yields the partitioning of $H$ into five distinct components, which are necessary for the quantum algorithm construction. 
\begingroup
\allowdisplaybreaks
\begin{align}
H_1&=\sum_{n=0}^{N/2-1}\frac{1}{4a}\left(X_{2n}X_{2n+1}+Y_{2n}Y_{2n+1}\right)\,,\label{Hamiltonia1}\\
H_2&=\sum_{n=1}^{N/2-1}\frac{1}{4a}\left(X_{2n-1}X_{2n}+Y_{2n-1}Y_{2n}\right)+\frac{(-1)^{N/2}}{4a}\left(X_{N-1}X_{0}+Y_{N-1}Y_{0}\right)\prod_{i=1}^{N-2}Z_i\,,\label{Hamiltonia2}\\
H_3&=\frac{m}{2}\sum_{n=0}^{N-1}(-1)^nZ_n\,,\label{Hamiltonia3}\\
H_4&=-\frac{g}{4a}\sum_{n,m=0}^{N-1}(-1)^{n+m}Z_n Z_m\,,\label{Hamiltonia4}\\
H_5&=-\frac{\mu}{2}\sum_{n=0}^{N-1}Z_n\label{Hamiltonia5}\,,\\
H_6&=-\frac{\mu_5}{2}\sum_{n=0}^{N/2-1}\left(X_{2n}Y_{2n+1}-Y_{2n}X_{2n+1}\right)\,.\label{Hamiltonia6}
\end{align}
\endgroup
Specifically, \cref{Hamiltonia2} includes the second term to enforce periodic boundary conditions. The resulting Hamiltonians outlined in \cref{Hamiltonia1}-\cref{Hamiltonia6} can be evolved using a quantum simulation through the application of the Suzuki-Trotter decomposition~\cite{trotter1959product,Suzuki}, which  enables us to examine the influence of the chiral chemical potential $\mu_5$ on the chiral condensate $\langle\bar{\psi}\psi\rangle$ and the chirality charge density $n_5$ of the $(1+1)$-dimensional NJL model at finite temperature on a quantum simulator.

\subsection{Quantum imaginary time evolution algorithm}\label{sec:QITE}
In the present subsection, we propose the use of the quantum imaginary time evolution (\texttt{QITE}) algorithm~\cite{Motta} to evaluate the temperature dependence of the NJL model for different values of the baryochemical potential $\mu$ and chiral chemical potential $\mu_5$. As highlighted in ~\cite{Motta}, the \texttt{QITE} algorithm possesses several advantages over other methods for quantum thermal averaging procedures~\cite{Terhal:1998yh,Temme:2009wa,https://doi.org/10.48550/arxiv.1603.02940,Brandao:2016mfe}. Specifically, it enables the computation of thermal averages of observables without any ancillary qubits or deep circuits. In addition, the \texttt{QITE} algorithm is highly resource-efficient, requiring exponentially less space and time for each iteration than classical counterparts.

\subsubsection{Trotterized evolution}
In general, the Suzuki-Trotter decomposition~\cite{trotter1959product,Suzuki} can be employed to approximate the (Euclidean) evolution operator $e^{-\beta H}$ for a given Hamiltonian $H$.
\begin{align}
e^{- \beta H} = \left(e^{-\Delta\beta H}\right)^{N} + O(\Delta\beta^2),
\end{align}
In this regard, the chosen imaginary time step $\Delta\beta$ determines the number of iterations $N=\beta/\Delta\beta$ required to attain the desired imaginary time $\beta=1/T$ at temperature $T$. Nevertheless, as the evolution operator $e^{- \Delta\beta H}$ is not unitary, it is not feasible to realize it as a sequence of unitary quantum gates. Therefore, to achieve the Euclidean time evolution of a state $\ket \Psi$ on a quantum computer, an approximation of the operator $e^{- \Delta\beta H}$ by some unitary operator is necessary. Fortunately, the \texttt{QITE} algorithm offers a scheme to tackle this issue.

The \texttt{QITE} algorithm employs the introduction of a Hermitian operator $A$ to approximate the non-unitary operator $e^{- \Delta\beta H}$ acting on a quantum state $\ket\Psi$ with a unitary operator $e^{-i \Delta\beta A}$. It should be noted that quantum states are represented by ${\alpha\ket\Psi : \alpha\in\mathbb C}$ in a Hilbert space, and not by the vectors themselves, as the normalization and phase of the state vectors are nonphysical.
\begin{align}
\frac{1}{\sqrt{c(\Delta\beta)}} e^{- \Delta\beta H} \ket{\Psi} \approx e^{-i \Delta\beta A} \ket{\Psi}\,.\label{eq:nonu_to_u}
\end{align}
Here the normalization is given by $c(\Delta\beta) = \langle\Psi|e^{-2\Delta\beta H}\ket{\Psi}$. 

In the regime where $\Delta\beta$ is very small, one may expand \cref{eq:nonu_to_u} up to the first nontrivial term, namely up to $\mathcal{O}(\Delta\beta)$, and truncate the higher-order terms. By doing so, one can approximate the change of quantum states under the operators $e^{-\Delta\beta H}$ and $e^{-i\Delta\beta A}$ over a small imaginary time interval $\Delta\beta$ at imaginary time $\beta$. This can be expressed as:
\begin{align}
\ket{\Delta\Psi_H(\beta)}&=\frac{1}{\Delta\beta}\left(\frac{1}{\sqrt{c(\Delta\beta)}} e^{- \Delta\beta H} \ket{\Psi(\beta)}-\ket{\Psi(\beta)}\right)\,,\\
\ket{\Delta\Psi_A(\beta)}&=\frac{1}{\Delta\beta}\bigg(e^{-i \Delta\beta A} \ket{\Psi(\beta)}-\ket{\Psi(\beta)}\bigg)\,,
\end{align}
In accordance with the approach presented in~\cite{Motta}, the Hermitian operator $A$ can be obtained by parameterizing it in terms of Pauli matrices as follows:
\begin{align}
A(\boldsymbol a)=\sum_{\mu}a_{\mu}\hat{\sigma}{\mu},.\label{eq:amu}
\end{align}
The various Pauli strings are labeled by the subscript $\mu$, and the corresponding Pauli string $\hat{\sigma}{\mu}=\prod{l}{\sigma}{\mu_l,l}$ is utilized. The evaluation of the Hermitian operator $A$ requires the minimization of an objective function $F(a)$, where $a{\mu}$ denotes the coefficients of the Pauli strings, and
\begin{align}
F(a)=&||\big(\ket{\Delta\Psi_H(\beta)}-\ket{\Delta_A\Psi(\beta)}\big)||^2\label{eq:fa}\\
=&||\,\ket{\Delta\Psi_H(\beta)}\,||^2+\sum_{\mu,\nu}a_\nu a_\mu\bra{\Psi(\beta)}\hat{\sigma}_\nu^\dagger\hat{\sigma}_\mu\ket{\Psi(\beta)}\nonumber\\
&+i\sum_\mu \frac{a_\mu}{\sqrt{c(\Delta\beta)}}\bra{\Psi(\beta)} \big(H\hat{\sigma}_\mu-\hat{\sigma}_\mu^\dagger H\big)\ket{\Psi(\beta)}\,.\nonumber
\end{align}
The first term $||\,\ket{\Delta\Psi_H(\beta)}\,||^2$ is independent of $a_\mu$. Therefore, we can differentiate with respect to $a_\mu$ and set the resulting expression equal to zero. This leads to a linear equation of the form $({\boldsymbol S}+{\boldsymbol S}^T)\,{\boldsymbol a}={\boldsymbol b}$, where the matrix ${\boldsymbol S}$ and vector ${\boldsymbol b}$ are defined as follows:
\begin{align}
S_{\mu\nu}&=\bra{\Psi(\beta)}\hat{\sigma}_\nu^\dagger\hat{\sigma}_\mu\ket{\Psi(\beta)}\,,
\\
b_\mu&=-\frac{i}{\sqrt{c(\Delta\beta)}}\bra{\Psi(\beta)} \big(H\hat{\sigma}_\mu+\hat{\sigma}_\mu^\dagger H\big)\ket{\Psi(\beta)}\,.
\end{align}
Given the equation $({\boldsymbol S}+{\boldsymbol S}^T)\,{\boldsymbol a}={\boldsymbol b}$ derived earlier by setting the derivative of the objective function $F(a)$ to zero, we can solve for $a_\mu$ and use it to evolve an initial quantum state to any imaginary time $\beta$ by Trotterization. Specifically, we can apply the Suzuki-Trotter formula to approximate the evolution operator $e^{-\beta H}$ and write it as a product of operators $e^{-i\Delta\beta A}$ and $e^{-\Delta\beta H}$. By iteratively applying these operators, we can evolve the initial quantum state to any desired imaginary time with high accuracy, without the need for ancilla qubits or deep quantum circuits.
\begin{align}
\ket{\Psi(\beta)}&=\left(e^{- i\Delta\beta A}\right)^N \ket{\Psi(0)}+\mathcal{O}(\Delta\beta)\,.
\end{align}
In \cref{fig:init}, we present the implementation of the procedure for evolving a quantum state under the unitary operator $e^{-i\Delta\beta A}\ $, where $U_j|\Psi(\beta_j)\rangle= e^{-i\Delta\beta A}|\Psi(\beta_j)\rangle=|\Psi(\beta_j+\Delta\beta)\rangle$. Specifically, each block in the figure performs the evolution of the state $\ket{\Psi(\beta_j)}$ to $\ket{\Psi(\beta_j+\Delta\beta)}$. By repeating this procedure for $n$ steps, we can obtain the state at the target total evolution time $\beta$.
\begin{figure}[h]
\centering
\includegraphics[width=0.7\textwidth]{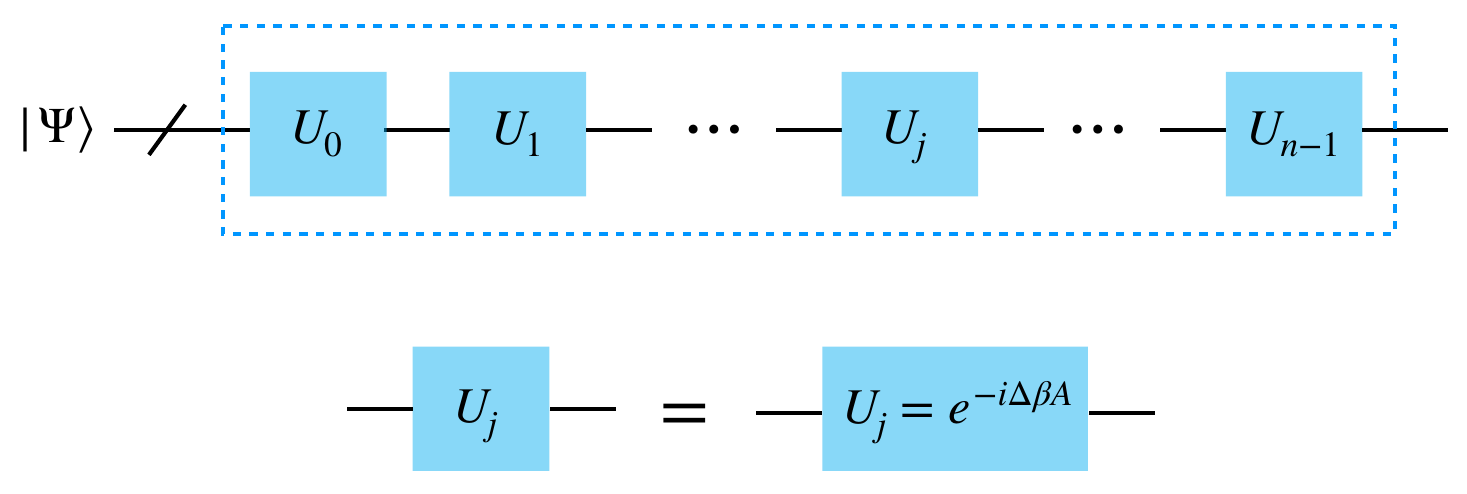}
\caption{The scheme of quantum circuit for generating the thermal state evolved from the initial state $\ket{\Psi}$ at imaginary time $\beta=n\Delta\beta$. The dashed-line box contains $n$ blocks, each of which evolves the state for a small imaginary time interval $\Delta\beta$. Together, these blocks constitute the \texttt{QITE} algorithm for thermal state preparation.}\label{fig:init}
\end{figure}
The QMETTS algorithm ~\cite{Motta} is employed to calculate the expectation value of an observable $\hat{O}$ at finite temperature $T=1/\beta$ by a Markov-like process that starts from a product state. The process involves the imaginary time evolution of the state using \texttt{QITE}, measurement of $\hat{O}$, and measurement in a product basis to collapse back to a product state.

Fig.~\ref{fig:sec2mu} demonstrates the imaginary time evolution of the chiral condensate $\langle\bar\psi\psi\rangle$ at chemical potentials $\mu=0$ MeV (left), $100$ MeV (middle), and $\mu=150$ MeV (right) using small-time steps $\Delta\beta=0.001$ MeV$^{-1}$ (solid line) and $\Delta\beta=0.005$ MeV$^{-1}$ (dashed line). The \texttt{QITE} algorithm is applied to an initial equal superposition state $|++++\rangle$ on a $(1+1)$-dimensional NJL model with coupling constant $g=1$ MeV$^{-1}$, quark mass $m=100$ MeV, and lattice spacing $a=1$ MeV$^{-1}$, followed by exact diagonalization of the NJL Hamiltonian given by \cref{Hamiltonia_all} for reference. \new{Here we chose the number of qubits to be 4, i.e. $N=4$ in our simulation, and thus each staggered fermion field locates on one qubit.} 
Smaller evolution steps result in more accurate quantum simulations. For the rest of this work, we use $\Delta\beta=0.001$ MeV$^{-1}$ for the \texttt{QITE} algorithm and compare the effective mass $M$ among quantum simulation, exact diagonalization, and theoretical analysis.
\begin{figure}[h]
\centering
\includegraphics[width=0.98\textwidth]{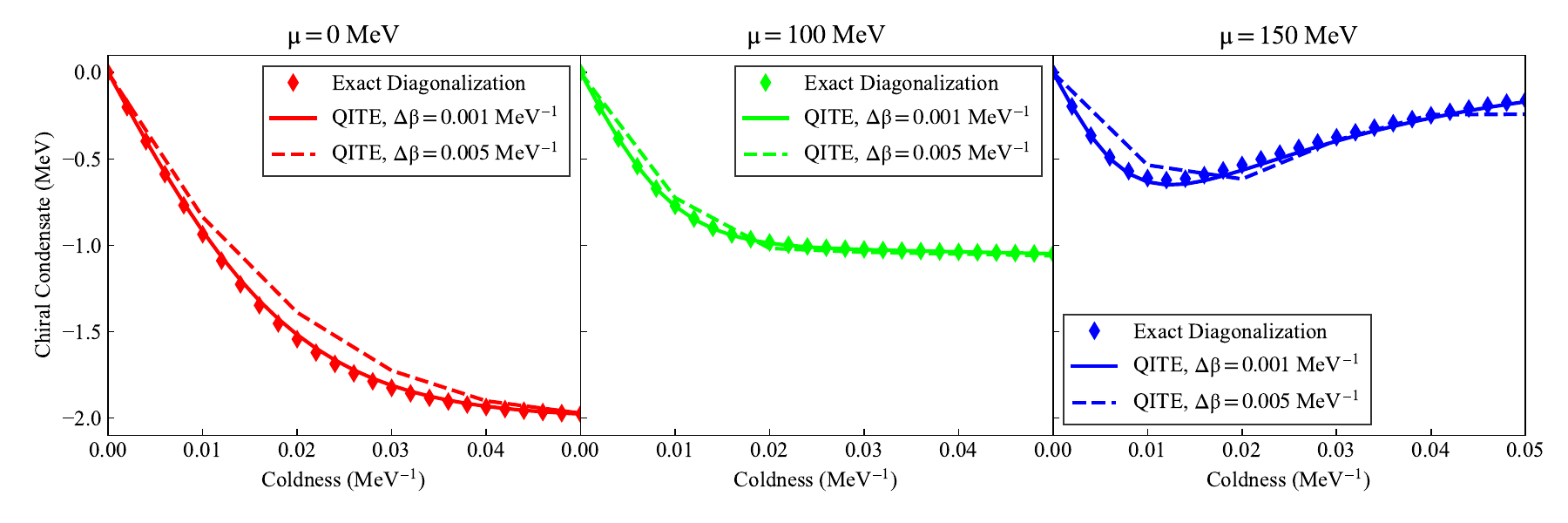}
\caption{The chiral condensate $\langle\bar{\psi}\psi\rangle$ at chemical potentials $\mu=0$ MeV, $\mu=100$ MeV, and $\mu=150$ MeV is simulated as a function of the coldness parameter $\beta=1/T$ (MeV$^{-1}$) for the $(1+1)$-dimensional NJL model with a coupling constant of $g=1$ MeV$^{-1}$, bare quark mass $m=100$ MeV, and lattice spacing $a=1$ MeV$^{-1}$. The simulation is performed using the equal superposition state $|\Psi\rangle=\ket{++++}$ as the initial state. The results obtained using the \texttt{QITE} algorithm with an imaginary time step of $\Delta\beta=0.001$ MeV$^{-1}$ and $\Delta\beta=0.005$ MeV$^{-1}$ are shown in solid and dashed lines, respectively. The exact diagonalization results are presented as points for comparison.}\label{fig:sec2mu}
\end{figure}

\subsubsection{Compare to other algorithms}
In the ensuing section, we undertake a comparative study of the efficacy of the \texttt{QITE} algorithm and the Variational Quantum Eigensolver (\texttt{VQE}) algorithms in evaluating the ground-state energy of the $(1+1)$-dimensional NJL model described earlier. As highlighted in ~\cite{Motta}, the \texttt{QITE} algorithm is notably efficient for ground-state energy calculations. Fig.~\ref{fig:vqe1} portrays the ground-state energy with bare mass $m=100$ MeV, chemical potentials $\mu=100$ MeV and $\mu_5=10$ MeV at $g=1$, as a function of the number of operation steps executed by the algorithms. The \texttt{VQE} algorithms have made considerable strides in obtaining numerous notable outcomes on NISQ hardware~\cite{Barison:2022drt,Johnson:2022lpl,Cao:2021uls,Omiya:2021vol,Johnson:2022lpl}, which have recently garnered significant attention. However, the efficacy of the algorithm is limited by the reliance on an ansatz, as the Hilbert space segment that the \texttt{VQE} algorithm can investigate is affected by the specific variational ansatz employed, and the classical component of the algorithm requires optimization. The \texttt{QITE} algorithm, in contrast, does not necessitate an ansatz and systematically evolves the prepared state toward the ground state after each time-step in a controlled manner. The state is expected to converge to the ground state provided that the initial state shares some overlap with it, with an error that can be accurately estimated.

As can be observed from the graph in \cref{fig:vqe1}, the performance of the \texttt{QITE} and Variational Quantum Eigensolver (\texttt{VQE}) algorithms are compared by plotting the ground-state energy with varying operation steps for the $(1+1)$-dimensional NJL model with bare mass $m=100$ MeV, chemical potentials $\mu=100$ MeV and $\mu_5=10$ MeV at $g=1$. The \texttt{QITE} algorithm, implemented using the \texttt{QFORTE} quantum algorithms library based on \texttt{PYTHON}, utilizes an imaginary time step of $\Delta\beta=0.001$. On the other hand, \texttt{VQE} algorithm, with results obtained using \texttt{QISKIT} of \texttt{IBMQ}, implements various optimizers for comparison with \texttt{QITE} simulations.

It is worth noting that the \texttt{QITE} algorithm achieves the ground-state energy with higher accuracy in fewer operation steps than the \texttt{VQE} implementations shown in the plot. The performance of the \texttt{VQE} algorithm is restricted by its reliance on a variational ansatz, which limits its ability to explore the entirety of the Hilbert space. Moreover, the classical optimization component of the algorithm also affects its efficiency. Conversely, the \texttt{QITE} algorithm evolves the prepared state closer to the ground state in a controlled manner without requiring an ansatz. Provided the initial state has some overlap with the ground state, the state should converge to the ground state with a quantifiable error.

One can also observe that the error of the \texttt{VQE} implementations begins to level off at around $1\%$, owing to the finite set of the variational ansatz scanning distance from the true vacuum. The graph in \cref{fig:vqe1} shows $500$ operation steps for both \texttt{QITE} and \texttt{VQE} algorithms, where the former is represented by blue points with a curve, and the latter by light-blue, green and orange curves for various optimizers.
\begin{figure}
\centering
\includegraphics[width=0.65\textwidth,trim={0cm 1cm 2cm 2cm},clip]{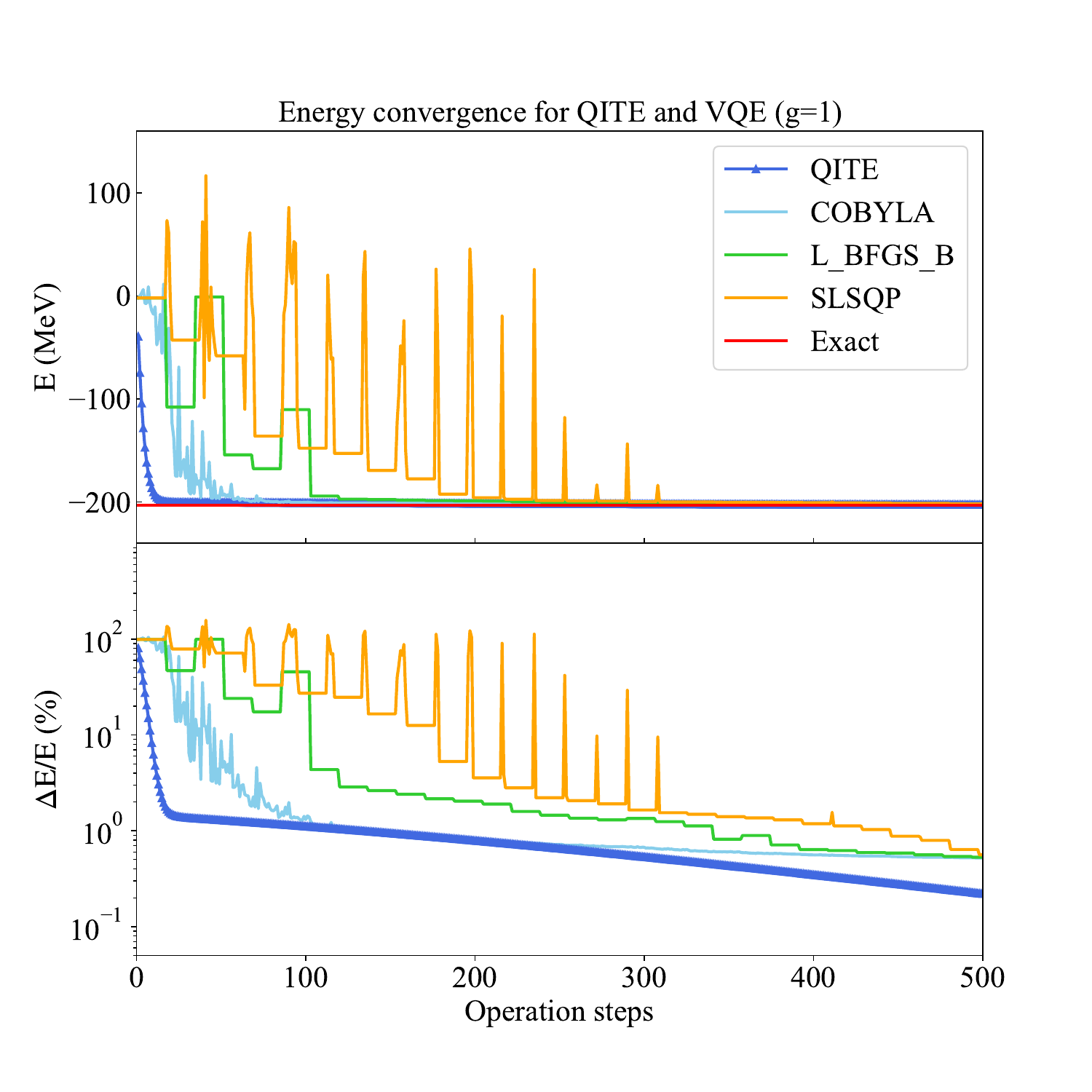}
\caption{Comparative analysis between the \texttt{QITE} and \texttt{VQE} algorithms utilizing various optimizers, namely \texttt{QOBYLA}, \texttt{L\_BFGS\_B}, and \texttt{SLSQP} for $g=1$ case. To investigate the efficacy of these algorithms in obtaining the ground-state energy of the NJL Hamiltonian, we employ a 4-qubit setup on $N=2$ lattice sites with $m=100$ MeV, $\mu=100$ MeV, and $\mu_5=10$ MeV. }\label{fig:vqe1}
\end{figure}

Notably, in \cref{fig:vqe1} we plot the number of optimization steps for the \texttt{VQE} algorithm and the number of thermal evolution steps $x=\beta/\Delta\beta$ ($\Delta\beta=0.001$) for the \texttt{QITE} algorithms on the $x$-axis. We implement the quantum circuit of the \texttt{QITE} algorithm in \texttt{QFORTE}, a Python-based quantum algorithms library. Conversely, we utilize \texttt{QISKIT} of \texttt{IBMQ} for the \texttt{VQE} algorithm, wherein the maximum optimization steps for all optimizers is set as 500. Our results, as depicted in \cref{fig:vqe1}, highlight that the \texttt{QITE} algorithm consistently achieves a higher degree of accuracy with fewer operation steps, thus outperforming the \texttt{VQE} algorithm with various optimizers. In contrast, the error in the \texttt{VQE} implementations seems to level off at around $1\%$, indicating that the variational ansatz scans a finite set of possibilities that are at a finite distance from the true vacuum.

In addition to its utility in ground state energy calculations, the \texttt{QITE} algorithm also exhibits great potential in the simulation of thermal processes at different temperatures. In this study, we aim to employ the \texttt{QITE} algorithm introduced earlier to generate the thermal state $|\Psi(\beta/2)\rangle$, and subsequently compute the thermal average of an observable $\hat{O}$ using the following relation:
\begin{align}
\langle \hat{O}\rangle_\beta=\frac{{\textrm Tr}(e^{-\beta \hat{H}}\hat{O})}{{\textrm Tr}(e^{-\beta \hat{H}})}=\frac{\sum_{i\in\mathcal{S}}\bra{i}e^{-\beta \hat{H}/2}\hat{O}e^{-\beta \hat{H}/2}\ket{i}}{\sum_{i\in\mathcal{S}}\bra{i}e^{-\beta \hat{H}}\ket{i}}\,.
\end{align}
In this context, $\mathcal{S}$ constitutes a full set of basis states for the ground state~\cite{Motta}. In this study, we will focus on the computation of the thermal mean of two observables, namely the chiral condensate $\langle\bar{\psi}\psi\rangle$ and the chirality charge density $n_5=\langle\bar{\psi}\gamma_0\gamma_5\psi\rangle$, by selecting $\hat{O}=\bar{\psi}\psi$ and $\hat{O}=\bar{\psi}\gamma_0\gamma_5\psi$, respectively.

\subsection{Results}\label{sec:pheno}
In the present subsection, we present a demonstration of the NJL chiral phase transition in $(1+1)$ dimensions through the depiction of the quark condensate $\langle\bar{\psi}\psi\rangle$ as a function of temperature $T$ and chemical potential $\mu$. The adoption of the $(1+1)$-dimensional model facilitates us to derive analytical results by solving the gap equation as explained in \cref{sec:theory}, thereby allowing for a comparison with quantum simulations. Moreover, the constraint imposed by the $(1+1)$ dimensionality enables the design of a quantum circuit that employs a relatively small number of qubits, rendering it feasible for implementation on presently available hardware for further research purposes.

Our computations employ the following set of parameters: lattice spacing $a=1$ MeV$^{-1}$, bare mass $m=100$ MeV, and coupling constant $g=1$ MeV$^{-1}$. Our outcomes have been derived via three distinct avenues:
\begin{enumerate}
\item Simulation of thermal states through the QITE algorithm;
\item Exact diagonalization of the Hamiltonian in spin representation;
\item Numerical solution of the gap equation, leading to an analytical calculation.
\end{enumerate}

\subsubsection{Chiral phase diagram at $\mu_5=0$}
We first explore the chiral condensate $\langle\bar{\psi}\psi\rangle$ as a function of temperature $T$ and chemical potential $\mu$ in the context of the $(1+1)$-dimensional NJL chiral phase transition with $\mu_5=0$. To achieve this, we employ the QITE algorithm with the QForte package, and modify it for calculating the properties of the NJL Hamiltonian in quantum simulation. The initial state is set to be the equal superposition state $|+\cdots+\rangle$, and the algorithm is utilized to compute the quark condensate at different values of $\beta=1/T$. The simulation results are presented alongside the theoretical calculations and exact diagonalization for comparison. Specifically, the diamond data points represent the results obtained from the QITE simulation, the solid curves indicate the theoretical calculation described in \cref{sec:theory}, and the dashed curves denote the exact diagonalization results based on the matrix form of the discretized NJL Hamiltonian given by Eqs.\cref{Hamiltonia1} - \cref{Hamiltonia5}. Our findings indicate that the quantum simulations align well with both the exact diagonalization and theoretical analyses.

In \cref{fig:pheno1}, we show the temperature dependence of the chiral condensate $\langle\bar{\psi}\psi\rangle$ for different values of the chemical potential $\mu$, ranging from 0 to 200 MeV. It is noteworthy that the quantum simulations are carried out with a lattice spacing of $a=1$ MeV$^{-1}$, a bare mass of $m=100$ MeV, and a coupling constant of $g=1$ MeV$^{-1}$. At chemical potentials below 100 MeV, the quark condensate increases with temperature. Conversely, for large chemical potentials, the quark condensate exhibits a contrasting behavior within a small temperature range before gradually increasing with temperature, as revealed by the figure. Moreover, the quark condensate at various finite chemical potentials $\mu$ tends to converge at high temperatures.
\begin{figure}[h]
\centering
\includegraphics[width=0.75\textwidth]{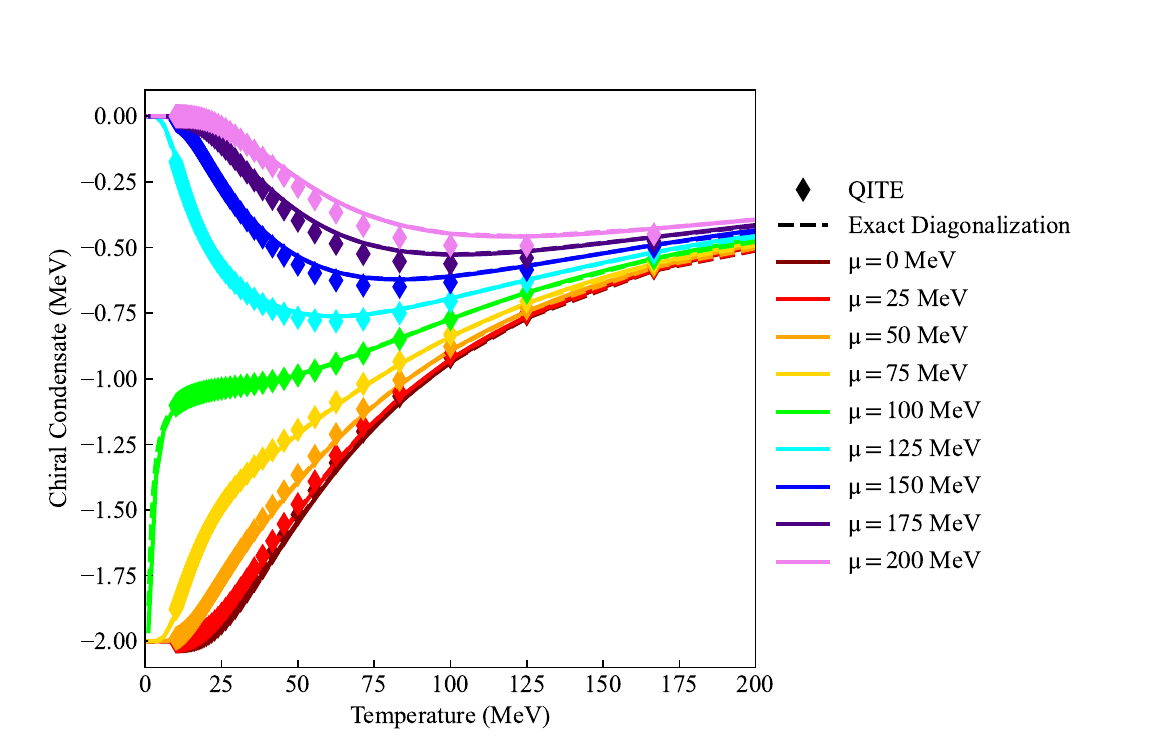}
\caption{Chiral Condensate $\langle
\bar\psi\psi\rangle$ as a function of temperature $T$ at finite chemical potentials $\mu$, ranging from 0 to 200 MeV. Dashed curves represent the results from exact diagonalization, while solid curves correspond to analytical calculations.}\label{fig:pheno1}
\end{figure}

\cref{fig:pheno2} depicts the temperature dependence of the quark condensate $\langle\bar{\psi}\psi\rangle$ for various ratios of the chemical potential and temperature, $\mu/T$, ranging from 0 to 8. The simulation is conducted using the QITE algorithm and the same set of parameters as \cref{fig:pheno1}. The diamond data points represent the quantum simulation results, whereas the solid and dashed lines represent the analytical calculation and exact diagonalization results, respectively. At low temperatures, the quark condensate reaches a minimum value of about $-2$ MeV. As the temperature increases, the quark condensate gradually increases to approach zero. It is observed from the curves that, for a larger ratio of $\mu/T$, the quark condensate increases more rapidly from $-2$ to 0 MeV as the temperature increases.
\begin{figure}[h]
\centering
\includegraphics[width=0.75\textwidth]{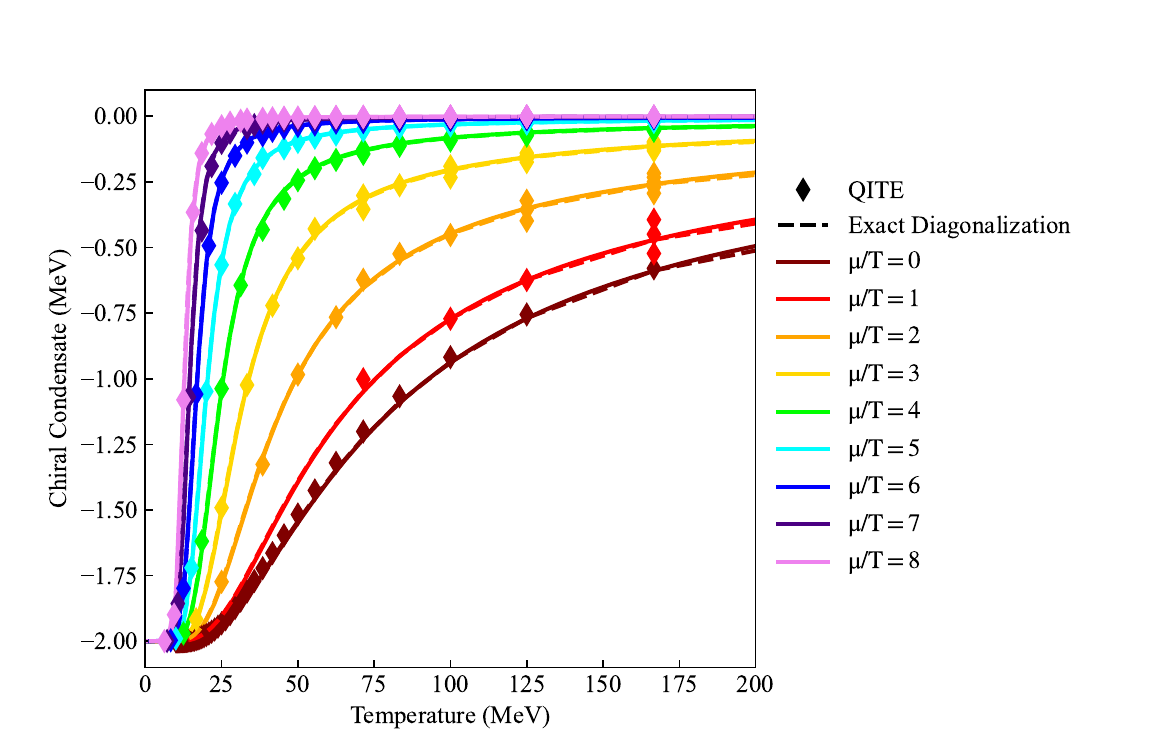}
\caption{This plot displays the temperature dependence of the Chiral Condensate $\langle
\bar\psi\psi\rangle$ for different ratios of the chemical potential and temperature, $\mu/T$, ranging from 0 to 8. The diamond points represent the results obtained from the QITE algorithm, while the solid curves correspond to the analytical calculation and the dashed curves to the exact diagonalization.}\label{fig:pheno2}
\end{figure}

We present the temperature dependence of the quark condensate $\langle\bar{\psi}\psi\rangle$ at various chemical potentials in the $\langle\bar{\psi}\psi\rangle-\mu$ plane in \cref{fig:pheno3}, where the temperatures are chosen as $T=25,\ 50,\ 100,\ 125,\ 250$ MeV. We observe that the absolute value of the quark condensate is larger at lower temperatures and chemical potentials. Moreover, as the chemical potential $\mu$ increases, the quark condensate experiences a rapid increase and approaches zero. Conversely, at higher temperatures, the quark condensate remains nearly unchanged for different chemical potentials. As such, the $\langle\bar{\psi}\psi\rangle-\mu$ plot exhibits a more pronounced phase transition at higher temperatures.
\begin{figure}[h]
\centering
\includegraphics[width=0.75\textwidth]{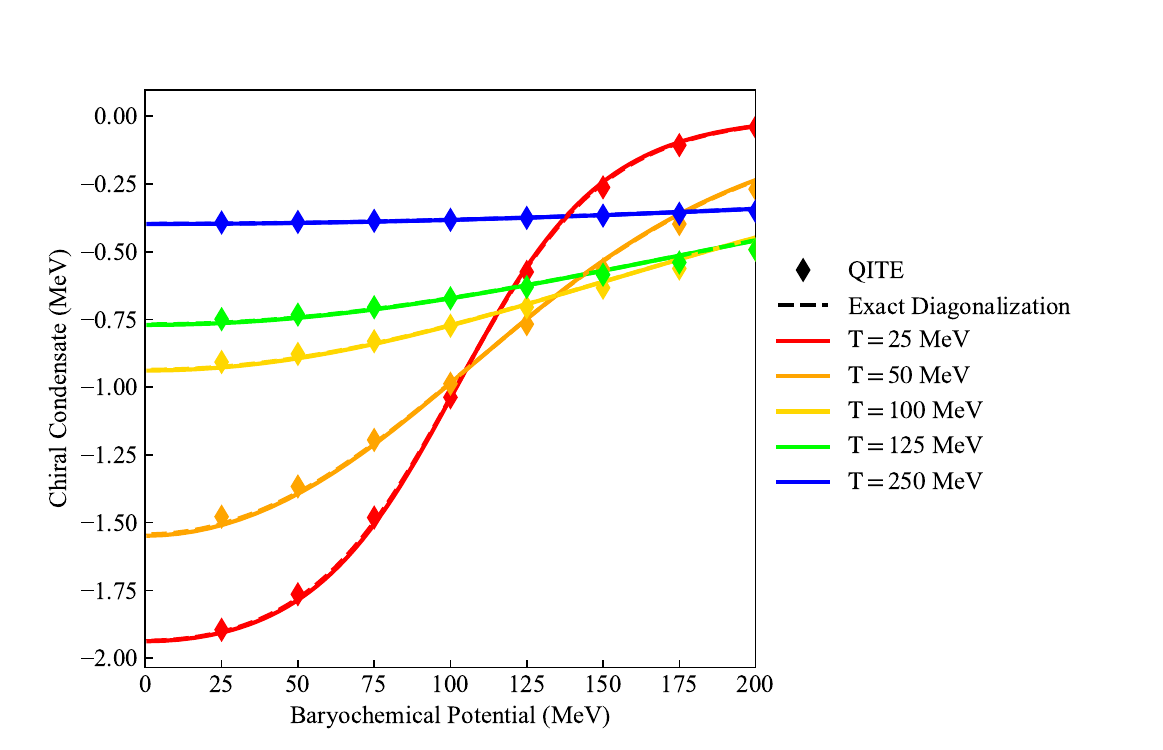}
\caption{In the $\langle\bar{\psi}\psi\rangle-\mu$ plane, we plot the quark condensate $\langle\bar{\psi}\psi\rangle$ at temperatures $T=25,\ 50,\ 100,\ 125,\ 250$ MeV for various chemical potentials. The solid curves represent the analytical calculation, while the dashed curves depict the results from exact diagonalization.}\label{fig:pheno3}
\end{figure}

In this study, we have successfully developed a quantum simulation for the chiral phase transition of the 1+1 dimensional NJL model at finite temperature and chemical potentials using the QITE algorithm. \new{Specifically, we use a 4-qubit quantum circuit, i.e. each staggered fermion field is located in one qubit. On one aspect, our work~\cite{Czajka:2021yll} serve as a demonstration using a limited number of lattice sites like the above examples at the frontier of this field. At the same time, we are aware of efforts in the field of quantum computing with larger lattice sizes ($>100$ sites) and actively engaging in endeavors aimed at augmenting the lattice size to approach the continuum limit and gain deeper insights into the NJL model.}

\new{Our results reveal a consistency between the digital quantum simulation and the standard method for the lattice NJL Hamiltionian with a limited lattice sites. Even though there is still a long way to go for comparing the quantum simulation with the continuum limit of the NJL model, this small illustration makes it promising for further development where the potential of quantum computing in simulating finite-temperature behaviors for QCD could be possible, and offers an exciting avenue for exploring finite density effects in QCD and other field theories.} With the promise of scalable quantum computer technology on the horizon, this study and previous ones showcase the potential for NISQ quantum computers to tackle physical problems that are challenging or impossible to solve using classical computing algorithms. 

\subsubsection{Chiral phase diagram at $\mu_5\ne0$}
In this subsection, we investigate the thermal behavior of the effective mass $M$ and chirality charge density $n_5=\langle\bar{\psi}\gamma_0\gamma_5\psi\rangle$ in the $(1+1)$-dimensional NJL model as given by \cref{eq:Hamiltonian05}. As we have highlighted in previous sections, our approach employs a quantum algorithm to simulate the thermal properties of physical observables. To demonstrate the accuracy and reliability of our simulations, we present the results with the same three approaches as mentioned in the previous subsection. For consistency among the three methods, we fix the bare mass $m=100$ MeV and lattice spacing $a=1$ MeV$^{-1}$, and consider the effects of the four-fermion interaction term in the Lagrangian by testing the coupling constant at $g=1$.

In order to implement quantum circuits, various quantum simulation packages have been developed and have shown consistent results for quantum simulations. These software libraries, including \texttt{PYQUILL}~\cite{https://doi.org/10.48550/arxiv.1608.03355} (Rigetti), \texttt{TEQUILA}~\cite{Kottmann_2021}, \texttt{Q\#}~\cite{qsharp} (Microsoft), \texttt{QISKIT}~\cite{gadi_aleksandrowicz_2019_2562111} (IBM), \texttt{QFORTE}~\cite{stair2021qforte}, \texttt{XACC}~\cite{McCaskey_2020}, \texttt{FQE}~\cite{Rubin:2021znj}, and \texttt{CIRQ}~\cite{cirq_developers_2021_5182845}, are coded in \texttt{PYTHON}, and generate expected outputs of an ideal quantum computer. General quantum simulation packages can be found in~\cite{Bharti:2021zez}, while specific implementations of quantum algorithms can be found in~\cite{Anand:2021xbq}. In our case, the \texttt{QITE} algorithm is executed using a quantum circuit constructed with the open-source software package \texttt{QFORTE}~\cite{stair2021qforte}, which has implemented many useful quantum algorithms.


\begin{figure}
\centering
\includegraphics[width=0.495\textwidth,trim={2.5cm 2.5cm 3cm 0},clip]{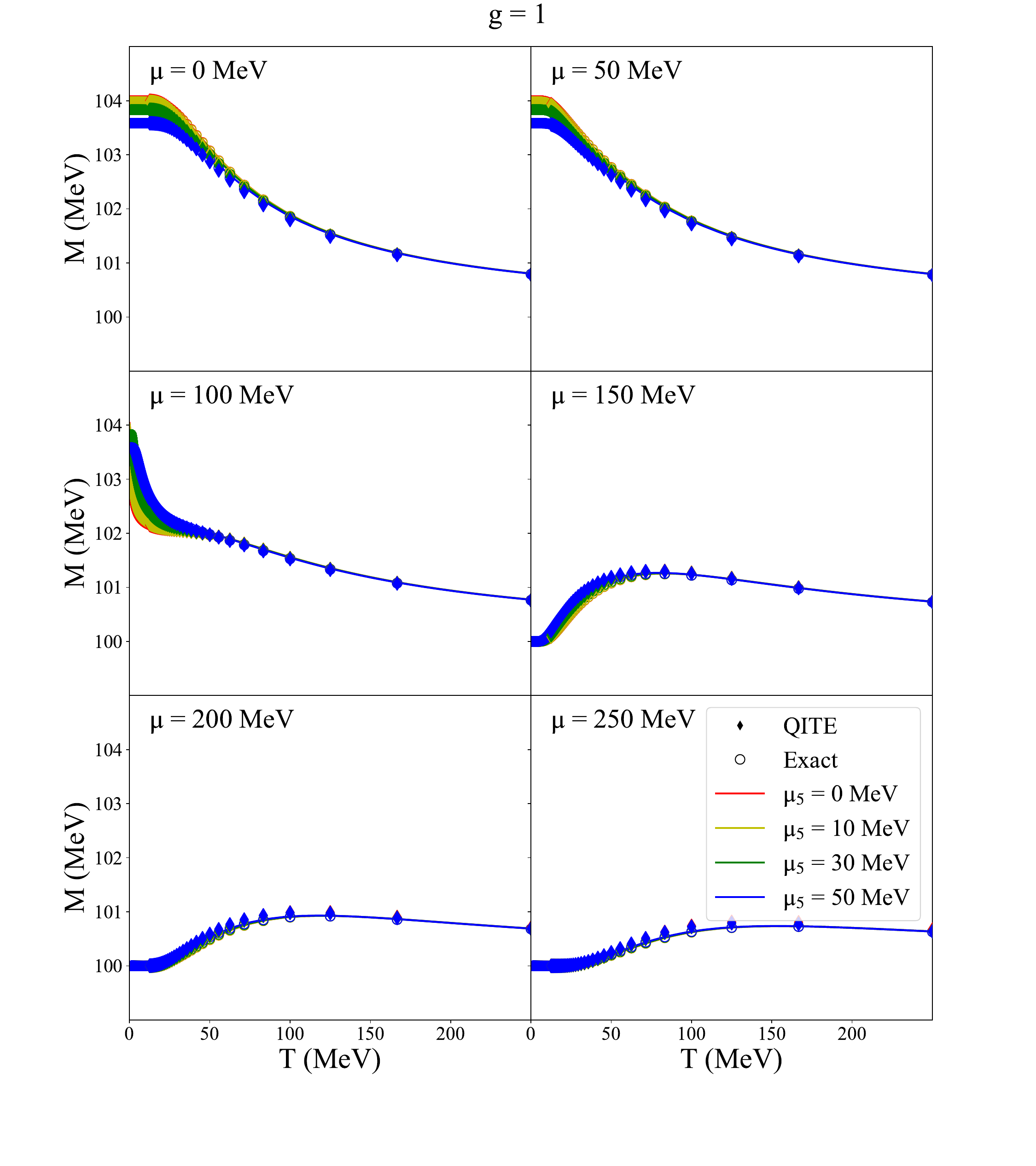}
\includegraphics[width=0.495\textwidth,trim={2.5cm 2.5cm 3cm 0},clip]{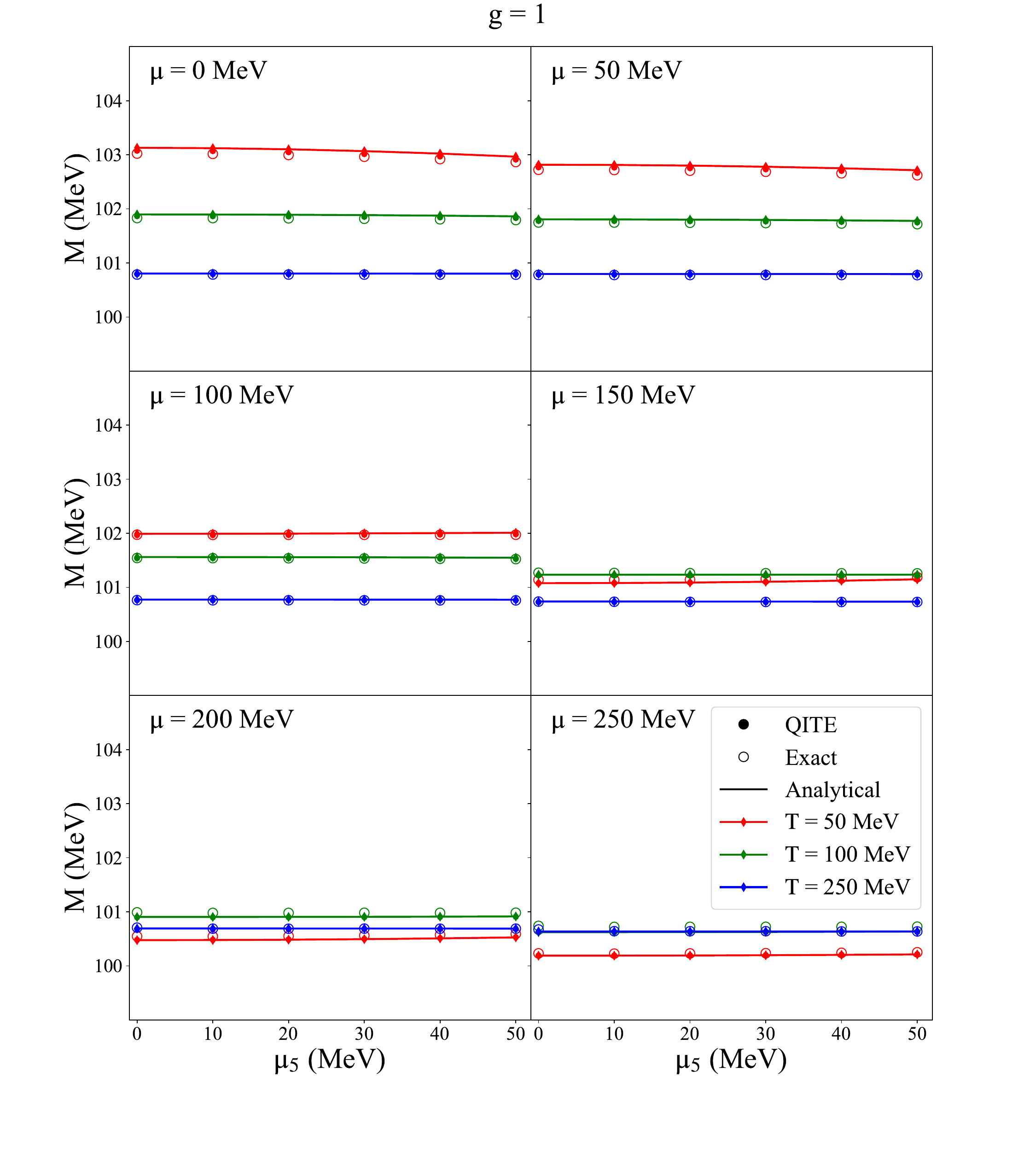}
\caption{Effective mass $M$ as a function of temperature $T$ (left panels) or $\mu_5$ (right panels) at coupling constant $g=1$. Each panel adopts a different value of chemical potential $\mu$.}\label{fig:Tmu5_g1}
\end{figure}

In this study, we investigate the effects of chiral imbalance on the chiral condensate at finite temperatures and chemical potentials. We present several plots of the effective mass $M$ at different chemical potentials $\mu$, $\mu_5$, and temperatures $T$. In \cref{fig:Tmu5_g1}, we compare the temperature (left panels) and chiral chemical potentials $\mu_5$ (right panels) dependence of the effective mass $M$ at coupling constant $g=1$ for various baryochemical potentials $\mu$. The results are obtained using the \texttt{QITE} algorithm, exact diagonalization, and analytical calculations. Notably, we observe a change in the pattern of the effective mass between $\mu=100$ MeV and $\mu=150$ MeV panels, from decreasing to increasing with temperature. Additionally, we find that the effective mass at smaller $\mu_5$ is larger for $\mu\leq 100$ MeV panels and smaller for $\mu\geq 150$ MeV panels. We also observe that at $\mu\leq 100$ MeV, $M$ is lower for higher temperature, as expected by asymptotic freedom, while at higher $\mu$, the effective mass $M$ becomes smaller at lower temperatures.

\begin{figure}
\centering
\includegraphics[width=0.65\textwidth,trim={2cm 1.5cm 2cm 0},clip]{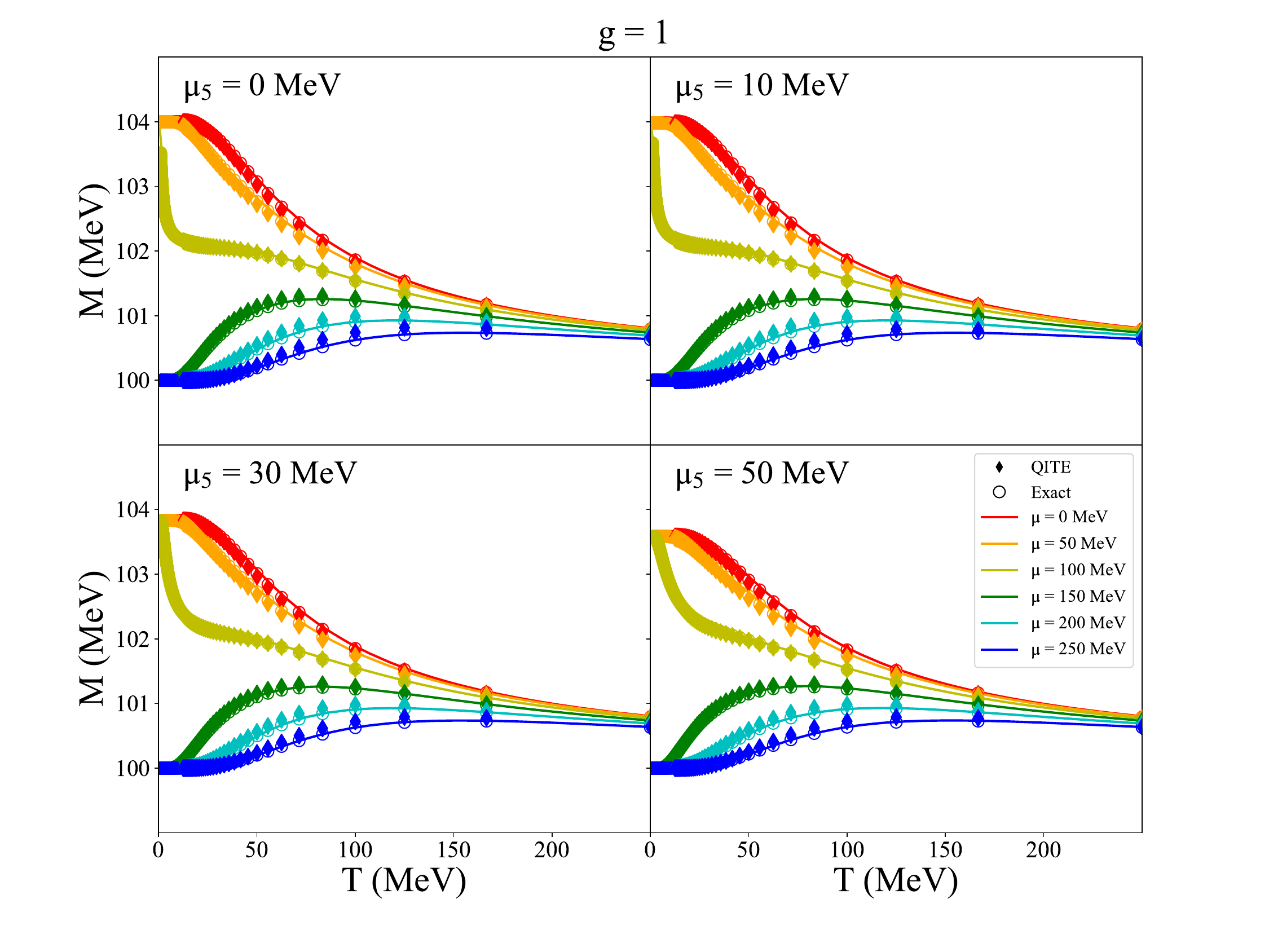}
\caption{Effective mass $M$ as a function of temperature $T$ MeV at coupling constant $g=1$ at a fixed chiral chemical potential $\mu_5$ in each panel.}\label{fig:mu5_g1}
\end{figure}

In \cref{fig:mu5_g1}, we fix the value of chiral chemical potentials $\mu_5$ and plot the $M\mathrm{-}T$ curves for various baryochemical potentials $\mu$ at $g=1$. We observe a non-trivial phase transition at $100<\mu<150$ MeV in all panels. The pattern of the $M\mathrm{-}T$ curves looks similar at various $\mu_5$, with the effective mass $M$ changing more rapidly as a function of $T$ at smaller $\mu_5$.

\begin{figure}[htp]
\centering
\includegraphics[width=0.495\textwidth,trim={2.5cm 2.5cm 3cm 0},clip]{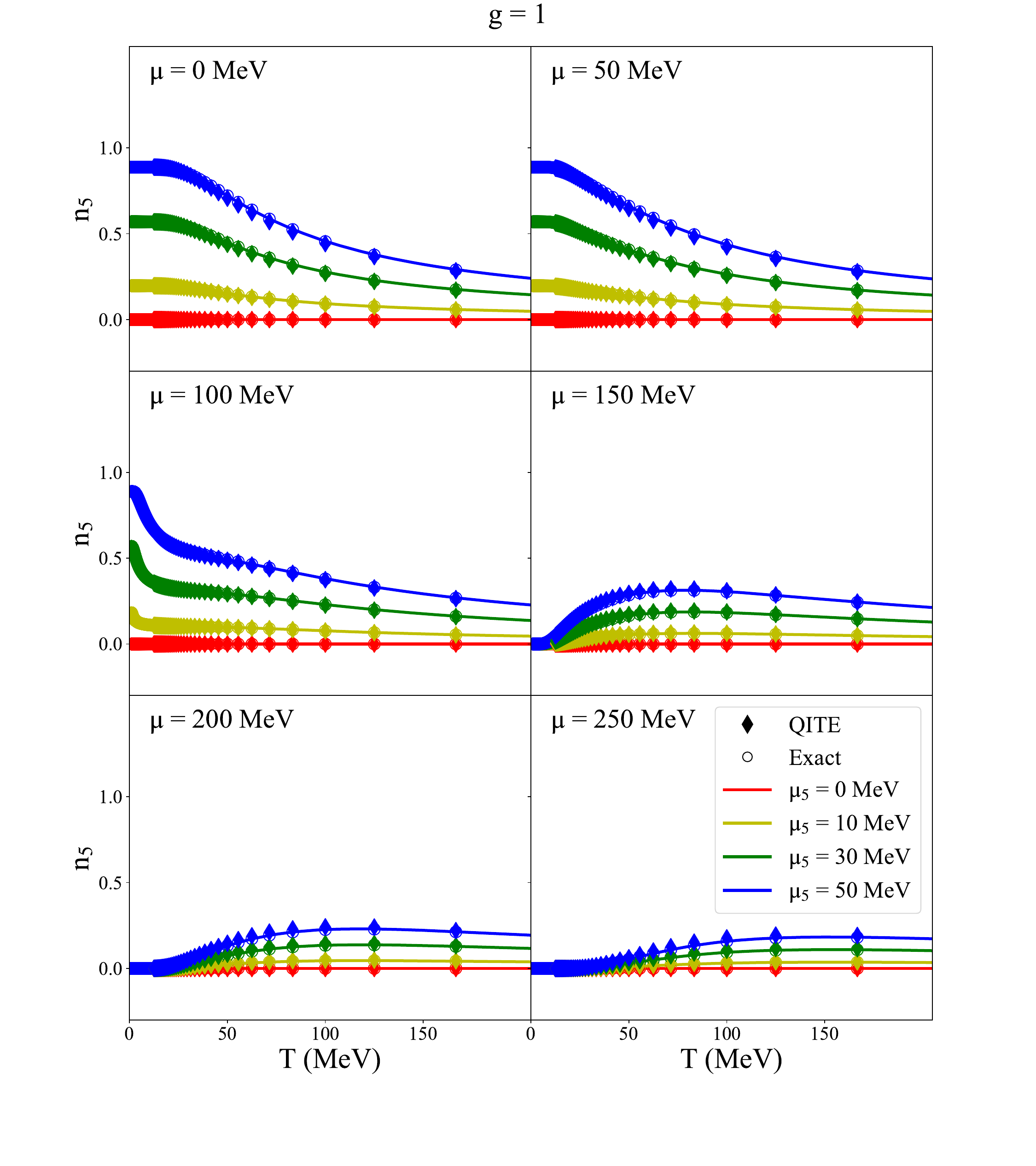}
\includegraphics[width=0.495\textwidth,trim={2.5cm 2.5cm 3cm 0},clip]{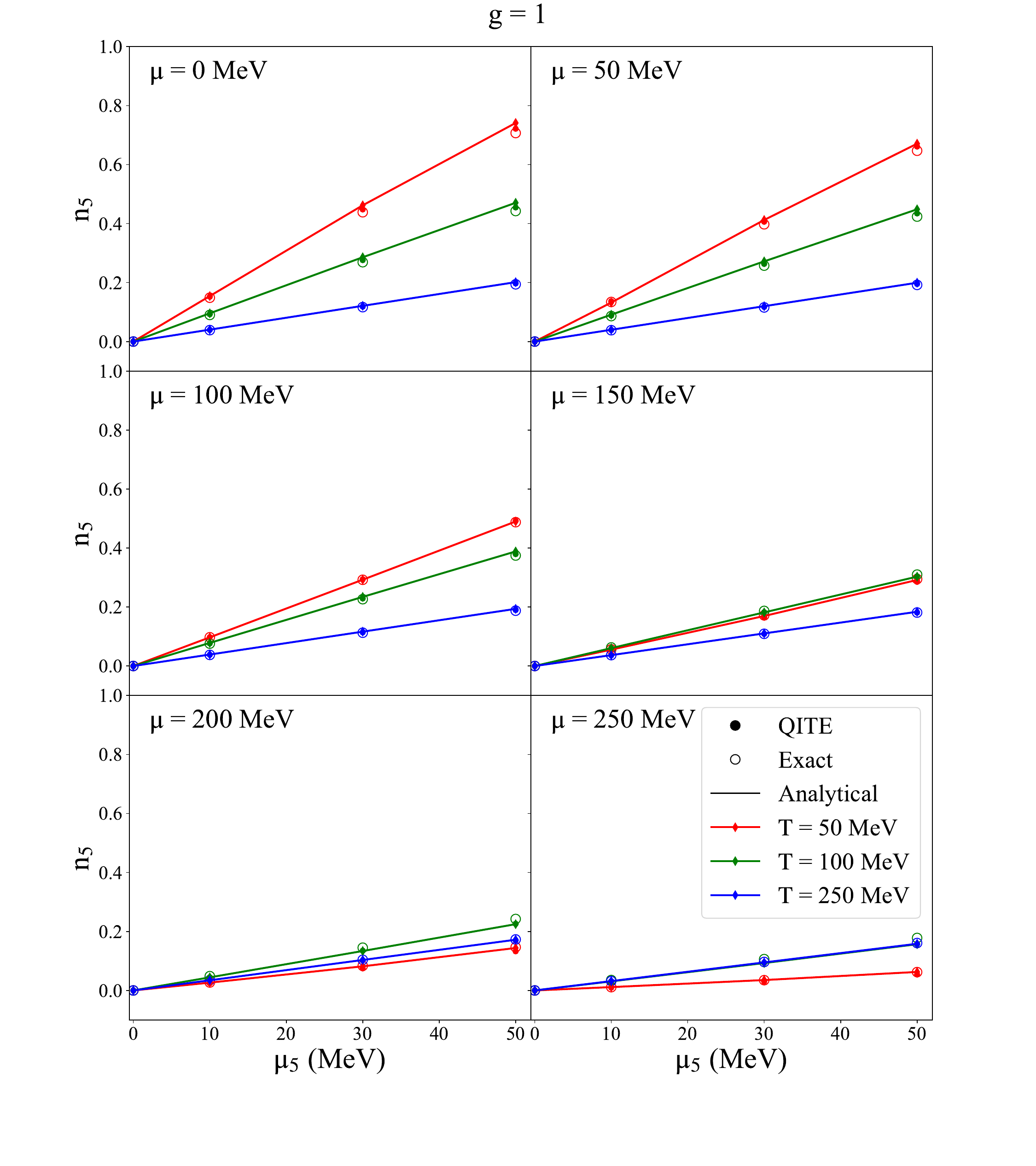}
\caption{Chirality charge density $n_5$ as a function of temperature $T$ (left panels) or $\mu_5$ (right panels) at coupling constant $g=1$. Each panel adopts a different value of chemical potential $\mu$.}\label{fig:n5_1}
\end{figure}

In \cref{fig:n5_1}, we present the chirality charge density as a function of temperature $T$ (left panels) or as a function of chiral chemical potential $\mu_5$ (right panels) for coupling constants $g=1$ and $g=5$. We adopt distinct hues to differentiate various chiral chemical potentials $\mu_5=0,\ 10,\ 30$ and $50$ MeV, while fixing the chemical potential $\mu$ at $0,\ 50,\ 100, \cdots $ or $250$ MeV for each panel. In the absence of any other mechanism for generating a non-zero $\mathcal{N}_5=\bar{\psi}\gamma_0\gamma_5\psi=\psi^\dagger_R\psi_R-\psi^\dagger_L\psi_L$, $n_5=0$ at $\mu_5=0$ MeV. Thus, $n_5=\langle\mathcal{N}_5\rangle$ only exists when $\mu_5\ne 0$.

We also observe that, across all chemical potentials $\mu$ and temperatures $T$, the chirality charge density $n_5$ is approximately proportional to the chiral chemical potential $\mu_5$. Specifically, $n_5$ starts at zero at $\mu_5 = 0$ MeV and increases linearly with $\mu_5$. Thus, in this model, the chiral chemical potential serves as a direct measure of the chiral imbalance present in the plasma. However, the rate of increase of $n_5$ with $\mu_5$ decreases at higher chemical potentials $\mu$. Notably, a phase transition can be discerned between $\mu \leq 100$ MeV and $\mu \geq 150$ MeV. In the former case, $n_5$ decreases with temperature for each $\mu_5$, while in the latter case, $n_5$ increases with temperature from $T = 50$ MeV to $T = 100$ MeV before subsequently decreasing from $T = 100$ MeV to $T = 200$ MeV.

\begin{figure}
\centering
\includegraphics[width=0.65\textwidth,trim={2cm 1.5cm 2cm 0cm},clip]{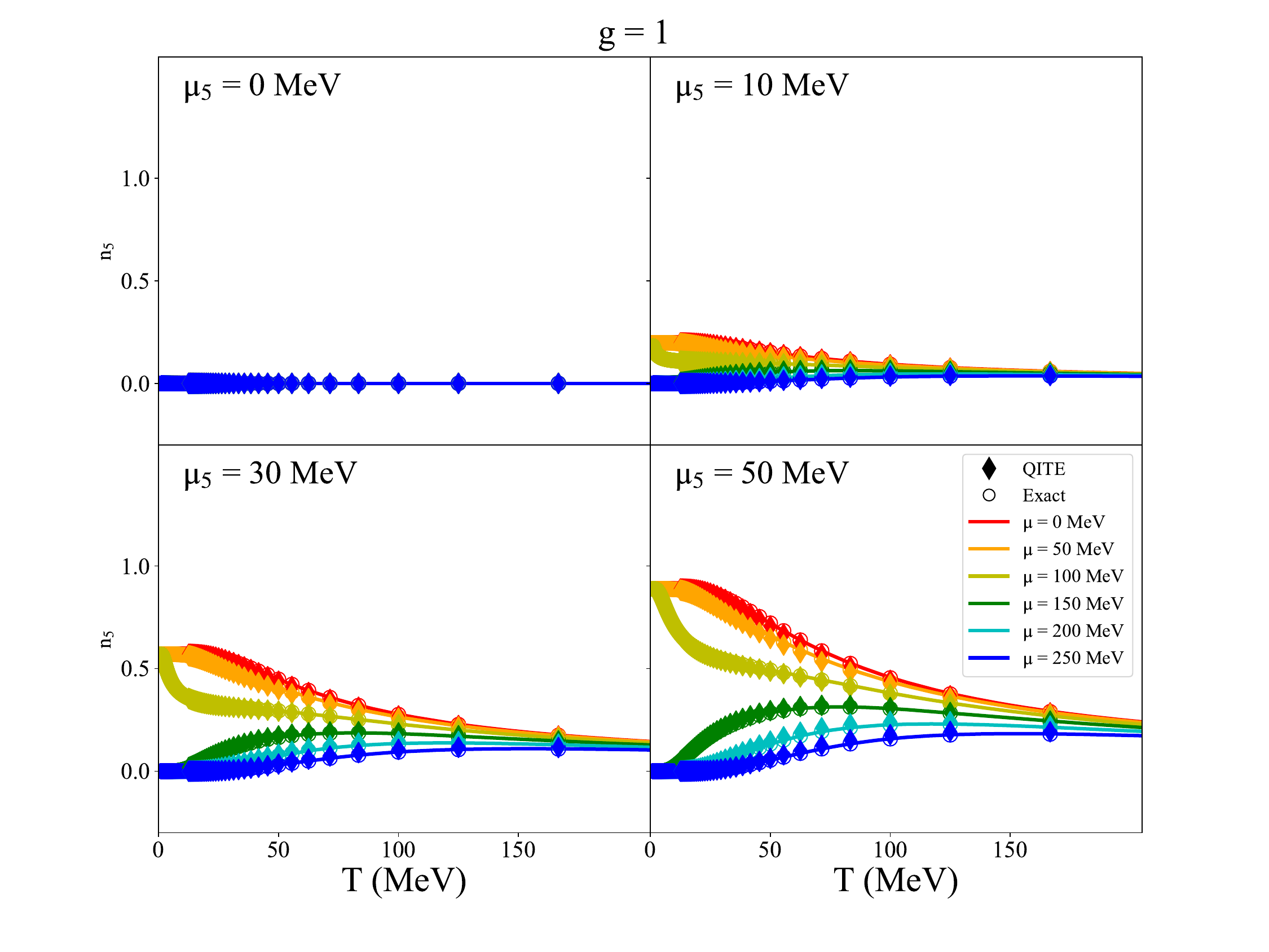}
\caption{Chirality charge density $n_5$ as a function of temperature at $g=1$ with a fixed chiral chemical potential $\mu_5$ in each panel.}\label{fig:n5_3}
\end{figure}

Similarly to the findings from the effective mass plots in \cref{fig:mu5_g1}, a non-trivial phase transition emerges between $\mu=100$ MeV and $\mu=150$ MeV. For $\mu\leq 100$ MeV, the chirality charge density decreases with increasing temperature, while for $\mu\geq 150$ MeV, it first increases before declining and ultimately approaching zero as temperature increases.

In \cref{fig:n5_3}, we depict the chirality charge density $n_5$ as a function of temperature $T$ for coupling constants $g=1$ and $g=5$. We utilize distinct hues to distinguish various chemical potentials $\mu\in{0,\ 50, \cdots, 250}$ MeV, while fixing the chiral chemical potential $\mu_5$ at $0,\ 10,\ 30$ or $50$ MeV for each panel. As expected from the preceding figure, the chirality charge density is $0$ for all temperatures and chemical potentials when $\mu_5 = 0$ MeV. At non-zero $\mu_5$ values, we also observe the phase transition between $\mu = 100$ MeV and $\mu = 150$ MeV. Specifically, for $\mu \leq 100$ MeV, the chirality charge density commences with some non-zero value at $T = 0$ MeV and diminishes as temperature increases, while for curves with $\mu \geq 150$ MeV, the chirality charge density starts at $0$ at $T=0$ MeV and initially increases before decreasing and converging to 0 with increasing temperature. Furthermore, for curves with greater $\mu_5$ values, those with $\mu \leq 100$ MeV start with higher chirality charge densities.

\end{part}

\chapter{Conclusion}\label{sec:conclude}
In this PhD dissertation, we have explored the fascinating world of particle physics, with a particular focus on Quantum Chromodynamics (QCD) and its pivotal role in describing the strong force. Our investigation revolved around two significant aspects of QCD studies, aiming to deepen our understanding of elementary particles and their behavior.

The first aspect of our research delved into the exciting realm of perturbative QCD methods for quantum 3D imaging of hadrons. Through QCD factorization and perturbative expansions, we harnessed these powerful tools to predict the behavior of high-energy interactions, specifically in inclusive jet and hadron-in-jet production at colliders. This allowed us to gain valuable insights into the intricate dynamics of quarks and gluons within hadrons, shedding light on phenomena like hadron production inside jets and spin asymmetries. This new concept ``polarized jet fragmentation functions'' allow us to probe multi-dimensional structure of the nucleons and hadrons, an important thrust for the future Electron-Ion Collider. 

Simultaneously, we recognized the significance of non-perturbative QCD studies, particularly in investigating the QCD phase diagram and understanding the properties of hadrons. Embracing the advancements in quantum computing and simulators, we explored the potential of enhancing finite-temperature behavior simulations. This breakthrough opens up avenues to scrutinize extreme temperatures and densities with unprecedented accuracy and detail, offering fresh perspectives on the nature of QCD.

Throughout this research, we pursued two approaches to enrich our comprehension of QCD. Firstly, we focused on investigating the nucleon structure by analyzing polarized jet fragmentation functions. Secondly, we harnessed quantum computing techniques to explore chiral phase transitions. These enabled us to delve into the non-perturbative facets of QCD and discern new applications for quantum computing in the realm of particle physics.

By undertaking these multifaceted studies, we have contributed to the ongoing efforts to decipher the intricate interactions between elementary particles. Our findings provide essential insights into the behavior of quarks and gluons within hadrons, advancing our knowledge of particle physics. Moreover, the integration of quantum computing methodologies demonstrates the immense potential it holds in unraveling the mysteries of QCD and enhancing the accuracy of simulations at extreme conditions.

As we conclude this thesis, we envision a promising future where continued research and innovation in QCD studies, combined with quantum computing advancements, will lead us towards new frontiers of knowledge. The pursuit of understanding the fundamental building blocks of our universe remains an ever-evolving and deeply rewarding journey, and we hope that this work serves as a stepping stone towards further discoveries in the captivating realm of particle physics.

\bibliographystyle{uclathes}
\bibliography{ch-ref}

\end {document}